\RequirePackage{fix-cm}
\documentclass[smallextended]{svjour3}       
\smartqed  
\usepackage{graphicx, amsmath, amssymb, nicefrac, hyperref, color}
\newcommand{\apj}{{\it{Astrophys. J. }}}
\newcommand{\aap}{{\it{Astron. Astrophys. }}}
\newcommand{\solphys}{{\it{Sol. Phys. }}}
\newcommand{\apjs}{{\it{Astrophys. J. Supp. }}}
\newcommand{\apjl}{{\it{Astrophys. J. Lett. }}}
\newcommand{\memsai}{{\it{Mem. Soc. Astron. Italiana }}}
\newcommand{\mnras}{{\it{Mon. Notices Royal Astron. Soc. }}}
\newcommand{\pasj}{{\it{Publ. Astron. Soc. Jpn }}}
\newcommand{\procspie}{{\it{Proc. SPIE }}}
\newcommand{\ssr}{{\it{Space Sci. Rev. }}}

\newcommand{\rsta}{{\it{Phil. Trans. R. Soc. A }}}
\newcommand{\nat}{{\it{Nature }}}
\newcommand{\araa}{{\it{Annu. Rev. Astron. Astrophys. }}}
\newcommand{\apss}{{\it{Astrophys. Space Sci. }}}
\newcommand{\aapr}{{\it{Astron. Astrophys. Rev. }}}
\newcommand{\aplett}{{\it{Astrophys. Lett. }}}
\newcommand{\zap}{{\it{Z. Astrophys. }}}
\newcommand{\skytel}{{\it{Sky \& Telesc. }}}
\newcommand{\aj}{{\it{Astronom. J. }}}
\newcommand{\ao}{{\it{Appl. Opt. }}}
\newcommand{\Alfven}{Alfv{\'{e}}n}
\newcommand*{\affaddr}[1]{#1} 
\newcommand*{\affmark}[1][*]{\textsuperscript{#1}}
\usepackage{threeparttable}

\usepackage{natbib}
\usepackage{array}
\usepackage[dvipsnames]{xcolor}
\newcolumntype{L}{>{\centering\arraybackslash}m{3cm}}
\usepackage{xcolor}

\usepackage{lastpage}
\usepackage{totcount}
\regtotcounter{figure}
\regtotcounter{table}
\regtotcounter{equation}
\newtotcounter{citnum} 
\def\oldbibitem{} \let\oldbibitem=\bibitem
\def\bibitem{\stepcounter{citnum}\oldbibitem}

\hypersetup{
  colorlinks   = true, 
  urlcolor     = blue, 
  linkcolor    = blue, 
  citecolor   = blue 
}

\begin{document}
\title{Waves in the Lower Solar Atmosphere}
\subtitle{The Dawn of Next-generation Solar Telescopes}


\author{David B. Jess\protect\affmark[1,2]         \and
        Shahin Jafarzadeh\affmark[3,4]      \and \\
        Peter H. Keys\affmark[1]          \and
        Marco Stangalini\affmark[5]       \and \\
        Gary Verth\affmark[6]             \and 
        Samuel D. T. Grant\affmark[1] 
}

\authorrunning{D.B. Jess et al.} 

\institute{
              \affaddr{\affmark[1]Astrophysics Research Centre, School of Mathematics and Physics, Queen's University Belfast, Belfast, BT7 1NN, United Kingdom} \\
              Tel.: (+44)(0)28 9097 3052 \\
              \email{d.jess@qub.ac.uk} \\
              \affaddr{\affmark[2]Department of Physics and Astronomy, California State University Northridge, Northridge, CA 91330, USA} \\
              \affaddr{\affmark[3]Max Planck Institute for Solar System Research, Justus-von-Liebig-Weg 3, 37077 G{\"{o}}ttingen, Germany} \\
              \affaddr{\affmark[4]Rosseland Centre for Solar Physics, University of Oslo, P.O. Box 1029 Blindern, 0315 Oslo, Norway} \\
              \affaddr{\affmark[5]Italian Space Agency (ASI), Via del Politecnico snc, 00133 Roma, Italy} \\
              \affaddr{\affmark[6]Plasma Dynamics Group, School of Mathematics and Statistics, University of Sheffield, Hicks Building, Hounsfield Road, Sheffield, S3 7RH, United Kingdom} \\
}

\date{Received: date / Accepted: date}

\maketitle

\begin{abstract}
Waves and oscillations have been observed in the Sun's atmosphere for over half a century. While such phenomena have readily been observed across the entire electromagnetic spectrum, spanning radio to gamma-ray sources, the underlying role of waves in the supply of energy to the outermost extremities of the Sun's corona has yet to be uncovered. Of particular interest is the lower solar atmosphere, including the photosphere and chromosphere, since these regions harbor the footpoints of powerful magnetic flux bundles that are able to guide oscillatory motion upwards from the solar surface. As a result, many of the current- and next-generation ground-based and space-borne observing facilities are focusing their attention on these tenuous layers of the lower solar atmosphere in an attempt to study, at the highest spatial and temporal scales possible, the mechanisms responsible for the generation, propagation, and ultimate dissipation of energetic wave phenomena. Here, we present a two-fold review that is designed to overview both the wave analyses techniques the solar physics community currently have at their disposal, as well as highlight scientific advancements made over the last decade. Importantly, while many ground-breaking studies will address and answer key problems in solar physics, the cutting-edge nature of their investigations will naturally pose yet more outstanding observational and/or theoretical questions that require subsequent follow-up work. This is not only to be expected, but should be embraced as a reminder of the era of rapid discovery we currently find ourselves in. We will highlight these open questions and suggest ways in which the solar physics community can address these in the years and decades to come. 
\keywords{shock waves \and Sun: chromosphere \and Sun: oscillations \and Sun: photosphere \and telescopes}
\end{abstract}



\newpage
\section{Introduction}
\label{sec:introduction}
Understanding the energy flow through the Sun's dynamic and tenuous atmosphere has long been a scientific interest for the global astrophysical community. The challenge of identifying the source(s) responsible for the elevated multi-million Kelvin temperatures in the solar corona has produced two main theoretical mechanisms. The first is via magnetic reconnection -- the so-called `DC' heating mechanism. Here, the continual re-configuration of the omnipresent magnetic fields that populate the Sun's atmosphere allow the production of intense thermal heating as the magnetic energy is converted through the process of reconnection, producing dramatic flares that often release energies in excess of $10^{31}$~ergs during a single event \citep{Priest1986, Priest1999, Shibata2011, Benz2017}. However, such large-scale solar flares are relatively rare, and hence cannot supply the global background heating required to continuously maintain the corona's elevated temperatures. Instead, there is evidence to suggest that the frequency of flaring events, as a function of their energy, is governed by a power-law relationship \citep{Shimizu1997, Krucker1998, Aschwanden2000, Parnell2000}, whereby smaller-scale micro- and nano-flares (with energies $\sim 10^{27}$~ergs and $\sim 10^{24}$~ergs, respectively) may occur with such regularity that they can sustain the thermal inputs required to maintain the hot corona. Many modern numerical and observational studies have been undertaken to try and quantify the ubiquity of these faint reconnection events, which often lie at (or below) the noise level of current-generation facilities \citep{Terzo2011}. Due to the difficulties surrounding the extraction of nanoflare characteristics embedded within the noise limitations of the data, only tentative evidence exists to support their global heating abilities of the outer solar atmosphere \citep[][to name but a few recent examples]{Viall2013, Viall2015, Viall2016, Viall2017, 2014ApJ...795..172J, 2019ApJ...871..133J, Bradshaw2015, Tajfirouze2016a, Tajfirouze2016b, Ishikawa2017}.

\begin{figure*}[!t]
\begin{center}
\includegraphics[width=1.0\textwidth, trim=0.8cm 0cm 0.8cm 0cm, clip]{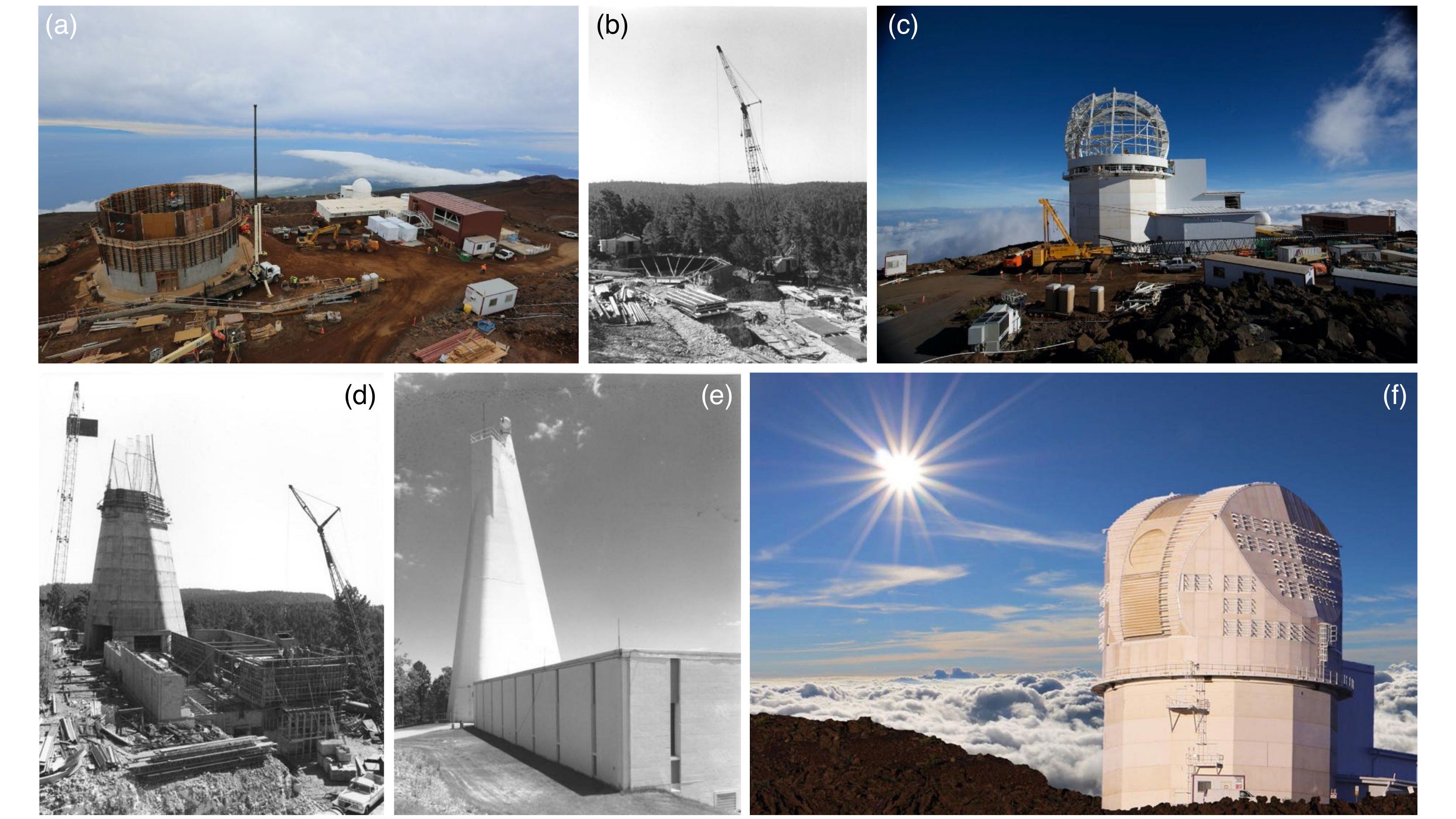}
\end{center}
\caption{Images depicting the construction of National Science Foundation facilities some 50~years apart. Panels (b), (d) and (e) display construction stages of the Dunn Solar Telescope, which was first commissioned in 1969 in the Sacramento Peak mountains of New Mexico, USA. Panels (a), (c) and (f) depict similar stages of construction for the Daniel K. Inouye Solar Telescope, which acquired first-light observations in 2019 at the Haleakal{$\bar{\mathrm{a}}$} Observatory on the Hawaiian island of Maui, USA. Images courtesy of Doug Gilliam (NSO) and Brett Simison (NSO).}
\label{fig:DST_DKIST_construction} 
\end{figure*}

The second energy-supplying mechanism for the Sun's outer atmosphere involves the creation, propagation, and ultimately dissipation of wave-related phenomena -- often referred to as the `AC' heating mechanism \citep{1948ApJ...107....1S}. The specific oscillatory processes responsible for supplying non-thermal energy to the solar atmosphere have come under scrutiny since wave motions were first discovered more than 60~years ago \citep{Leighton1960, 1962ApJ...135..474L, Noyes1963}. Of course, such early observations were without the modern technological improvements that enhance image quality, such as adaptive optics \citep[AO;][]{Rimele2011} and image reconstruction techniques, including speckle \citep{2008A&A...488..375W} and multi-object multi-frame blind deconvolution \citep[MOMFBD;][]{2005SoPh..228..191V}. As a result, many pioneering papers documenting the characteristics of wave phenomena in the lower solar atmosphere relied upon the study of large-scale features that would be less effected by seeing-induced fluctuations, including sunspots and super-granular cells, captured using premiere telescope facilities of the time such as the McMath-Pierce Solar Telescope \citep{1964ApOpt...3.1337P} at the Kitt Peak Solar Observatory, USA, and the National Science Foundation's Dunn Solar Telescope \citep[DST;][]{Dunn1969}, situated in the Sacramento Peak mountains of New Mexico, USA (see Figure~{\ref{fig:DST_DKIST_construction}}).

\begin{figure*}
\begin{center}
\includegraphics[width=0.66\textwidth]{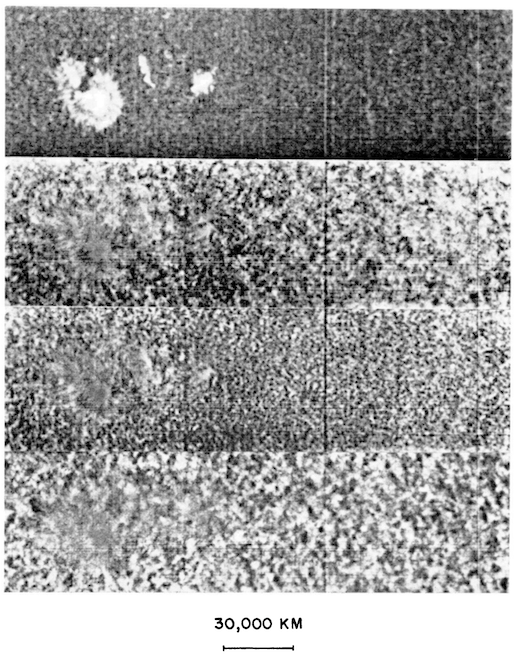}
\end{center}
\caption{Observations of the photospheric Fe~{\sc{i}} absorption line, showing the sum of blue- and red-wing intensities (displayed in a negative color scale; top), the total measured Doppler velocities across the field-of-view (middle-top), the slowly varying component of the plasma flows (middle-bottom), and the Doppler velocity map arising purely from oscillatory motion (bottom). The region of interest includes a large sunspot structure (left-hand side), and shows ubiquitous oscillatory signatures with periods $\sim$$300$~s and velocity amplitudes $\sim$$0.6$~km{\,}s$^{-1}$. Image taken from \citet{Sheeley1971}.}
\label{fig:Sheeley1971} 
\end{figure*}

Even at large spatial scales, Doppler velocity and intensity time series from optical spectral lines, including Fe~{\sc{i}} \citep{Deubner1967}, H$\alpha$ \citep{Deubner1969}, Ca~{\sc{ii}} \citep{Musman1970}, C~{\sc{i}} \citep{Deubner1971}, and Na~{\sc{i}} \citep{Slaughter1972} demonstrated the ubiquitous nature of oscillations throughout the photosphere and chromosphere. Through segregation of slowly-varying flows and periodic velocity fluctuations, \citet{Sheeley1971} were able to map the spatial structuring of wave power in the vicinity of a sunspot (see Figure~\ref{fig:Sheeley1971}), and found clear evidence for ubiquitous photospheric oscillatory motion with periods $\sim$$300$~s and velocity amplitudes $\sim$$0.6$~km{\,}s$^{-1}$. Such periodicities and amplitudes were deemed observational manifestations of the pressure-modulated global $p$-mode spectrum of the Sun \citep{1970ApJ...162..993U, Leibacher1971, 1975A&A....44..371D, Rhodes1977}, where internal acoustic waves are allowed to leak upwards from the solar surface, hence producing the intensity and velocity oscillations synonymous with the compressions and rarefactions of acoustic waves.

Difficulties arose in subsequent work, when the measured phase velocities of the waves between two atmospheric heights were too large to remain consistent with a purely acoustic wave interpretation \citep{Osterbrock1961, Mein1976}. It was not yet realized that the 5-minute oscillations are not propagating acoustic waves, but instead are evanescent in character since their frequency was lower than the associated acoustic cut-off value (see Section~{\ref{sec:globalmodes}} for further details). Researchers further hypothesized that the magnetic fields, which were often synonymous with the observed oscillations, needed to be considered in order to accurately understand and model the wave dynamics \citep[][to name but a few examples]{1973SoPh...30...47M, Nakagawa1973b, Nakagawa1973a, 1974ARA&A..12..407S, 1977SoPh...52..283M, Mein1978}. The field of magnetohydrodynamics (MHD) was introduced to effectively link the observed wave signatures to the underlying magnetic configurations, where the strong field strengths experienced in certain locations \citep[e.g., field strengths that can approach approximately $6000$~G in sunspot umbrae;][]{Livingston2006, Okamoto2018} produce wave modes that are highly modified from their purely acoustic counterparts. 

The importance of the magnetic field in the studies of wave phenomena cannot be overestimated, since both the alignment of the embedded magnetic field, $B_0$, with the wavevector, $k$, and the ratio of the kinetic pressure, $p_0$, to the magnetic pressure, $B_{0}^{2}/2\mu_{0}$, play influential roles in the characteristics of any waves present \citep[see the reviews by, e.g.,][]{1974ARA&A..12..407S, Bogdan2000, 2013SSRv..175....1M, 2015SSRv..190..103J, 2016GMS...216..449J}. Commonly, the ratio of kinetic to magnetic pressures is referred to as the plasma-$\beta$, defined as, 
\begin{equation}
\label{eqn:plasmabeta}
    \beta = \frac{2\mu_{0}p_{0}}{B_{0}^{2}} \ ,
\end{equation} 
where $\mu_{0}$ is the magnetic permeability of free space \citep{Wentzel1979, 1983SoPh...88..179E, Spruit1983}. Crucially, by introducing the local hydrogen number density, $n_{\mathrm{H}}$, the plasma-$\beta$ can be rewritten (in cgs units) in terms of the Boltzmann constant, $k_{B}$, and the temperature of the plasma, $T$, giving the relation, 
\begin{equation}
\label{eqn:plasmabetacgs}
    \beta = \frac{8{\pi}n_{\mathrm{H}}Tk_{B}}{B_{0}^{2}} \ .
\end{equation} 

In the lower regions of the solar atmosphere, including the photosphere and chromosphere, temperatures are relatively low ($T \lesssim 15{\,}000$~K) when compared to the corona. This, combined with structures synonymous with the solar surface, including sunspots, pores, and magnetic bright points \citep[MBPs;][]{Berger1995, SanchezAlmeida2004, Ishikawa2007, Utz2009, Utz2010, Utz2013a, 2013A&A...554A..65U, 2011ApJ...740L..40K, 2013MNRAS.428.3220K, 2014A&A...566A..99K}, all of which possess strong magnetic field concentrations ($B_{0} \gtrsim 1000$~G), presents wave conduits that are inherently `low-$\beta$' (i.e., are dominated by magnetic pressure; $\beta \ll 1$). \citet{Gary2001} has indicated how such structures (particularly for highly-magnetic sunspots) can maintain their low-$\beta$ status throughout the entire solar atmosphere, even as the magnetic fields begin to expand into the more volume-filling chromosphere \citep{Gudiksen2006, 2013A&A...553A..73B}. Using non-linear force-free field \citep[NLFFF;][]{Wiegelmann2008, Aschwanden2016a, 2021LRSP...18....1W} extrapolations, \citet{Aschwanden2016} and \citet{2018NatPh..14..480G} provided further evidence that sunspots can be best categorized as low-$\beta$ wave guides, spanning from the photosphere through to the outermost extremities of the corona. As can be seen from Equation~{\ref{eqn:plasmabetacgs}}, the hydrogen number density ($n_{\mathrm{H}}$) also plays a pivotal role in the precise local value of the plasma-$\beta$. As one moves higher in the solar atmosphere, a significant drop in the hydrogen number density is experienced \citep[see, e.g., the sunspot model proposed by][]{1981phss.conf..235A}, often with an associated scale-height on the order of $150-200$~km \citep{10.3389/fspas.2020.574460}. As a result, the interplay between the number density and the expanding magnetic fields plays an important role in whether the environment is dominated by magnetic or plasma pressures.

Of course, not all regions of the Sun's lower atmosphere are quite so straightforward. Weaker magnetic elements, including small-scale MBPs \citep{2020A&A...633A..60K}, are not able to sustain dominant magnetic pressures as their fields expand with atmospheric height. This results in the transition to a `high-$\beta$' environment, where the plasma pressure dominates over the magnetic pressure (i.e., $\beta > 1$), which has been observed and modeled under a variety of highly magnetic conditions \citep[e.g.,][]{2011LRSP....8....4B, 2013ApJ...779..168J, 2017ApJ...850L..29B, 2018NatPh..14..480G}. This transition has important implications for the embedded waves, since the allowable modes become effected as the wave guide passes through the $\beta \sim 1$ equipartition layer. Here, waves are able to undergo mode conversion/transmission \citep{2006MNRAS.372..551S, Cally2007, 2016MNRAS.456.1826H}, which has the ability to change the properties and observable signatures of the oscillations. However, we note that under purely quiescent conditions (i.e., related to quiet Sun modeling and observations), the associated intergranular lanes \citep{Lin1999} and granules themselves \citep{Lites2008} will already be within the high plasma-$\beta$ regime at photospheric heights.

\begin{figure*}
\begin{center}
\includegraphics[width=\textwidth]{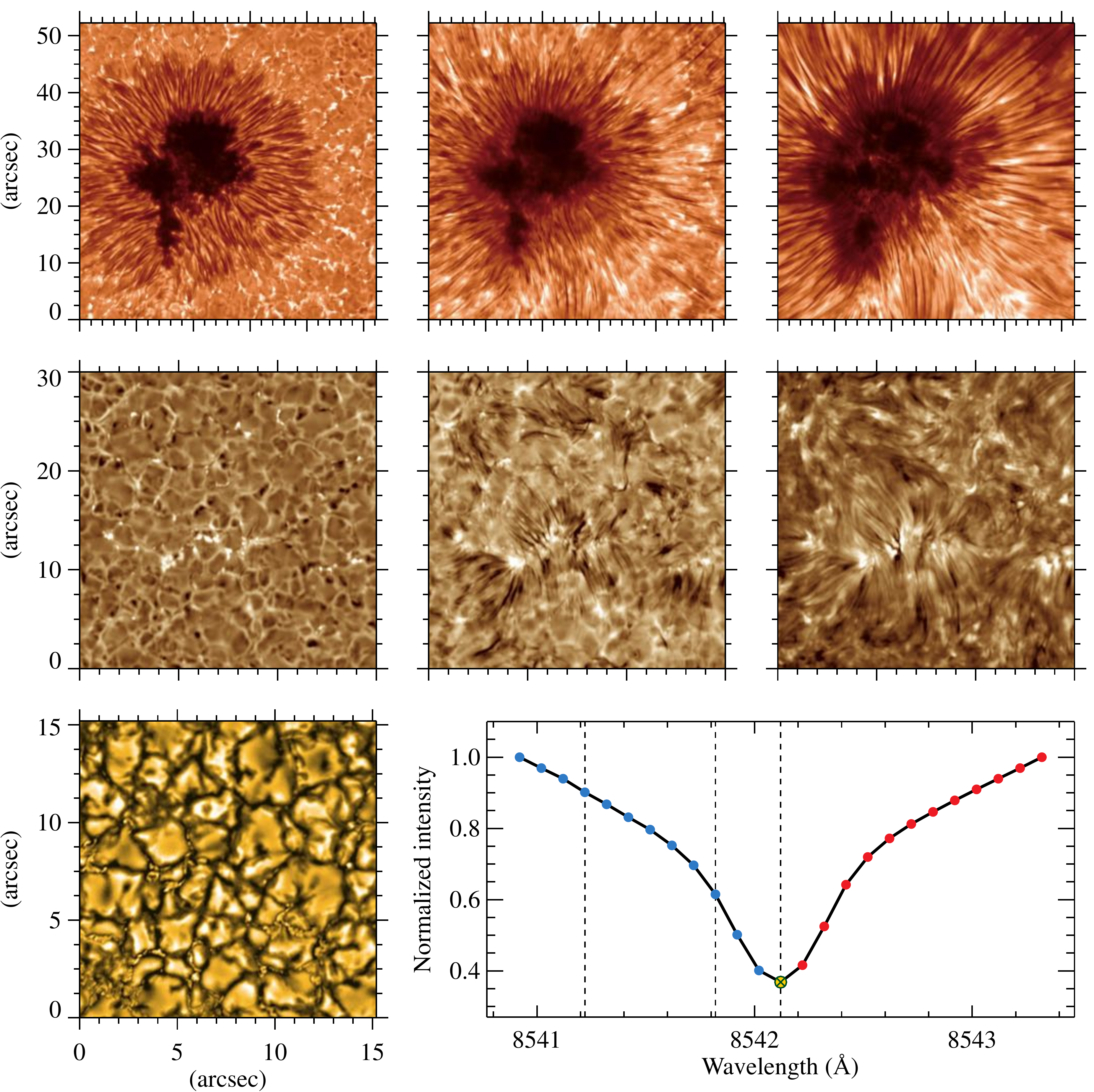}
\end{center}
\caption{Observations of a sunspot (top row) and a quiet-Sun region (middle row) in the lower solar atmosphere, sampled at three wavelength positions in the Ca~{\sc ii}~8542{\,}{\AA} spectral line from the 1m Swedish Solar Telescope (SST). The wavelength positions, from left to right, correspond to $-900${\,}{m\AA}, $-300${\,}{m\AA}, and $0${\,}{m\AA} from the line core, marked with vertical dashed lines in the bottom-right panel, where the average spectral line and all sampled positions are also depicted. The bottom-left panel illustrates a photospheric image sampled with a broadband filter (centered at 3950{\,}{\AA}; filter width $\approx13.2${\,}{\AA}). For better visibility, a small portion of the observed images are presented. All images are squared. Images courtesy of the Rosseland Centre for Solar Physics, University of Oslo.}
\label{fig:sstobservations} 
\end{figure*}

Since the turn of the century, there has been a number of reviews published in the field of MHD waves manifesting in the outer solar atmosphere, including those linked to standing \citep{VanDoorsselaere2009, Wang2011}, quasi-periodic \citep{Nakariakov2005}, and propagating \citep{DeMoortel2009, 2009SSRv..149..355Z, Lin2011} oscillations. Many of these review articles focus on the outermost regions of the solar atmosphere (i.e., the corona), or only address waves and oscillations isolated within a specific layer of the Sun's atmosphere, e.g., the photosphere \citep{2016GMS...216..449J} or the chromosphere \citep{2015SSRv..190..103J, 2016GMS...216..431V}. As such, previous reviews have not focused on the coupling of MHD wave activity between the photosphere and chromosphere, which has only recently become possible due to the advancements made in multi-wavelength observations and data-driven MHD simulations. Here, in this review, we examine the current state-of-the-art in wave propagation, coupling, and damping/dissipation within the lower solar atmosphere, which comprises of both the photosphere and chromosphere, which are the focal points of next-generation ground-based telescopes, such as DKIST.

In addition, we would also like this review to be useful for early career researchers (PhD students and post-doctoral staff) who may not necessarily be familiar with all of the wave-based analysis techniques the solar physics community currently have at their disposal, let alone the wave-related literature currently in the published domain. As a result, we wish this review to deviate from traditional texts that focus on summarizing and potential follow-up interpretations of research findings. Instead, we will present traditional and state-of-the-art methods for detecting, isolating, and quantifying wave activity in the solar atmosphere. This is particularly important since modern data sequences acquired at cutting-edge observatories are providing us with incredible spatial, spectral, and temporal resolutions that require efficient and robust analyses tools in order to maximize the scientific return. Furthermore, we will highlight how the specific analysis methods employed often strongly influence the scientific results obtained, hence it is important to ensure that the techniques applied are fit for purpose. To demonstrate the observational improvements made over the last $\sim50$~years we draw the readers attention to Figures~{\ref{fig:Sheeley1971}} \& {\ref{fig:sstobservations}}. Both Figures~{\ref{fig:Sheeley1971}} \& {\ref{fig:sstobservations}} show sunspot structures captured using the best techniques available at that time. However, with advancements made in imaging (adaptive) optics, camera architectures, and post-processing algorithms, the drastic improvements are clear to see, with the high-quality data sequences shown in Figure~{\ref{fig:sstobservations}} highlighting the incredible observations of the Sun's lower atmosphere we currently have at our disposal. 

After the wave detection and analysis techniques have been identified, with their strengths/weaknesses defined, we will then take the opportunity to summarize recent theoretical and observational research focused on the generation, propagation, coupling, and dissipation of wave activity spanning the base of the photosphere, through to the upper echelons of the chromosphere that couples into the transition region and corona above. Naturally, addressing a key question in the research domain may subsequently pose two or three more, or pushing the boundaries of observational techniques and/or theoretical modeling tools may lead to ambiguities or caveats in the subsequent interpretations. This is not only to be expected, but should be embraced as a reminder of the era of rapid discovery we currently find ourselves in. The open questions we will pose not only highlight the challenges currently seeking solution with the dawn of next-generation ground-based and space-borne telescopes, but will also set the scene for research projects spanning decades to come.  

\begin{figure*}[!t]
\begin{center}
\includegraphics[width=1.0\textwidth]{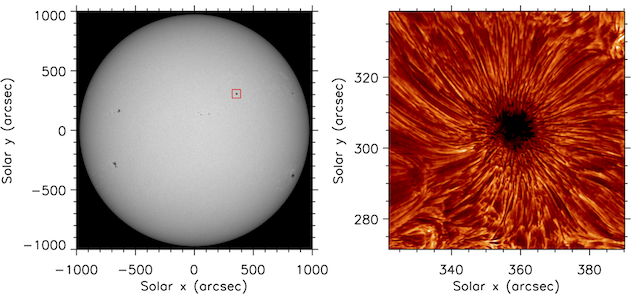}
\end{center}
\caption{An SDO/HMI full-disk continuum image (left), with a red box highlighting the HARDcam field-of-view captured by the DST facility on 2011 December 10. An H$\alpha$ line core image of active region NOAA~11366, acquired by HARDcam at 16:10~UT, is displayed in the right panel. Axes represent heliocentric coordinates in arcseconds.}
\label{fig:10Dec2011_FOV} 
\end{figure*}

\section{Wave Analysis Tools}
\label{sec:waveanalysistools}
Identifying, extracting, quantifying, and understanding wave-related phenomena in astrophysical time series is a challenging endeavor. Signals that are captured by even the most modern charge-coupled devices (CCDs) and scientific complementary metal-oxide-semiconductor (sCMOS) detectors are accompanied by an assortment of instrumental and noise signals that act to mask the underlying periodic signatures. For example, the particle nature of the incident photons leads to Poisson-based shot noise, resulting in randomized intensity fluctuations about the time series mean \citep{1977ApJ...213L..93T, 2008SoPh..248..441D}, which can reduce the clarity of wave-based signatures. Furthermore, instrumental and telescope effects, including temperature sensitivity and pointing stability, can lead to mixed signals either swamping the signatures of wave motion, or artificially creating false periodicities in the resulting data products. Hence, without large wave amplitudes it becomes a challenge to accurately constrain weak wave signals in even the most modern observational time series, especially once the wave fluctuations become comparable to the noise limitations of the data sequence. In the following sub-sections we will document an assortment of commonly available tools available to the solar physics community that can help quantify wave motion embedded in observational data. 

\subsection{Observations}
\label{sec:observations}
In order for meaningful comparisons to be made from the techniques presented in Section~{\ref{sec:waveanalysistools}}, we will benchmark their suitability using two observed time series. We would like to highlight that the algorithms described and demonstrated below can be applied to any form of observational data product, including intensities, Doppler velocities, and spectral line-widths. As such, it is important to ensure that the input time series are scientifically calibrated before these wave analysis techniques are applied.

\begin{figure*}[!t]
\begin{center}
\includegraphics[width=1.0\textwidth]{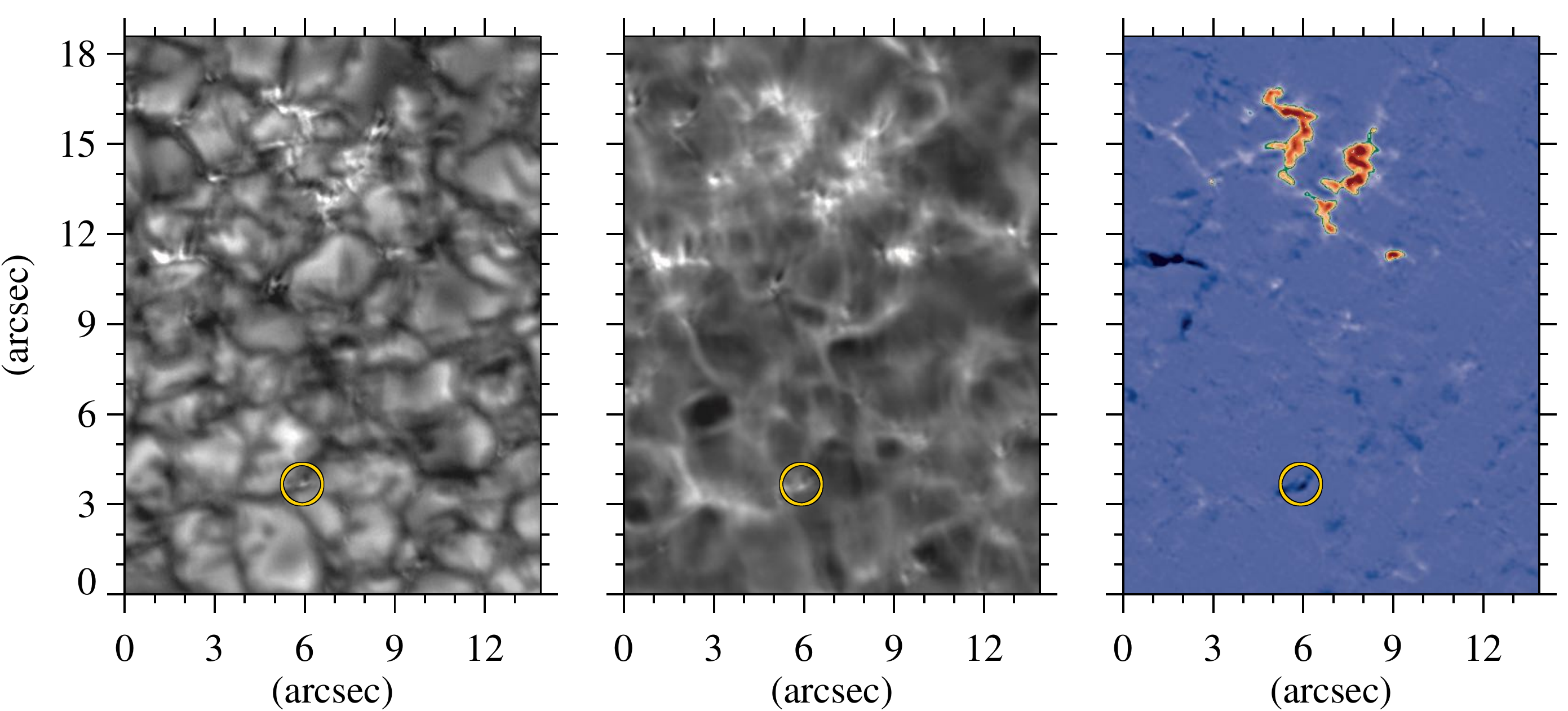}
\end{center}
\caption{A small region of an image acquired in 300~nm (left) and in Ca~{\sc ii}~H  spectral lines (middle) from SuFI/{\sc Sunrise}, along with their corresponding line-of-sight magnetic fields from IMaX/{\sc Sunrise} (right). The latter ranges between $-1654$~G and $2194$~G. The circle includes a small-scale magnetic feature whose oscillatory behavior is shown in Figure~\ref{fig:SUNRISE_wavelet_phaselag}.}
\label{fig:SUNRISE_09June2009} 
\end{figure*}

\subsubsection{HARDcam -- 2011 December 10}
\label{sec:HARDcam}
The Hydrogen-alpha Rapid Dynamics camera \citep[HARDcam;][]{2012ApJ...757..160J} is an sCMOS instrument designed to acquire high-cadence H$\alpha$ images at the DST facility. The data captured by HARDcam on 2011~December~10 consists of 75~minutes (16:10 -- 17:25~UT) of H$\alpha$ images, acquired through a narrowband 0.25{\,}{\AA} Zeiss filter, obtained at 20 frames per second. Active region NOAA~11366 was chosen as the target, which was located at heliocentric coordinates ($356''$, $305''$), or N17.9W22.5 in the more conventional heliographic coordinate system. A non-diffraction-limited imaging platescale of $0{\,}.{\!\!}{''}138$ per pixel was chosen to provide a field-of-view size equal to $71''\times71''$. During the observing sequence, high-order adaptive optics \citep{2004SPIE.5490...34R, Rimele2011} and speckle reconstruction algorithms \citep{2008A&A...488..375W} were employed, providing a final cadence for the reconstructed images of 1.78~s. The dataset has previously been utilized in a host of scientific studies \citep{2013ApJ...779..168J, 2016NatPh..12..179J, 2017ApJ...842...59J, 2015ApJ...812L..15K, 2021RSPTA.37900181A} due to the excellent seeing conditions experienced and the fact that the sunspot observed was highly circularly symmetric in its shape. A sample image from this observing campaign is shown in the right panel of Figure~{\ref{fig:10Dec2011_FOV}}, alongside a simultaneous continuum image captured by the Helioseismic and Magnetic Imager \citep[HMI;][]{2012SoPh..275..229S}, onboard the Solar Dynamics Observatory \citep[SDO;][]{2012SoPh..275....3P}.

In addition to the HARDcam data of this active region, we also accessed data from the Atmospheric Imaging Assembly \citep[AIA;][]{2012SoPh..275...17L} onboard the SDO. Here, we obtained 1700{\,}{\AA} continuum (photospheric) images with a cadence of 24~s and spanning a 2.5~hour duration. The imaging platescale is $0{\,}.{\!\!}{''}6$ per pixel, with a $350\times350$~pixel$^{2}$ cut-out providing a $210''\times210''$ field-of-view centered on the NOAA~11366 sunspot. The SDO/AIA images are used purely for the purposes of comparison to HARDcam information in Section~{\ref{sec:threedimensionalFourierfiltering}}.

\subsubsection{SuFI -- 2009 June 9}
\label{sec:SuFI}
The {\sc Sunrise} Filter Imager \citep[SuFI;][]{2011SoPh..268...35G} onboard the {\sc Sunrise} balloon-borne solar observatory \citep{2010ApJ...723L.127S,2011SoPh..268....1B,2011SoPh..268..103B} sampled multiple photospheric and chromospheric heights, with a 1~m telescope, in distinct wavelength bands during its first and second flights in 2009 and 2013, respectively \citep{2017ApJS..229....2S}. High quality, seeing-free time-series of images at 300~nm and 397~nm (Ca~{\sc ii}~H) bands (approximately corresponding to the low photosphere and low chromosphere, respectively) were acquired by SuFI/{\sc Sunrise} on 2009 June 9, between 01:32 UTC and 02:00 UTC, at a cadence of 12~sec after phase-diversity reconstructions \citep{2010ApJ...723L.154H,2011A&A...529A.132H}. The observations sampled a quiet region located at solar disk center with a field of view of $14''\times40''$ and a spatial sampling of $0{\,}.{\!\!}{''}02$ per pixel. Figure~\ref{fig:SUNRISE_09June2009} illustrates sub-field-of-view sample images in both bands \citep[with an average height difference of $\approx450$~km;][]{2017ApJS..229...10J}, along with magnetic-field strength map obtained from Stokes inversions of the Fe~{\sc i}~525.02~nm spectral line from the {\sc Sunrise} Imaging Magnetograph eXperiment (IMaX; \citealt{2011SoPh..268...57M}). A small magnetic bright point is also marked on all panels of Figure~\ref{fig:SUNRISE_09June2009} with a circle. Wave propagation between these two atmospheric layers in the small magnetic element is discussed in Section~\ref{WaveletPhaseMeasurements}. 

\subsection{One-dimensional Fourier Analysis}
\label{sec:1Dfourieranalysis}
Traditionally, Fourier analysis \citep{1824AnP....76..319F} is used to decompose time series into a set of cosines and sines of varying amplitudes and phases in order to recreate the input lightcurve. Importantly, for Fourier analysis to accurately benchmark embedded wave motion, the input time series must be comprised of both {\it{linear}} and {\it{stationary}} signals. Here, a purely linear signal can be characterized by Gaussian behavior (i.e., fluctuations that obey a Gaussian distribution in the limit of large number statistics), while a stationary signal has a constant mean value and a variance that is independent of time \citep{doi:10.1111/j.1467-9892.1989.tb00037.x, doi:10.1080/0740817X.2014.999180}. If non-linear signals are present, then the time series displays non-Gaussian behavior \citep{2019ApJ...871..133J}, i.e., it contains features that cannot be modeled by linear processes, including time-changing variances, asymmetric cycles, higher-moment structures, etc. In terms of wave studies, these features often manifest in solar observations in the form of sawtooth-shaped structures in time series synonymous with developing shock waves \citep{1993A&A...273..671F, 2003A&A...403..277R, 2009A&A...494..269V, 2013A&A...556A.115D, 2018ApJ...860...28H}. Of course, it is possible to completely decompose non-linear signals using Fourier analysis, but the subsequent interpretation of the resulting amplitudes and phases is far from straightforward and needs to be treated with extreme caution \citep{10.1117/12.960516}.

\begin{figure*}[!t]
\begin{center}
\includegraphics[width=0.48\textwidth]{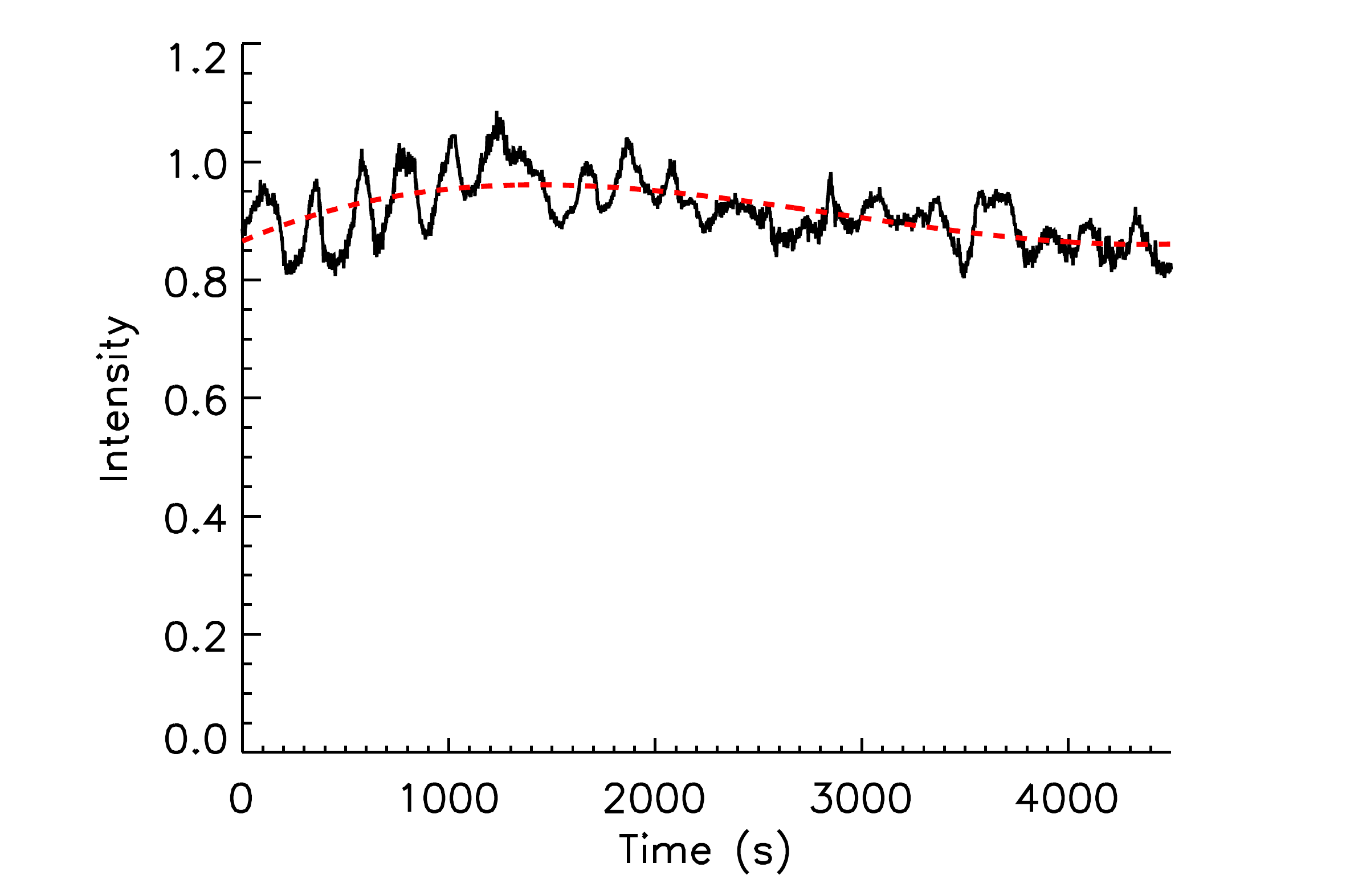}
\includegraphics[width=0.49\textwidth]{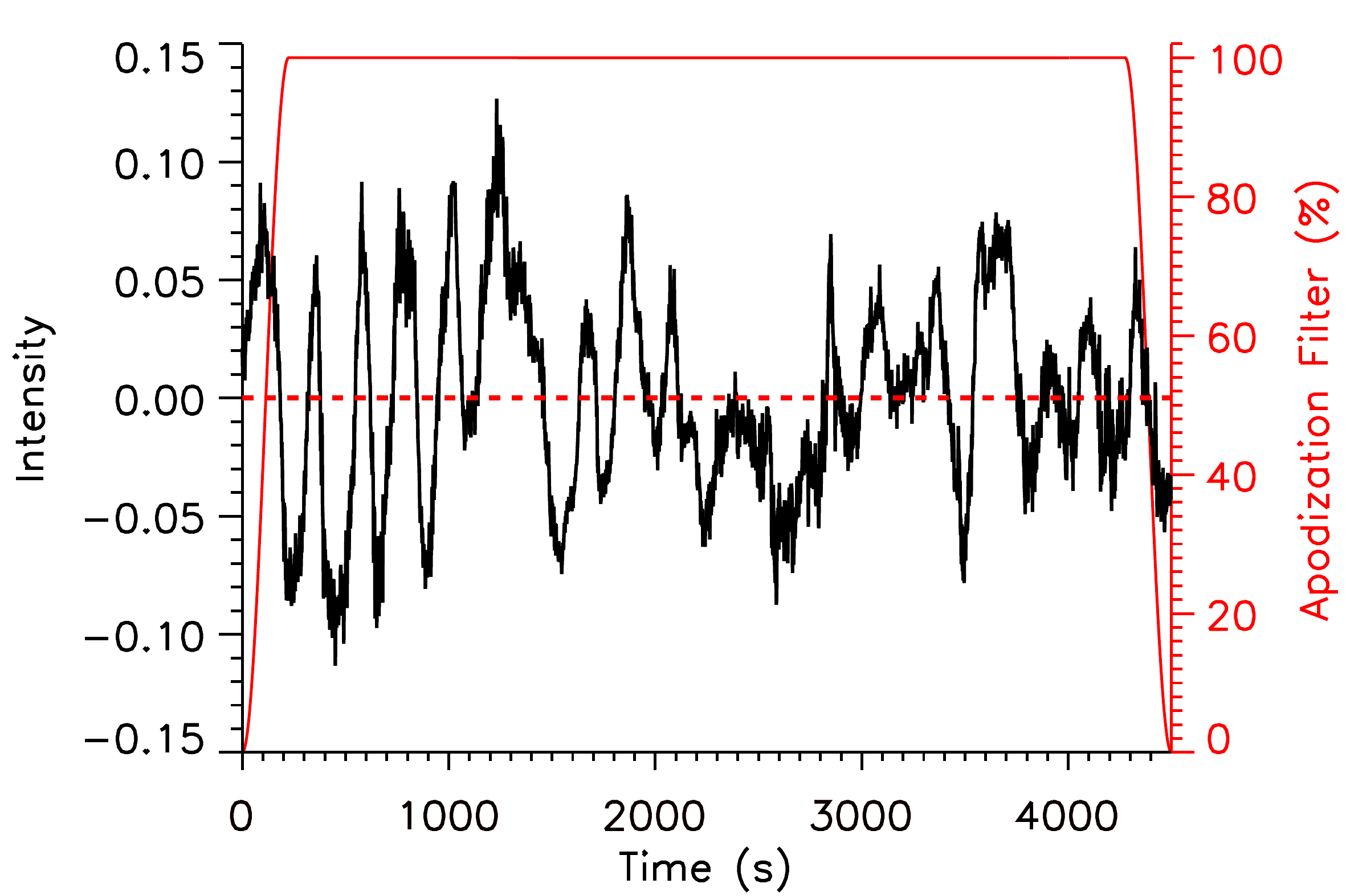}
\includegraphics[width=0.49\textwidth]{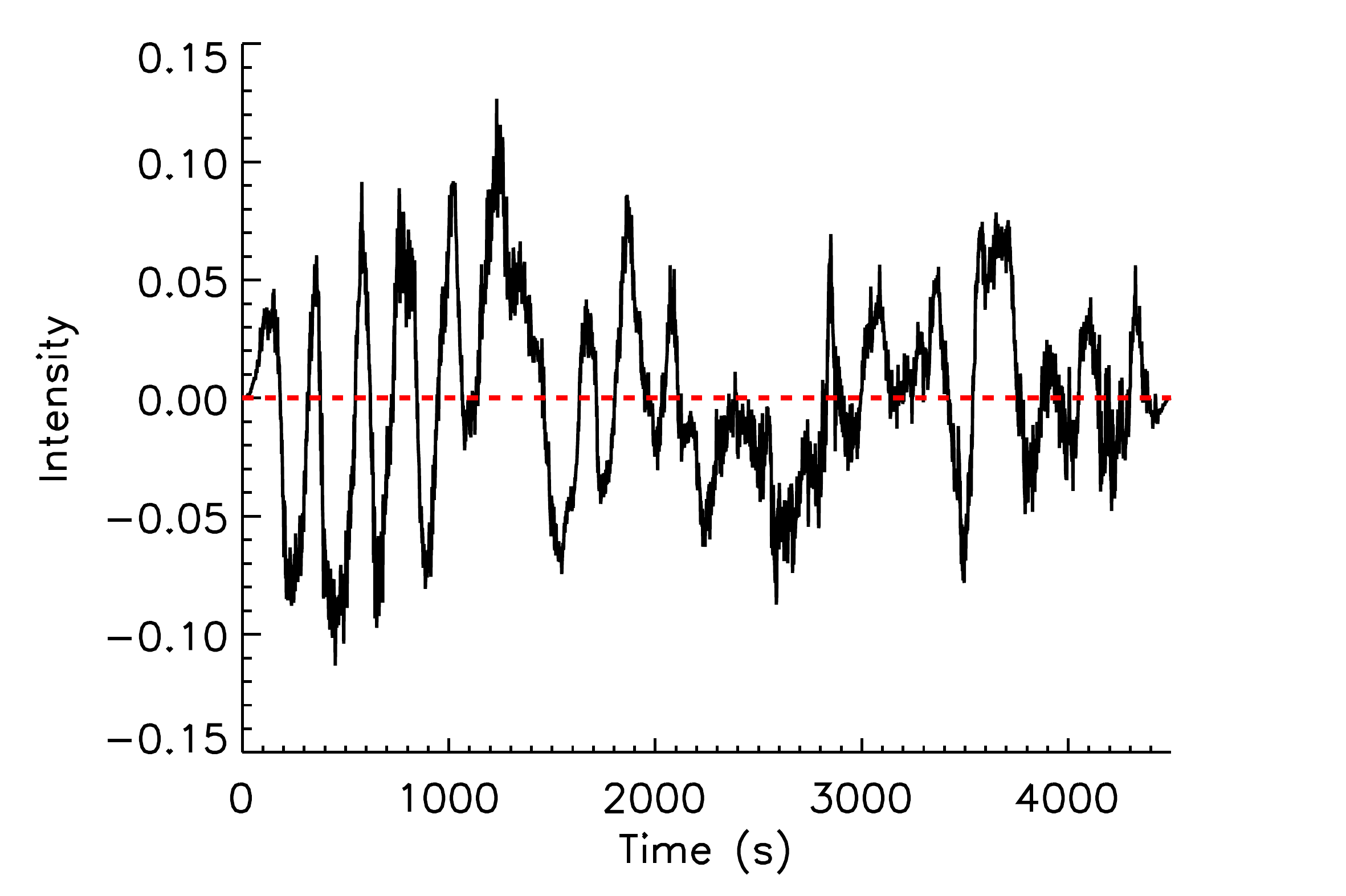}
\end{center}
\caption{An H$\alpha$ line core intensity time series (upper left; solid black line) extracted from a penumbral location of the HARDcam data described in Section~{\ref{sec:HARDcam}}. The intensities shown have been normalized by the time-averaged H$\alpha$ intensity established in a quiet Sun region within the field-of-view. A dashed red line shows a third-order polynomial fitted to the lightcurve, which is designed to detrend the data to provide a stationary time series. The upper-right panel displays the resulting time series once the third-order polynomial trend line has been subtracted from the raw intensities (black line). The solid red line depicts an apodization filter designed to preserve 90\% of the original lightcurve, but gradually reduce intensities to zero towards the edges of the time series to help alleviate any spurious signals in the resulting FFT. The lower panel reveals the final lightcurve that is ready for FFT analyses, which has been both detrended and apodized to help ensure the resulting Fourier power is accurately constrained. The horizontal dashed red lines signify the new mean value of the data, which is equal to zero due to the detrending employed.}
\label{fig:HARDcam_lightcurves} 
\end{figure*}

On the other hand, non-stationary time series are notoriously difficult to predict and model \citep{doi:10.1111/j.1467-9892.1989.tb00037.x}. A major challenge when applying Fourier techniques to non-stationary data is that the corresponding Fourier spectrum incorporates numerous additional harmonic components to replicate the inherent non-stationary behavior, which artificially spreads the true time series energy over an uncharacteristically wide frequency range \citep{2004ApJ...614..435T}. Ideally, non-stationary data needs to be transformed into stationary data with a constant mean and variance that is independent of time. However, understanding the underlying systematic (acting according to a fixed plan or system; methodical) and stochastic (randomly determined; having a random probability distribution or pattern that may be analyzed statistically but may not be predicted precisely) processes is often very difficult \citep{doi:10.1007/s00521-013-1386-y}. In particular, differencing can mitigate stochastic (i.e., non-systematic) processes to produce a difference-stationary time series, while detrending can help remove deterministic trends (e.g., time-dependent changes), but may struggle to alleviate stochastic processes \citep{doi:10.5815/ijisa.2015.09.07}. Hence, it is often very difficult to ensure observational time series are truly linear and stationary. 

The upper-left panel of Figure~{\ref{fig:HARDcam_lightcurves}} displays an intensity time series (lightcurve) that has been extracted from a penumbral pixel in the chromospheric HARDcam H$\alpha$ data. Here, the intensities have been normalized by the time-averaged quiescent H$\alpha$ intensity. It can be seen in the upper-left panel of Figure~{\ref{fig:HARDcam_lightcurves}} that in addition to sinusoidal wave-like signatures, there also appears to be a background trend (i.e., moving average) associated with the intensities. Through visual inspection, this background trend does not appear linear, thus requiring higher order polynomials to accurately model and remove. It must be remembered that very high order polynomials will likely begin to show fluctuations on timescales characteristic of the wave signatures wishing to be studied. Hence, it is important that the lowest order polynomial that best fits the data trends is chosen to avoid contaminating the embedded wave-like signatures with additional fluctuations arising from high-order polynomials. Importantly, the precise method applied to detrend the data can vary depending upon the signal being analyzed \citep[e.g.,][]{1972SoPh...22..276E, 1972SoPh...23...47E, 2001A&A...379.1052K, 2003A&A...407..735R, 2005A&A...441.1183D, 2005A&A...430.1119D}. For example, some researchers choose to subtract the mean trend, while others prefer to divide by the fitted trend then subtract `1' from the subsequent time series. Both approaches result in a more stationary time series with a mean value of `0'. However, subtracting the mean preserves the original unit of measurement and hence the original shape of the time series (albeit with modified numerical axes labels), while dividing by the mean provides a final unit that is independent of the original measurement and thus provides a method to more readily visualize fractional changes to the original time series. It must be noted that detrending processes, regardless of which approach is selected, can help remove deterministic trends (e.g., time-dependent changes), but often struggle to alleviate stochastic processes from the resulting time series.

The dashed red line in the upper-left panel of Figure~{\ref{fig:HARDcam_lightcurves}} displays a third-order polynomial trend line fitted to the raw H$\alpha$ time series. The line of best fit is relatively low order, yet still manages to trace the global time-dependent trend. Subtracting the trend line from the raw intensity lightcurve provides fluctuations about a constant mean equal to zero (upper-right panel of Figure~{\ref{fig:HARDcam_lightcurves}}), helping to ensure the resulting time series is stationary. It can be seen that wave-like signatures are present in the lightcurve, particularly towards the start of the observing sequence, where fluctuations on the order of $\approx$8\% of the continuum intensity are visible. However, it can also be seen from the right panel of Figure~{\ref{fig:HARDcam_lightcurves}} that between times of approximately $300 - 1300$~s there still appears to be a local increase in the mean (albeit no change to the global mean, which remains zero). To suppress this local change in the mean, higher order polynomial trend lines could be fitted to the data, but it must be remembered that such fitting runs the risk of manipulating the true wave signal. Hence, for the purposes of this example, we will continue to employ third-order polynomial detrending, and make use of the time series shown in the upper-right panel of Figure~{\ref{fig:HARDcam_lightcurves}}. 

For data sequences that are already close to being stationary, one may question why the removal of such background trends is even necessary since the Fourier decomposition with naturally put the trend components into low-frequency bins. Of course, the quality and/or dynamics of the input time series will have major implications regarding what degree of polynomial is required to accurately transform the data into a stationary time series. However, from the perspective of wave investigations, non-zero means and/or slowly evolving backgrounds will inappropriately apply Fourier power across low frequencies, even though these are not directly wave related, which may inadvertently skew any subsequent frequency-integrated wave energy calculations performed. The sources of such non-stationary processes can be far-reaching, and include aspects related to structural evolution of the feature being examined, local observing conditions (e.g., changes in light levels for intensity measurements), and/or instrumental effects (e.g., thermal impacts that can lead to time-dependent variances in the measured quantities). As such, some of these sources (e.g., structural evolution) are dependent on the precise location being studied, while other sources (e.g., local changes in the light level incident on the telescope) are global effects that can be mapped and removed from the entire data sequence simultaneously. Hence, detrending the input time series helps to ensure that the resulting Fourier power is predominantly related to the embedded wave activity.

Another step commonly taken to ensure the reliability of subsequent Fourier analyses is to apply an apodization filter to the processed time series \citep{1976JOSA...66..259N}. An Fourier transform assumes an infinite, periodically repeating sequence, hence leading to a looping behavior at the ends of the time series. Hence, an apodization filter is a function employed to smoothly bring a measured signal down to zero towards the extreme edges (i.e., beginning and end) of the time series, thus mitigating against sharp discontinuities that may arise in the form of false power (edge effect) signatures in the resulting power spectrum. 

Typically, the apodization filter is governed by the percentage over which the user wishes to preserve the original time series. For example, a 90\% apodization filter will preserve the middle 90\% of the overall time series, with the initial and final 5\% of the lightcurve being gradually tapered to zero \citep{1987ApJ...314L..15D}. There are many different forms of the apodization filter shape that can be utilized, including tapered cosines, boxcar, triangular, Gaussian, Lorentzian, and trapezoidal profiles, many of which are benchmarked using solar time series in \citet{2015SoPh..290.1135L}. A tapered cosine is the most common form of apodization filter found in solar physics literature \citep[e.g.,][]{1998A&A...329..276H}, and this is what we will employ here for the purposes of our example dataset. The upper-right panel of Figure~{\ref{fig:HARDcam_lightcurves}} reveals a 90\% tapered cosine apodization filter overplotted on top of the detrended H$\alpha$ lightcurve. Multiplying this apodization filter by the lightcurve results in the final detrended and apodized time series shown in the bottom panel of Figure~{\ref{fig:HARDcam_lightcurves}}, where the stationary nature of this processed signal is now more suitable for Fourier analyses. It is worth noting that following successful detrending of the input time series, the apodization percentage chosen can often be reduced, since the detrending process will suppress any discontinuities arising at the edges of the data sequence \citep[i.e., helps to alleviate spectral leakage;][]{1163506}. As such, the apodization percentage employed may be refined based on the ratio between the amplitude of the (primary) oscillatory signal and the magnitude of the noise present within that signal \citep[i.e., linked to the inherent signal-to-noise ratio;][]{stoica2005spectral, alma991002695602804901}.

\begin{figure*}[!t]
\begin{center}
\includegraphics[trim = 1cm 0cm 2.5cm 0cm, clip, width=0.32\textwidth]{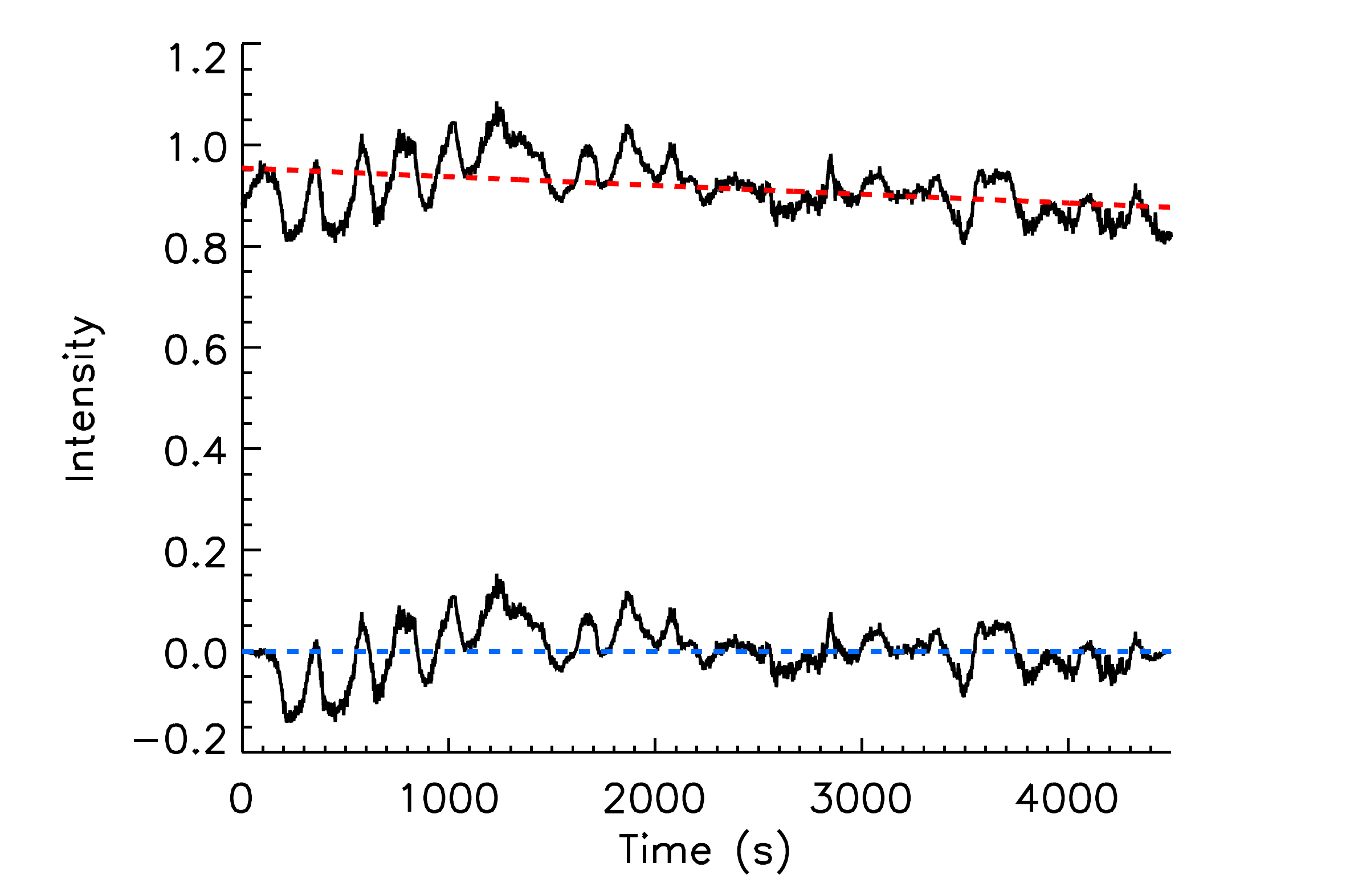}
\includegraphics[trim = 1cm 0cm 2.5cm 0cm, clip, width=0.32\textwidth]{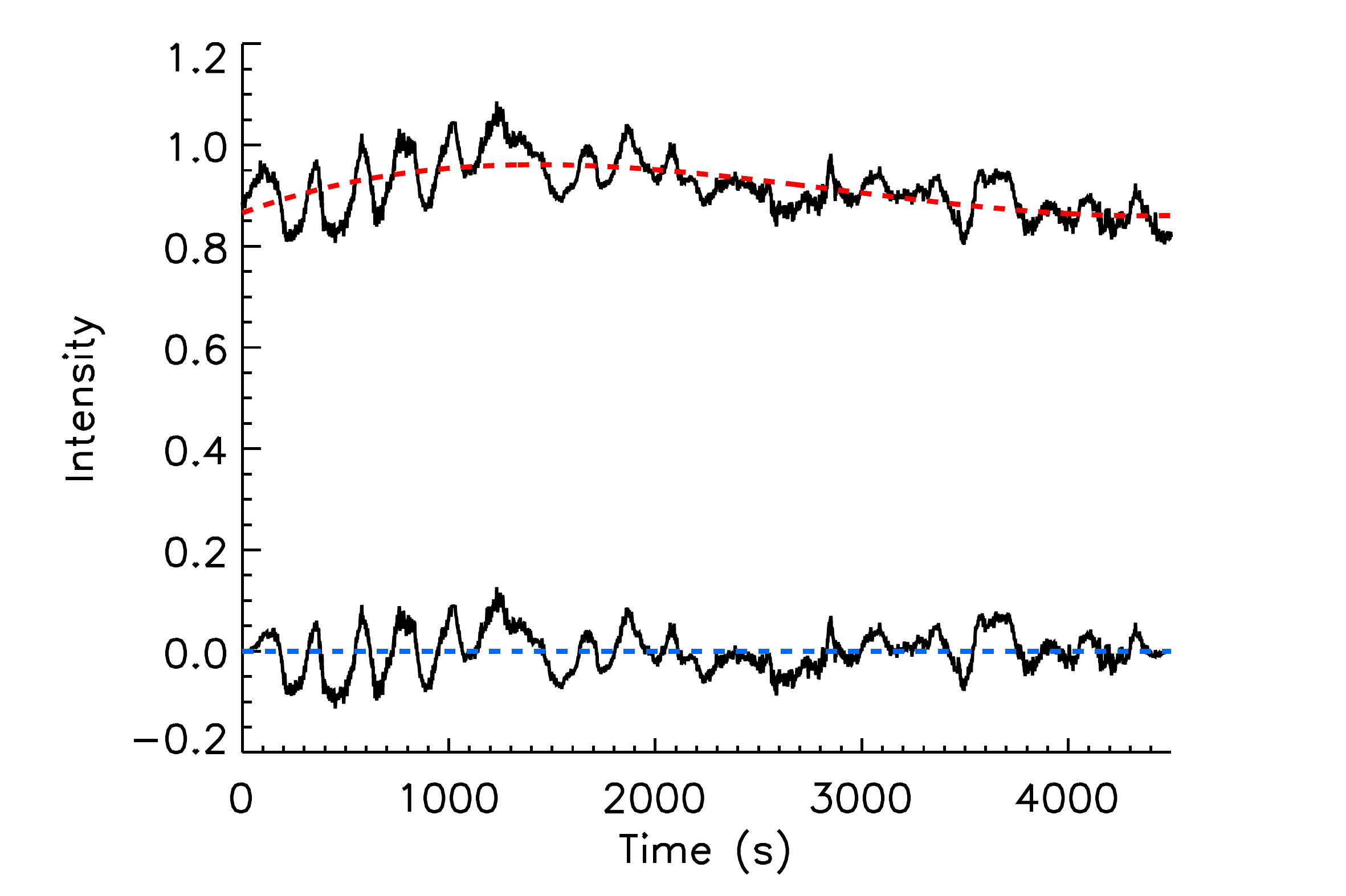}
\includegraphics[trim = 1cm 0cm 2.5cm 0cm, clip, width=0.32\textwidth]{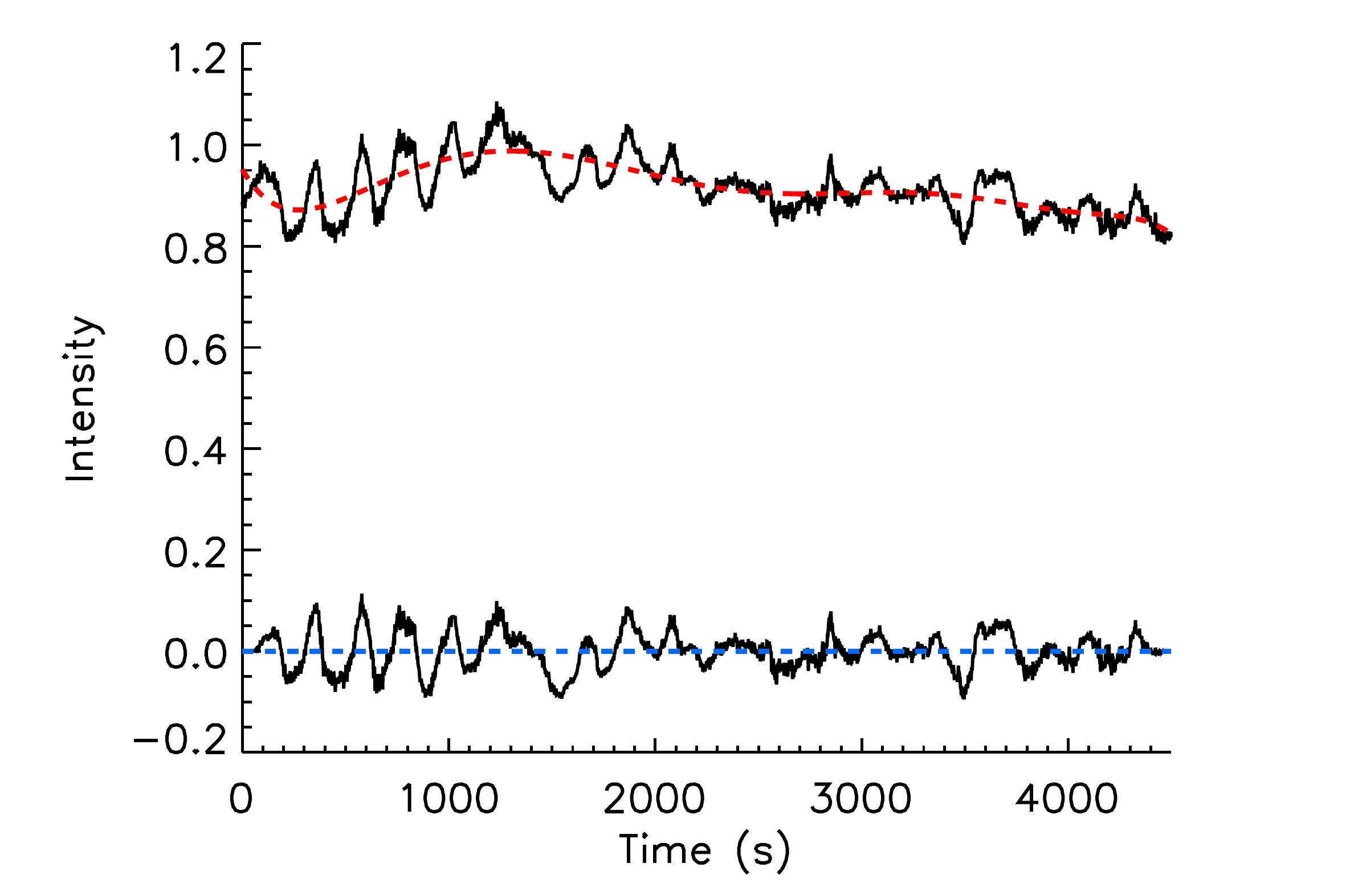}
\includegraphics[trim = 0cm 0cm 2.5cm 0cm, clip, width=0.32\textwidth]{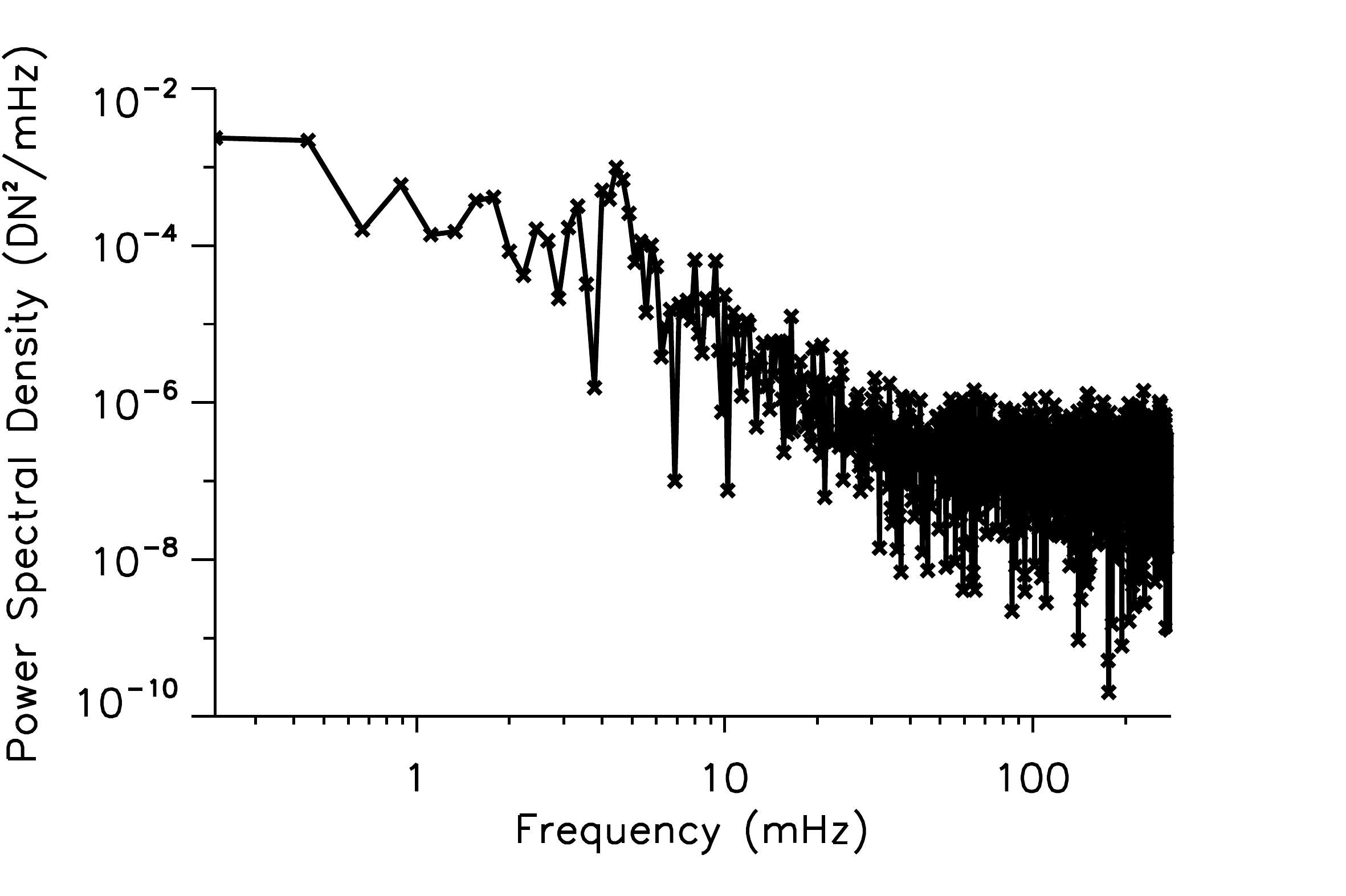}
\includegraphics[trim = 0cm 0cm 2.5cm 0cm, clip, width=0.32\textwidth]{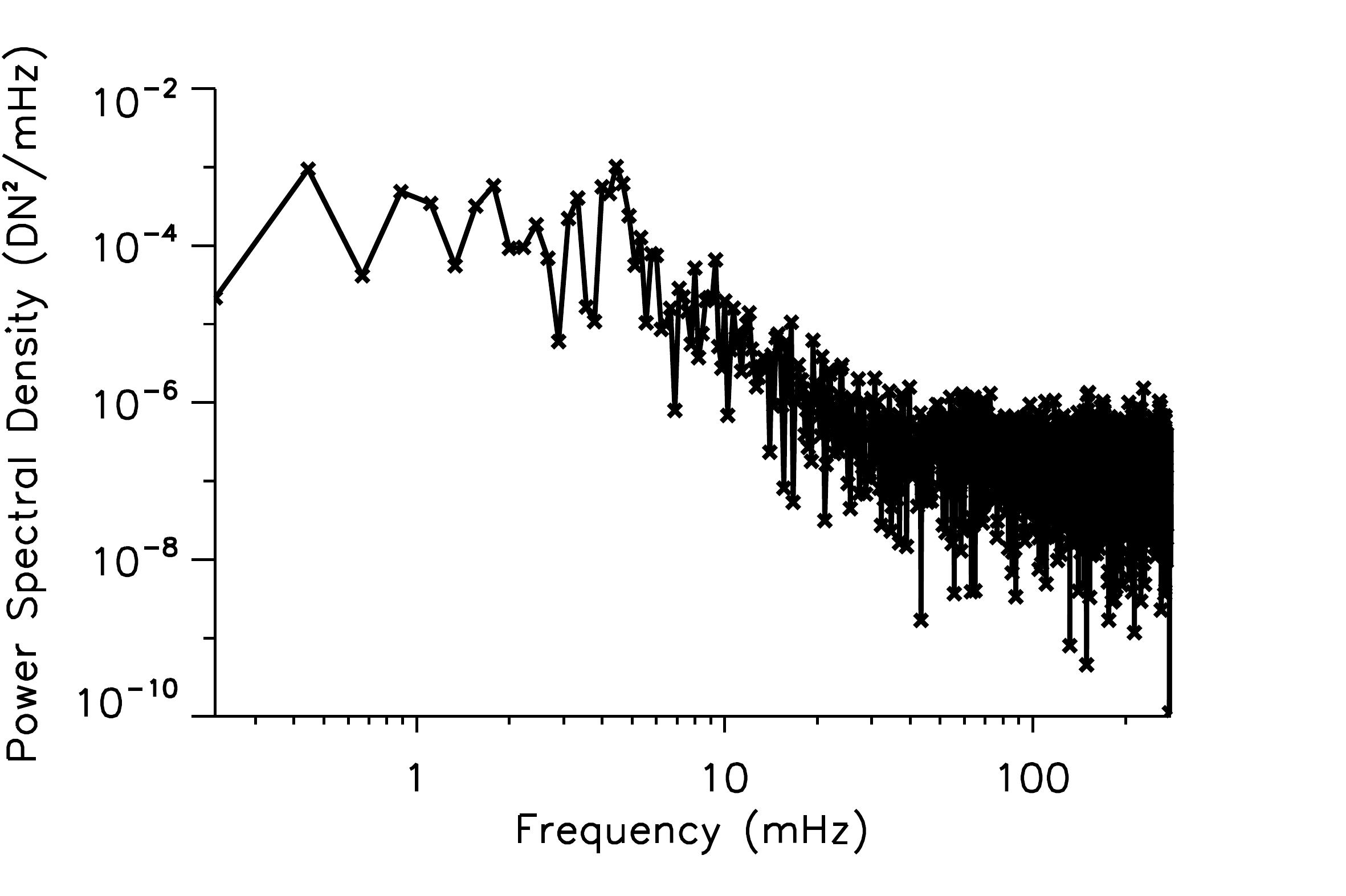}
\includegraphics[trim = 0cm 0cm 2.5cm 0cm, clip, width=0.32\textwidth]{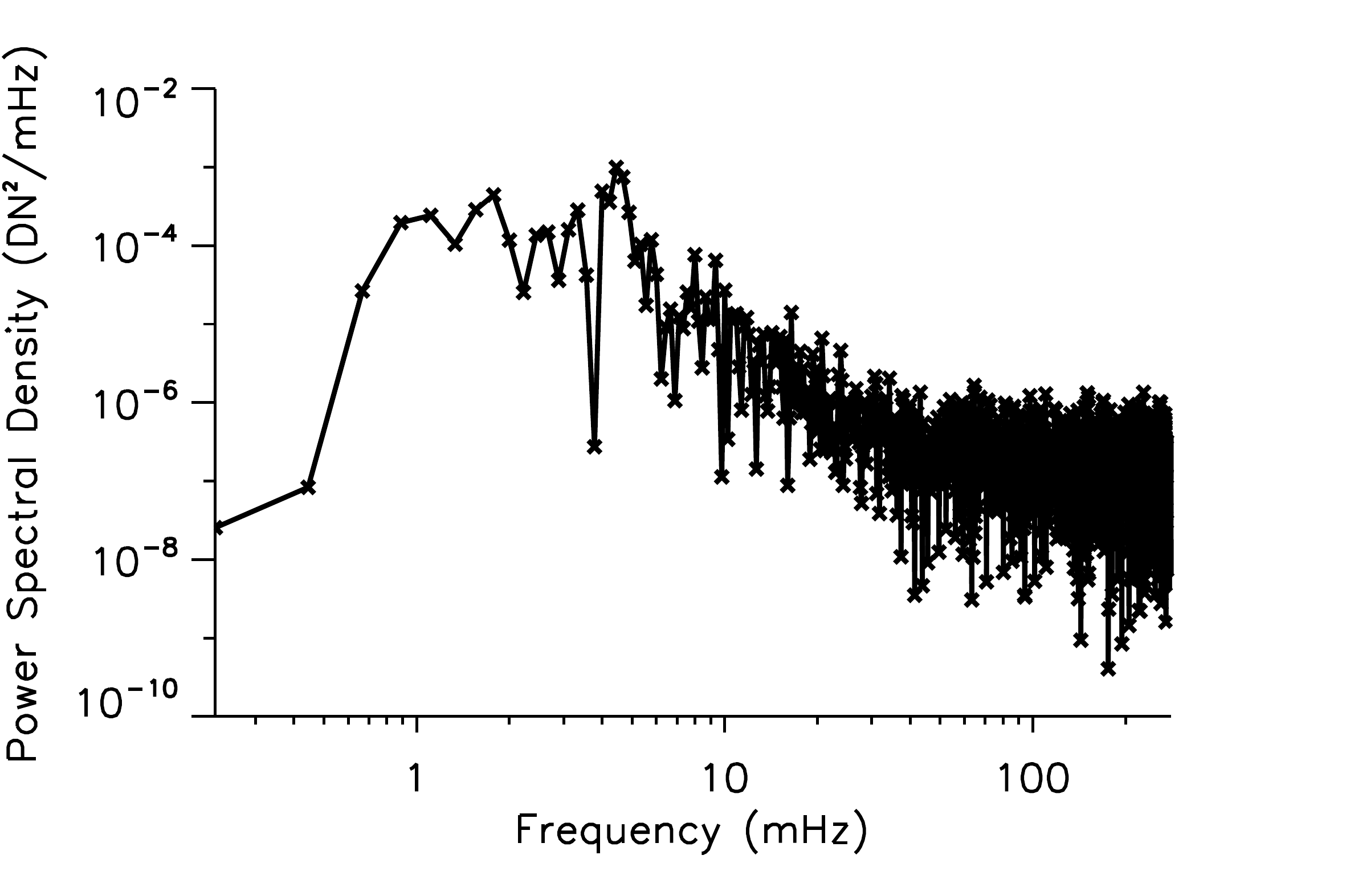}
\end{center}
\caption{Taking the raw HARDcam H$\alpha$ lightcurve shown in the upper-left panel of Figure~{\ref{fig:HARDcam_lightcurves}}, the upper row displays the resultant detrended time series utilizing linear (left), third-order polynomial (middle), and nineth-order polynomial (right) fits to the data. In each panel the dashed red line highlights the line of best fit, while the dashed blue line indicates the resultant data mean that is equal to zero following detrending. The lower row displays the corresponding Fourier power spectral densities for each of the linear (left), third-order polynomial (middle), and nineth-order polynomial detrended time series. Changes to the power spectral densities are particularly evident at low frequencies.}
\label{fig:HARDcam_power_different_detrends} 
\end{figure*}

Performing a fast Fourier transform \citep[FFT;][]{doi:10.2307/2003354} of the detrended time series provides a Fourier amplitude spectrum, which can be displayed as a function of frequency. An FFT is a computationally more efficient version of the discrete Fourier transform \citep[DFT;][]{GRUNBAUM1982355}, which only requires $N\log{N}$ operations to complete compared with the $N^{2}$ operations needed for the DFT, where $N$ is the number of data points in the time series, which can be calculated by dividing the time series duration by the acquisition cadence. Following a Fourier transform of the input data, the number of (non-negative) frequency bins, $N_{f}$, can be computed by adding one to the number of samples \citep[to account for the zeroth frequency representing the time series mean;][]{10.5555/1795494}, $N+1$, dividing the result by a factor of two, before rounding to the nearest integer. The Nyquist frequency is the highest constituent frequency of an input time series that can be evaluated at a given sampling rate \citep{grenander1959probability}, and is defined as $f_{\mathrm{Ny}} = {\mathrm{sampling~rate}}/2 = 1/(2 \times {\mathrm{cadence}})$. To evaluate the frequency resolution, $\Delta{f}$, of an input time series, one must divide the Nyquist frequency by the number of non-zero frequency bins (i.e., the number of steps between the zeroth and Nyquist frequencies, $N/2$), providing,
\begin{equation}
\label{eqn:frequencyresolution}
    \Delta{f} = \frac{f_{\mathrm{Ny}}}{N/2} ~=~ \frac{\frac{1}{2 \times {\mathrm{cadence}}}}{\frac{{\mathrm{time~series~duration}}}{2 \times {\mathrm{cadence}}}} ~=~ \frac{1}{{\mathrm{time~series~duration}}} \ . 
\end{equation}
As a result, it is clear to see that the observing duration plays a pivotal role in the corresponding frequency resolution \citep[see, e.g.,][for considerations in the helioseismology community]{1985ESASP.235..199H, 1997SoPh..170...63D, 2017A&A...600A..35G}. It is also important to note that the frequency bins remain equally spaced across the lowest (zeroth frequency or mean) to highest (Nyquist) frequency that is resolved in the corresponding Fourier spectrum. See Section~{\ref{sec:commonmisconceptionsinvolvingfourierspace}} for a more detailed comparison between the terms involved in Fourier decomposition.

The HARDcam dataset utilized has a cadence of 1.78~s, which results in a Nyquist frequency of $f_{\text{\tiny{Ny}}}\approx280$~mHz $\left(\frac{1}{2\times1.78}\right)$. It is worth noting that only the positive frequencies are displayed in this review for ease of visualization. Following the application of Fourier techniques, both negative and positive frequencies, which are identical except for their sign, will be generated for the corresponding Fourier amplitudes. This is a consequence of the Euler relationship that allows sinusoidal wave signatures to be reconstructed from a set of positive and negative complex exponentials \citep{smith2007mathematics}. Since input time series are real valued (e.g., velocities, intensities, spectral line widths, magnetic field strengths, etc.) with no associated imaginary terms, then the Fourier amplitudes associated with the negative and positive frequencies will be identical. This results in the output Fourier transform being Hermitian symmetric \citep{NAPOLITANO2020497}. As a result, the output Fourier amplitudes are often converted into a power spectrum (a measure of the square of the Fourier wave amplitude), or following normalization by the frequency resolution, into a power spectral density. This approach is summarized by \citet{StullRolandB1988AItB}, where the power spectral density, PSD, can be calculated as,
\begin{equation}
\label{eqn:PSD}
    \mathrm{PSD}(n) = \frac{2 \cdot |\mathcal{F}_{A}(n)|^{2}}{\Delta f} = \frac{2 \cdot \left( \left[ \mathcal{F}_{\mathrm{real~part}}(n) \right]^{2} + \left[ \mathcal{F}_{\mathrm{imaginary~part}}(n) \right]^{2} \right)}{\Delta f} \ .
\end{equation}

\begin{figure*}[!t]
\begin{center}
\includegraphics[width=0.65\textwidth]{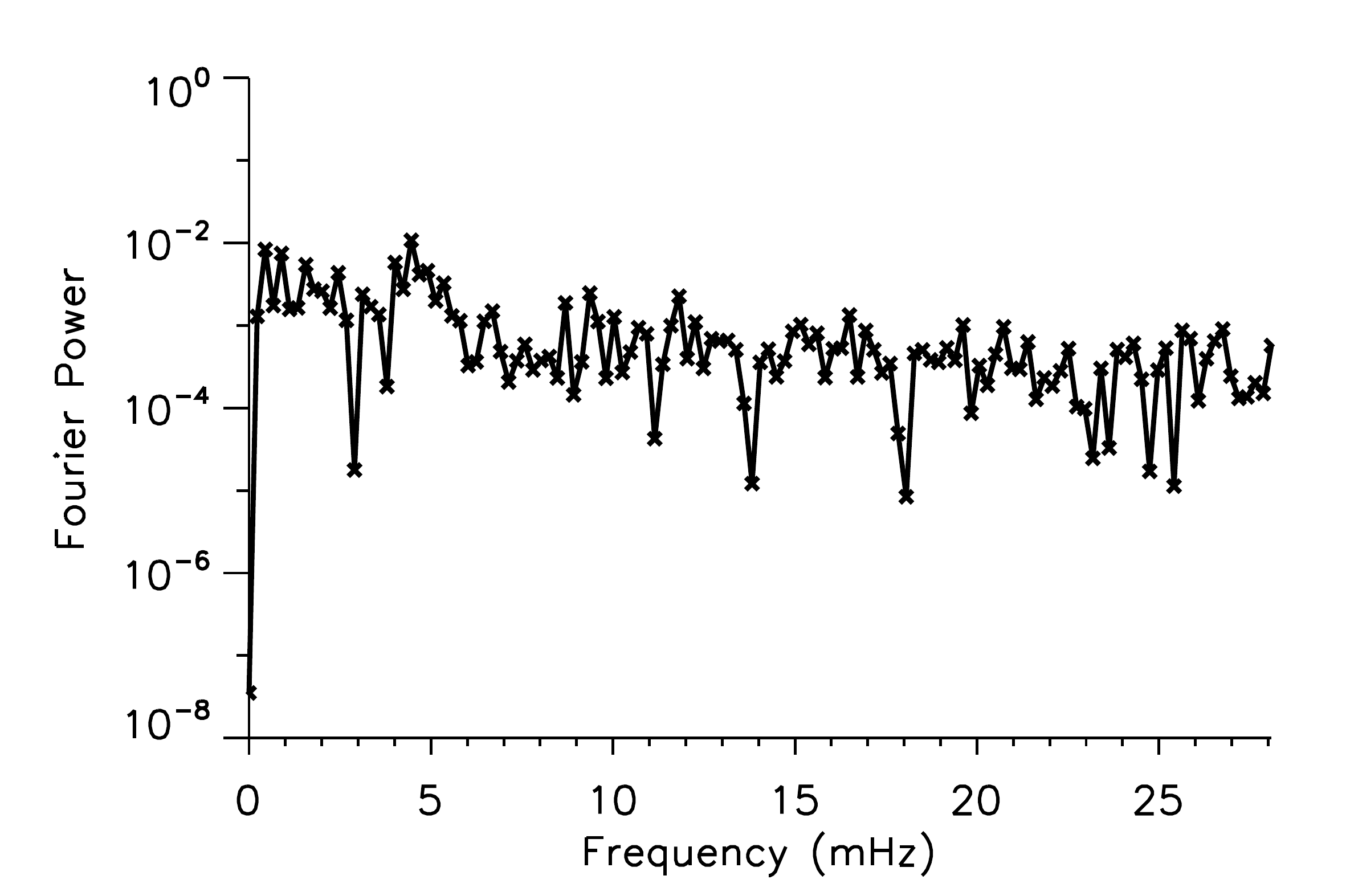}
\includegraphics[width=0.49\textwidth]{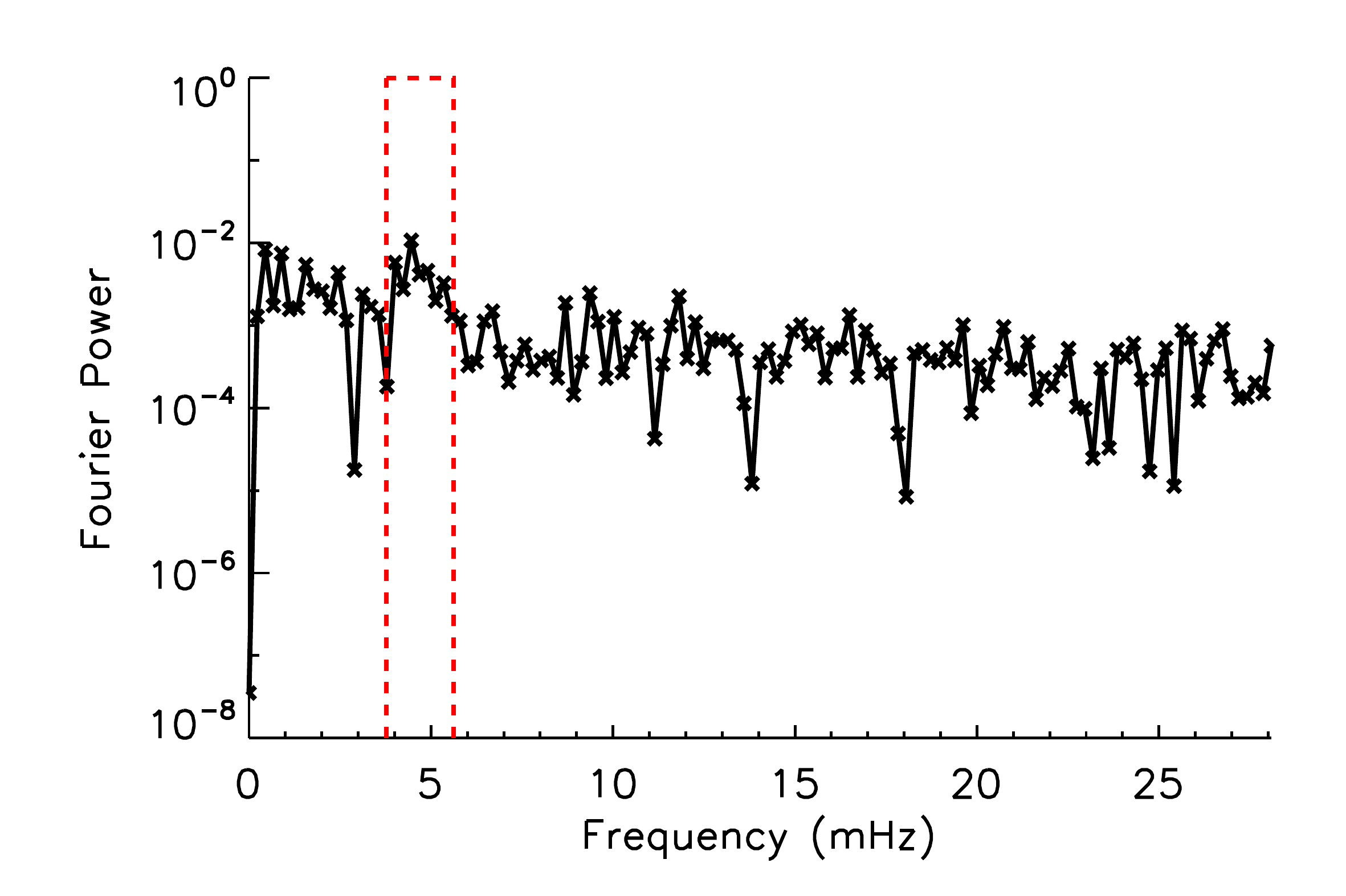}
\includegraphics[width=0.49\textwidth]{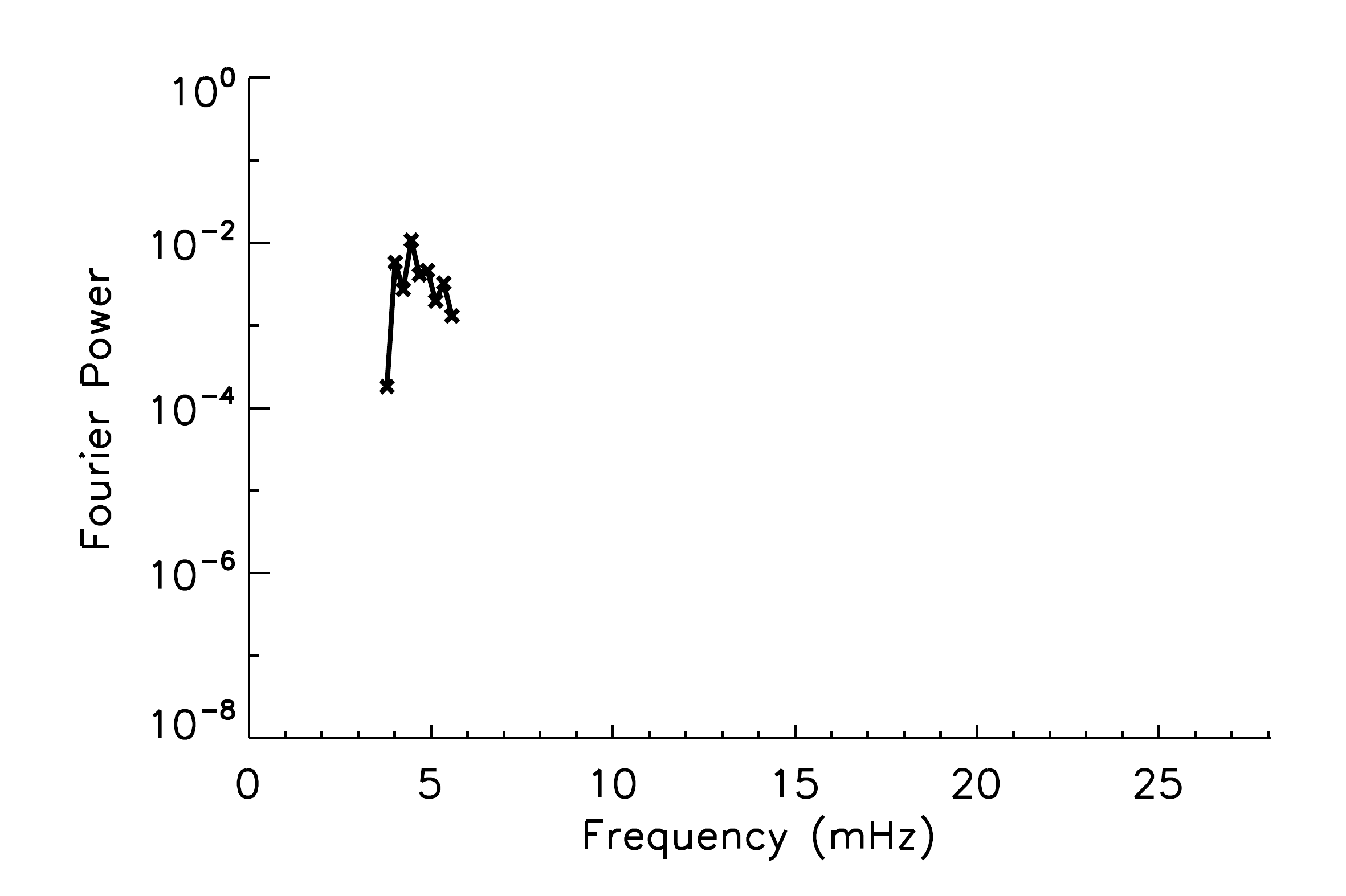}
\includegraphics[width=0.49\textwidth]{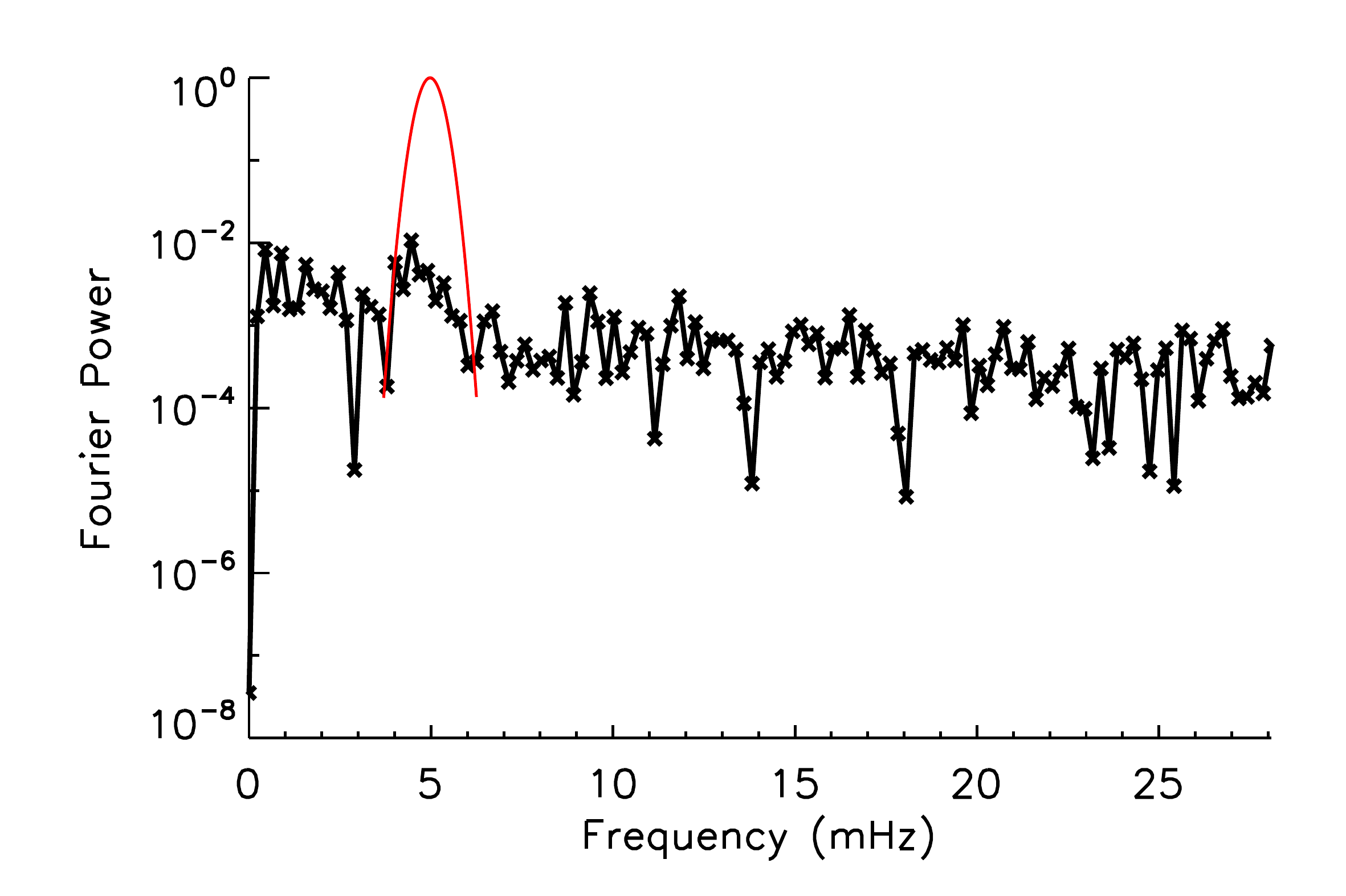}
\includegraphics[width=0.49\textwidth]{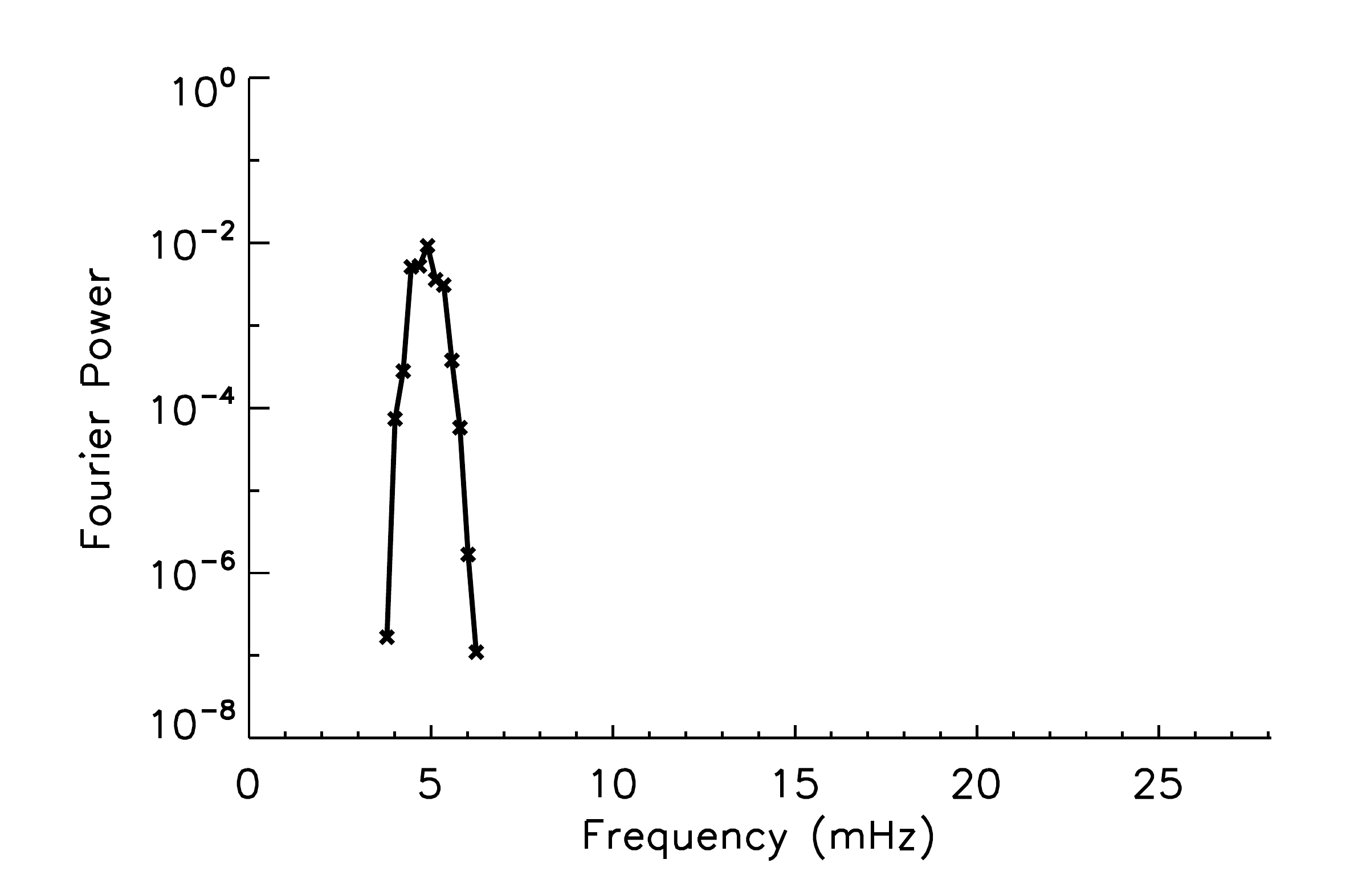}
\end{center}
\caption{Fourier power spectrum of the HARDcam H$\alpha$ detrended lightcurve shown in the lower panel of Figure~{\ref{fig:HARDcam_lightcurves}} (top). For the purposes of wave filtering, a step function is shown on the Fourier spectrum using a dashed red line (middle left), where the step function equals unity between frequencies spanning $3.7 - 5.7$~mHz (i.e., $4.7\pm1.0$~mHz). Multiplying the Fourier power spectrum by this step function results in isolated power features, which are displayed in the middle-right panel. Alternatively, a Gaussian function centered on 4.7~mHz, with a FWHM of 2.0~mHz, is overplotted on top of the Fourier power spectrum using a red line in the lower-left panel. Multiplying the power spectrum by the Gaussian function results in similar isolated power features, shown in the lower-right panel, but with greater apodization of edge frequencies to help reduce aliasing upon reconstruction of the filtered time series. }
\label{fig:HARDcam_power} 
\end{figure*}

In Equation~{\ref{eqn:PSD}}, $\mathcal{F}_{A}(n)$ is the Fourier amplitude for any given positive frequency, $n$, while $\Delta f$ is the corresponding frequency resolution of the Fourier transform (see definition above and further discussion points in Section~{\ref{sec:commonmisconceptionsinvolvingfourierspace}}). Note that the factor of `2' is required due to the wrapping of identical Fourier power at negative frequencies into the positive domain. The normalization of the power spectrum by the frequency resolution is a best practice to ensure that the subsequent plots can be readily compared against other data sequences that may be acquired across shorter or longer observing intervals, hence affecting the intrinsic frequency resolution (see Section~{\ref{sec:commonmisconceptionsinvolvingfourierspace}}). As an example, the power spectral density of an input velocity time series, with units of km/s, will have the associated units of km$^{2}$/s$^{2}$/mHz \citep[e.g.,][]{Stangalini2021_bomega}. The power spectral density for the detrended HARDcam H$\alpha$ time series is depicted in the lower-middle panel of Figure~{\ref{fig:HARDcam_power_different_detrends}}. Here, the intensity time series is calibrated into normalized data number (DN) units, which are often equally labeled as
`counts' in the literature. Hence, the resulting power spectral density has units of DN$^{2}$/mHz. 

An additional step often employed following the calculation of the PSD of an input time series is to remove the Fourier components associated with noise. It can be seen in the lower panels of Figure~{\ref{fig:HARDcam_power_different_detrends}} that there is a flattening of power towards higher frequencies, which is often due to the white noise that dominates the signal at those frequencies \citep{1976A&A....47..449H, 2017ApJ...847....5K}. Here, white noise is defined as fluctuations in a time series that give rise to equal Fourier power across all frequencies, hence giving rise to a flat PSD \citep{bendat2011random}. Often, if white noise is believed to be the dominant source of noise in the data (i.e.,  the signal is well above the detector background noise, hence providing sufficient photon statistics so that photon noise is the dominant source of fluctuations), then its PSD can be estimated by applying Equation~{\ref{eqn:PSD}} to a random light curve generated following a Poisson distribution, with an amplitude equivalent to the square root of the mean intensity of the time series \citep{2005Natur.435..919F, 2011ApJ...743L..24L}. Subtraction of the background noise is necessary when deriving, for example, the total power of an oscillation isolated in a specific frequency window \citep{doi:https://doi.org/10.1002/9780470740156.ch13}. Other types of noise exist that have discernible power-law slopes associated with their PSDs as a function of frequency. For example, while white noise has a flat power-law slope, pink and red noise display $1/f$ and $1/f^{2}$ power-law slopes, respectively, resulting in larger amplitudes at lower frequencies \citep{2016A&A...592A.153K, 2018Ge&Ae..58..893S}. The specific dominant noise profile must be understood before it is subtracted from the relevant data PSDs.

As a result of the detrending employed in Figure~{\ref{fig:HARDcam_lightcurves}}, the absolute Fourier wave amplitude related to a frequency of 0~Hz (i.e., representing the time series mean; upper panel of Figure~{\ref{fig:HARDcam_power}}) is very low; some 4 orders-of-magnitude lower than the power associated with white noise signatures at high frequencies. Of course, if the processed time series mean is exactly zero, then the Fourier wave amplitude at 0~Hz should also be zero. In the case of Figure~{\ref{fig:HARDcam_power}}, the detrended time series does have a zero mean. However, because the time series is not antisymmetric about the central time value, it means that the application of the tapered cosine apodization function results in a very small shift in the time series mean away from the zero value. As a result, the subsequent Fourier amplitudes are fractionally (e.g., at the $10^{-8}$ level for the upper panel of Figure~{\ref{fig:HARDcam_power}}) above the zero point. Once the processes of detrending and apodization are complete, it is possible to re-calculate the time series mean and subtract this value to ensure the processed mean remains zero before the application of Fourier analyses. However, for the purposes of Figures~{\ref{fig:HARDcam_power_different_detrends}} \& {\ref{fig:HARDcam_power}}, this additional mean subtraction has not been performed to better highlight this potential artifact at the lowest temporal frequencies.

Note that Figure~{\ref{fig:HARDcam_power}} does not have the frequency axis displayed on a log-scale in order to reveal the 0~Hz component. As such, the upper frequency range is truncated to $\approx28$~Hz to better reveal the signatures present at the lower frequencies synonymous with wave activity in the solar atmosphere. The suppression of Fourier wave amplitudes at the lowest frequencies suggests that the third-order polynomial trend line fitted to the raw H$\alpha$ intensities is useful at removing global trends in the visible time series. However, as discussed above, care must be taken when selecting the polynomial order to ensure that the line of best fit does not interfere with the real wave signatures present in the original lightcurve. To show the subtle, yet important impacts of choosing a suitable trend line, Figure~{\ref{fig:HARDcam_power_different_detrends}} displays the resultant detrended time series of the original HARDcam H$\alpha$ lightcurve for three different detrending methods, e.g., the subtraction of a linear, a third-order polynomial, and a nineth-order polynomial line of best fit. It can be seen from the upper panels of Figure~{\ref{fig:HARDcam_power_different_detrends}} that the resultant (detrended) lightcurves have different perturbations away from the new data mean of zero. This translates into different Fourier signatures in the corresponding power spectral densities (lower panels of Figure~{\ref{fig:HARDcam_power_different_detrends}}), which are most apparent at the lowest frequencies (e.g., $<3$~mHz). Therefore, it is clear that care must be taken when selecting the chosen order of the line of best fit so that it doesn't artificially suppress true wave signatures that reside in the time series. It can be seen in the lower-middle panel of Figure~{\ref{fig:HARDcam_power_different_detrends}} that the largest Fourier power signal is at a frequency of $\approx 4.7$~mHz, corresponding to a periodicity of $\approx 210$~s, which is consistent with previous studies of chromospheric wave activity in the vicinity of sunspots \citep[e.g.,][to name but a few examples]{2010ApJ...722..131F, 2013ApJ...779..168J, 2016A&A...591A..63L}. 

\subsubsection{Common Misconceptions involving Fourier Space}
\label{sec:commonmisconceptionsinvolvingfourierspace}
Translating a time series into the frequency-dependent domain through the application of a Fourier transform is a powerful diagnostic tool for analyzing the frequency content of (stationary) time series. However, when translating between the temporal and frequency domains it becomes easy to overlook the importance of the sampling cadence and the time series duration in the corresponding frequency axis. For example, one common misunderstanding is the belief that increasing the sampling rate of the data (e.g., increasing the frame rate of the observations from 10~frames per second to 100~frames per second) will improve the subsequent frequency resolution of the corresponding Fourier transform. Unfortunately, this is not the case, since increasing the frame rate raises the Nyquist frequency (highest frequency component that can be evaluated), but does not affect the frequency resolution of the Fourier transform. Instead, to improve the frequency resolution one must obtain a longer-duration time series or employ `padding' of the utilized lightcurve to increase the number of data points spanning the frequency domain \citep{10.5555/524406}. 

To put these aspects into better context, we will outline a worked example that conveys the importance of both time series cadence and duration. Let us consider two complementary data sequences, one from the Atmospheric Imaging Assembly \citep[AIA;][]{2012SoPh..275...17L} onboard the SDO spacecraft, and one from the 4m ground-based Daniel K. Inouye Solar Telescope \citep[DKIST;][]{2016AN....337.1064T,2020SoPh..295..172R,2021arXiv200808203R}. Researchers undertaking a multi-wavelength investigation of wave activity in the solar atmosphere may choose to employ these types of complementary observations in order to address their science objectives. Here, the AIA/SDO observations consist of 3~hours (10{\,}800~s) of 304{\,}{\AA} images taken at a cadence of 12.0~s, while the DKIST observations comprise of 1~hour (3600~s) of H$\alpha$ observations taken by the Visual Broadband Imager \citep[VBI;][]{2014SPIE.9147E..9IW} at a cadence of 3.2~s. 

\begin{figure*}[!t]
\begin{center}
\includegraphics[width=\textwidth]{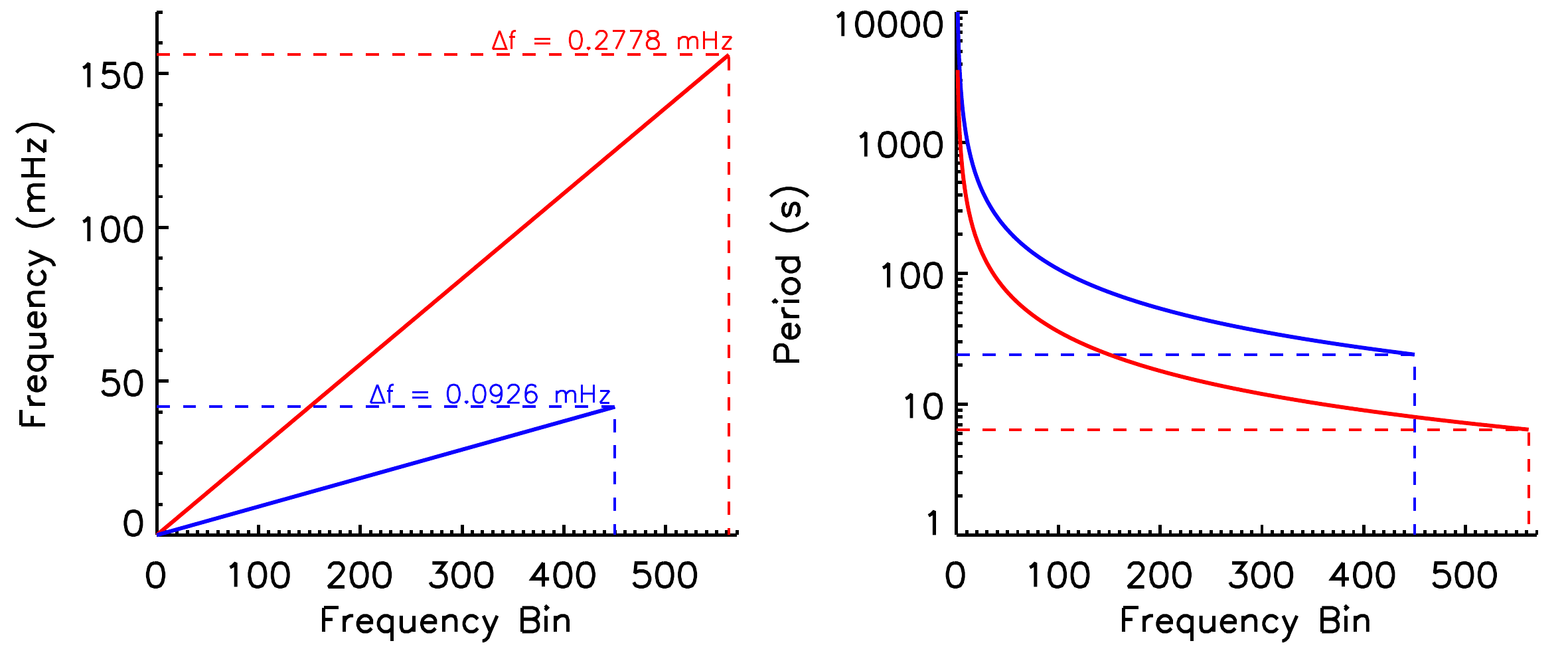}
\end{center}
\caption{The frequencies (left panel) and corresponding periodicities (right panel) that can be measured through the application of Fourier analysis to an input time series. Here, the solid blue lines depict AIA/SDO observations spanning a 3~hour duration and acquired with a temporal cadence of 12.0~s, while the solid red lines highlight VBI/DKIST observations spanning a 1~hour window and acquired with a temporal cadence of 3.2~s. It can be seen that both the cadence and observing duration play pivotal roles in the resulting frequencies/periodicities achievable, with the longer duration AIA/SDO observations providing a better frequency resolution, $\Delta{f}$, while the higher cadence VBI/DKIST data results in a better Nyquist frequency that allows more rapid wave fluctuations to be studied. In the left and right panels, the dashed blue and red lines depict the Nyquist frequencies and corresponding periodicities for the AIA/SDO and VBI/DKIST data sequences, respectively (see text for more information).} 
\label{fig:Fourierfrequencyperiodbins} 
\end{figure*}

The number of samples, $N$, for each of the time series can be calculated as $N_{\mathrm{AIA}} = 10800 / 12.0 = 900$ and $N_{\mathrm{VBI}} = 3600 / 3.2 = 1125$. Therefore, it is clear that even though the AIA/SDO observations are obtained over a longer time duration, the higher cadence of the VBI/DKIST observations results in more samples associated with that data sequence. The number of frequency bins, $N_{f}$, can also be computed as $N_{f({\mathrm{AIA}})} = (900+1) / 2 = 451$, while ${N_{f({\mathrm{VBI}})} = (1125+1) / 2 = 563}$. Hence, the frequency axes of the corresponding Fourier transforms will be comprised of $451$ and $563$ positive real frequencies (i.e., $\ge 0$~Hz) for the AIA/SDO and VBI/DKIST data, respectively. The increased number of frequency bins for the higher cadence VBI/DKIST observations sometimes leads to the belief that this provides a higher frequency resolution. However, we have not yet considered the effect of the image cadence on the corresponding frequency axes.

In the case of the AIA/SDO and VBI/DKIST observations introduced above, the corresponding Nyquist frequencies can be computed as $f_{\mathrm{Ny(AIA)}} = 1/(2 \times 12.0) \approx 42$~mHz and ${f_{\mathrm{Ny(VBI)}} = 1/(2 \times 3.2) \approx 156}$~mHz, respectively. As a result, it should become clear that while the VBI/DKIST observations result in a larger number of corresponding frequency bins (i.e., $N_{f({\mathrm{VBI}})} > N_{f({\mathrm{AIA}})}$), these frequency bins are required to cover a larger frequency interval up to the calculated Nyquist value. Subsequently, for the case of the AIA/SDO and VBI/DKIST observations, the corresponding frequency resolutions can be calculated as ${\Delta{f}_{\mathrm{AIA}} = 1/10800 = 0.0926}$~mHz and ${\Delta{f}_{\mathrm{VBI}} = 1/3600 = 0.2778}$~mHz, respectively. Note that while the frequency resolution is constant, the same cannot be said for the period resolution due to the reciprocal nature between these variables. For example, at a frequency of $3.3$~mHz ($\approx5$~min oscillation), the period resolution for VBI/DKIST is $\approx25$~s (i.e., $\approx303\pm25$~s), while for AIA/SDO the period resolution is $\approx8$~s (i.e., $\approx303\pm8$~s). Similarly, at a frequency of $5.6$~mHz ($\approx3$~min oscillation), the period resolutions for VBI/DKIST and AIA/SDO are $\approx9$~s (i.e., $\approx180\pm9$~s) and $\approx3$~s (i.e., $\approx180\pm3$~s), respectively.

Figure~{\ref{fig:Fourierfrequencyperiodbins}} depicts the Fourier frequencies (left panel), and their corresponding periodicities (right panel), as a function of the derived frequency bin. It can be seen from the left panel of Figure~{\ref{fig:Fourierfrequencyperiodbins}} that the AIA/SDO observations produce a lower number of frequency bins (i.e., a result of less samples, $N_{\mathrm{AIA}} < N_{\mathrm{VBI}}$), alongside a smaller peak frequency value (i.e., a lower Nyquist frequency, ${f_{\mathrm{Ny(AIA)}}} < {f_{\mathrm{Ny(VBI)}}}$, caused by the lower temporal cadence). However, as a result of the longer duration observing sequence for the AIA/SDO time series (i.e., 3~hours for AIA/SDO versus 1~hour for VBI/DKIST), the resulting frequency resolution is better (i.e., ${\Delta{f}_{\mathrm{AIA}}} < {\Delta{f}_{\mathrm{VBI}}}$), allowing more precise frequency-dependent phenomena to be uncovered in the AIA/SDO observations. Of course, due to the AIA/SDO cadence being longer than that of VBI/DKIST (i.e., 12.0~s for AIA/SDO versus 3.2~s for VBI/DKIST), this results in the inability to examine the fastest wave fluctuations, which can be seen more clearly in the right panel of Figure~{\ref{fig:Fourierfrequencyperiodbins}}, whereby the VBI/DKIST observations are able to reach lower periodicities when compared to the complementary AIA/SDO data sequence. The above scenario is designed to highlight the important interplay between observing cadences and durations with regards to the quantitative parameters achievable through the application of Fourier transforms. For example, if obtaining the highest possible frequency resolution is of paramount importance to segregate closely matched wave frequencies, then it is the overall duration of the time series ({\it{not the observing cadence}}) that facilitates the necessary frequency resolution.

Another important aspect to keep in mind is that the Fourier spectrum is only an estimate of the real power spectrum of the studied process. The finite-duration time series, noise, and distortions due to the intrinsic covariance within each frequency bin may lead to spurious peaks in the spectrum, which could be wrongly interpreted as real oscillations. As a result, one may believe that by considering longer time series the covariance of each frequency bin will reduce, but this is not true since the bin width itself becomes narrower. One way forward is to divide the time series into different segments and average the resulting Fourier spectra calculated from each sub-division -- the so-called Welch method \citep{welch}, at the cost of reducing the resolution of frequencies explored. However, data from ground-based observatories are generally limited to $1-2$~hours each day, and it is not always possible to obtain such long time series. Therefore, special attention must be paid when interpreting the results.

\begin{figure*}[!t]
\begin{center}
\includegraphics[width=\textwidth]{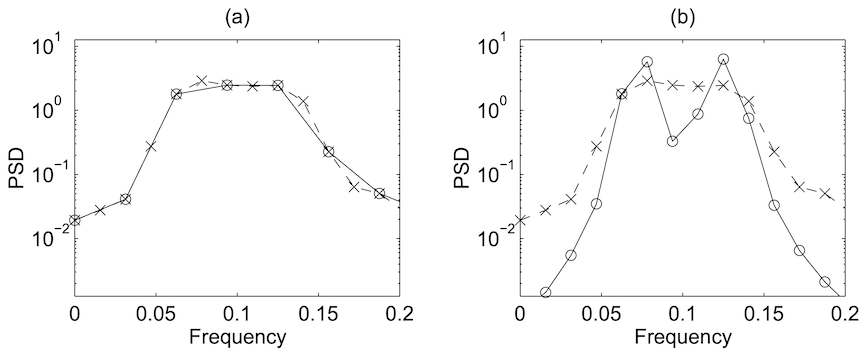}
\end{center}
\caption{Panels revealing the effect of padding an input time series on the resulting Fourier transform. For this example, two sinusoids are superimposed with normalized frequencies equal to 0.075 and 0.125 of the sampling frequency. Panels (a) and (b) show the resulting power spectral densities (PSDs) following the Fourier transforms of 32 input data points (solid black line with circular data points; left) and 64 input data points (solid black line with circular data points; right), respectively. In both panels, the dashed black lines with crosses represent the Fourier transforms of 32 input data points that have been padded to a total of 64 data points. It can be seen that the increased number of data points associated with the padded array results in more samples along the frequency axis, but this does not improve the frequency resolution to the level consistent with supplying 64 genuine input samples (solid black line in the right panel). Image reproduced from \citet{1998ISSIR...1....5E}.} \label{fig:Erikssonzeropaddingexample} 
\end{figure*}

It is also possible to artificially increase the duration of the input time series through the process known as `padding' \citep{2002AJ....124.1788R}, which has been employed across a wide range of solar studies incorporating the photosphere, chromosphere, and corona \citep[e.g.,][]{2002ApJ...566..505B, 2016ApJ...825..110A, 2021ApJ...909...66H, 2021RSPTA.37900174J}. Here, the beginning and/or end of the input data sequence is appended with a large number of data points with values equal to the mean of the overall time series. The padding adds no additional power to the data, but it acts to increase the fine-scale structure present in the corresponding Fourier transform since the overall duration of the data has been artificially increased. Note that padding with the data mean is preferable to padding with zeros since this alleviates the introduction of low-frequency power into the subsequent Fourier transform. Of course, if the input time series had previously been detrended (see Section~{\ref{sec:1Dfourieranalysis}}) so that the resulting mean of the data is zero, then zero-padding and padding with the time series mean are equivalent. 

\begin{figure*}[!t]
\begin{center}
\includegraphics[width=\textwidth]{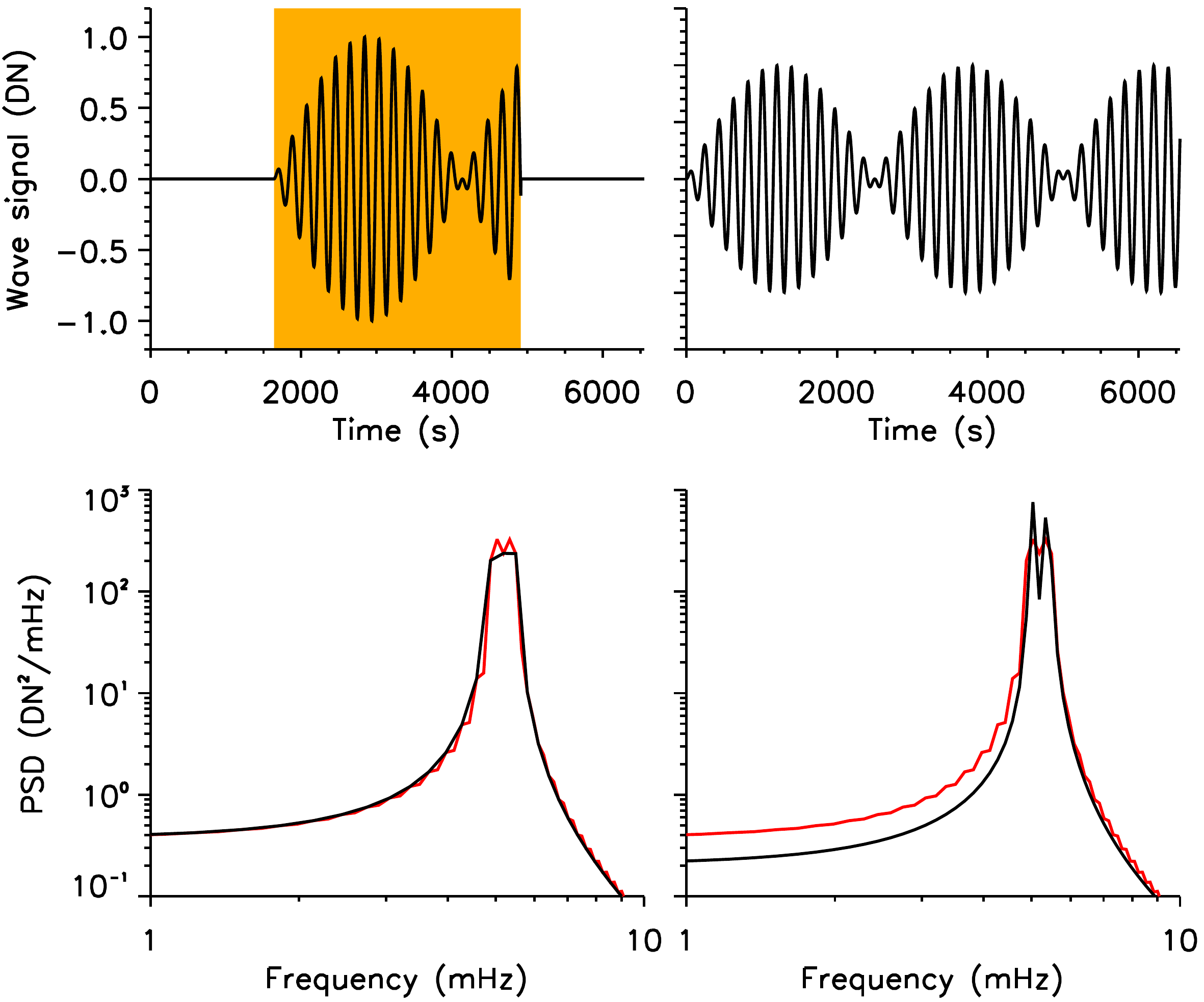}
\end{center}
\caption{{\it{Upper left:}} Inside the shaded orange region is a synthetic lightcurve created from the superposition of 5.0~mHz and 5.4~mHz waves, which are generated with a 3.2~s cadence (i.e., from VBI/DKIST) over a duration of $\approx3275$~s. This time series is zero-padded into a $\approx6550$~s array, which is displayed in its entirety in the upper-left panel using a solid black line. {\it{Upper right:}} The same resultant waveform created from the superposition of 5.0~mHz and 5.4~mHz waves, only now generated for the full $\approx6550$~s time series duration (i.e., no zero-padding required). {\it{Lower left:}} The power spectral density (PSD) of the original (un-padded) lightcurve is shown using a solid black line, while the solid red line reveals the PSD of the full zero-padded time series. It is clear that the padded array offers better visual segregation of the two embedded wave frequencies. {\it{Lower right:}} The PSDs for both the full $\approx6550$~s time series (solid black line) and the zero-padded original lightcurve (solid red line; same as that depicted in the lower-left panel). It can be seen that while the padded array provides some segregation of the 5.0~mHz and 5.4~mHz wave components, there is no better substitute at achieving high frequency resolution than obtaining long-duration observing sequences. Note that both PSD panels have the frequency axis truncated between $1-10$~mHz for better visual clarity.} 
\label{fig:datapaddingexample} 
\end{figure*}

Note that the process of padding is often perceived to increase the usable Fourier frequency resolution of the dataset, which is unfortunately incorrect. The use of padded time series acts to reveal small-scale structure in the output Fourier transform, but as it does not add any real signal to the input data sequence, the frequency resolution remains governed by the original time series characteristics \citep{1998ISSIR...1....5E}. As such, padding cannot recover and/or recreate any missing information in the original data sequence. This effect can be visualized in Figure~{\ref{fig:Erikssonzeropaddingexample}}. Here, a resultant wave consisting of two sinusoids with normalized frequencies 0.075 and 0.125 of the sampling frequency is cropped to 32 and 64 data points in length. Figure~{\ref{fig:Erikssonzeropaddingexample}}(a) shows the corresponding power spectral density (PSD) following Fourier transformation on both the raw 32 data samples array (solid black line with circular data points) and the original 32 data point array that has been padded to a total of 64 data points (dashed black line with crosses). In addition, Figure~{\ref{fig:Erikssonzeropaddingexample}}(b) shows another PSD for the data array containing 64 input samples (solid black line with circular data points), alongside the same PSD for the original 32 data point array that has been padded to a total of 64 data points (dashed black line with crosses; same as Figure~{\ref{fig:Erikssonzeropaddingexample}}a). From Figure~{\ref{fig:Erikssonzeropaddingexample}}(a) it can be seen that while the padding increases the number of data points along the frequency axis (and therefore creates some additional small-scale fluctuations in the resulting PSD), it does not increase the frequency resolution to a value needed to accurately identify the two sinusoidal components. This is even more apparent in Figure~{\ref{fig:Erikssonzeropaddingexample}}(b), where the Fourier transform of the time series containing 64 data points now contains sufficient information and frequency resolution to begin to segregate the two sinusoidal components. The padded array (32 data points plus 32 padded samples) contains the same number of elements along the frequency axis, but does not increase the frequency resolution to allow the quantification of the two embedded wave frequencies. The use of padding is often employed to decrease the computational time. Indeed, FFT algorithms work more efficiently if the number of samples is an integer power of 2. 

Of course, while data padding strictly does not add usable information into the original time series, it can be utilized to provide better visual segregation of closely spaced frequencies. To show an example of this application, Figure~{\ref{fig:datapaddingexample}} displays the effects of padding and time series duration in a similar format to Figure~{\ref{fig:Erikssonzeropaddingexample}}. In Figure~{\ref{fig:datapaddingexample}}, the upper-left panel shows an intensity time series that is created from the superposition of two closely spaced frequencies, here 5.0~mHz and 5.4~mHz. The resultant time series is $\approx3275$~s ($\sim55$~minutes) long, and constructed with a cadence of 3.2~s to remain consistent with the VBI/DKIST examples shown earlier in this section. The absolute extent of this 3275~s time series is bounded in the upper-left panel of Figure~{\ref{fig:datapaddingexample}} by the shaded orange background. In order to pad this lightcurve, a new time series is constructed that has twice as many data points in length, making the time series duration now $\approx6550$~s ($\sim110$~minutes). The original $\approx3275$~s lightcurve is placed in the middle of the new (expanded) array, thus providing zero-padding at the start and end of the time series. The corresponding power spectral densities (PSDs) for both the original and padded time series are shown in the lower-left panel of Figure~{\ref{fig:datapaddingexample}} using black and red lines, respectively. Note that the frequency axis is cropped to the range of $1-10$~mHz for better visual clarity. It is clear that the original input time series creates a broad spectral peak at $\approx5$~mHz, but the individual 5.0~mHz and 5.4~mHz components are not visible in the corresponding PSD (solid black line in the lower-left panel of Figure~{\ref{fig:datapaddingexample}}). On the other hand, the PSD from the padded array (solid red line in the lower-left panel of Figure~{\ref{fig:datapaddingexample}}) does show a double peak corresponding to the 5.0~mHz and 5.4~mHz wave components, highlighting how such padding techniques can help segregate multi-frequency wave signatures. 

Of course, padding cannot be considered a universal substitute for a longer duration data sequence. The upper-right panel of Figure~{\ref{fig:datapaddingexample}} shows the same input wave frequencies (5.0~mHz and 5.4~mHz), only with the resultant wave now present throughout the full $\sim110$~minute time sequence. Here, the beat pattern created by the superposition of two closely spaced frequencies can be readily seen, which is a physical manifestion of wave interactions also studied in high-resolution observations of the lower solar atmosphere \citep[e.g.,][]{2015ApJ...812L..15K}. The resulting PSD of the full-duration time series is depicted in the lower-right panel of Figure~{\ref{fig:datapaddingexample}} using a solid black line. For comparison, the PSD constructed from the padded original lightcurve is also overplotted using a solid red line (same as shown using a solid red line in the lower-left panel of Figure~{\ref{fig:datapaddingexample}}). It is clearly seen that the presence of the wave signal across the full time series provides the most prominent segregation of the 5.0~mHz and 5.4~mHz spectral peaks. While these peaks are also visible in the padded PSD (solid red line), they are less well defined, hence reiterating that while time series padding can help provide better isolation of closely spaced frequencies, there is no better candidate for high frequency resolution than long duration observing sequences. 

On the other hand, if rapidly fluctuating waveforms are wanting to be studied, then achieving a high Nyquist frequency is necessary to achieve these objectives, which the duration of the observing sequence is unable to assist with. Hence, it is important to tailor the observing strategy to ensure the frequency requirements are met. This, of course, can present challenges for particular facilities. For example, if a frequency resolution of $\Delta{f} \approx 35~\mu$Hz is required \citep[e.g., to probe the sub-second timescales of physical processes affecting frequency distributions in the aftermath of solar flares;][]{2019ApJ...886...32W}, this would require an observing duration of approximately 8 continuous hours, which may not be feasible from ground-based observatories that are impacted by variable weather and seeing conditions. Similarly, while space-borne satellites may be unaffected by weather and atmospheric seeing, these facilities may not possess a sufficiently large telescope aperture to probe the wave characteristics of small-scale magnetic elements \citep[e.g.,][]{2012ApJ...752...48C, 2017ApJ...850...64V, Keys2018} and naturally have reduced onboard storage and/or telemetry restrictions, thus creating difficulties obtaining 8 continuous hours of observations at maximum acquisition cadences. Hence, complementary data products, including ground-based observations at high cadence and overlapping space-borne data acquired over long time durations, are often a good compromise to help provide the frequency characteristics necessary to achieve the science goals. Of course, next-generation satellite facilities, including the recently commissioned Solar Orbiter \citep{2013SoPh..285...25M, 2020A&A...642A...1M} and the upcoming Solar-C \citep{2020SPIE11444E..0NS} missions, will provide breakthrough technological advancements to enable longer duration {\em{and}} higher cadence observations of the lower solar atmosphere than previously obtained from space. Another alternative to achieve both long time-series and high-cadence observations is the use of balloon-borne observatories, including the {\sc {Sunrise}} \citep{2010ApJ...723L.134B} and Flare Genesis \citep{1996SPIE.2804..141M, 2000SPD....31.0289B} experiments, where the data are stored in onboard discs. Such missions, however, have their own challenges and are limited to only a couple of days of observations during each flight.

\subsubsection{Calculating Confidence Levels}
\label{sec:calculatingconfidencelevels}
After displaying Fourier spectra, it is often difficult to pinpoint exactly what features are significant, and what power spikes may be the result of noise and/or spurious signals contained within the input time series. A robust method of determining the confidence level of individual power peaks is to compare the Fourier transform of the input time series with the Fourier transform of a large number (often exceeding 1000) of randomized lightcurves based on the original values \citep[i.e., ensuring an identical distribution of intensities throughout the new randomized time series;][]{2001A&A...368.1095O}. Following the randomization and computation of FFTs of the new time series, the probability, $p$, of randomized fluctuations being able to reproduce a given Fourier power peak in the original spectrum can be calculated. To do this, the Fourier power at each frequency element is compared to the power value calculated for the original time series, with the proportion of permutations giving a Fourier power value greater than, or equal to, the power associated with the original time series providing an estimate of the probability, $p$. Here, a small value of $p$ suggests that the original lightcurve contains real oscillatory phenomena, while a large value of $p$ indicates that there is little (or no) real periodicities contained within the data \citep{2001A&A...371.1137B, 2001A&A...368.1095O}. Indeed, it is worth bearing in mind that probability values of $p=0.5$ are consistent with noise fluctuations \citep[i.e., the variance of a binomial distribution is greatest at $p=0.5$;][]{Lyden2019}, hence why the identification of real oscillations requires small values of $p$.

Following the calculation of the probability, $p$, the value can be reversed to provide a percentage probability that the detected oscillatory phenomenon is real, through the relationship,
\begin{equation}
\label{eqn:preal}
    p_{\text{real}} = (1 - p) \times 100 \ .
\end{equation}
Here, $p_{\text{real}}=100\%$ would suggest that the wave motion present in the original time series is real, since no (i.e., $p=0$) randomized time series provided similar (or greater) Fourier power. Contrarily, $p_{\text{real}}=0\%$ would indicate a real (i.e., statistically significant) power deficit at that frequency, since all (i.e., $p=1$) randomized time series provided higher Fourier power at that specific frequency. Finally, a value of $p_{\text{real}} = 50\%$ would indicate that the power peak is not due to actual oscillatory motions.
A similar approach is to calculate the means and standard deviations of the Fourier power values for each independent frequency corresponding to the randomized time series. This provides a direct estimate of whether the original measured Fourier power is within some number of standard deviations of the mean randomized-data power density. As a result, probability estimations of the detected Fourier peaks can be estimated providing the variances and means of the randomized Fourier power values are independent \citep[i.e., follow a normal distribution;][]{10.1093/mnras/sty2731}.

\begin{figure*}[!t]
\begin{center}
\includegraphics[width=0.49\textwidth]{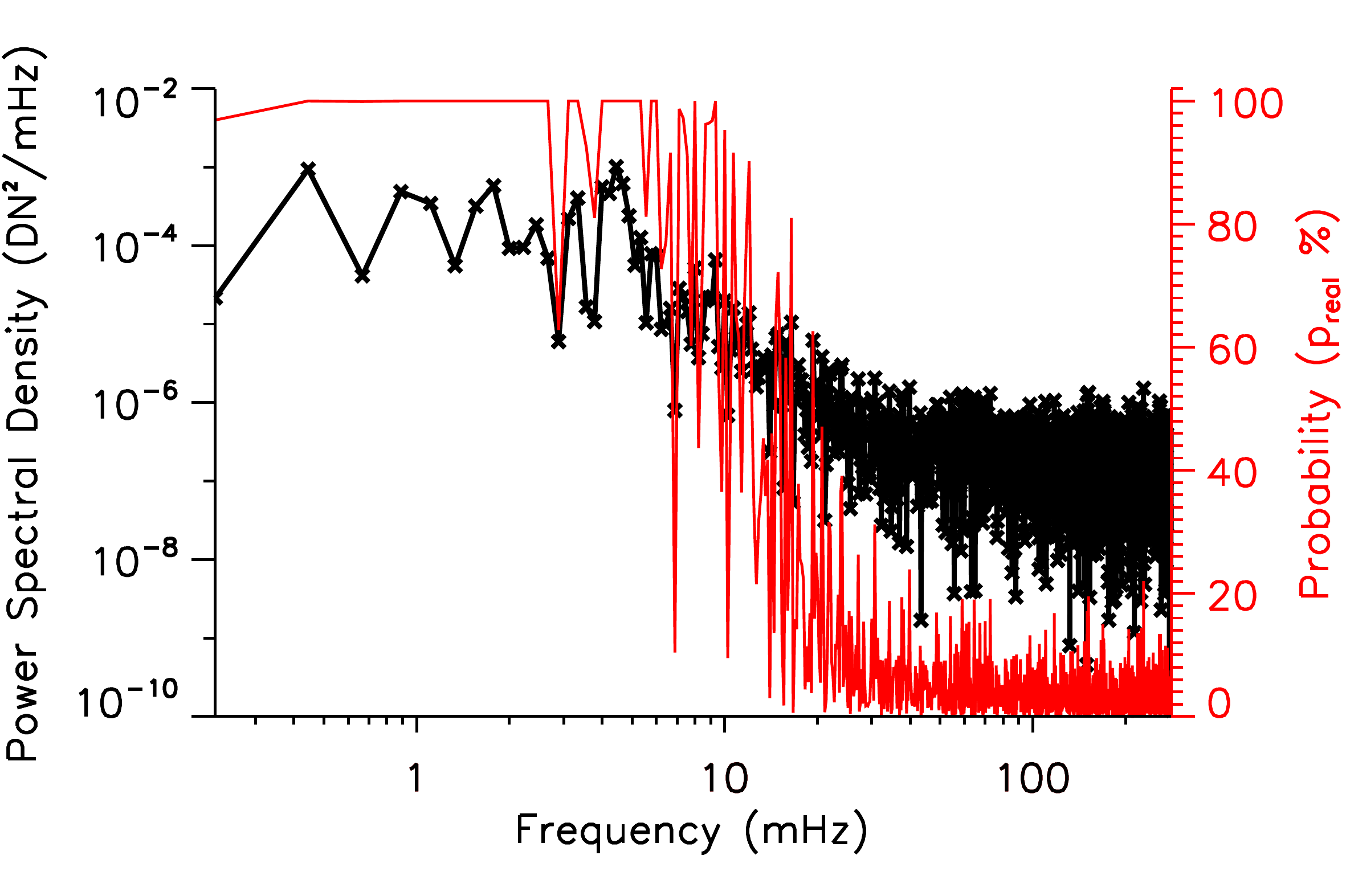}
\includegraphics[width=0.49\textwidth]{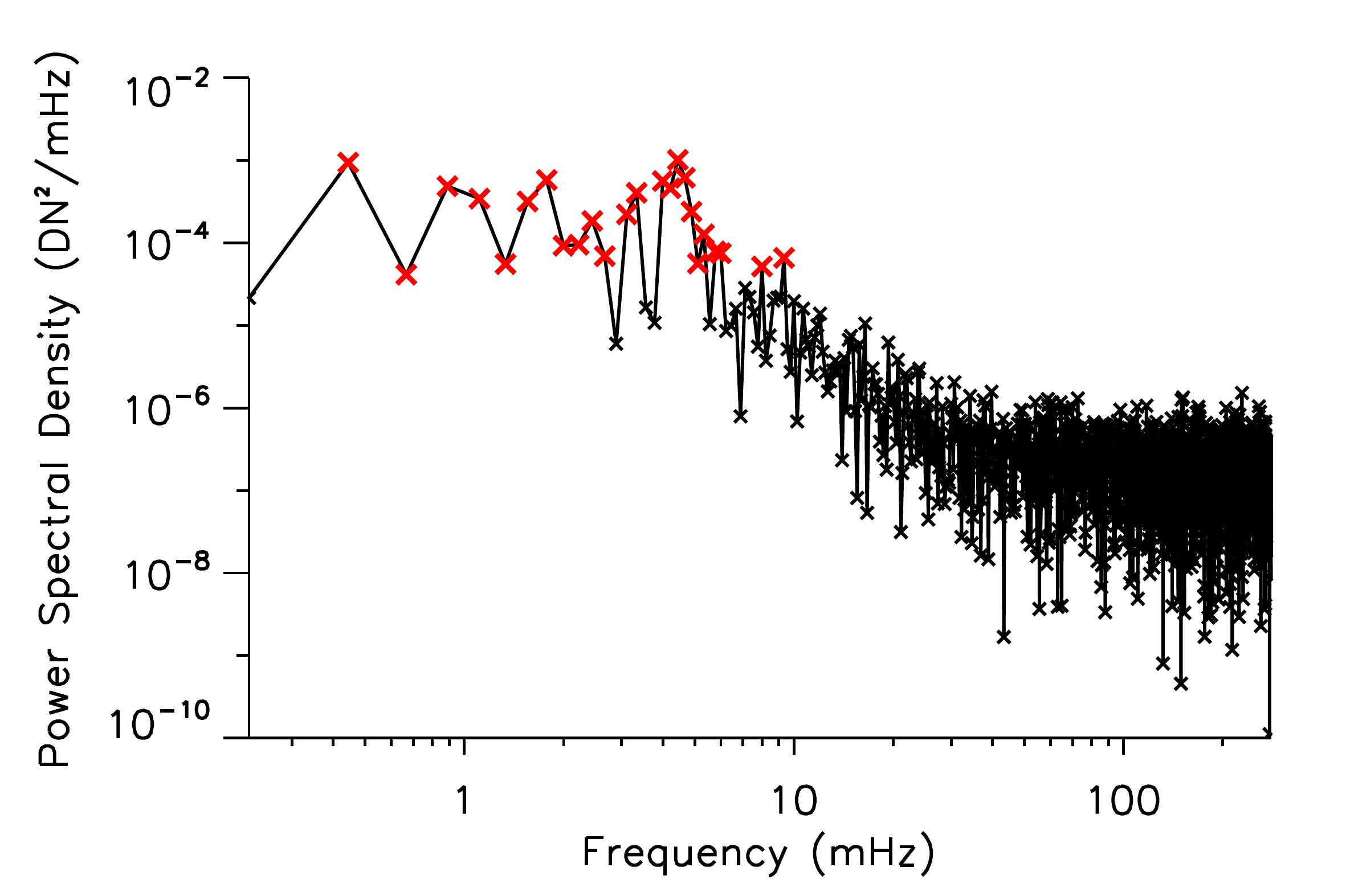}
\end{center}
\caption{The full frequency extent of the Fourier power spectral densities shown in the lower-middle panel of Figure~{\ref{fig:HARDcam_power_different_detrends}}, displayed using a log-log scale for better visual clarity (left panel). Overplotted using a solid red line are the percentage probabilities, $p_{\text{real}}$, computed over 1000 randomized permutations of the input lightcurve. Here, any frequencies with $p_{\text{real}} \ge 99\%$ correspond to a statistical confidence level in excess of 95\%. The same Fourier power spectral density is shown in the right panel, only now with red crossed symbols highlighting the locations where the Fourier power provides confidence levels greater than 95\%.} 
\label{fig:FFTprobabilities} 
\end{figure*}

If a large number ($\ge 1000$) of randomized permutations are employed, then the fluctuation probabilities will tend to Gaussian statistics \citep{1985AJ.....90.2317L, 2008SoPh..248..441D, 2019ApJ...871..133J}. In this case, the confidence level can be obtained using a standardized Gaussian distribution. For many solar applications \citep[e.g.,][to name but a few examples]{2002ApJ...567L.165M, 2003ApJ...587..806M, 2008SerAJ.177...87A, 2009A&A...508..941B, 2012A&A...539L...4S, 2014A&A...563A..12D, 2016ApJ...817...44F, 2017ApJS..229...10J}, a confidence level of 95\% is typically employed as a threshold for reliable wave detection. In this case, $99\% \le p_{\text{real}} \le 100\%$ (or $0.00 \le p \le 0.01$) is required to satisfy the desired 95\% confidence level. 

To demonstrate a worked example, we utilize the HARDcam H$\alpha$ time series shown in the left panel of Figure~{\ref{fig:HARDcam_lightcurves}}, which consists of 2528~individual time steps. This, combined with 1000 randomized permutations of the lightcurve, provides 1000 FFTs with 1000 different measures in each frequency bin; more than sufficient to allow the accurate use of Gaussian number statistics \citep{Montgomery.Runger2003}. For each randomization, the resulting Fourier spectrum is compared to that depicted in the upper panel of Figure~{\ref{fig:HARDcam_power}}, with the resulting percentage probabilities, $p_{\text{real}}$, calculated according to Equation~{\ref{eqn:preal}} for each of the temporal frequencies. The original Fourier power spectrum, along with the percentage probabilities for each corresponding frequency, are shown in the left panel of Figure~{\ref{fig:FFTprobabilities}}. It can be seen that the largest power signal at $\approx4.7$~mHz ($\approx 210$~s) has a high probability, suggesting that this is a detection of a real oscillation. Furthermore, the neighboring frequencies also have probabilities above 99\%, further strengthening the interpretation that wave motion is present in the input time series. It should be noted that with potentially thousands of frequency bins in the high-frequency regime of an FFT, having some fraction of points that exceed a 95\% (or even 99\%) confidence interval is to be expected. Therefore, many investigations also demand some degree of coherency in the frequency and/or spatial distributions to better verify the presence of a real wave signal \citep[similar to the methods described by][]{1982A&A...111..272D, https://doi.org/10.1002/2017JA024922}.
To better highlight which frequencies demonstrate confidence levels exceeding 95\%, the right panel of Figure~{\ref{fig:FFTprobabilities}} overplots (using bold red crosses) those frequencies containing percentage probabilities in excess of 99\%. 

\subsubsection{Lomb-Scargle Techniques}
\label{sec:lombscargletechniques}
A requirement for the implementation of traditional Fourier-based analyses is that the input time series is regularly and evenly sampled. This means that each data point of the lightcurve used should be obtained using the same exposure time, and subsequent time steps should be acquired with a strict, uniform cadence. Many ground-based and space-borne instruments employ digital synchronization triggers for their camera systems that can bring timing uncertainties down to the order of $10^{-6}$~s \citep{2010SoPh..261..363J}, which is often necessary in high-precision polarimetric studies \citep{10.1117/12.2309508}. This helps to ensure the output measurements are sufficiently sampled for the application of Fourier techniques. 

However, often it is not possible to obtain time series with strict and even temporal sampling. For example, raster scans using slit-based spectrographs can lead to irregularly sampled observations due to the physical times required to move the spectral slit\footnote{Note: while individual slit positions on a raster scan can be irregularly sampled in the time domain, often the same slit position has a regular temporal cadence between successive complete rasters.}. Also, some observing strategies interrupt regularly sampled data series for the measurement of Stokes~$I/Q/U/V$ signals every few minutes, hence introducing data gaps during these times \citep[e.g.,][]{2016ApJ...828...23S}. Furthermore, hardware requiring multiple clocks to control components of the same instrument \citep[e.g., the mission data processor and the polarization modulator unit on board the Hinode spacecraft;][]{2007SoPh..243....3K} may have a tendency to drift away from one another, hence effecting the regularity of long-duration data sequences \citep{2007PASJ...59S.637S}. In addition, some facilities including the Atacama Large Millimeter/submillimeter Array \citep[ALMA;][]{2009IEEEP..97.1463W, 2016SSRv..200....1W} require routine calibrations that must be performed approximately every 10~minutes \citep[with each calibration taking $\sim2.5$~minutes;][]{2020A&A...635A..71W}, hence introducing gaps in the final time series \citep{2021RSPTA.37900174J}. Finally, in the case of ground-based observations, a period of reduced seeing quality or the passing of a localized cloud will result in a number of compromised science frames, which require removal and subsequent interpolation \citep{2016ApJ...823...45K}. 

If the effect of data sampling irregularities is not believed to be significant (i.e., is a fraction of the wave periodicities expected), then it is common to interpolate the observations back on to a constant cadence grid \citep[e.g.,][]{2012ApJ...746..183J, 2016A&A...585A.110K}. Of course, how the data points are interpolated (e.g., linear or cubic fitting) may effect the final product, and as a result, care should be taken when interpolating time series so that artificial periodicities are not introduced to the data through inappropriate interpolation. This is particularly important when the data sequence requires subsequent processing, e.g., taking the derivative of a velocity time series to determine the acceleration characteristics of the plasma. Under these circumstances, inappropriate interpolation of the velocity information may have drastic implications for the derived acceleration data. For this form of analysis, the use of 3-point Lagrangian interpolation is often recommended to ensure the endpoints of the time series remain unaffected due to the use of error propagation formulae \citep{2008ApJ...681L.113V}. However, in the case for very low cadence data, 3-point Lagrangian interpolation may become untrustworthy due to the large temporal separation between successive time steps \citep{2013A&A...557A..96B}. For these cases, a Savtizky-Golay \citep{doi:10.1021/ac60214a047} smoothing filter can help alleviate sharp (and often misleading) kinematic values \citep{2015JSWSC...5A..19B}. 

If interpolation of missing data points and subsequent Fourier analyses is not believed to be suitable, then Lomb-Scargle techniques \citep{1976Ap&SS..39..447L, 1982ApJ...263..835S} can be implemented. As overviewed by \citet{2009A&A...496..577Z}, the Lomb-Scargle algorithms are useful for characterizing periodicities present in unevenly sampled data products. Often, least-squares minimization processes assume that the data to be fitted are normally distributed \citep{2012ApJ...746..131B}, which may be untrue since the spectrum of a linear, stationary stochastic process naturally follows a $\chi_{2}^{2}$ distribution \citep{1975ApJS...29..285G, 1993MNRAS.261..612P}. However, a benefit of implementing the Lomb-Scargle algorithms is that the noise at each individual frequency can be represented by a $\chi^{2}$ distribution, which is equivalent to a spectrum being reliably derived from more simplistic least-squares analysis techniques \citep{2018ApJS..236...16V}. 

Crucially, Lomb-Scargle techniques differ from conventional Fourier analyses by the way in which the corresponding spectra are computed. While Fourier-based algorithms compute the power spectrum by taking dot products of the input time series with pairs of sine- and cosine-based waveforms, Lomb-Scargle techniques attempt to first calculate a delay timescale so that the sinusoidal pairs are mutually orthogonal at discrete sample steps, hence providing better power estimates at each frequency without the strict requirement of evenly sampled data \citep{10.5555/1403886}. In the field of solar physics, Lomb-Scargle techniques tend to be more commonplace in investigations of long-duration periodicities spanning days to months \citep[i.e., often coupled to the solar cycle;][]{NI2012282, 2017JSWSC...7A..34D}, although they can be used effectively in shorter duration observations where interpolation is deemed inappropriate \citep[e.g.,][]{2013SoPh..288...73M}.

\subsubsection{One-dimensional Fourier Filtering}
\label{sec:onedimensionalFourierfiltering}
Often, it is helpful to filter the time series in order to isolate specific wave signatures across a particular range of frequencies. This is useful for a variety of studies, including the identification of beat frequencies \citep{2015ApJ...812L..15K}, the more reliable measurement of phase variations between different wavelengths/filters \citep{2017ApJ...847....5K}, and in the identification of various wave modes co-existing within single magnetic structures \citep{2018ApJ...857...28K}. From examination of the upper panel of Figure~{\ref{fig:HARDcam_power}} and Figure~{\ref{fig:FFTprobabilities}}, it is clear that the frequency associated with peak Fourier power is $\approx 4.7$~mHz, and is accompanied by high confidence levels exceeding 95\%. 

\begin{figure*}[!t]
\begin{center}
\includegraphics[width=0.9\textwidth]{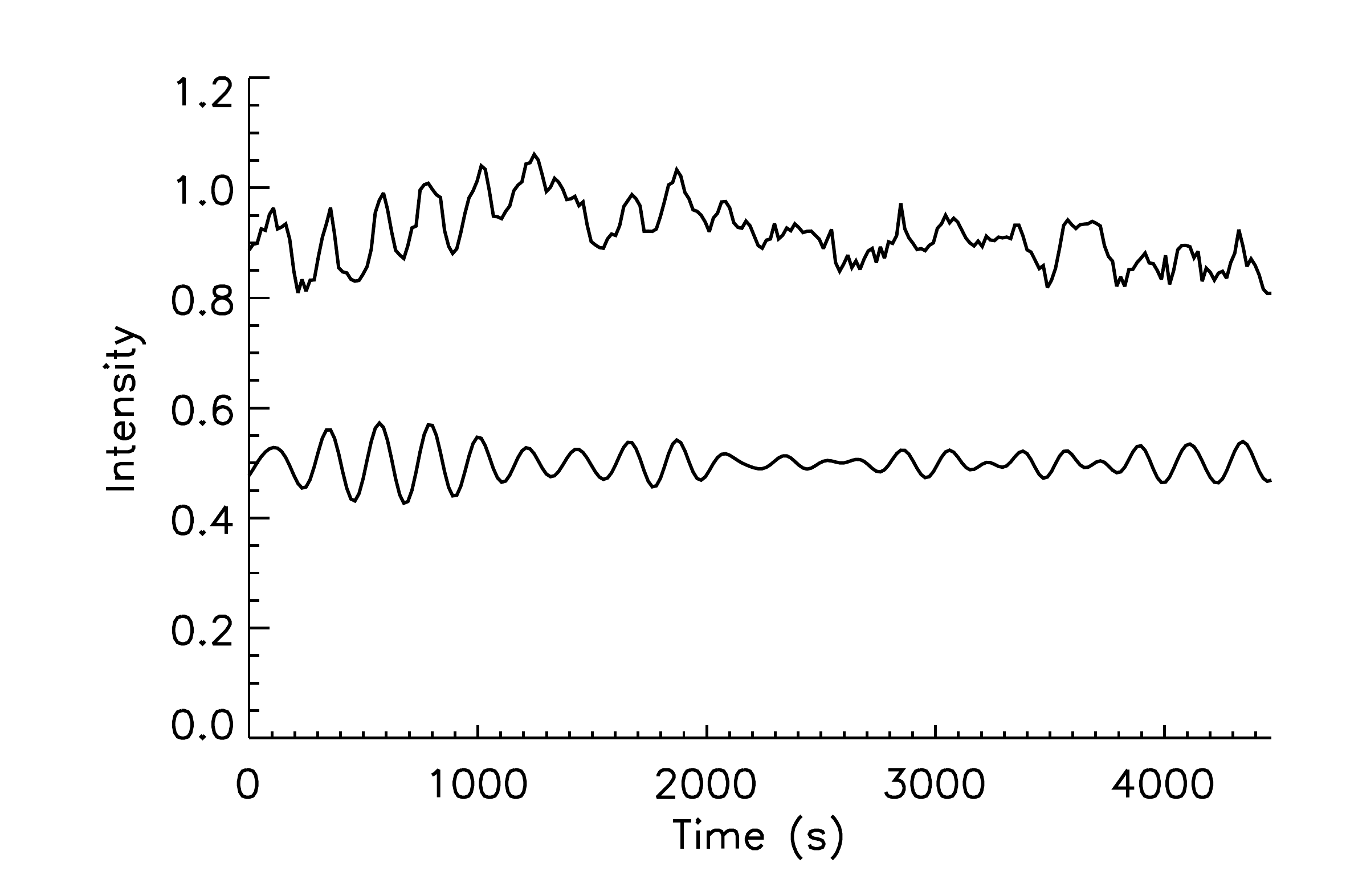}
\end{center}
\caption{The original HARDcam time series (upper solid black line), normalized by the quiescent H$\alpha$ continuum intensity, and displayed as a function of time. The lower solid black line is a Fourier filtered lightcurve, which has been detrended using a third-order polynomial (right panel of Figure~{\ref{fig:HARDcam_lightcurves}}), convolved with a Gaussian frequency filter centered on 4.7~mHz with a FWHM of 2.0~mHz (lower-right panel of Figure~{\ref{fig:HARDcam_power}}), before applying an inverse FFT to reconstruct the filtered time series. For visual clarity, the filtered lightcurve has been offset to bring it closer to the original time series intensities.}
\label{fig:HARDcam_reconstruction} 
\end{figure*}

If we wish to reconstruct a filtered time series centered on this dominant frequency, then we have a number of options available. 
The dashed red line in the middle-left panel of Figure~{\ref{fig:HARDcam_power}} depicts a step function frequency range of $4.7\pm1.0$~mHz, whereby the filter is assigned values of `1' and `0' for frequencies inside and outside, respectively, this chosen frequency range. Multiplying the Fourier power spectrum by this step function frequency filter results in the preserved power elements that are shown in the middle-right panel of Figure~{\ref{fig:HARDcam_power}}, which can be passed through an inverse FFT to create a Fourier filtered time series in the range of $4.7\pm1.0$~mHz. However, by employing a step function frequency filter, there is a sharp and distinct transition between elevated power signals and frequencies with zero Fourier power. This abrupt transition can create aliasing artifacts in the reconstructed time series \citep{doi:10.1029/2005WR004374}. Alternatively, to help mitigate against aliasing (i.e., sharp Fourier power transitions at the boundaries of the chosen frequency range), the Fourier power spectrum can be multiplied by a filter that peaks at the desired frequency, before gradually reducing in transmission towards the edges of the frequency range. 
An example of such a smoothly varying filter is documented in the lower panels of Figure~{\ref{fig:HARDcam_power}}, where a Gaussian centered at 4.7~mHz, with a full-width at half-maximum (FWHM) of 2~mHz, is overplotted on top of the Fourier spectrum using a solid red line, which can be multiplied by the original Fourier spectrum to gradually decrease the power down to zero at the edges of the desired frequency range (lower-right panel of Figure~{\ref{fig:HARDcam_power}}).
Performing an inverse FFT on this filtered Fourier power spectrum results in the reconstruction of an H$\alpha$ lightcurve containing dominant periodicities of $\approx 210$~s, which can be seen in Figure~{\ref{fig:HARDcam_reconstruction}}. This process is identical to convolving the detrended intensity time series with the given Gaussian frequency filter, but we perform this process step-by-step here for the purposes of clarity.

It must be noted that here we employ a Gaussian frequency filter to smoothly transition the Fourier power to values of zero outside of the desired frequency range. However, other filter shapes can also be chosen, including Lorentzian, Voigt, or even custom profile shapes depending upon the level of smoothing required by the investigators. At present, there is no firm consensus regarding which filter profile shape is best to use, so it may be necessary to choose the frequency filter based upon the specifics of the data being investigated, e.g., the frequency resolution, the amplitude of the spectral components wishing to be studied, the width of the documented Fourier peaks, etc. Of course, we must remind the reader that isolating a relatively limited range of frequencies in Fourier space and transforming these back into real (temporal) space will always result in the appearance of a periodic signal at the frequency of interest, even if the derived Fourier transform was originally noise dominated. Therefore, it is necessary to combine confidence interval analysis (see Section~{\ref{sec:calculatingconfidencelevels}}) with such Fourier filtering techniques to ensure that only statistically significant power is being considered in subsequent analyses.

\subsubsection{Fourier Phase Lag Analysis}
\label{sec:fourierphaselaganalysis}
Many observational datasets will be comprised of a combination of multi-wavelength and/or multi-component spectral measurements. For example, the Rapid Oscillations in the Solar Atmosphere \citep[ROSA;][]{2010SoPh..261..363J} instrument at the DST is able to observe simultaneously in six separate bandpasses. It is common practice to acquire contemporaneous imaging observations through a combination of G-band, 3500{\,}{\AA} and 4170{\,}{\AA} broadband continuum filters, in addition to Ca~{\sc{ii}}~K, Na~{\sc{i}}~D$_{1}$, and H$\alpha$ narrowband filters, which allows wave signatures to be studied from the depths of the photosphere through to the base of the transition region \citep[e.g.,][]{2011ApJ...729L..18M, 2012NatCo...3.1315M, 2012ApJ...757..160J, 2012ApJ...744L...5J, 2012ApJ...746..183J, 2012ApJ...750...51K, 2015ApJ...806..132G, 2015ApJ...812L..15K, 2016ApJ...823...45K, 2017ApJ...847....5K, 2018ApJ...857...28K}. On the other hand, Fabry-P{\'{e}}rot spectral imaging systems such as the Crisp Imaging Spectropolarimeter \citep[CRISP;][]{2008ApJ...689L..69S} and the Interferometric Bi-dimensional Spectrometer \citep[IBIS;][]{2006SoPh..236..415C}, are able to capture two-dimensional spatial information (often including spectropolarimetric Stokes~$I/Q/U/V$ measurements) across a single or multiple spectral lines. This allows a temporal comparison to be made between various spectral parameters of the same absorption line, such as the full-width at half-maximum (FWHM), intensity, Doppler velocity, and magnitudes of circular/linear polarization (providing spectropolarimetric measurements are made). As a result, harnessing multi-wavelength and/or multi-component observations provides the ability to further probe the coupling of wave activity in the lower solar atmosphere. 

\begin{figure*}[!t]
\begin{center}
\includegraphics[width=0.8\textwidth]{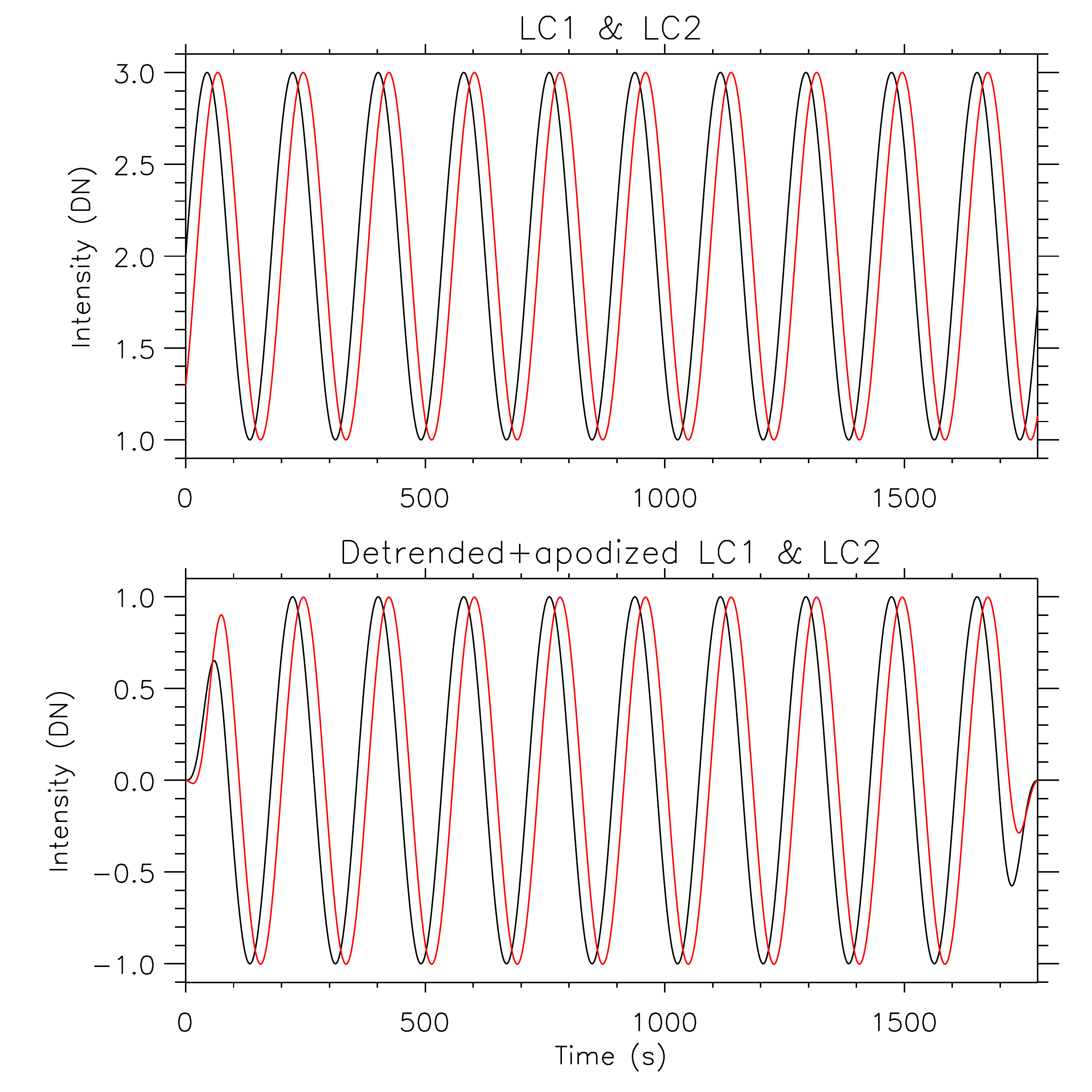}
\end{center}
\caption{Synthetic time series (upper panel), each with a cadence of 1.78~s, displaying a frequency of 5.6~mHz ($\approx$180~s periodicity) and a mean intensity equal to 2. The red lightcurve is delayed by 45~degrees (0.785~radians) with respect to the black lightcurve. The lower panel displays the detrended and apodized time series, which are now suitable for subsequent FFT analyses. }
\label{fig:samplelightcurves} 
\end{figure*}

The upper panel of Figure~{\ref{fig:samplelightcurves}} displays two synthetic intensity time series generated with a cadence of 1.78~s (consistent with the HARDcam H$\alpha$ data products overviewed in Section~{\ref{sec:HARDcam}}), each with a frequency of 5.6~mHz ($\approx$180~s periodicity) and a mean intensity equal to 2. However, the red lightcurve (LC2) is delayed by 45 degrees, and hence lags behind the black lightcurve (LC1) by 0.785~radians. As part of the standard procedures prior to the implementation of Fourier analysis (see, e.g., Section~{\ref{sec:1Dfourieranalysis}}), each of the time series are detrended (in this case by subtracting a linear line of best fit) and apodized using a 90\% tapered cosine apodization filter. The final intensity time series are shown in the lower panel of Figure~{\ref{fig:samplelightcurves}}, and are now suitable for subsequent Fourier analyses. 

\begin{figure*}[!t]
\begin{center}
\includegraphics[width=0.8\textwidth]{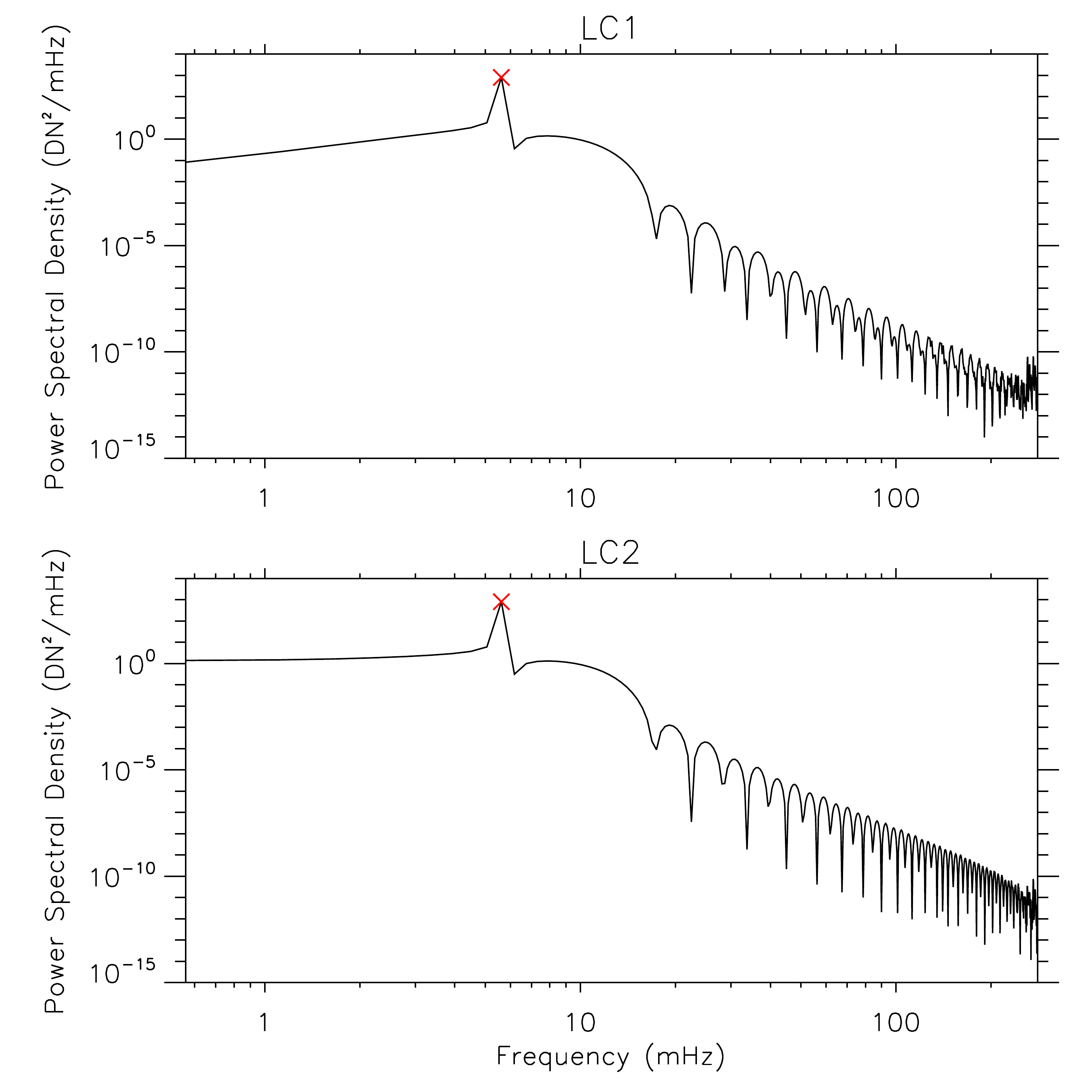}
\end{center}
\caption{FFT power spectral densities for LC1 (upper panel) and LC2 (lower panel), corresponding to the solid black and red lines in the lower panel of Figure~{\ref{fig:samplelightcurves}}, respectively. The red crosses highlight frequencies where the calculated Fourier power is above the 95\% confidence level. It can be seen that the synthetic 5.6~mHz input signal is accurately identified in both corresponding power spectra, with its associated Fourier power being in excess of the 95\% confidence threshold. The oscillatory behavior at high frequencies is due to the selected apodization filter.}
\label{fig:sampleFFTs} 
\end{figure*}

\begin{figure*}[!t]
\begin{center}
\includegraphics[width=0.8\textwidth]{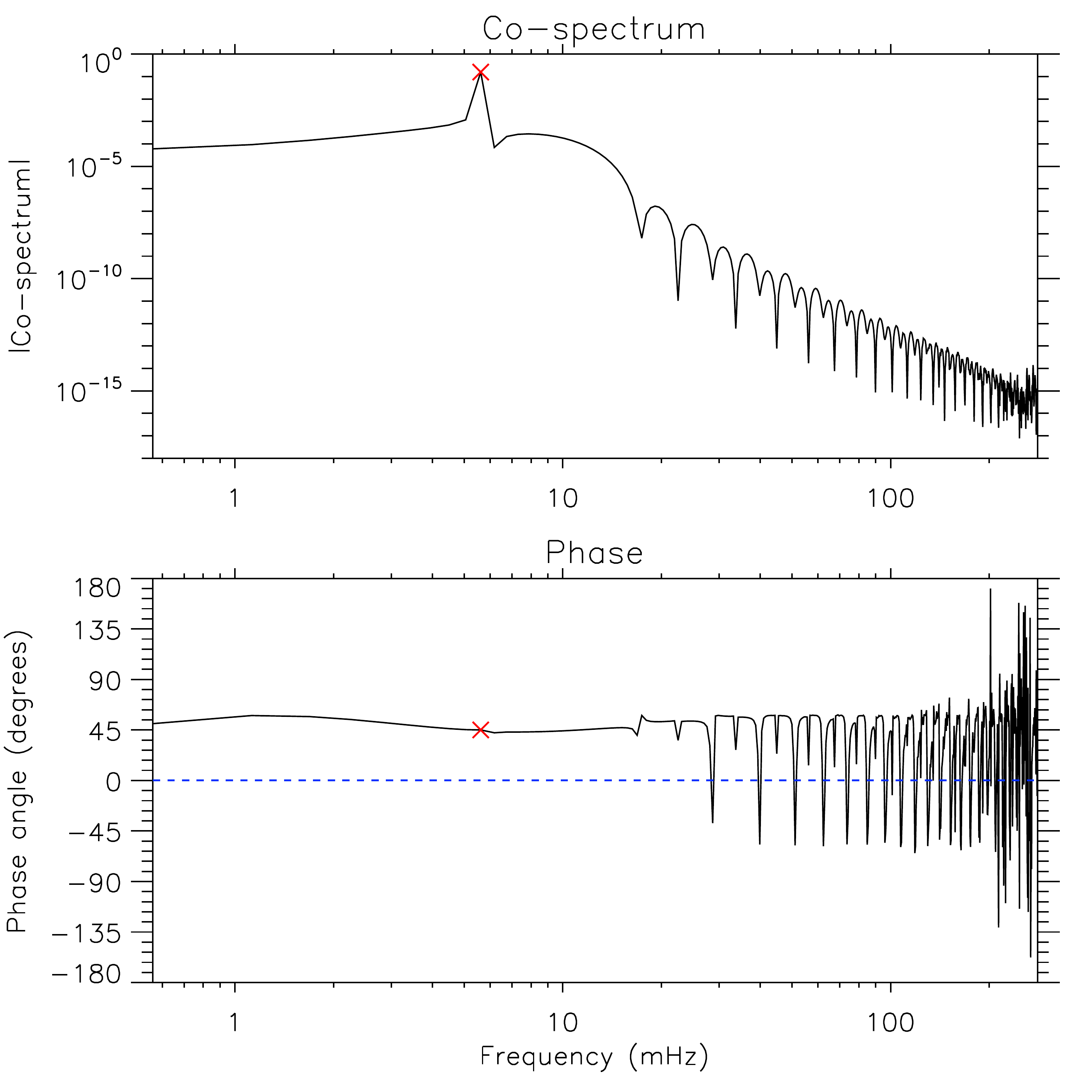}
\end{center}
\caption{Co-spectrum (upper panel; real part of the cross-power spectrum) of the input time series LC1 and LC2 shown in the lower panel of Figure~{\ref{fig:samplelightcurves}}. The lower panel displays the phase angle between the input time series LC1 and LC2, which corresponds to the phase of the complex cross-spectrum. Here, a positive phase angle indicates that LC1 leads LC2 (i.e., LC2 lags behind LC1), which can be seen visually through examination of the individual lightcurves depicted in Figure~{\ref{fig:samplelightcurves}}. The red crosses indicate the frequency where the calculated Fourier power for LC1 and LC2 both exceed the 95\% confidence levels (see Figure~{\ref{fig:sampleFFTs}}). The horizontal dashed blue line in the lower panel highlights a phase angle of 0~degrees.}
\label{fig:samplecpsphase} 
\end{figure*}

Following the approaches documented in Section~{\ref{sec:calculatingconfidencelevels}}, FFTs of the detrended and apodized time series are taken, with 95\% confidence levels calculated. The resulting FFT power spectral densities are shown in Figure~{\ref{fig:sampleFFTs}}, where the red crosses indicate frequencies where the associated power is in excess of the calculated 95\% confidence levels for each respective time series. It can be seen in both the upper and lower panels of Figure~{\ref{fig:sampleFFTs}} that the input 5.6~mHz signal is above the 95\% confidence threshold for both LC1 and LC2. Next, the cross-power spectrum, $\Gamma_{12}(\nu)$, between the FFTs of LC1 and LC2 is calculated following the methods described by \citet{Piersol_2000} as;
\begin{equation}
    \Gamma_{12}(\nu) = F(LC1) * \overline{F(LC2)} \ ,
\end{equation}
with $F$ denoting an FFT and $\overline{F}$ the complex conjugate of the FFT. The cross-power spectrum is a complex array (just like the FFTs from which it is computed), and therefore has components representative of its co-spectrum ($d(\nu)$; real part of the cross-power spectrum) and quadrature spectrum ($c(\nu)$; imaginary part of the cross-power spectrum). The co-spectrum from the input time series LC1 and LC2 is shown in the upper panel of Figure~{\ref{fig:samplecpsphase}}. The red cross signifies the frequency where the Fourier power exceeded the 95\% confidence level in both FFTs, namely 5.6~mHz, which is consistent with the synthetic lightcurves shown in Figure~{\ref{fig:samplelightcurves}}.

Finally, the co-spectrum and quadrature spectrum can be utilized to calculate the phase lag between the input lightcurves LC1 and LC2 as a function of frequency, defined by \citet{2011ApJ...734...47P} as,
\begin{equation}
\label{eqn:phase_angle}
    \phi(\nu) = \arctan\left(\frac{\langle c(\nu) \rangle}{\langle d(\nu) \rangle}\right) \ .
\end{equation}
Here, the phase angle, commonly chosen to span the interval $-180^{\circ} \rightarrow +180^{\circ}$, is simply the phase of the complex cross-spectrum \citep[see the nomenclature of][]{1997ApJ...474L..43V}. The lower panel of Figure~{\ref{fig:samplecpsphase}} displays the calculated phase angles, again with the red cross highlighting the phase value at the frequency where the Fourier power exceeds the 95\% confidence level in both FFTs corresponding to LC1 and LC2. In this example, the phase angle corresponding to a frequency of $\approx$5.6~mHz is equal to $45^{\circ}$, which is consistent with the input lightcurves depicted in Figure~{\ref{fig:samplelightcurves}}. Here, a positive phase angle indicates that LC1 leads LC2 (i.e., LC2 lags behind LC1), which can be visually confirmed in Figure~{\ref{fig:samplelightcurves}} with LC1 (solid black line) leading LC2 (solid red line). 

It must be noted that phase angles can be computed for all possible frequencies (see, e.g., the lower panel of Figure~{\ref{fig:samplecpsphase}}). However, it is important to determine which of these phase values are reliable before they are used in subsequent scientific interpretations. For the purposes of the example shown here, we selected that frequency at which both times series LC1 and LC2 demonstrated Fourier power exceeding the 95\% confidence levels in both of their corresponding FFTs. However, a common alternative is to calculate the coherence level for each constituent frequency, which can then be employed (independently of the confidence levels) to pinpoint reliable frequencies in the corresponding cross-power spectrum. The coherence level is estimated from the normalized square of the amplitude of the complex cross-spectrum \citep[see, e.g.,][]{storch_zwiers_1999}, providing a measure, ranging between `0' and `1', of the linear correlation between the two input time series. Under this regime, values of `0' and `1' indicate no and perfect correlation, respectively. For the purposes of solar physics research, it is common to adopt a coherence value $>0.80$ to signify robust and reliable phase measurements \citep{2003ApJ...587..806M, 2004ApJ...604..936B, 2004ApJ...617..623B, 2013A&A...554A.115S, 2018ApJ...869..110S, 2016A&A...585A.110K}. 

Therefore, the cross-power spectrum and coherence are both used to examine the relationship between two time series as a function of frequency. The cross spectrum identifies common large power (i.e., significant peaks) at the same frequencies in the power spectra of the two time series, and whether such frequencies are related to each other (the relationship is quantified by phase differences). Such correlations cannot, however, be revealed if one or both time series do not have significant power enhancements at particular frequencies, e.g., if the power spectra at those frequencies are indistinguishable from red noise. Nonetheless, there still may be coherent modes at such frequencies, that can be identified in the coherence spectrum, i.e., two time series can have a large coherence at a frequency even though both or one of the power spectra do not show large power at that frequency. Thus, the coherence is a measure of the degree of linear correlation between the two time series at each frequency. In solar physics, the coherence is particularly useful when the two signals are associated to, e.g., different solar atmospheric heights (with, e.g., different amplitudes) and/or two different physical parameters. An example, from real observations, where oscillatory power (at specific time-frequency locations) appears only in one of the signals is demonstrated in Figure~\ref{fig:SUNRISE_wavelet_phaselag}. Hence, no significant power is detected in the cross-power spectrum, whereas a large coherence level, exceeding 0.8, is identified. The significance of phase measurements for reliable coherence values can be evaluated by either introducing a coherence floor level (e.g., the 0.8 threshold mentioned above) or estimating confidence levels. To approximate a floor level, \citet{2004ApJ...617..623B} randomized both time series for a very large number of realizations and calculated the coherence for each, from which, the threshold was estimated as an average over all realizations plus some multiples of the standard deviation of the coherence values. For the confidence levels, the coherence values should be tested against the null hypothesis of zero population coherence, i.e., whether the coherence exceeds expected values from arbitrary colored (e.g., white or red) noise backgrounds. While various methods have been employed for this statistical test, one common approach is to estimate the confidence levels by means of Monte Carlo simulations \citep{1998BAMS...79...61T,2004NPGeo..11..561G,2016CG.....91...11B}.

With reliable phase angles calculated, it then becomes possible to estimate a variety of key wave characteristics. 
If $T$ is the period of the wave, then the phase lag, $\phi$ (in degrees), can be converted into a physical time delay through the relationship,
\begin{equation}
\label{eqn:time_delay}
    \text{time~delay~(s)}~=~ \frac{\phi}{360} \times T \ .
\end{equation}
The time delay value (arising from the measured phase lag) corresponds to a wave propagating between the different atmospheric layers. Of course, phase angles deduced from the co-spectrum and quadrature spectrum (see Equation~{\ref{eqn:phase_angle}}) inherently have phase wrapping at $\pm180^{\circ}$, hence introducing a $360^{\circ}$ ambiguity associated with the precise phase angle \citep[as discussed in][]{2006ApJ...640.1153C, 2008A&A...479..213B, 10.1111/j.1365-2966.2009.14708.x}. Hence, the true time delay may need to include multiples of the period to account for the $360^{\circ}$ ambiguity, hence transforming Equation~{\ref{eqn:time_delay}} into,
\begin{equation}
\label{eqn:time_delay2}
    \text{time~delay~(s)}~=~ \frac{\phi}{360} \times nT \ ,
\end{equation}
where $n$ is a non-zero integer. Many studies to date have examined the propagation of relatively long-period oscillations (e.g., $100-300$~s), which permit the assumption of $n=1$ without violating theoretical considerations \citep[e.g., sound speed restrictions;][]{2012ApJ...746..183J}, hence allowing direct use of Equation~{\ref{eqn:time_delay}}. However, as future studies examine higher-frequency (lower-period) wave propagation, then more careful consideration of Fourier phase wrapping will need to be taken into consideration to ensure the derived time delay is consistent with the observations. As part of a phase `unwrapping' process, the identification of quasi-periodic waves and/or those with modulated amplitudes will allow phase ambiguities to be practically alleviated. For example, by tracking the commencement of a wave, and hence the time delay as it propagates between closely-spaced atmospheric layers, the phase angle can be computed without the $\pm$360$^{\circ}$ phase wrapping uncertainty. Alternatively, a modulated waveform will provide secondary peaks associated with the propagating group, which supplies additional information to better establish the precise value of $n$ in Equation~{\ref{eqn:time_delay2}}, hence assisting with the phase unwrapping of the data, which will enable much more precise tracking of wave energy flux through the solar atmosphere.

Finally, if the geometric height separation, $d$ (in km), between the two layers is known or can be estimated \citep{2020A&A...634A..19G}, then the average phase velocity, $v_{\text{ph}}$, of the wave propagating between these two distinct layers can be deduced via,
\begin{equation}
\label{eqn:phase_speed}
    v_{\text{ph}}~\text{(km/s)}~=~\frac{360d}{T\phi} \ .
\end{equation}
Similar estimations of the phase velocities of embedded waves have been made by \citet{1977SoPh...52..283M}, \citet{1979ApJ...229.1147A}, \citet{1979ApJS...39..347W}, \citet{2006ApJ...640.1153C}, \citet{2010A&A...522A..31B}, \citet{2012ApJ...746..183J}, \citet{2015ApJ...806..132G}, \citet{2017ApJS..229....9J}, to name but a few examples. Importantly, Equation~{\ref{eqn:phase_speed}} can also be rearranged to estimate the atmospheric height separation between two sets of observations. For example, the acoustic sound speed is approximately constant in the lower photosphere, hence this value, alongside the derived time lag, can be utilized to provide an estimate of the height separation, $d$ \citep[e.g.,][]{1989A&A...213..423D}.

\subsection{Three-dimensional Fourier Analysis}
\label{sec:3Dfourieranalysis}
Telescope facilities deployed in a space-borne environment, which benefit from a lack of day/night cycles and atmospheric aberrations, have long been able to harness three-dimensional Fourier analyses to examine the temporal ($t \leftrightarrow \omega$) {\it{and}} spatial ($[x,y] \leftrightarrow [k_{x},k_{y}]$) domains. Here, $t$ and $\omega$ represent the coupled time and frequency domains, respectively, while the $[x,y]$ and $[k_{x},k_{y}]$ terms represent the coupled spatial distances and spatial wavenumbers in orthogonal spatial directions, respectively. Such three-dimensional Fourier analyses has been closely coupled with the field of helioseismology, which is employed to study the composition and structure of the solar interior by examining large-scale wave patterns on the solar surface \citep{1997SoPh..170...43K, 2001NuPhS..91...73T, 2000SoPh..192..285B, 2010ARA&A..48..289G, 2011LNP...832....3K, 10.3389/fspas.2019.00042}, which often give rise to patterns consistent with `rings' and `trumpets' when viewed in Fourier space \citep{1988ApJ...333..996H}.

\begin{figure*}[!t]
\begin{center}
\includegraphics[width=\textwidth]{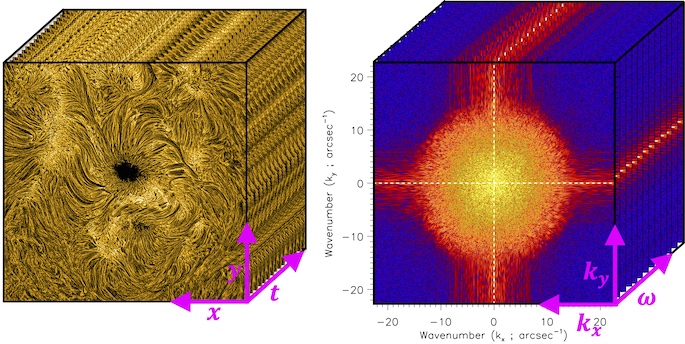}
\end{center}
\caption{An example application of an FFT to a three-dimensional datacube, converting [$x,y,t$] (left) into its frequency counterparts [$k_{x}, k_{y}, \omega$] (right). The HARDcam H$\alpha$ dataset presented here is taken from the work of \citet{2018NatPh..14..480G}.}
\label{fig:HARDcam_komega_cube} 
\end{figure*}

Up until recently, it has been challenging to apply the same three-dimensional Fourier techniques to high-resolution datasets from ground- and space-based observatories \citep{1963ARA&A...1...19L, 1990ARA&A..28..263S}. These techniques have been applied with ground-based observations to study convective phenomena \citep{1991ApJ...372..314C, 1992A&A...256..652S} and plage \citep{1992ApJ...393..782T}. With the advent of high image pointing stability, brought to fruition through a combination of high-order AO, photometrically accurate image reconstruction algorithms, precise telescope control hardware, and sub-pixel cross-correlation image co-alignment software, it is now possible to achieve long-duration image and/or spectral sequences that are stable enough to allow Fourier analyses in both temporal {\it{and}} spatial domains. The benefit of using high-resolution facilities
is that they offer unprecedented Nyquist temporal frequencies ($\omega$) and spatial wavenumbers ($[k_{x},k_{y}]$) due to their high temporal and spatial sampling, respectively. For example, the HARDcam H$\alpha$ dataset described in Section~{\ref{sec:HARDcam}} has a temporal cadence of 1.78~s and a spatial sampling of $0{\,}.{\!\!}{''}138$ per pixel, providing a Nyquist frequency of $\omega_{\text{\tiny{Ny}}}\approx280$~mHz $\left(\frac{1}{2\times1.78}\right)$ and a Nyquist wavenumber of $k_{\text{\tiny{Ny}}}\approx22.8$~arcsec$^{-1}$ $\left(\frac{2\pi}{2\times0.138}\right)$. This allows for the examination of the smallest and most rapidly varying phenomena currently visible in such high-resolution datasets.

\begin{figure*}[!t]
\begin{center}
\includegraphics[width=0.41\textwidth]{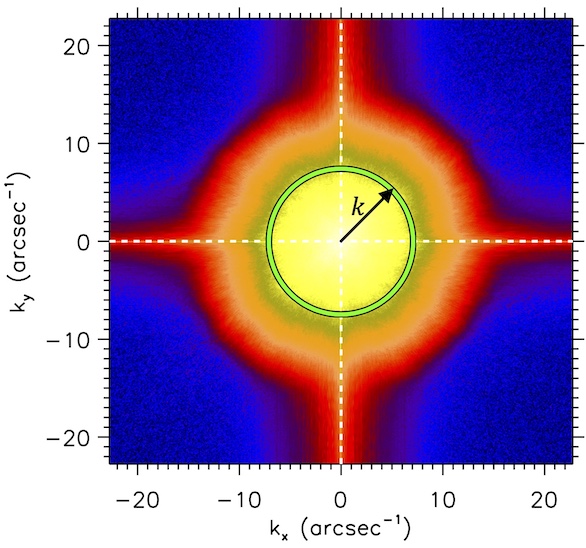}
\includegraphics[width=0.56\textwidth]{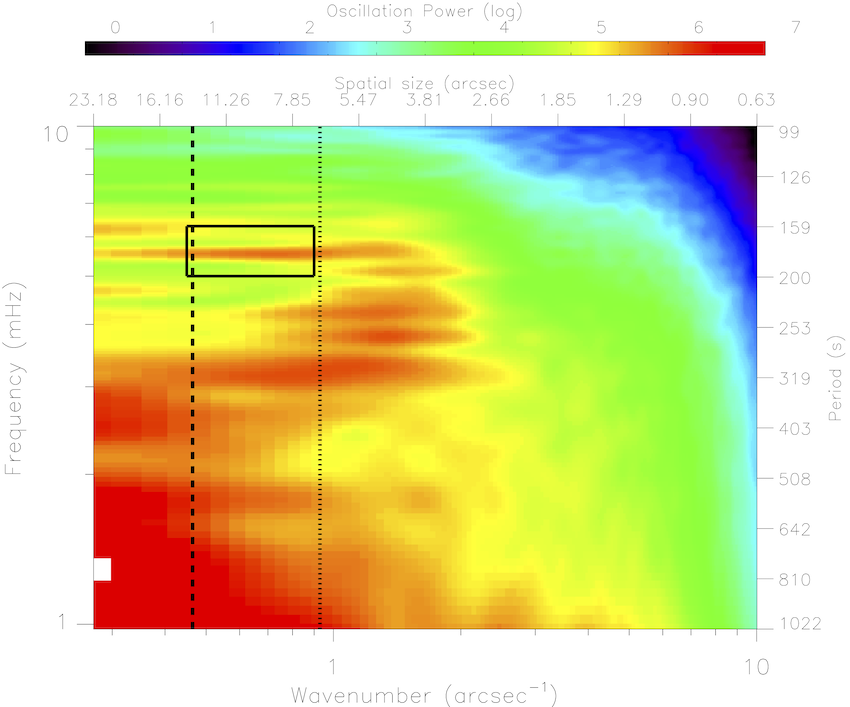}
\end{center}
\caption{A two-dimensional $[k_{x},k_{y}]$ cross-cut for a single temporal frequency, $\omega$, corresponding to the HARDcam H$\alpha$ data acquired on 2011 December 10 and described in Section~{\ref{sec:HARDcam}} (left panel). Due to the symmetries often found between $k_{x}$ and $k_{y}$, it is common to perform azimuthal averaging (e.g., along the solid green contour) to collapse this two-dimensional information into a single dimension, i.e. $[k_{x}, k_{y}] \rightarrow [k]$. This allows the three-dimensional FFT cube (see, e.g., the right panel of Figure~{\ref{fig:HARDcam_komega_cube}}) to be simplified into a standardized two-dimensional image, forming a $k$-$\omega$ diagram (right panel). Here, the $k$-$\omega$ diagram is cropped between approximately $1<\omega<10$~mHz and $0.3<k<10.0$~arcsec$^{-1}$, and displayed on a log-log scale to assist visual clarity. The colors represent oscillatory power that is displayed on a log-scale, while the vertical dashed and dotted lines correspond to the spatial size of the umbral diameter ($\approx13{\,}.{\!\!}{''}50$) and the radius of the umbra ($\approx6{\,}.{\!\!}{''}75$), respectively. The solid black box indicates a region of excess wave power at $\approx5.9$~mHz ($\approx170$~s) over the entire spatial extent of the sunspot umbra. Image reproduced from \citet{2017ApJ...842...59J}.}
\label{fig:komega_azimuthal_average} 
\end{figure*}

Applying an FFT to a three-dimensional dataset converts the spatial/temporal signals, [$x, y, t$], into its frequency counterparts, [$k_{x}, k_{y}, \omega$]. An example of this process can be seen in Figure~{\ref{fig:HARDcam_komega_cube}}, whereby an FFT has been applied to the HARDcam H$\alpha$ dataset documented by \citet{2018NatPh..14..480G}. It can be seen in the right panel of Figure~{\ref{fig:HARDcam_komega_cube}} that the Fourier power signatures are approximately symmetric in the $k_{x}/k_{y}$ plane. As a result, it is common for [$k_{x}, k_{y}$] cross-cuts at each frequency, $\omega$, to be azimuthally averaged providing a more straightforward two-dimensional representation of the Fourier power in the form of a $k$-$\omega$ diagram \citep{1988ApJ...324.1158D, 2001A&A...379.1052K, 2003A&A...407..735R, 2011A&A...532A.111K, 2012ApJ...746..183J, 2017ApJ...842...59J}.

\begin{figure*}[!t]
\begin{center}
\includegraphics[width=0.49\textwidth]{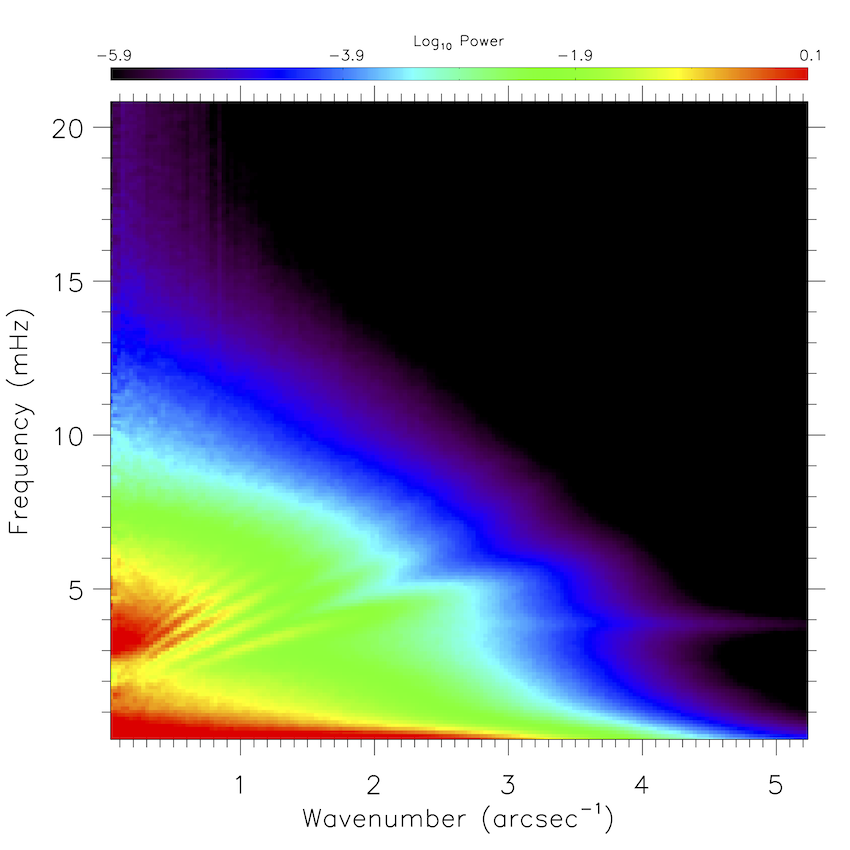}
\includegraphics[width=0.49\textwidth]{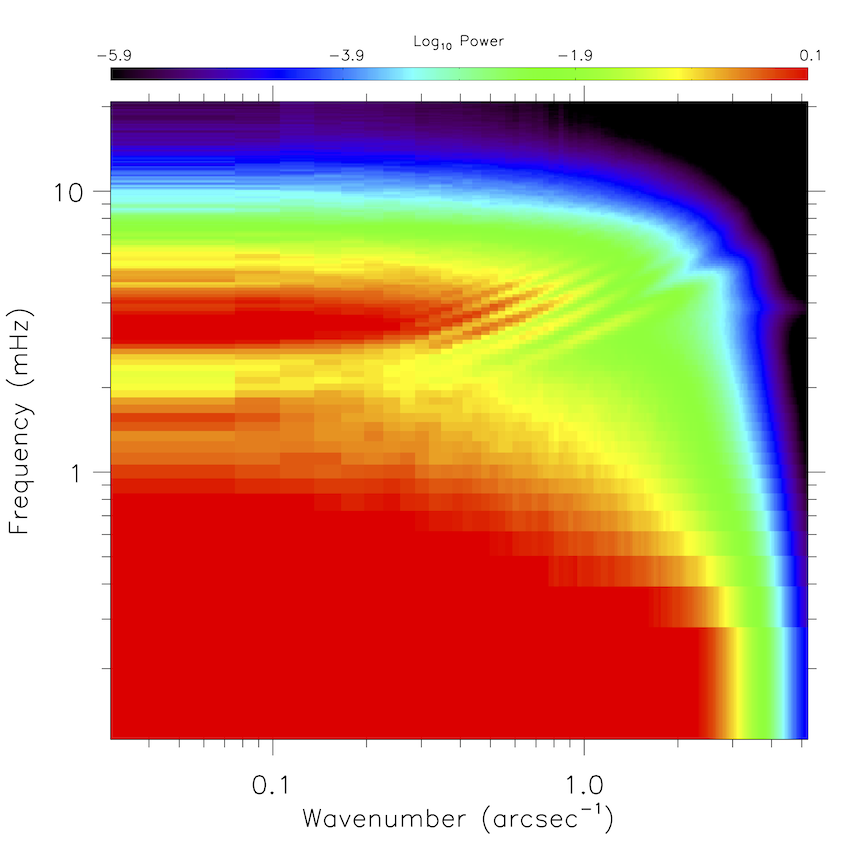}
\end{center}
\caption{A set of $k$-$\omega$ diagrams, derived from the photospheric SDO/AIA~1700{\,}{\AA} time series of active region NOAA~11366, which is co-spatial (and overlaps temporally) with the chromospheric HARDcam measurements presented in Figure~{\ref{fig:komega_azimuthal_average}}. Both $k$-$\omega$ diagrams are identical, however, the left panel is displayed on linear wavenumber ($k$) and frequency ($\omega$) scales, while the right panel is displayed on log-log axes. It is clear from inspection of the two panels that each have their merit when presenting results, with the linear axes giving less visual emphasis to the lower wavenumbers/frequencies, while the log-log axes allowing power-law trends in the power spectral densities to be modeled more easily through straight-line fitting. }
\label{fig:komega_SDO} 
\end{figure*}

An azimuthally averaged $k$-$\omega$ diagram for the HARDcam H$\alpha$ sunspot observations described in Section~{\ref{sec:HARDcam}} is shown in the right panel of Figure~{\ref{fig:komega_azimuthal_average}}. A number of important features are present in this diagram, including consistency with many quiet-Sun and internetwork Fourier power peaks documented by \citet{2001A&A...379.1052K}, \citet{2011A&A...532A.111K} and \citet{2012ApJ...746..183J}, whereby high power observed at larger spatial wavenumbers tends to be correlated with higher temporal frequencies. 
This can be visualized in the right panel of Figure~{\ref{fig:komega_azimuthal_average}}, whereby the dominant Fourier power is associated with the smallest spatial wavenumbers and temporal frequencies. However, as the wavenumber is increased to $>1$~arcsec$^{-1}$, the temporal frequencies corresponding to maximal Fourier power are concentrated within the $3-6$~mHz interval. This is consistent with the general trends observed in classical photospheric $k$-$\omega$ diagrams, such as that shown in Figure~{\ref{fig:komega_SDO}}. Here, two $k$-$\omega$ diagrams from the photospheric SDO/AIA~1700{\,}{\AA} time series that is co-spatial (and overlaps temporally) with the HARDcam H$\alpha$ observations (used to produce Figure~{\ref{fig:komega_azimuthal_average}}) are displayed. The information displayed in both panels of Figure~{\ref{fig:komega_SDO}} is identical, however, the left panel is displayed on a linear wavenumber ($k$) and frequency ($\omega$) scales, while the right panel is displayed on log-log axes. In both panels, similar trends (e.g., heightened Fourier power with increasing temporal frequency in the interval of $3-6$~mHz is linked to larger spatial wavenumbers) can be identified, which is consistent with the overall trends depicted in the right panel of Figure~{\ref{fig:komega_azimuthal_average}}. However, as discussed in \citet{2017ApJ...842...59J}, within the region highlighted by the solid black box in the right panel of Figure~{\ref{fig:komega_azimuthal_average}}, there is evidence of elevated Fourier power that spans a large range of spatial scales, yet remains confined to a temporal frequency on the order of 5.9~mHz ($\approx170$~s). This suggests that the embedded wave motion has strong coherency across a broad spectrum of spatial scales, yet can be represented by a very narrow range of temporal frequencies. Looking closely at the right panel of Figure~{\ref{fig:komega_azimuthal_average}}, it can be seen that elevated levels of Fourier power extend down to the smallest spatial wavenumbers allowable from the HARDcam dataset. This implies that the 5.9~mHz frequency is still significant on spatial scales much larger than the field of view captured by HARDcam. 

However, there are a number of key points related to Figures~{\ref{fig:komega_azimuthal_average}} \& {\ref{fig:komega_SDO}} that are worth discussing. First, Figure~{\ref{fig:komega_SDO}} highlights the merits of utilizing either linear or log-log axes depending on the features being examined. For example, the use of a linear scale (left panel of Figure~{\ref{fig:komega_SDO}}) results in less visual emphasis being placed on the lowest spatial waveneumbers and temporal frequencies. This can help prevent (visual) over-estimations of the trends present in the $k$-$\omega$ diagram since all of the frequency bins occupy identical sizes within the corresponding figure. However, as spatial and temporal resolutions dramatically improve with next generation instrumentation, the corresponding spatial/temporal Nyquist frequencies continue to become elevated, often spanning multiple orders-of-magnitude. If these heightened Nyquist frequencies are plotted on a purely linear scale, then many of the features of interest may become visually lost within the vast interval occupied by the $k$-$\omega$ diagram. An option available to counter this would be to crop the $k$-$\omega$ diagram to simply display the spatial wavenumbers and temporal frequencies of interest, although this comes at the price of discarding information that may be important within the remainder of the frequency space. Alternatively, it is possible to use log-log axes for the $k$-$\omega$ diagram, which can be visualized in the right panels of Figures~{\ref{fig:komega_azimuthal_average}} \& {\ref{fig:komega_SDO}}. This type of log-log display also benefits the fitting of any power-law trends that may be present within the $k$-$\omega$ diagram, since they will manifest as more straightforward (to fit) linear slopes in the plot. Finally, the right panel of Figure~{\ref{fig:komega_azimuthal_average}} reveals some horizontal banding of power that appears slightly different than the diagonal `arms' of Fourier power visible in Figure~{\ref{fig:komega_SDO}}. This may be a consequence of the reduced spatial wavenumber and temporal frequency resolutions achievable with large-aperture ground-based observatories, which naturally have a reduced field-of-view size (causing a relatively low spatial wavenumber resolution when compared to large field-of-view observations from, e.g., SDO) and limited time series durations (creating relatively low temporal frequency resolutions when compared to space-borne satellite missions that are unaffected by day/night cycles and/or atmospheric effects). Therefore, it is imperative that the investigative team examines the merits of each type of $k$-$\omega$ display and selects the use of either linear or log-log axes to best represent the physical processes at work in their dataset.


\begin{figure*}[!t]
\begin{center}
\includegraphics[width=\textwidth]{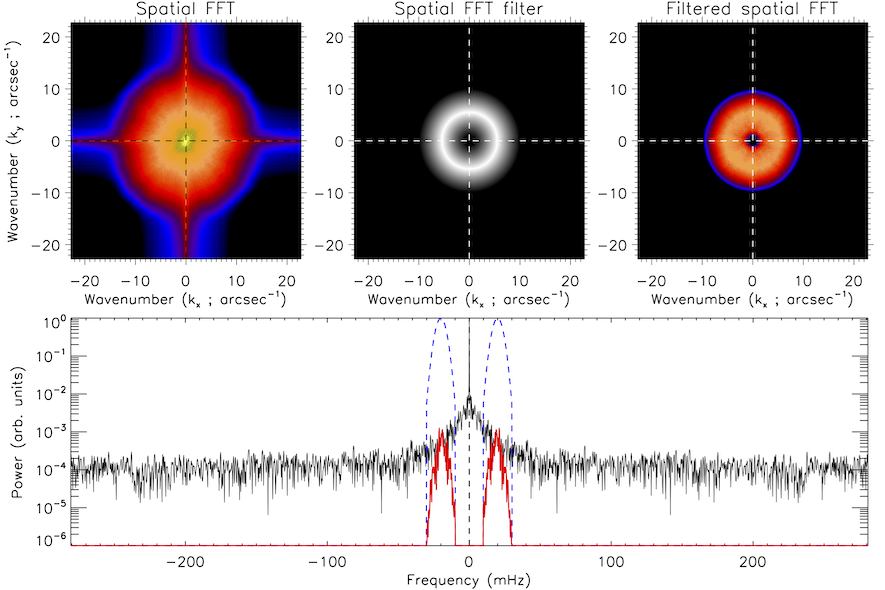}
\end{center}
\caption{Outputs provided by a commonly available three-dimensional Fourier filtering code \citep[QUEEFF;][]{2017ApJ...842...59J}, showing a frequency-averaged wavenumber spectrum (upper-left), a Gaussian (with $2<k<10$~arcsec$^{-1}$) wavenumber filter that resembles a torus shape when viewed in the $[k_{x}, k_{y}]$ plane (upper-middle), and the resulting transmitted wavenumber spectra once multiplied by the chosen filter (upper-right). The lower panel displays the wavenumber-averaged frequency spectrum (solid black line), where the Fourier power is displayed (using a log-scale) as a function of the temporal frequency, $\omega$. The dashed blue line highlights a chosen frequency filter, $20\pm10$~mHz, with a Gaussian shape to more smoothly reduce Fourier power at the edges of the chosen spectral range to reduce aliasing. The solid red line displays the resulting transmitted frequency spectrum once multiplied by the chosen Gaussian filter. In each panel, dashed black or white lines highlight the $k_{x}/k_{y}=0$~arcsec$^{-1}$ or $\omega=0$~mHz locations.}
\label{fig:QUEEFFoutput} 
\end{figure*}

\subsubsection{Three-dimensional Fourier Filtering}
\label{sec:threedimensionalFourierfiltering}
Taking the one-dimensional Fourier filtering methodology described in Section~{\ref{sec:onedimensionalFourierfiltering}} a step further, it is often useful to filter an input three-dimensional dataset ($[x,y,t]$) in terms of both its temporal frequencies, $\omega$, and its spatial wavenumbers, $k$. While it is common for the frequency to be defined as the reciprocal of the period, i.e., $\omega = 1/T$, where $T$ is the period of oscillation, the wavenumber is often defined as $k = 2\pi/\lambda$ \citep{2001A&A...379.1052K}, where $\lambda$ is the wavelength of the oscillation in the spatial domain (i.e., [$x,y$]). Hence, it is often important to bear in mind this additional factor of $2\pi$ when translating between wavenumber, $k$, and spatial wavelength, $\lambda$. Figures~{\ref{fig:komega_azimuthal_average}} \& {\ref{fig:QUEEFFoutput}} employ this form of frequency/wavenumber notation, meaning that the spatial wavelengths can be computed as $\lambda = 2\pi/k$, while the period is simply $T = 1/\omega$ \citep[similar to that shown in][]{1992A&A...256..652S, 2012ApJ...746..183J}. However, some research programs, particularly those adopting helioseismology nomenclature, utilize the factor of $2\pi$ in both the wavenumber and frequency domains (e.g., $T = 2\pi/\omega$ \citep{1981ApJ...249..349M}. As a result, it is important to select an appropriate scaling to ensure consistency across a piece of work. An example code capable of doing three-dimensional Fourier filtering is the QUEEn's university Fourier Filtering \citep[QUEEFF;][]{2017ApJ...842...59J} algorithm, which is based around the original techniques put forward by \citet{1988ESASP.286..315T}, \citet{1989ApJ...336..475T}, \citet{2003A&A...407..735R}, \citet{2010ApJ...723L.175R}, and \citet{2001A&A...379.1052K}, but now adapted into a publicly available Interactive Data Language \citep[{\sc{idl}};][]{doi:10.1061/40479(204)125} package\footnote{QUEEFF code download link -- \href{https://bit.ly/37mx9ic}{https://bit.ly/37mx9ic}}$^{,}$\footnote{WaLSA online wave analysis software repository available at \href{https://walsa.team/codes}{https://walsa.team/codes}}.

\begin{figure*}[!t]
\begin{center}
\includegraphics[width=\textwidth]{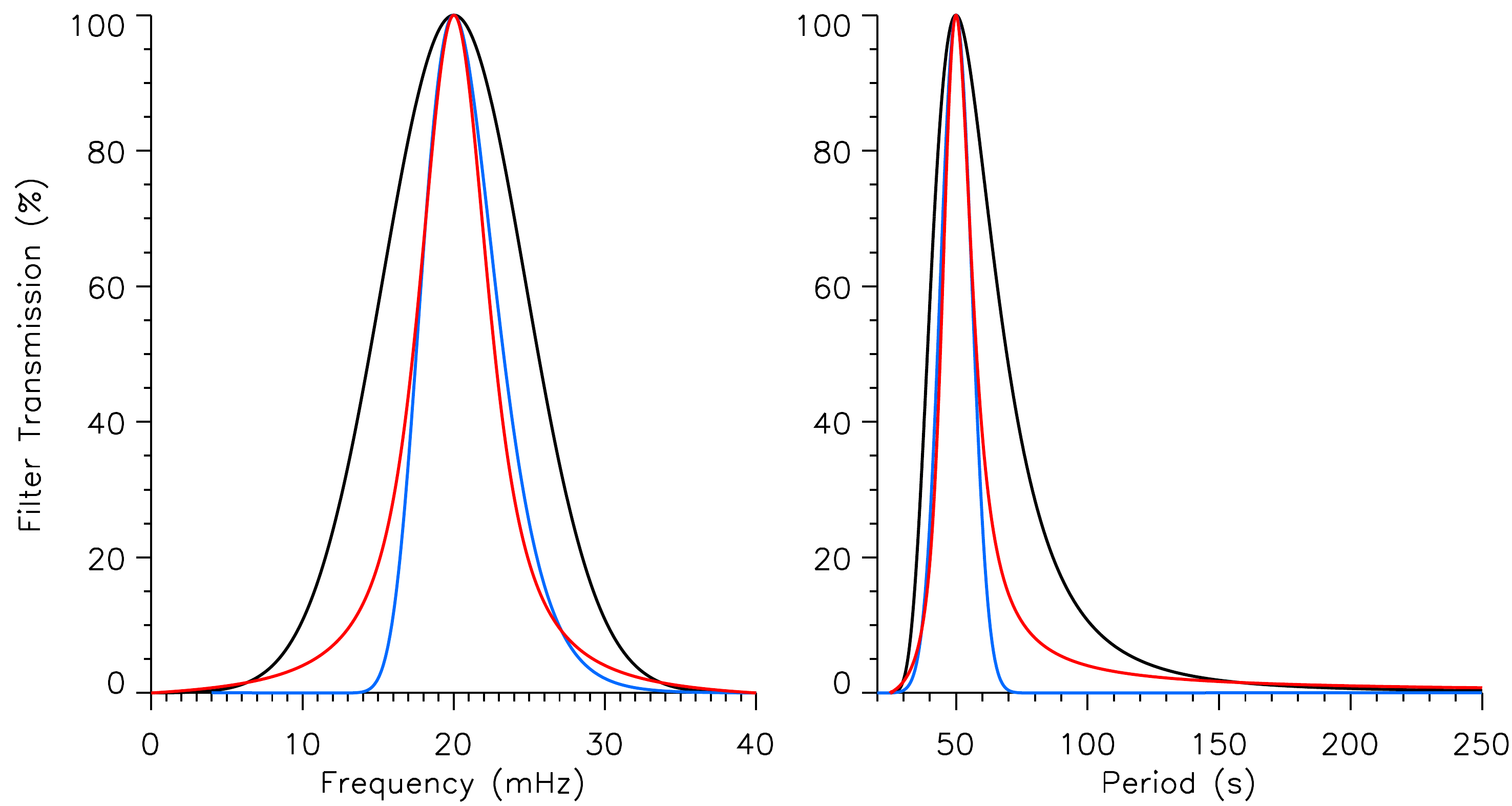}
\end{center}
\caption{Different types of frequency ($\omega$) filter that can be applied to time-resolved data products. The left panel displays the filter transmission (as a percentage) in terms of the frequency, while the right panel displays the same filters as a function of the oscillatory period. Presented using a solid black line is a Gaussian-shaped filter in the frequency domain with a FWHM equal to 10~mHz, while the solid red line indicates a Voigt-shaped filter in the frequency domain, both centered on 20~mHz. Contrarily, a Gaussian-shaped filter in the period domain, with a FWHM equal to 10~s, is shown using a solid blue line, again centered on 50~s to remain consistent with the 20~mHz profiles shown using red and black lines. It is clearly evident that the filter profile shape changes dramatically between the time and frequency domains, and hence it is important to select the correct filter based upon the science requirements.}
\label{fig:Gaussian_filter} 
\end{figure*}

Importantly, the QUEEFF code provides the user with the ability to apply Gaussian smoothing windows to both frequency and wavenumber regions of interest in order to help mitigate against elements of aliasing during subsequent dataset reconstruction. Figure~{\ref{fig:QUEEFFoutput}} shows an example figure provided by the QUEEFF code, which displays the frequency-averaged wavenumber power (upper-left panel), the chosen wavenumber filter (upper-middle panel) utilizing a Gaussian structure providing a torus-shaped filter spanning ${2-10}$~arcsec$^{-1}$, alongside the resulting filtered wavenumber spectra (upper-right panel). The lower panel of Figure~{\ref{fig:QUEEFFoutput}} displays the spatially-averaged frequency spectrum of the HARDcam H$\alpha$ dataset, where the Fourier power is displayed as a function of the frequency, $\omega$, using a solid black line. A Gaussian frequency filter, spanning $20\pm10$~mHz, is overplotted using a dashed blue line. The preserved temporal frequencies (i.e., once the original frequency spectrum has been multiplied by the chosen frequency filter) is shown using a solid red line. This filtered three-dimensional Fourier cube can then be passed through an inverse FFT to reconstruct an intensity image cube that contains the wavenumbers and frequencies of interest to the user.

Again, as discussed in Section~{\ref{sec:onedimensionalFourierfiltering}}, the QUEEFF three-dimensional Fourier filtering code constructs a Gaussian-shaped filter, which is applied in the Fourier domain. This ensures that the filter is symmetric about the chosen peak frequency (see, e.g., the black line in the left panel of Figure~{\ref{fig:Gaussian_filter}}). Of course, due to the oscillation period having a reciprocal relationship with the temporal frequency (i.e., $1/\omega$), this results in asymmetric sampling about the desired peak period (see, e.g., the solid black line in the right panel of Figure~{\ref{fig:Gaussian_filter}}). Depending upon the science requirements of the user, it may be more advantageous to apply a Gaussian-shaped filter in the period domain (e.g., the solid blue line in the right panel of Figure~{\ref{fig:Gaussian_filter}}), which ensures less inclusion of lower frequency (higher period) terms that may be undesirable in the final reconstructed time series. This is highlighted by the more rapid truncation of the filter (solid blue line in the left panel of Figure~{\ref{fig:Gaussian_filter}}) towards lower frequencies. Additionally, the user may select alternative frequency filters, such as a Voigt profile \citep{2007MNRAS.375.1043Z}, which is shown in Figure~{\ref{fig:Gaussian_filter}} using a solid red line. Furthermore, Figure~{\ref{fig:Gaussian_filter}} shows possible filtering combinations that can be applied to the temporal domain, yet similar options are available when filtering the spatial wavenumbers ($[k_{x}, k_{y}]$) too. Ultimately, it is the science objectives that drive forward the wave filtering protocols, so possible options need to be carefully considered before applying to the input data. 

Combination Fourier filters (i.e., that are functions of $k_{x}$, $k_{y}$ and $\omega$) have been utilized in previous studies to extract unique types of wave modes manifesting in the lower solar atmosphere. For example, specific Fourier filters may be employed to extract signatures of $f$- and $p$-mode oscillations manifesting in photospheric observations \citep[e.g.,][]{1988ApJ...333..996H, 1998ApJ...505..390S, 2004ApJ...614..472G, 2021ApJ...915...36B}. Another example of a well-used Fourier filter is the `sub-sonic filter', which can be visualized as a cone in $k-\omega$ space \citep{1989ApJ...336..475T}, 
\begin{equation}
    v_{\mathrm{ph}} = \frac{\omega}{k} \ ,
\end{equation}
where $v_{\mathrm{ph}}$ is the phase velocity of the wave. Here, all Fourier components inside the cone, where propagation velocities are less than the typical sound speed (i.e., $v_{\mathrm{ph}} < c_{s}$), are retained while velocities outside the cone are set to zero. An inverse Fourier transform of this filtered spectrum provides a dataset that is embodied by the convective part of the solar signal since the non-convective phenomena (e.g., solar $p$-modes) have been removed \citep{1997A&A...324..704S, 2003A&A...407..735R}. Alternatively, modification of the sub-sonic filter to include only those frequencies above the Lamb mode, $\omega = c_{s}k$ \citep{2021RSPTA.37900170F}, provides a reconstructed dataset containing oscillatory parts of the input signal. As highlighted above, it is the science objectives that define the filtering sequences required to extract the underlying time series of interest. However, well-proven examples of these exist for common phenomena (e.g., solar $f$- and $p$-modes), hence providing an excellent starting point for the community.

\subsection{Wavelet Analyses}
\label{waveletanalyses}
\begin{figure*}[!t]
\begin{center}
\includegraphics[width=0.49\textwidth]{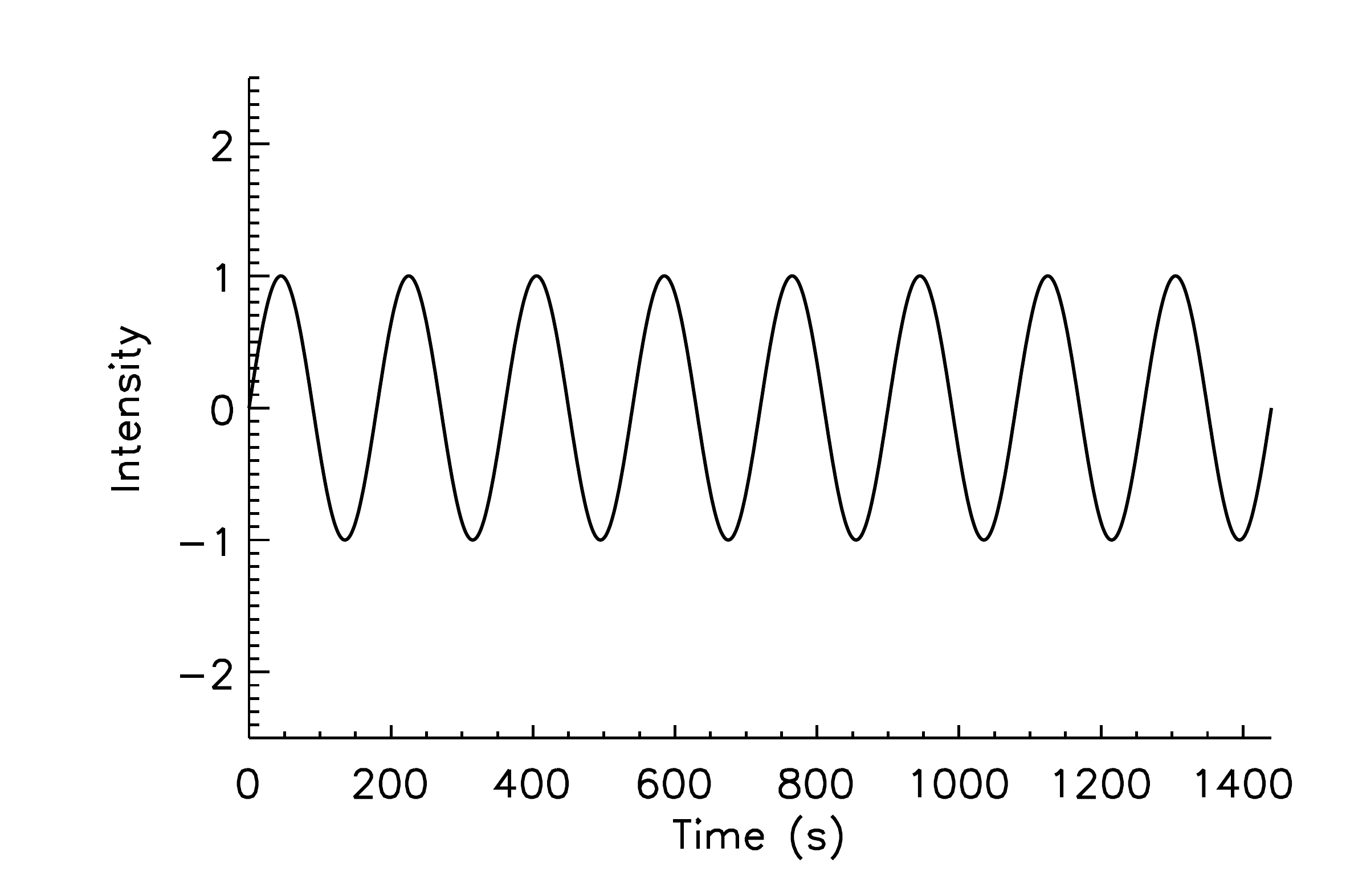}
\includegraphics[width=0.49\textwidth]{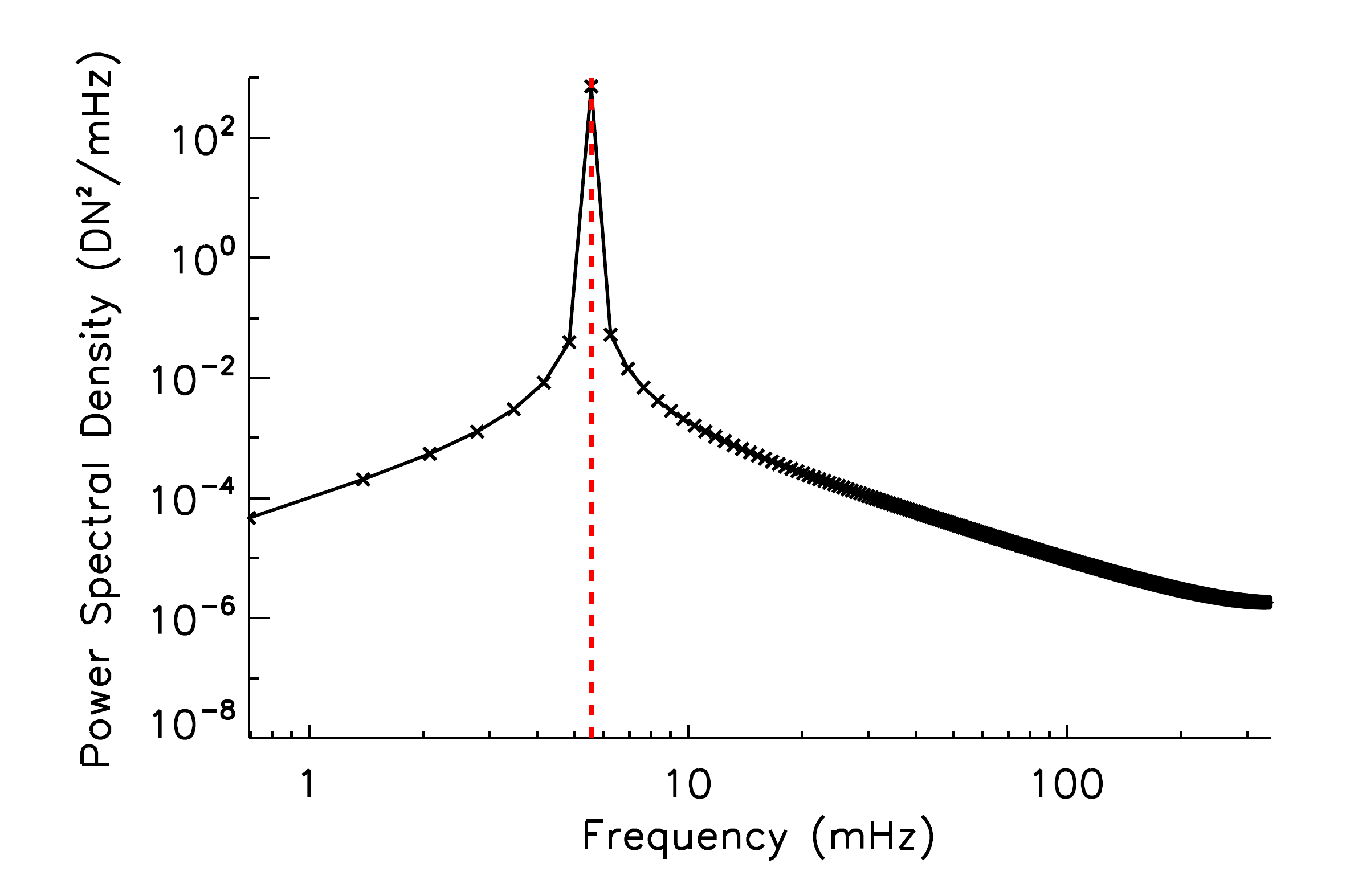}
\includegraphics[width=0.49\textwidth]{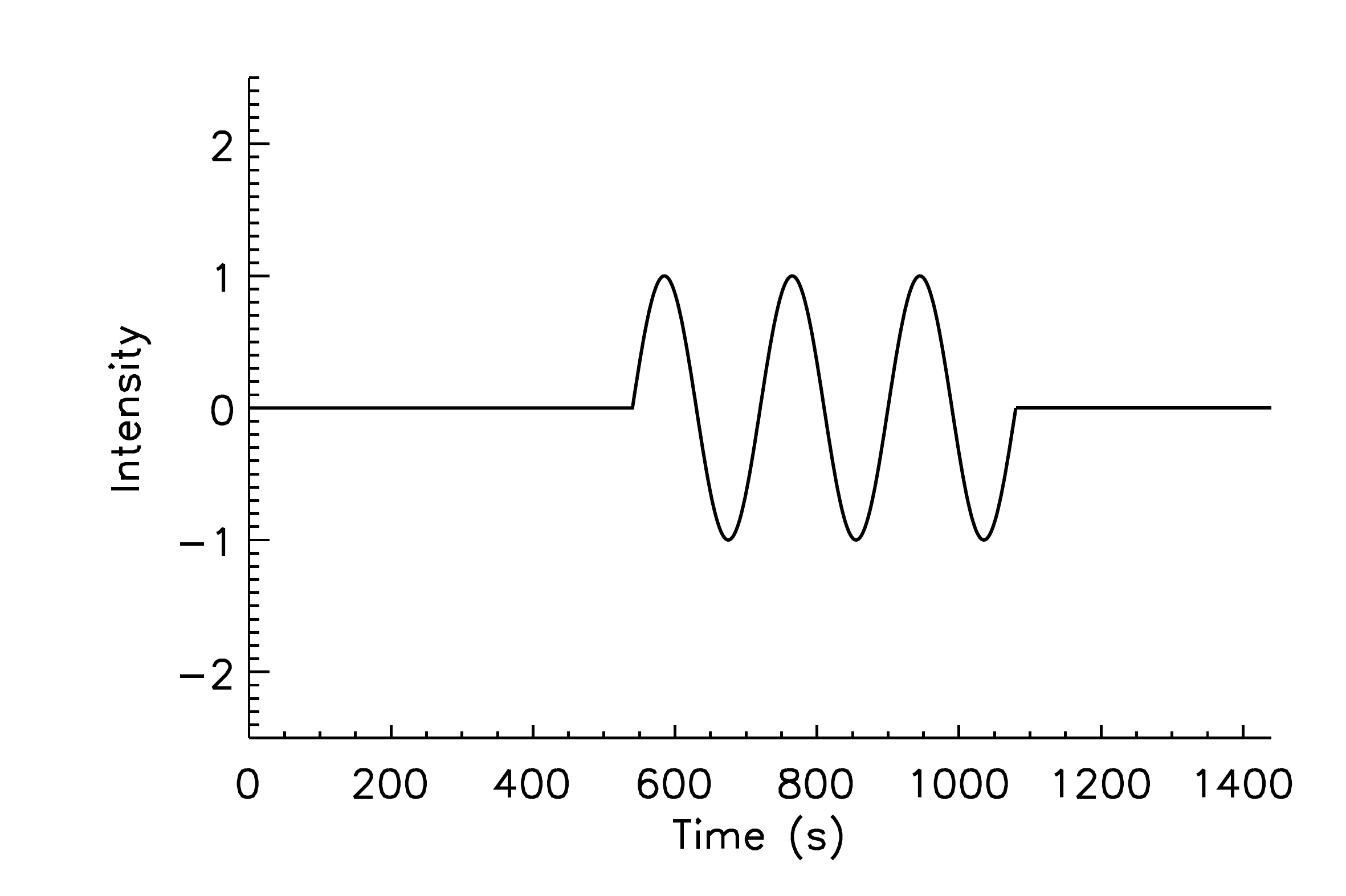}
\includegraphics[width=0.49\textwidth]{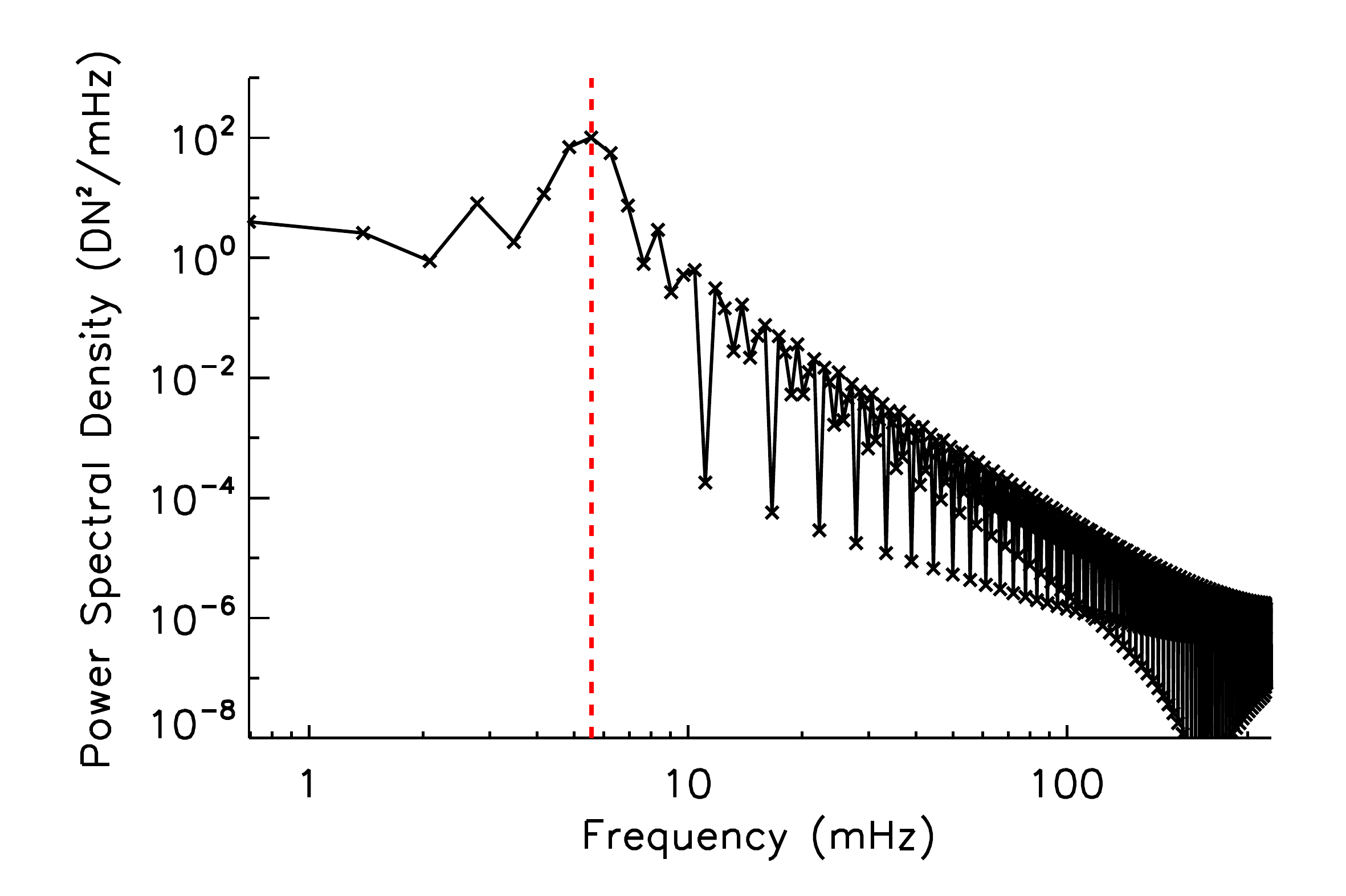}
\includegraphics[width=0.49\textwidth]{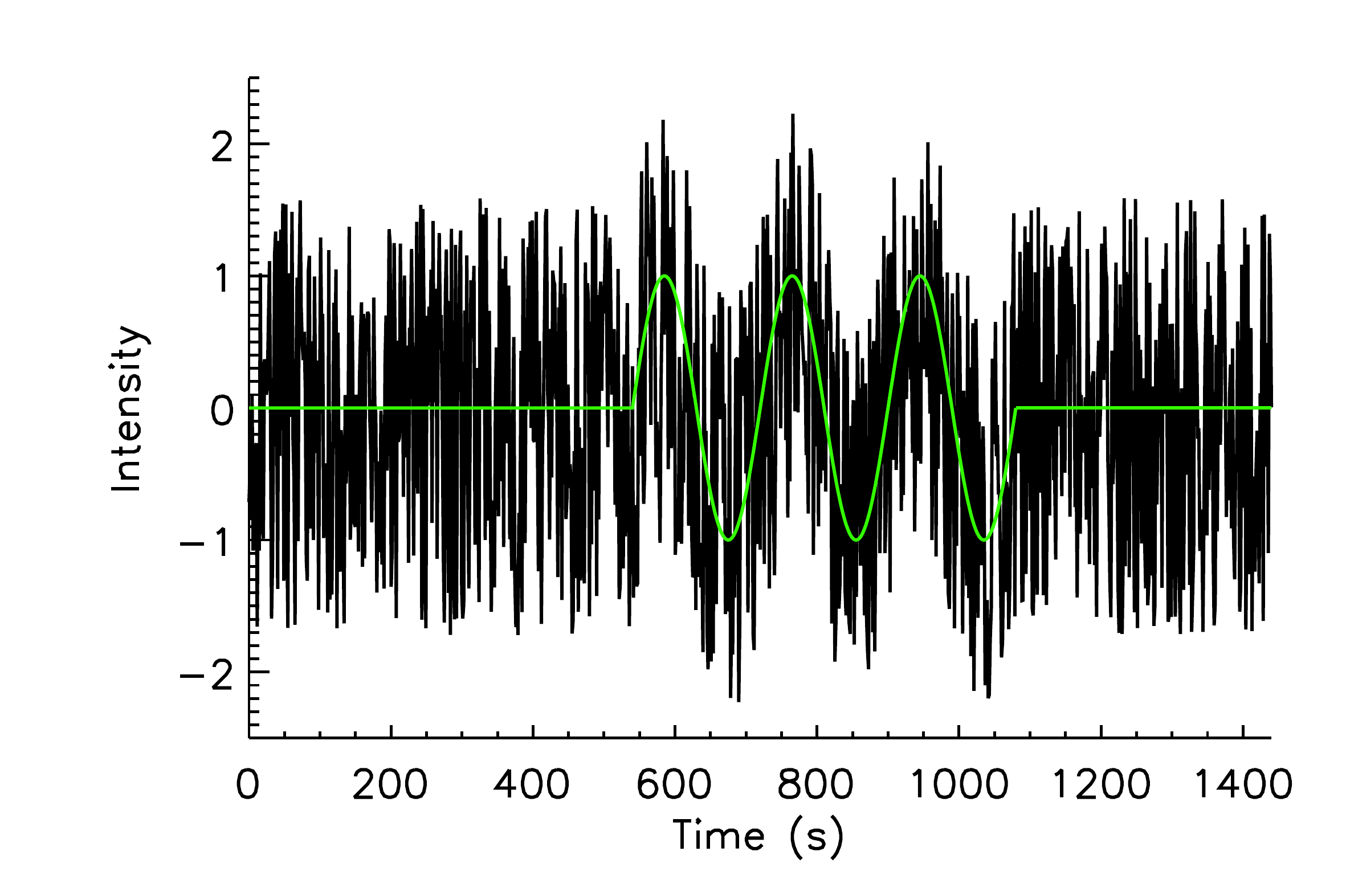}
\includegraphics[width=0.49\textwidth]{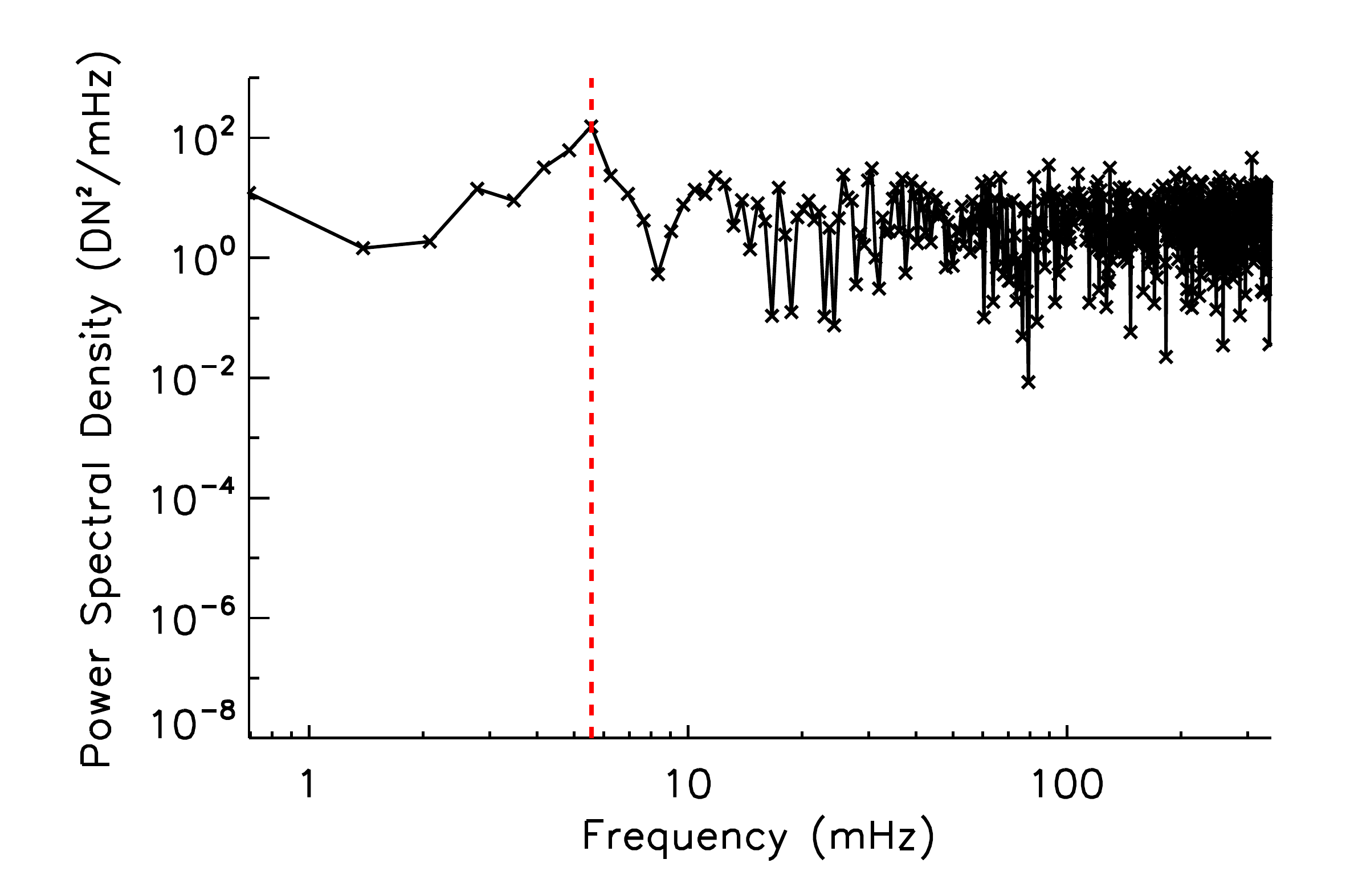}
\end{center}
\caption{An example time series consisting of a pure 180~s periodicity (5.56~mHz) signal, which is sampled at a cadence of 1.44~s to remain consistent with modern instrument capabilities (upper left). The middle-left panel shows the same example time series, only now with the first three and last two complete wave cycles suppressed, hence making a quasi-periodic wave signal. The lower-left panel shows the same quasi-period wave signal shown in the middle-left panel (solid green line), only now with superimposed Poisson (shot) noise added on top of the signal. Each of the right panels display the corresponding FFT-generated Fourier spectra, with the frequency and Fourier power values plotted on log-scales for better visual clarity. The vertical dashed red lines highlight the input 5.56~mHz signal.}
\label{fig:Fourierdrawbacks} 
\end{figure*}

While FFT analyses is very useful for identifying and characterizing persistent wave motion present in observational datasets, it begins to encounter difficulties when the time series contains weak signals and/or quasi-periodic signatures. Figure~{\ref{fig:Fourierdrawbacks}} shows example time series containing a persistent wave signal with a 180~s periodicity (5.56~mHz) and no embedded noise (top-left panel), a quasi-periodic 5.56~mHz wave signal with no noise (middle-left panel), and a qausi-periodic 5.56~mHz wave signal embedded in high-amplitude noise (lower-left panel). It can be seen for each of the corresponding right-hand panels, which reveal the respective Fourier power spectral densities, that the detected 5.56~mHz Fourier peak becomes progressively less apparent and swamped by noise, even becoming significantly broadened in the lower-right panel of Figure~{\ref{fig:Fourierdrawbacks}}. As a result, the application of Fourier analyses to solar time series often displaying quasi-periodic wave motion \citep[e.g., spicules, fibrils, rapid blueshift excursions (RBEs), etc.;][]{1968SoPh....3..367B, 2004Natur.430..536D, 2007ApJ...655..624D, 2007PASJ...59S.655D, 2009SSRv..149..355Z, 2013ApJ...764..164S, 2013ApJ...769...44S, 2015ApJ...802...26K} may not be the most appropriate as a result of the limited lifetimes associated with these features.

Wavelet techniques, pioneered by \citet{1998BAMS...79...61T}, employ a time-localized oscillatory function that is continuous in both time and frequency \citep{2004ApJ...604..936B}, which allows them to be applied in the search for dynamic transient oscillations. The time resolution of the input dataset is preserved through the modulation of a simple sinusoid (synonymous with standard FFT approaches) with a Gaussian envelope, providing the Morlet wavelet commonly used in studies of waves in the solar atmosphere \citep{2004ApJ...617..623B, 2007A&A...473..943J, 2012A&A...539L...4S, 2013A&A...554A.146K, 2015SoPh..290..363K, 2017ApJS..229...10J}. As a result, a wavelet transform is able to provide high frequency resolution at low frequencies and high time resolution at high frequencies, which is summarized by \citet{KEHTARNAVAZ2008175}.

\begin{figure*}[!t]
\begin{center}
\includegraphics[width=0.6\textwidth]{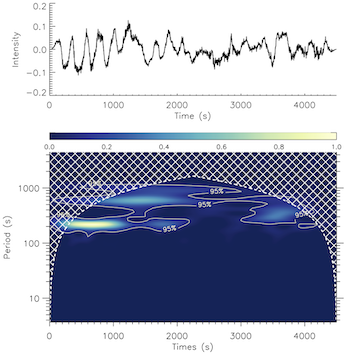}
\end{center}
\caption{The detrended and apodized HARDcam H$\alpha$ lightcurve shown in the lower panel of Figure~{\ref{fig:HARDcam_lightcurves}} (top). The bottom panel shows the corresponding wavelet transform, where the wave power is displayed as a function of the oscillatory period ($y$-axis) and observing time ($x$-axis). The color bar displays the normalized wavelet power, while the cross-hatched region (bounded by a dashed white line) highlights locations of the wavelet transform that may be untrustworthy due to edge effects. Solid white lines contour regions where the wavelet power exceeds the 95\% confidence level (i.e., significant at the 5\% level).}
\label{fig:wavelethardcam} 
\end{figure*}

Figure~{\ref{fig:wavelethardcam}} displays the wavelet power spectrum (lower panel) resulting from the application of a Morlet wavelet transform on the detrended and apodized HARDcam H$\alpha$ lightcurve (upper panel). Here, it is possible to see the effects of quasi-periodic wave phenomena, where there is clear evidence of a large-amplitude periodicity between times of $0-2200$~s at a period of $\approx210$~s ($\approx4.7$~mHz). This wave activity is highlighted in the wavelet transform by being bounded by the 95\% confidence level isocontours across these times and periods, which is equivalent to the oscillatory behavior being significant at the 5\% level \citep{1998BAMS...79...61T}. To calculate the wavelet power thresholds corresponding to the 95\% confidence isocontours, the wavelet background spectrum (i.e., the output theoretical background spectrum that has been smoothed by the wavelet function) is multiplied by the 95$^{\text{th}}$ percentile value for a $\chi_{2}^{2}$ distribution \citep{1963JAtS...20..182G}. Please note that for some considerations, including expensive computation times, the Monte Carlo randomization method is not preferred for wavelet transform \citep{1995BAMS...76.2391L,1998BAMS...79...61T}. The $\approx210$~s wavelet power signatures shown in the lower panel of Figure~{\ref{fig:wavelethardcam}} are consistent with the standardized FFT approach documented in Section~{\ref{sec:1Dfourieranalysis}}, although the quasi-periodic nature of the wave motion is likely a reason why the corresponding power in the traditional FFT spectrum (upper panel of Figure~{\ref{fig:HARDcam_power}}) is not as apparent. Importantly, with the wavelet transform it is possible to identify more clearly the times when this periodicity appears and disappears from the time series, which is seen to correlate visibly with the clear sinusoidal fluctuations present at the start of the H$\alpha$ time series (upper panel of Figure~{\ref{fig:wavelethardcam}}). Also, the lack of significant wavelet power at very long periods (low frequencies) suggests that the lightcurve detrending applied is working adequately.

\begin{figure*}[!t]
\begin{center}
\includegraphics[width=0.8\textwidth]{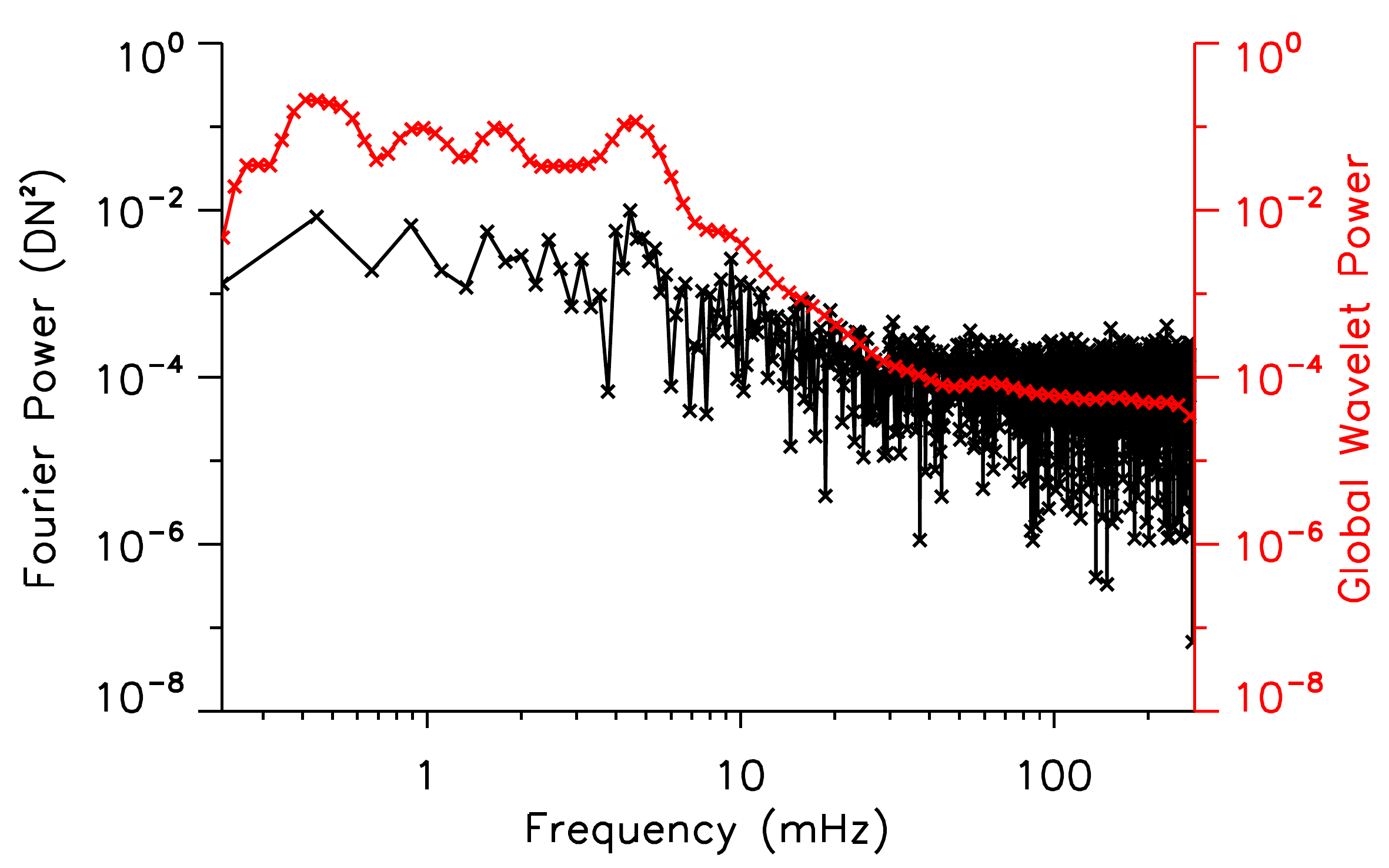}
\end{center}
\caption{Fourier (black line) and global wavelet (red line) power spectra of the HARDcam H$\alpha$ detrended lightcurve shown in the lower panel of Figure~{\ref{fig:HARDcam_lightcurves}}. It can be seen that at larger scales (lower frequencies) the global wavelet spectrum has increased power over that calculated from traditional Fourier techniques, due to the increased wavelet frequency resolution in this regime. Contrarily, at smaller scales (higher frequencies) the global wavelet spectrum appears as a smoothed Fourier spectrum due to the reduced frequency resolution at these smaller scales. While the global wavelet spectrum is a good estimation of the Fourier power spectrum, these biases need to be carefully considered when interpreting the embedded wave motion.}
\label{fig:Fourier_globalWS_comparison} 
\end{figure*}

Due to wavelet analyses preserving the time domain of the original input signal, care must be taken to ensure that any power visible in wavelet transforms is the result of wave motion and not an instantaneous spike in intensity. To achieve this, it is typical to exclude oscillations from subsequent analysis that last, in duration, for less than $\sqrt{2}$~wave cycles. This requirement is often referred to as the decorrelation time \citep{1998BAMS...79...61T}, which involves comparing the width of a peak in the wavelet power spectrum (defined as the time interval over which the wavelet power exceeds the 95\% confidence level -- see Section~{\ref{sec:calculatingconfidencelevels}}) with the period itself to determine the number of complete wave cycles \citep{1999A&A...347..355I, 2004ApJ...602..436M}. Oscillations that last for less time than $\sqrt{2}$~wave cycles are subsequently discarded as they may be the result of spikes and/or instrumental abnormalities in the data. In addition, periodicities manifesting towards the extreme edges of the lightcurve need to be considered carefully due to the possible presence of edge effects arising due to the finite duration of the time series \citep{1993MWRv..121.2858M}. This region where caution is required is highlighted in the lower panel of Figure~{\ref{fig:wavelethardcam}} using the cross-hatched solid white lines. Here, the ``cone of influence'' (COI) is defined as as the $e$-folding time for the autocorrelation of wavelet power at each scale, and for the traditional Morlet wavelet this is equal to $\sqrt{2}$~wave cycles \citep{1998BAMS...79...61T}, hence why longer periods are more heavily effected (in the time domain) than their shorter (high-frequency) counterparts.

Finally, many research studies employ the global wavelet spectrum to characterize the frequencies present in the input time series. Here, the global wavelet spectrum is defined as the average spectrum across all local wavelet spectra along the entire input time axis \citep{1998BAMS...79...61T}. Essentially, the global wavelet spectrum can be considered as an estimation of the true Fourier spectrum. For example, a time series comprised of mixed wave frequencies that are superimposed on top of a white noise background should produce Fourier spectral peaks equal to $2\sigma_{\epsilon}^{2} + NA_{i}^{2}/2$, where $A_{i}$ are the amplitudes of the oscillatory components, $\sigma_{\epsilon}^{2}$ is the variance of the noise, and $N$ is the number of steps in the time series \citep{nla.cat-vn2888327}. However, the corresponding peaks in the global wavelet spectrum will usually be higher at larger scales when compared to smaller scales, which is a consequence of the wavelet transform having better frequency resolution at long periods, albeit with worse time localization \citep{https://doi.org/10.1029/2004WR003843}. 

\begin{figure*}[!t]
\begin{center}
\includegraphics[width=1.0\textwidth]{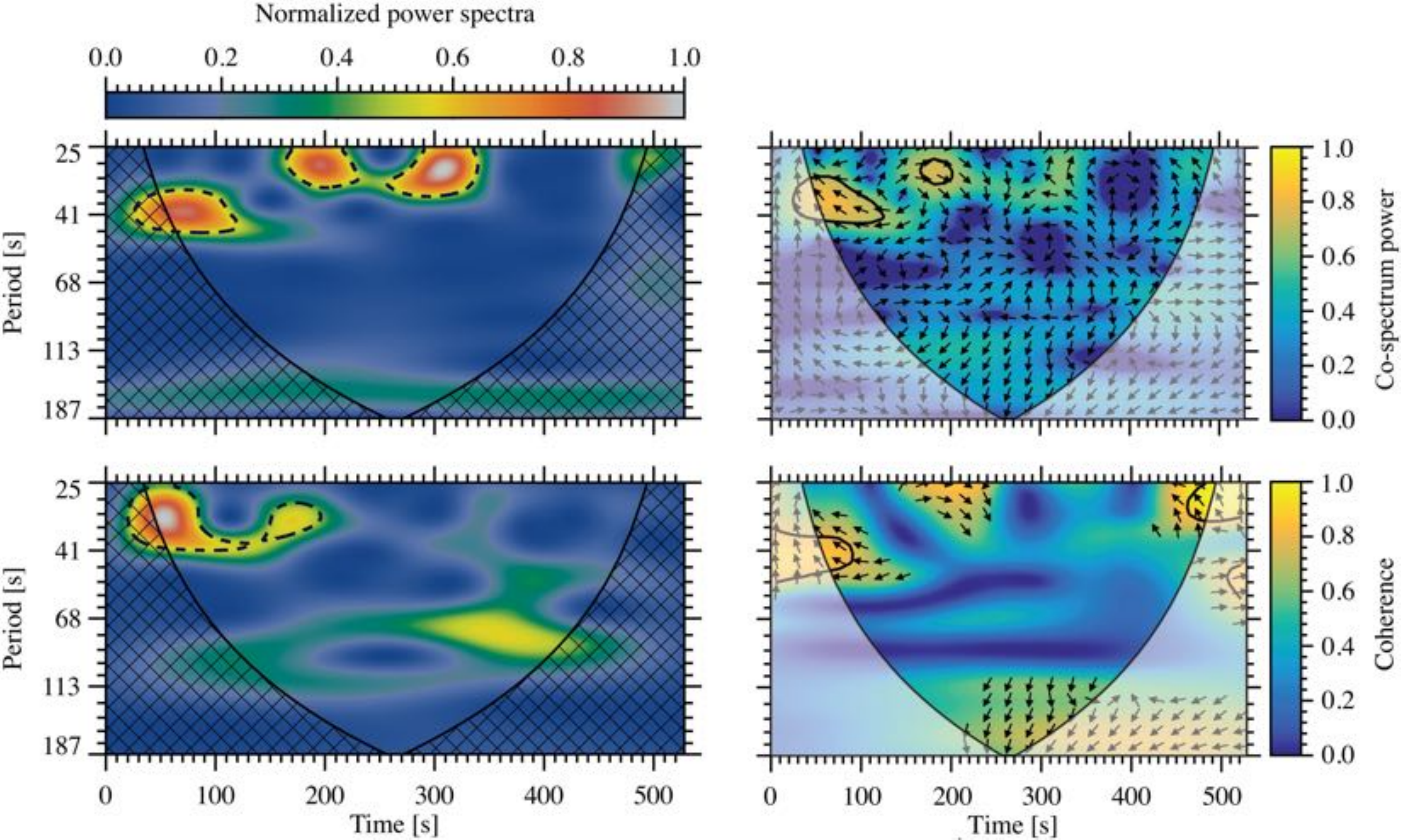}
\end{center}
\caption{Wavelet power spectra of transverse oscillations in a small magnetic element (marked with circles in Figure~\ref{fig:SUNRISE_09June2009}), from time-series of images acquired in 300~nm (lower left) and in Ca~{\sc ii}~H (upper left) bands from SuFI/{\sc Sunrise}. The right panels display the wavelet co-spectrum power (on the top) and coherence map (on the bottom). The 95\% confidence levels are identified with dashed/solid contours in all panels and the COIs are marked with the cross-hatched/shaded regions. The arrows on the right panels show the phase angles between oscillations at the two atmospheric heights, with in-phase oscillations depicted by arrows pointing right and fluctuations in Ca~{\sc ii}~H leading those in 300~nm by 90 degrees marked by arrows pointing straight up. Images reproduced from \cite{2017ApJS..229...10J}.}
\label{fig:SUNRISE_wavelet_phaselag} 
\end{figure*}

As such, the global wavelet spectrum is often considered a biased estimation of the true Fourier spectrum \citep{https://doi.org/10.1007/s11802-005-0062-y}. This effect can be clearly seen in Figure~{\ref{fig:Fourier_globalWS_comparison}}, which displays both the Fourier and global wavelet power spectra for the same HARDcam H$\alpha$ time series shown in the lower panel of Figure~{\ref{fig:HARDcam_lightcurves}}. In Figure~{\ref{fig:Fourier_globalWS_comparison}}, the higher power at larger scales (lower frequencies) is visible in the global wavelet spectrum (red line), when compared to that derived through traditional Fourier techniques (black line). However, at smaller scales (higher frequencies), both the global wavelet and Fourier spectra are in close agreement with one another, with the global wavelet spectrum appearing as a smoothed Fourier spectrum. The reason for these effects is due to the width of the wavelet filter in Fourier space. At large scales (low frequencies), the wavelet is narrower in frequency, resulting in sharper peaks that have inherently larger amplitudes. Contrarily, at small scales (high frequencies), the wavelet is more broad in frequency, hence causing any peaks in the spectrum to become smoothed \citep{1998BAMS...79...61T}. As such, it is important to take such biases into consideration when interpreting any embedded wave motion. Indeed, \citet{2001A&A...371.1137B}, \citet{2003ApJ...591..416C}, \citet{2016ApJ...828...23S}, \citet{2018MNRAS.479.5512K}, and \citet{2019ApJ...883...72C} have discussed the implementation of global wavelet and Fourier power spectra in the context of solar oscillations.

\subsubsection{Wavelet Phase Measurements}
\label{WaveletPhaseMeasurements}
Similar to the Fourier phase lag analysis described in Section~\ref{sec:fourierphaselaganalysis}, it is also useful to obtain phase angles, cross-power spectrum, and coherence between wavelet power spectra at different wavelengths, spatial locations, and/or multi-component spectral measurements. Hence, the phase angles are determined not only as a function of frequency, but also as a function of time. These phase angles are usually demonstrated as small arrows on a wavelet co-spectrum (or wavelet coherence) map, where their directions indicate the phase angles at different time-frequency locations. The convention with which an arrow direction represents, e.g., zero and 90 degrees phase angles (and which lightcurve leads or lags behind) should be specified.

\begin{figure*}[!t]
\begin{center}
\includegraphics[width=0.5\textwidth]{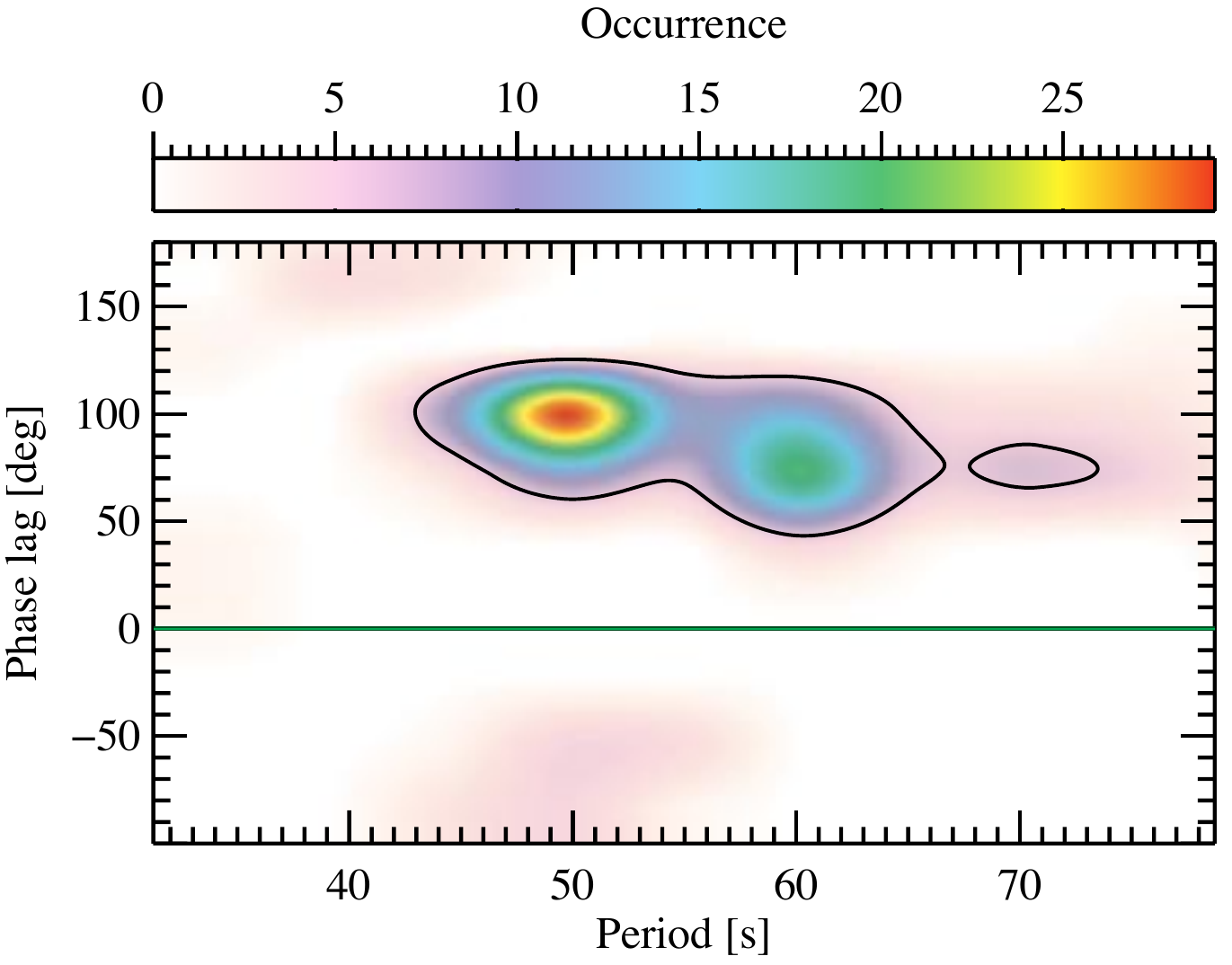}
\end{center}
\caption{Phase diagram of transverse oscillations in 7 small magnetic elements observed in two layers of the lower solar atmosphere (with $\approx450$~km height difference) from SuFI/{\sc Sunrise}. Image reproduced from \cite{2017ApJS..229...10J}.}
\label{fig:SUNRISE_wavelet_phasediagram} 
\end{figure*}

Reproduced from \citet{2017ApJS..229...10J}, the lower- and upper-left panels of Figure~\ref{fig:SUNRISE_wavelet_phaselag} display two wavelet power spectra (from a Morlet wavelet transform) of transverse oscillations in a small magnetic element (marked with circles in Figure~\ref{fig:SUNRISE_09June2009}) at two atmospheric heights sampled by the SuFI/{\sc Sunrise} 300~nm and Ca~{\sc ii}~H bands (with an average height difference of $\approx450$~km), respectively. Islands of high power, particularly those marked by the 95\% confidence level contours, are evident in both wavelet power spectra. The wavelet co-spectrum and coherence maps of these two power spectra are shown in the upper- and lower-right panels of Figure~\ref{fig:SUNRISE_wavelet_phaselag}, respectively. The phase-lag arrows are over plotted on the entire cross-power spectrum, while the same arrows are depicted on the latter map only where the coherence exceeds 0.7. Here, the arrows pointing right represent in-phase oscillations and those pointing straight up identify 90 degrees phase lags where the oscillations in 300~nm lag behind those observed in the Ca~{\sc ii}~H time series. Note here the changes of phase lags from one time-frequency region to another, particularly, in regions with confidence levels larger than 95\%, and/or areas with coherence exceeding 0.7 (or 0.8). However, most of the arrows point upwards (with different angles) in this example, implying an upward wave propagation in the lower solar atmosphere (i.e., from the low photosphere, sampled by the 300~nm band, to the heights corresponding to the temperature minimum/low chromosphere, sampled by the Ca~{\sc ii}~H images). A slight downward propagation is also observed in a small area. These may associate to various wave modes and/or oppositely propagating waves at different frequencies and times. We note that such phase changes with time could not be identified using a Fourier phase lag analysis (see Section~\ref{sec:fourierphaselaganalysis}), where phase angles are computed as a function of frequency only.

Whether the cross-power spectrum or coherence should be used for the wave identification greatly depends on the science and the types of data employed. While the co-spectrum (which is obtained through multiplying the wavelet power spectrum of a time series by the complex conjugate of the other) identifies regions with large power in common (between the two time series), the coherence (i.e., square of the cross-spectrum normalized by the individual power spectra; \citealt{2004NPGeo..11..561G}) highlights areas where the two time series co-move, but not necessarily sharing a high common power. An example is the area around the time and period of 70 and 47~s, respectively, that is associated to a coherence level exceeding 0.8 (and within the 95\% confidence levels), but with no significant power in the co-spectrum (only one of the power spectra, i.e., that from the Ca~{\sc ii}~H data, show large power at that time and period location).

As a working example, from the right panels of Figure~\ref{fig:SUNRISE_wavelet_phaselag}, the phase lag at the time and period of 75 and 41~s, respectively, reads about 140~degrees, which is translated to a time lag of $\approx16$~s. Given the average height difference of 450~km between the two atmospheric layers, it results in a wave propagation speed of $\approx28$~km/s (due to the transverse oscillations in the small-scale magnetic element). A similar analysis for intensity oscillations in the same small-scale magnetic element has also been presented in \citet{2017ApJS..229...10J}. Of course, as highlighted in Section~{\ref{sec:fourierphaselaganalysis}}, phase measurements are always subject to an associated uncertainty of $\pm$360$^{\circ}$ ($\pm2\pi$), which arises via phase wrapping. As a consequence, to alleviate ambiguities in phase angles, in addition to subsequently derived phase velocities, care must be taken to select observational time series where the atmospheric height separation is not too substantial (see Section~{\ref{sec:fourierphaselaganalysis}} for more discussion), which helps to minimize the ambiguities associated with phase wrapping.

Depending on science objectives, it may be helpful to inspect the variation of phase lags with frequency (or period). To this end, a statistical phase diagram can be created, where all reliable phase angles (e.g., those associated to power significant at 5\%, and/or with a coherence exceeding 0.8) are plotted as a function of frequency \citep{2012ApJ...746..183J}. Such a phase diagram can provide information about the overall wave propagation in, e.g., similar magnetic structures. Figure~\ref{fig:SUNRISE_wavelet_phasediagram} illustrates a phase diagram (i,e, a 2D histogram of phase angle versus period; from \citealt{2017ApJS..229...10J}) constructed from all the reliable phase angles obtained from the transverse oscillations in 7 small magnetic elements, similar to that discussed above. The background colors represent the occurrence frequency and the contours mark regions which are statistically significant (i.e., compared to the extreme outliers). From this phase diagram, it is evident that the upward propagating waves (i.e., the positive phase angles in the convention introduced here) appear preferential.

\begin{figure*}[!t]
\begin{center}
\includegraphics[width=0.45\textwidth]{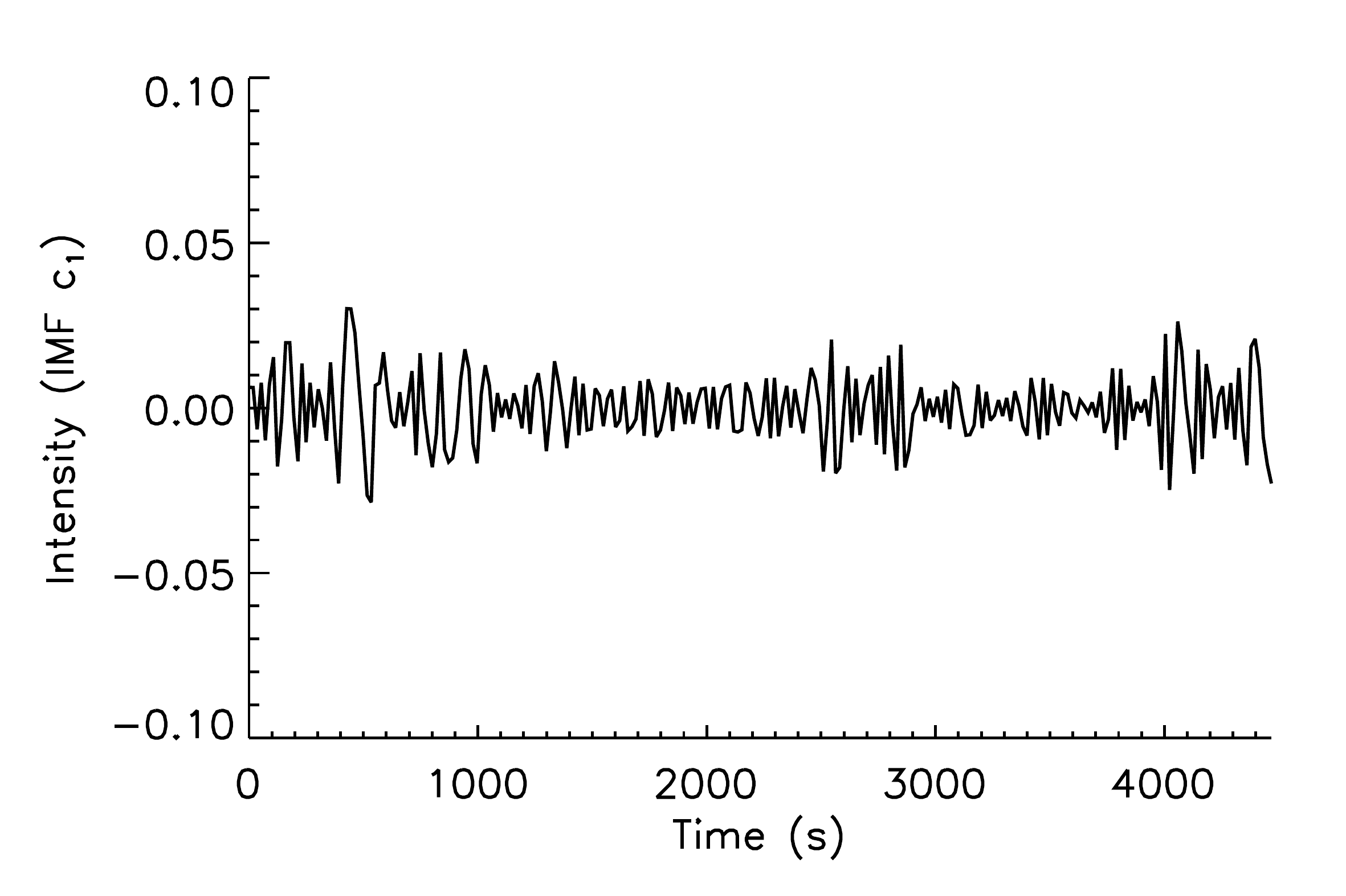}
\includegraphics[width=0.45\textwidth]{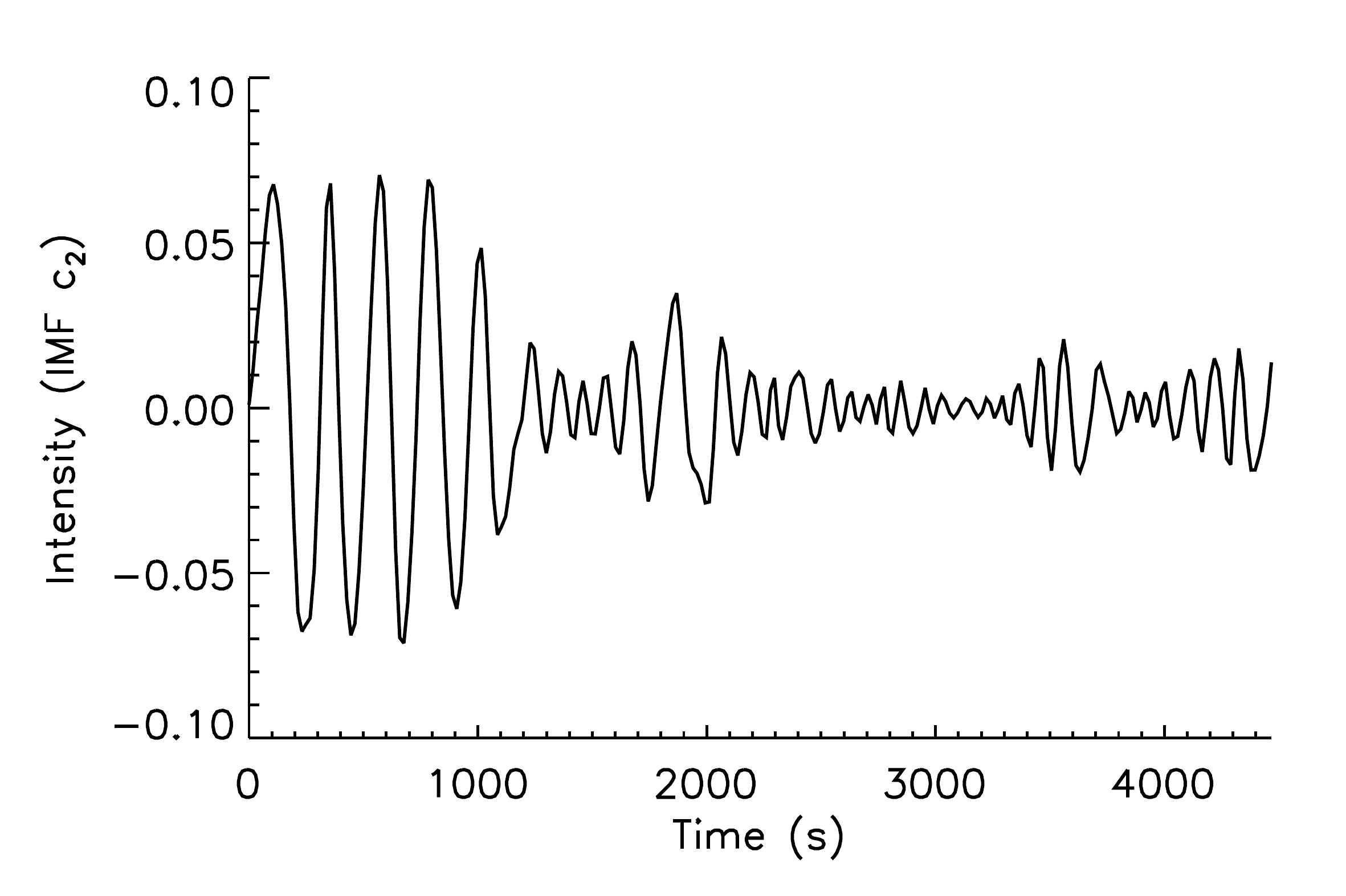}
\includegraphics[width=0.45\textwidth]{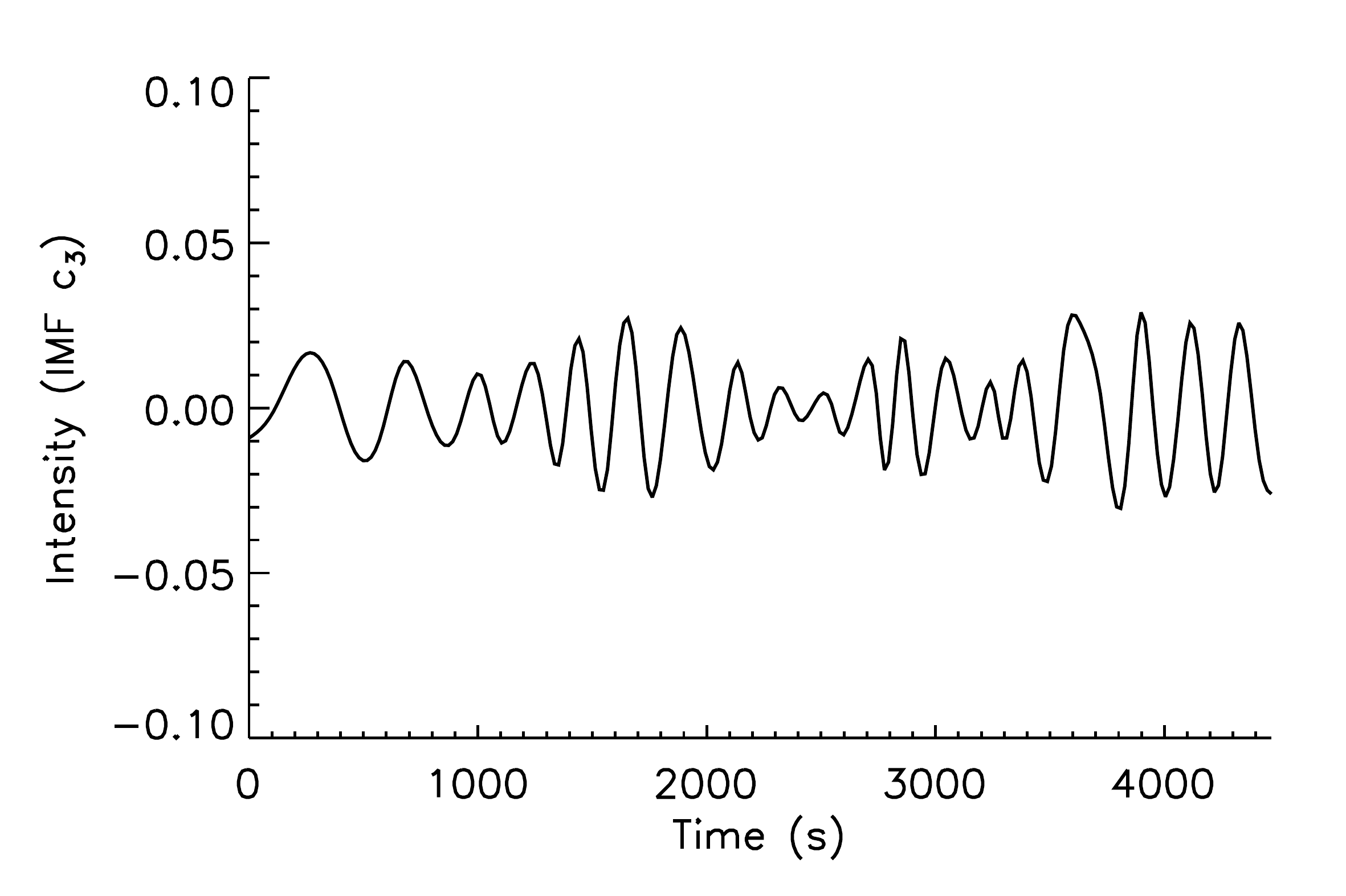}
\includegraphics[width=0.45\textwidth]{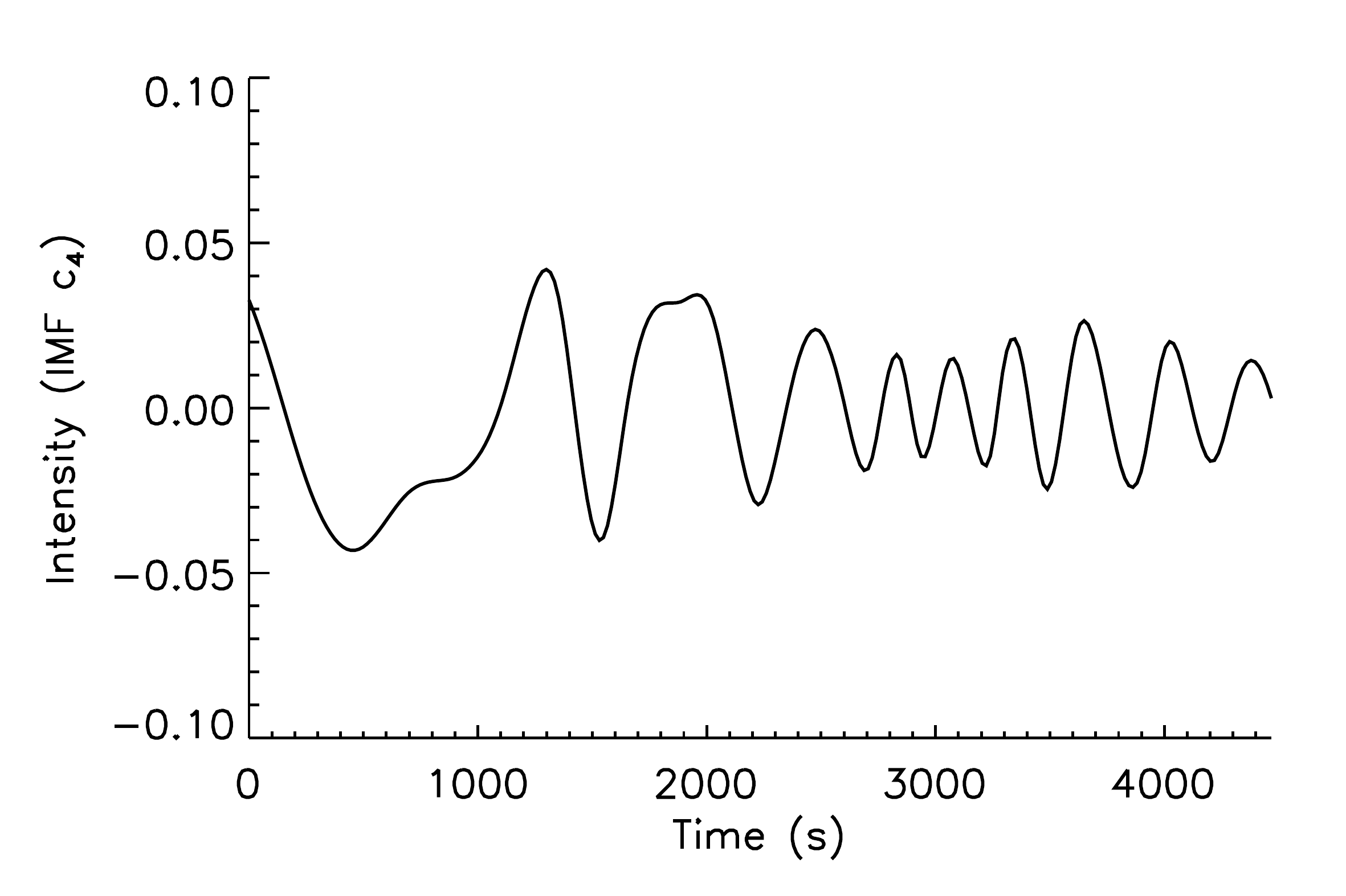}
\includegraphics[width=0.45\textwidth]{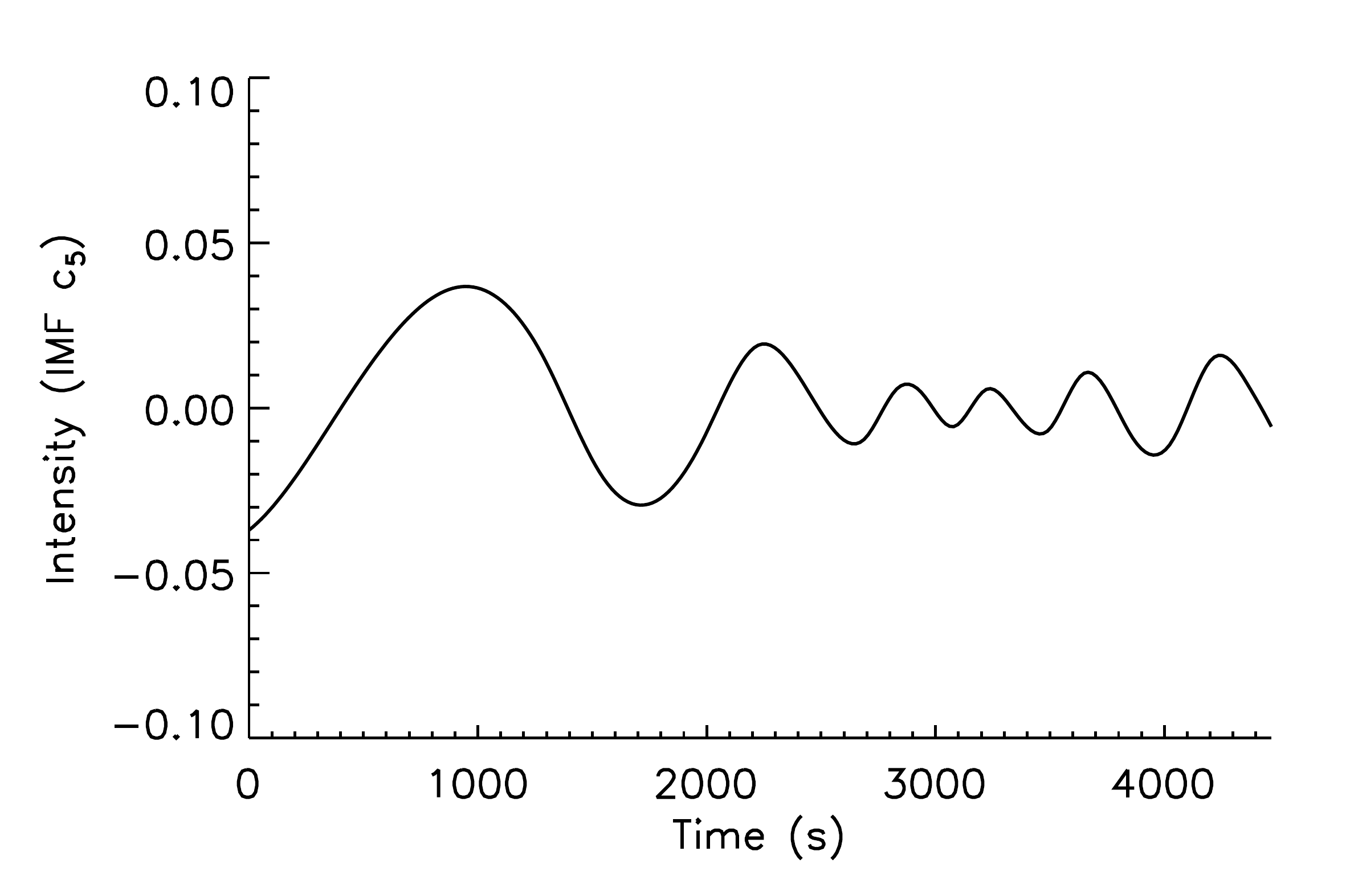}
\includegraphics[width=0.45\textwidth]{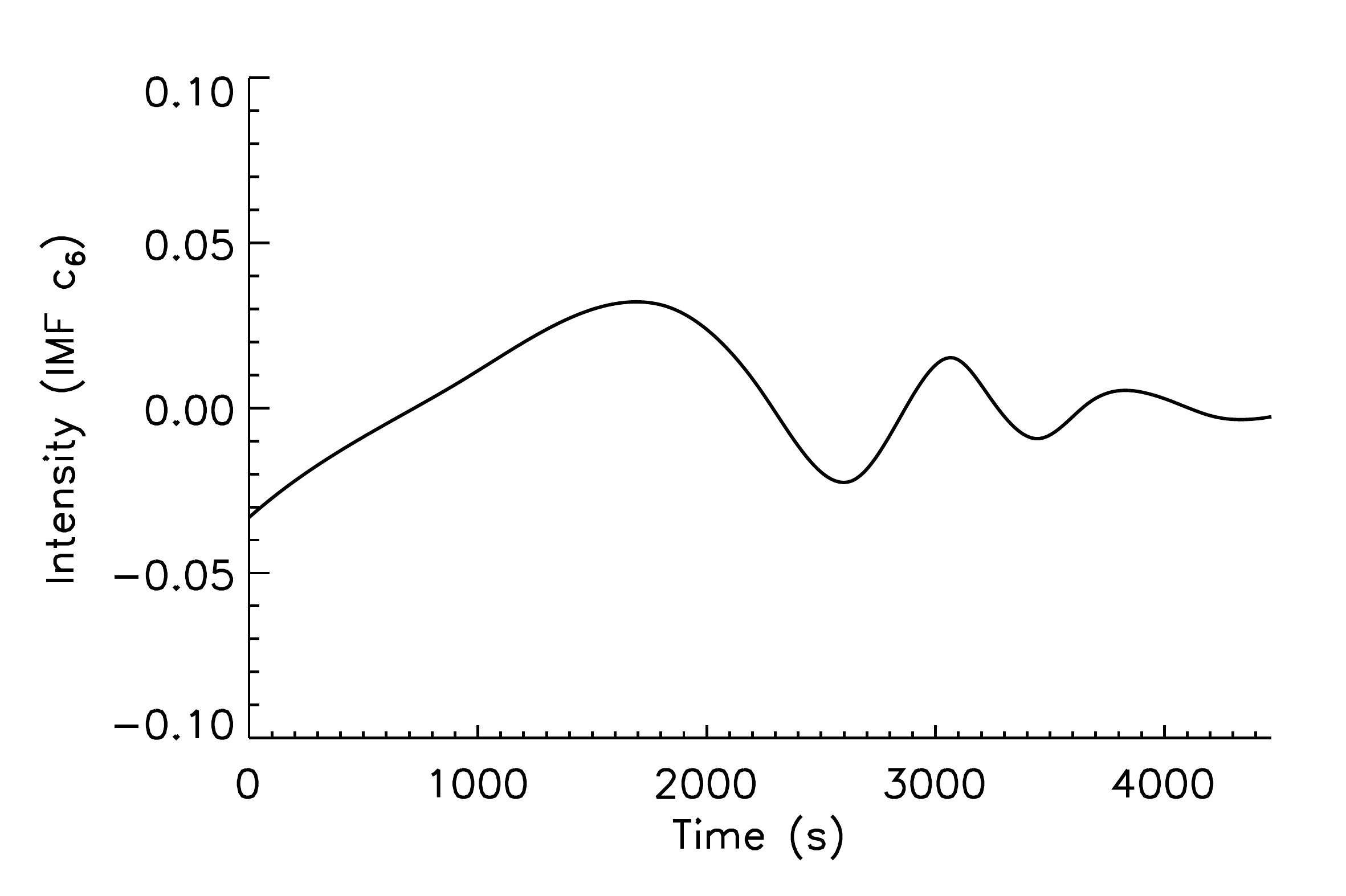}
\includegraphics[width=0.45\textwidth]{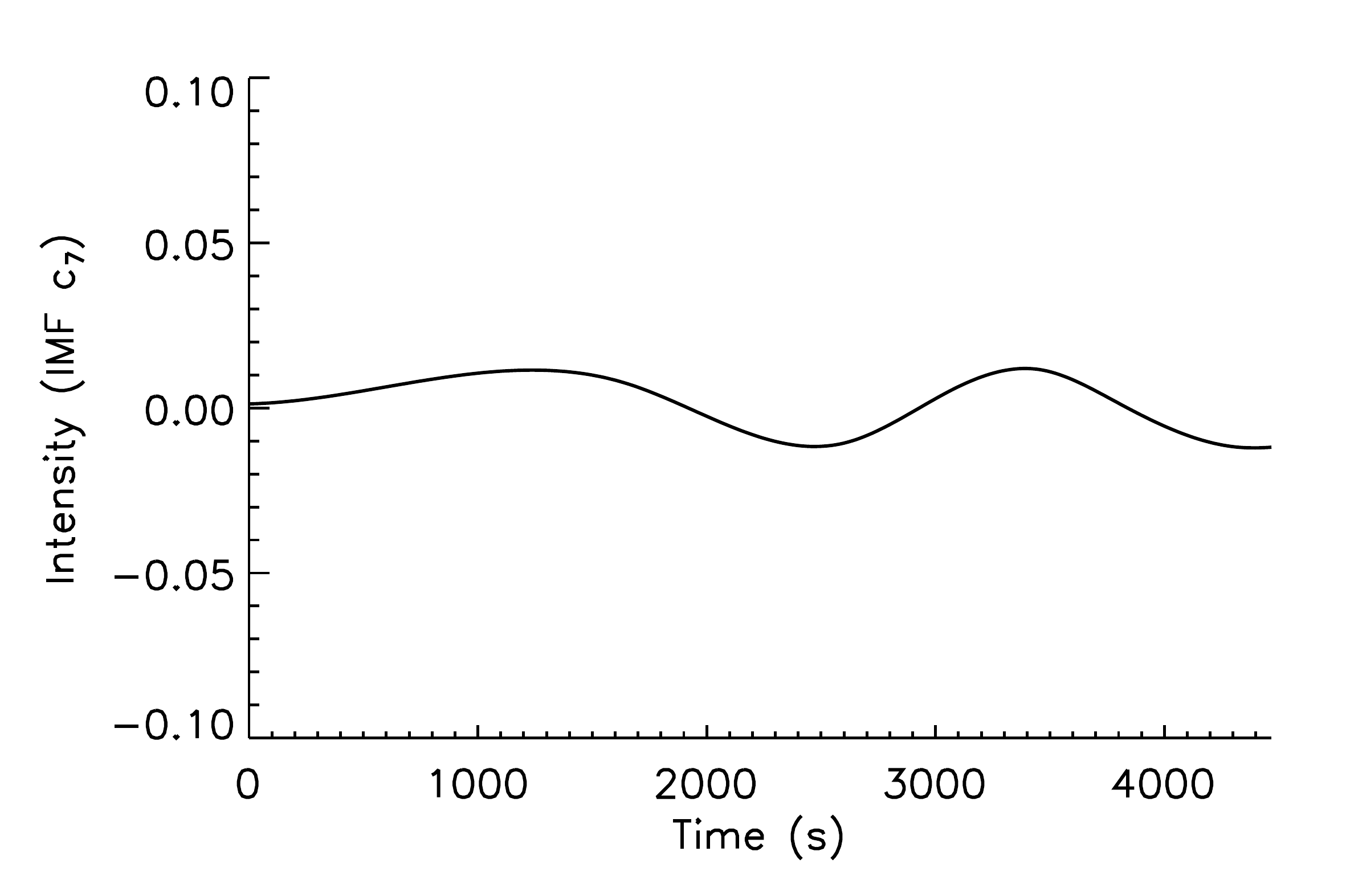}
\includegraphics[width=0.45\textwidth]{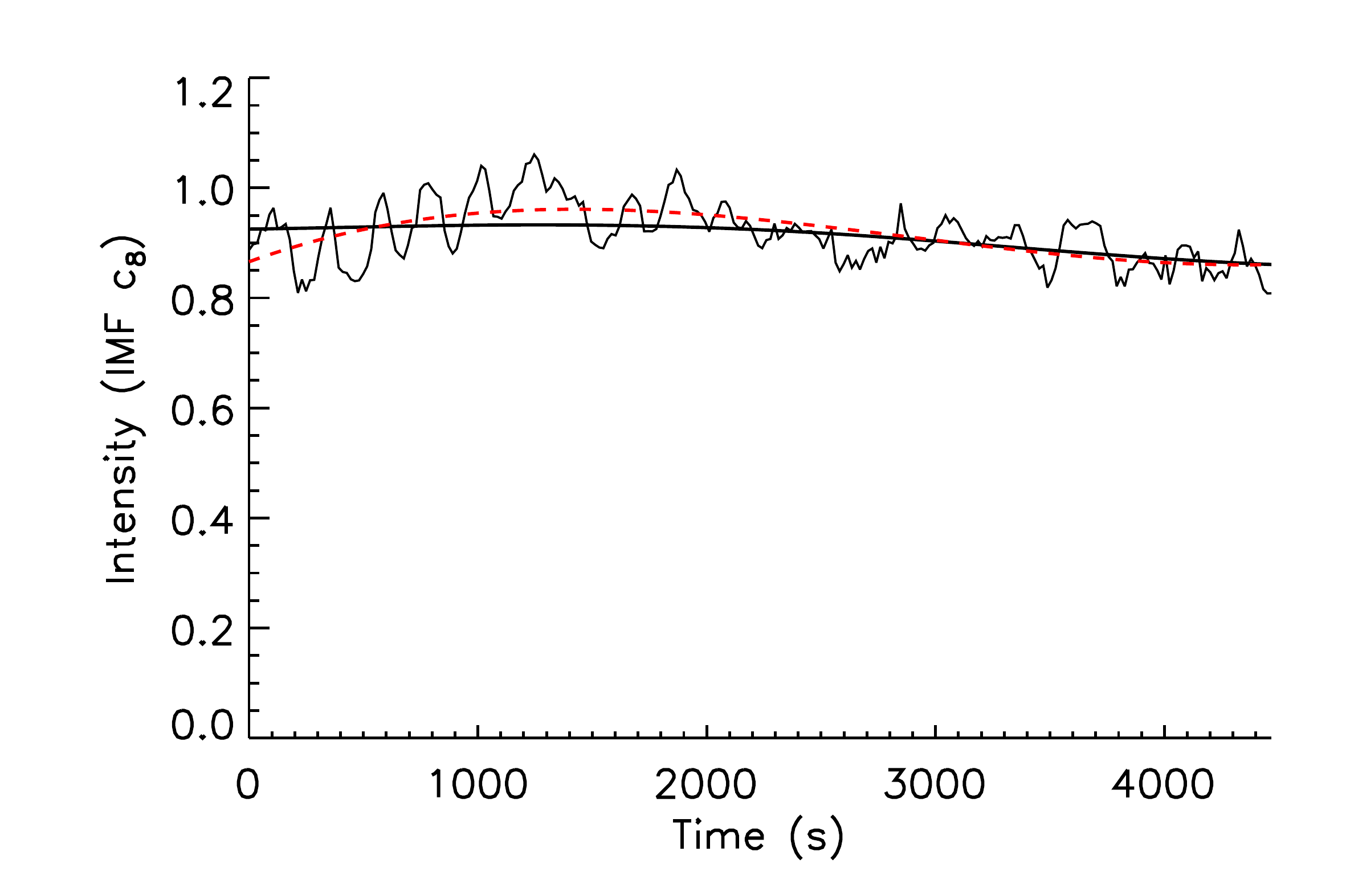}
\end{center}
\caption{IMFs $c_{1} \rightarrow c_{8}$, extracted from the original (non-detrended and non-apodized) HARDcam H$\alpha$ time series overplotted in the lower-right panel. In addition, the lower-right panel also shows the polynomial best-fit line (dashed red line) used to detrend the data prior to FFT/wavelet analyses. It can be seen that the longest period fluctuations making up IMF $c_{8}$ are similar to the global trend line calculated in Section~{\ref{sec:1Dfourieranalysis}}.Note that a summation of IMFs $c_{1} \rightarrow c_{8}$ will return the original signal.}
\label{fig:EMD} 
\end{figure*}

\vspace{5mm}
\subsection{Empirical Mode Decomposition}
\label{sec:EMD}
Empirical Mode Decomposition \citep[EMD;][]{1998RSPSA.454..903H, 1999AnRFM..31..417H} is a statistical tool developed to decompose an input time series into a set of intrinsic timescales. Importantly, EMD is a contrasting approach to traditional FFT/wavelet analyses since it relies on an empirical approach rather than strict theoretical tools to decompose the input data. Due to the decomposition being based on the local characteristic timescales of the data, it may be applied to non-linear and non-stationary processes without the detrending often applied before the application of Fourier-based techniques (i.e., under the assumption that such detrending is able to accurately characterize any non-stationary and/or non-periodic fluctuations in the time series with a low-order polynomial). As such, it is possible for EMD to overcome some of the limitations of FFT/wavelet analyses, including aspects of wave energy leakage across multiple harmonic frequencies \citep{2004ApJ...614..435T}. 
non-stationary/non-period fluctuations that can be characterized by a low-order polynomial

\begin{figure*}[!t]
\begin{center}
\includegraphics[width=\textwidth]{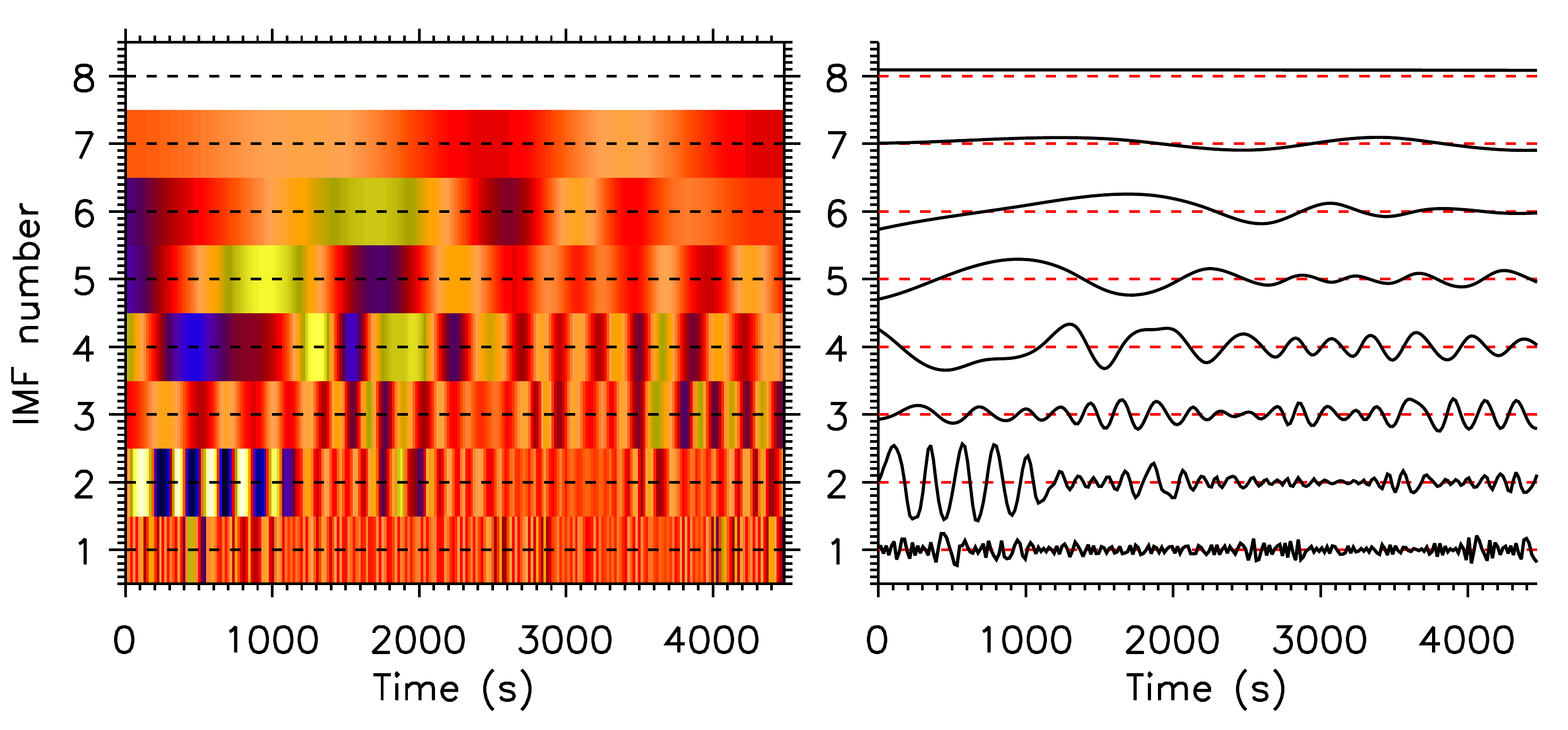}
\end{center}
\caption{IMFs $c_{1} \rightarrow c_{8}$ displayed as a two-dimensional map (left), where yellow and blue colors represent the peaks and troughs, respectively, of the IMF intensity fluctuations. The horizontal dashed black lines represent cuts through each IMF, with the corresponding intensity time series displayed in the right panel. The horizontal dashed red lines represent the zero value corresponding to each IMF.}
\label{fig:EMDmap} 
\end{figure*}

Following the methodology described by \citet{2004ApJ...614..435T}, we apply EMD techniques to the HARDcam H$\alpha$ time series depicted in the upper-left panel of Figure~{\ref{fig:HARDcam_lightcurves}}. To begin, extrema in the lightcurve are identified, and are then connected by a cubic spline fit to provide an upper envelope of the positive intensity fluctuations (i.e., fluctuations above the mean). Next, the same process is applied to find the lower envelope corresponding to negative intensity fluctuations (i.e., fluctuations below the mean). The mean value between the upper and lower envelopes, at each time step, is denoted $m_{1}(t)$. The difference between the original input data and the mean function is called the first component, $h_{1}(t)$. Providing the input time series contains no undershoots, overshoots, and/or riding waves \citep{1998RSPSA.454..903H}, then the first intrinsic mode function (IMF) is equal to $h_{1}(t)$. 

Unfortunately, many input time series contain signal blemishes, and removal of the first component, $h_{1}(t)$, from the original lightcurve will generate additional (false) extrema. Hence, to mitigate against these potential issues, the above procedure is repeated numerous times until the first true IMF is constructed \citep[see][for more information]{1998RSPSA.454..903H}. The first IMF constructed, $c_{1}(t)$, is comprised of the most rapid fluctuations of the signal. This can then be subtracted from the original time series, producing a residual lightcurve made up of longer duration fluctuations. The process can subsequently be repeated numerous times to extract additional IMFs until the amplitude of the residual lightcurve falls below a predetermined value, or becomes a function from which no more IMFs can be extracted \citep{2004ApJ...614..435T}.

\begin{figure*}[!t]
\begin{center}
\includegraphics[width=\textwidth]{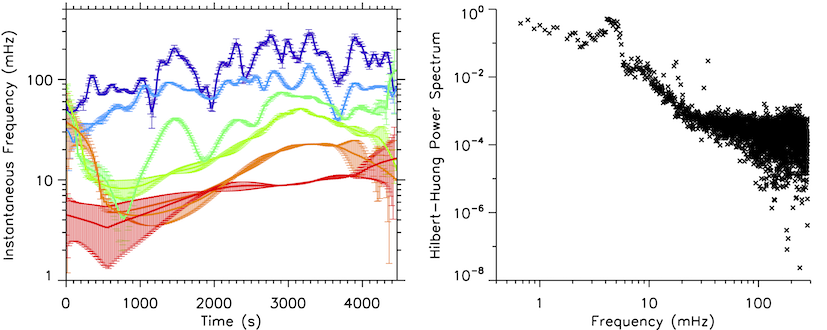}
\end{center}
\caption{Instantaneous frequencies computed from applying a Hilbert-Huang transform to the HARDcam H$\alpha$ lightcurve shown in the lower-right panel of Figure~{\ref{fig:EMD}} and displayed as a function of time (left panel). The solid purple, blue, dark green, light green, orange, and red lines correspond to moving average frequencies (computed over a 30~s window) for the IMFs~${c_{2} \rightarrow c_{7}}$, respectively. The vertical error bars correspond to the standard deviations of frequencies found within the 30~s moving average windows. The right panel displays the corresponding Hilbert-Huang power spectrum, calculated by integrating the instantaneous frequency spectra over time and normalized to the largest power value computed, hence providing a plot of relative changes in spectral energy as a function of frequency. Features within the power spectrum are consistent with the FFT and wavelet outputs shown in Figures~{\ref{fig:FFTprobabilities}} \& {\ref{fig:wavelethardcam}}.}
\label{fig:HHfreqspectrum} 
\end{figure*}

Figure~{\ref{fig:EMD}} shows a collection of IMFs extracted from the HARDcam H$\alpha$ time series depicted in the upper-left panel of Figure~{\ref{fig:HARDcam_lightcurves}}. It is clear that the most rapid fluctuations are present in IMF $c_{1}$, with IMF $c_{8}$ documenting the slowest evolving intensity variations. Plotted on top of IMF $c_{8}$ is the original H$\alpha$ time series, along with the polynomial best-fit line (dashed red line) used to detrend the lightcurve in Section~{\ref{sec:1Dfourieranalysis}} before the application of FFT/wavelet techniques. The global trends highlighted by IMF $c_{8}$ and the polynomial best-fit line are similar, again highlighting the appropriate use of detrending in Section~{\ref{sec:1Dfourieranalysis}}, but now compared with generalized empirical methods. Figure~{\ref{fig:EMDmap}} displays the 8 extracted IMFs in the form of a two-dimensional map, which can often be used to more readily display the corresponding interplay between the various amplitudes and variability timescales. 

\begin{figure*}[]
\begin{center}
\includegraphics[trim = 0cm 0cm 0cm 0cm, clip, width=\textwidth]{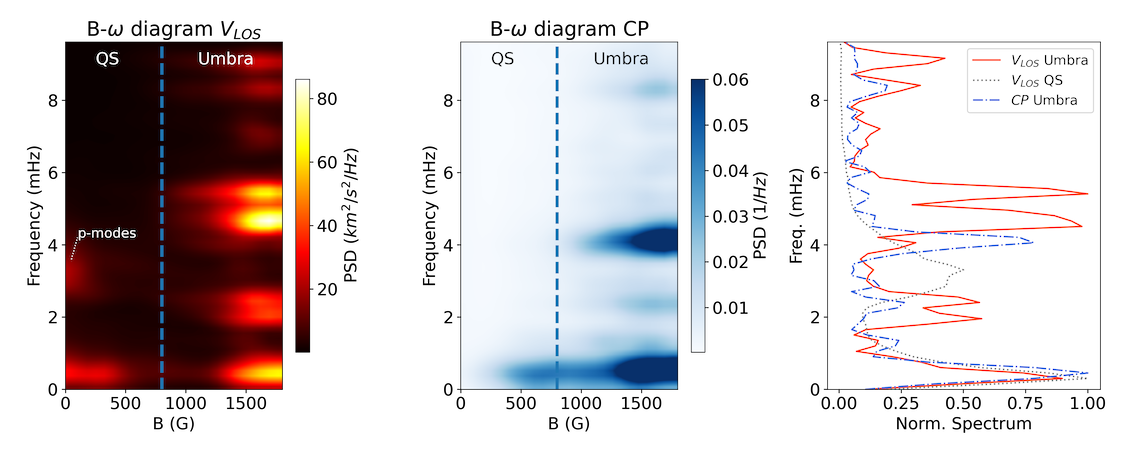}
\end{center}
\caption{Doppler velocity (left) and circular polarization (CP; center) $B$-$\omega$ diagrams of a magnetic pore observed in the photospheric Fe~{\sc{i}}~6173~{\AA} spectral line. The vertical blue dashed lines represents the boundary of the umbral region as inferred from intensity images. The right panel shows the average spectra outside and inside the magnetic umbra of the pore. The five-minute ($p$-mode) oscillations dominate the quiet Sun, but their amplitude is progressively reduced absorbed as one approaches the concentrated magnetic fields of the pore, until a series of eigenmodes are excited within the magnetic tube itself. Image reproduced from \citet{Stangalini2021_bomega}.}
\label{fig:BOmega} 
\end{figure*}

Once the IMFs have been extracted from the input time series, it is possible to employ Hilbert spectral analysis \citep{1998RSPSA.454..903H, 10.5555/1795494} to examine the instantaneous frequencies with time for each IMF. The combined application of EMD and Hilbert spectral analysis is often referred to as the Hilbert-Huang transformation \citep{2008RvGeo..46.2006H}. From the outputs of the Hilbert-Huang transformation, it is possible to display the instantaneous frequencies for each of the extracted IMFs as a function of time. The left panel of Figure~{\ref{fig:HHfreqspectrum}} displays the instantaneous frequencies corresponding to IMFs $c_{2} \rightarrow c_{7}$ using the purple, blue, dark green, light green, orange, and red lines, respectively. IMFs~$c_{1}$ and $c_{8}$ have been removed from the plot as these correspond to very high and low frequency fluctuations, respectively, which clutter the figure if included. The solid colored lines represent the running mean values of the instantaneous frequencies (calculated over a 30~s window), while the vertical colored error bars indicate the standard deviations of the frequency fluctuations found within the running mean sampling timescale. As already shown in Figures~{\ref{fig:EMD}} \& {\ref{fig:EMDmap}}, the frequencies associated with higher IMFs are naturally lower as a result of the residual time series containing less rapid fluctuations. It can be seen the in left panel of Figure~{\ref{fig:HHfreqspectrum}} that IMF~$c_{2}$ contains frequencies in the range of $50-300$~mHz ($3-20$~s), while IMF~$c_{7}$ displays lower frequencies spanning $1-30$~mHz ($33-1000$~s). We must note that the left panel of Figure~{\ref{fig:HHfreqspectrum}} is simply a representation of the instantaneous frequencies present in the time series as a function of time and does not contain information related to their relative importance (e.g., their respective amplitudes), although this information is indeed present in the overall Hilbert-Huang transform.

Finally, it is possible to integrate the instantaneous frequency spectra (including their relative amplitudes) across time, producing the Hilbert-Huang power spectrum shown in the right panel of Figure~{\ref{fig:HHfreqspectrum}}. The features of the Hilbert-Huang power spectrum are very similar to those depicted in the FFT spectrum shown in the right panel of Figure~{\ref{fig:FFTprobabilities}}. Notably, there is a pronounced power enhancement at $\approx4.7$~mHz, which is consistent with both the FFT power peaks (right panel of Figure~{\ref{fig:FFTprobabilities}}) and the heightened wave amplitudes found at $\approx210$~s in the wavelet transform shown in the bottom panel of Figure~{\ref{fig:wavelethardcam}}. This shows the consistency between FFT, wavelet, and EMD approaches, especially when visible wave activity is evident.

\vspace{5mm}
\subsection{Proper Orthogonal Decomposition and Dynamic Mode Decomposition}
\label{sec:PODDMD}

\begin{figure}[t!]
 \centering
\includegraphics[width=11cm]{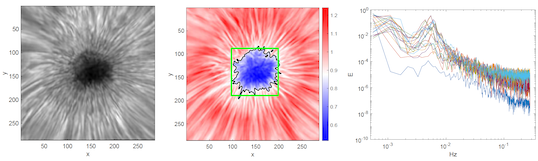}

\caption{The left panel shows a sunspot from \citet{2017ApJ...842...59J} in H$\alpha$ intensity using data from HARDcam (one pixel has a width of 0.138''). The middle panel shows the mean intensity of the time series, the colourbar displays the magnitude of the mean time series, the solid black line shows umbra/penumbra boundary (intensity threshold level 0.85) and the green box (101 $\times$ 101 pixels) shows the region where \citet{2021RSPTA.37900181A} applied POD and DMD. The right panel displays the PSD of the time coefficients of the first 20 POD modes (in log scale). The PSD shows peaks between frequencies $4.3-6.5$~mHz (corresponding to periods of $153-232$~s). Image adapted from \citet{2021RSPTA.37900181A}.} 
\label{fig:HARDcam_POD}
 \end{figure}

Recently, for the first time, \citet{2021RSPTA.37900181A} and \citet{2022ApJ...927..201A} applied the methods of Proper Orthogonal Decomposition \citep[POD; see e.g.,][]{doi:10.1080/14786440109462720,lumley1967} and Dynamic Mode Decomposition \citep[DMD; see e.g.,][]{schmid_2010} to identify MHD wave modes in a sunspot umbrae. The POD method defines the eigenfunctions to be orthogonal in space but places no constraints on their temporal behaviour. On the other hand, DMD puts no constraints on the spatial structure of the eigenfunctions but defines them to be orthogonal in time, meaning that each DMD mode has a distinct frequency. Hence, POD modes are permitted to have broadband frequency spectra but DMD modes are not. This is shown in the right panel of Figure~\ref{fig:HARDcam_POD}, which shows a broadband power spectral density (PSD) of 8 POD modes detected in a sunspot umbra by \citet{2021RSPTA.37900181A} using HARDcam H$\alpha$ intensity observations from \citet{2017ApJ...842...59J}.

\begin{figure}[t!]
  \centering
  \begin{tabular}{ccc}
    
    \includegraphics[trim=1mm 0mm 0mm 0mm, clip, width=0.31\textwidth]{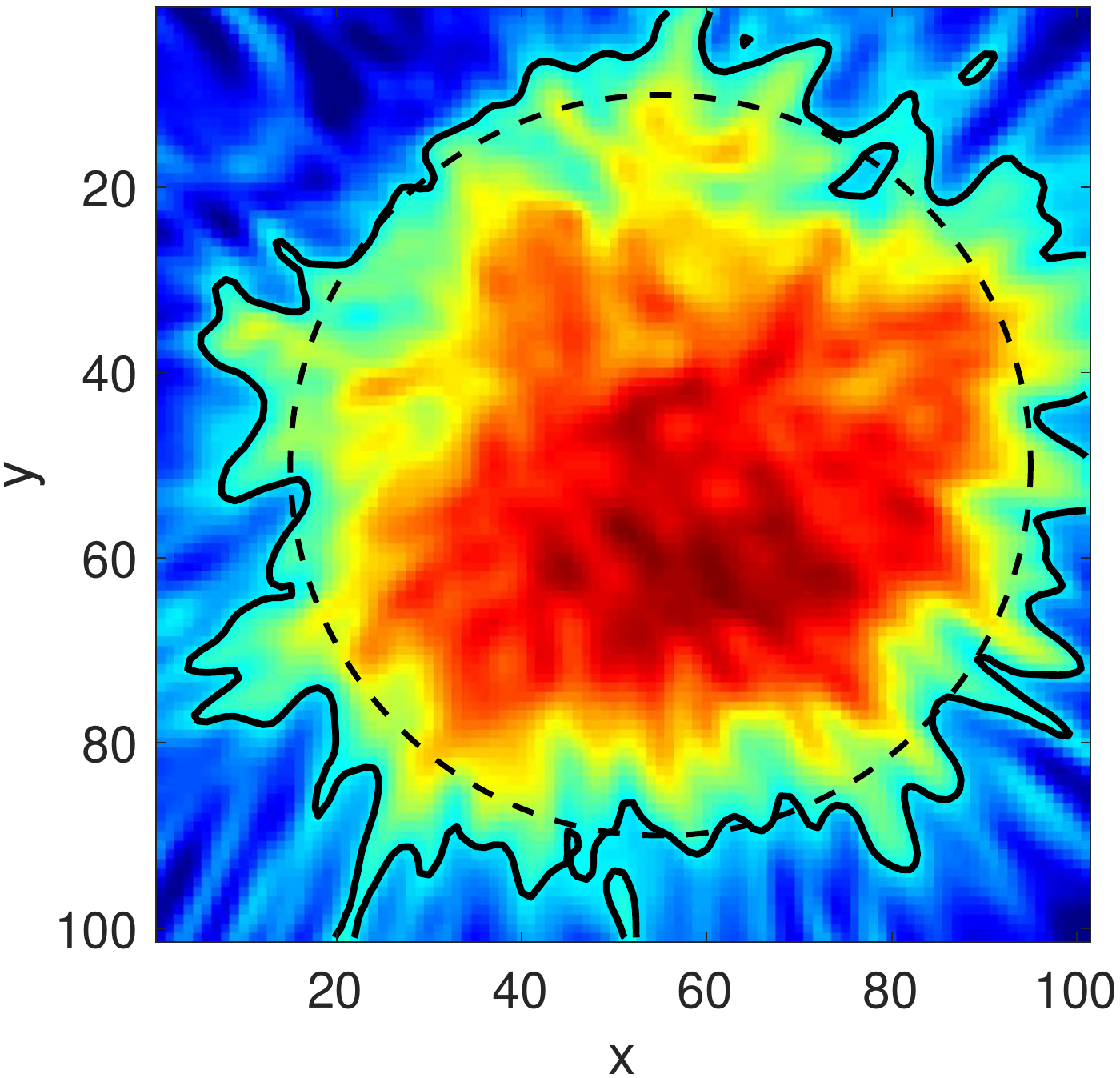}&

     \includegraphics[trim=1mm 0mm 0mm 0mm, clip, width=0.31\textwidth]{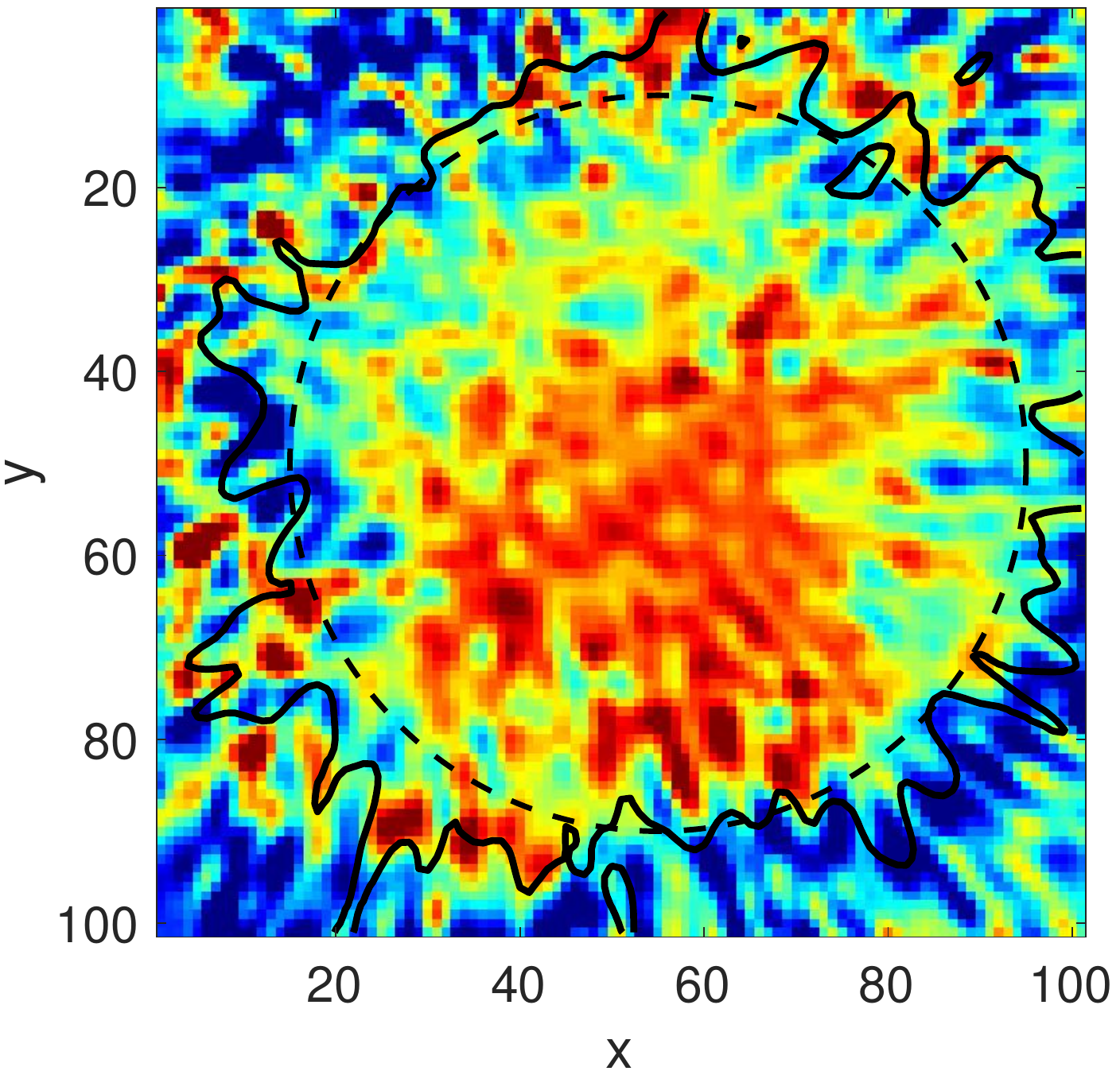}&

     \includegraphics[trim=1mm 0mm 0mm 0mm, clip, width=0.35\textwidth]{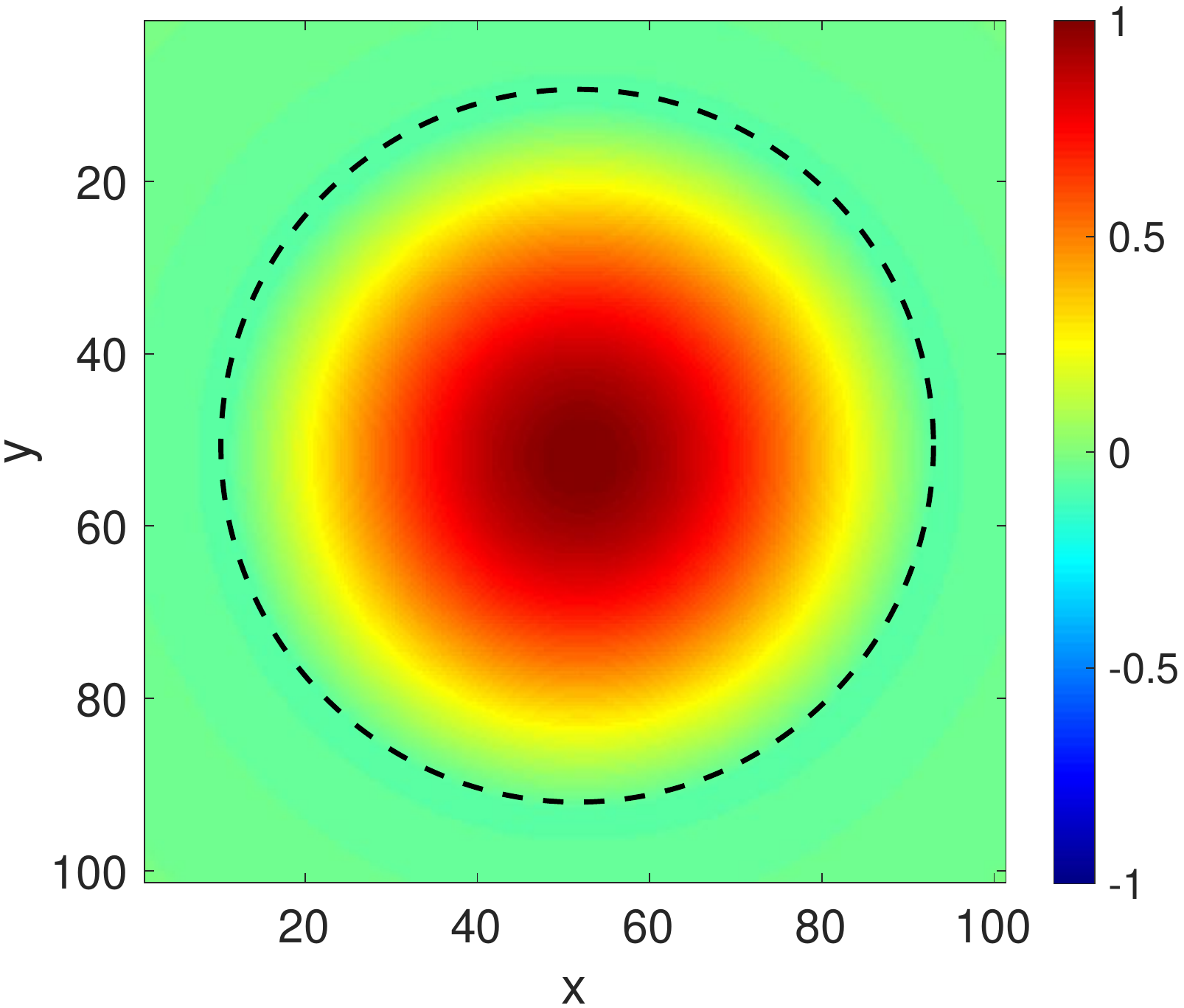}\\

      \includegraphics[trim=1mm 0mm 0mm 0mm, clip, width=0.31\textwidth]{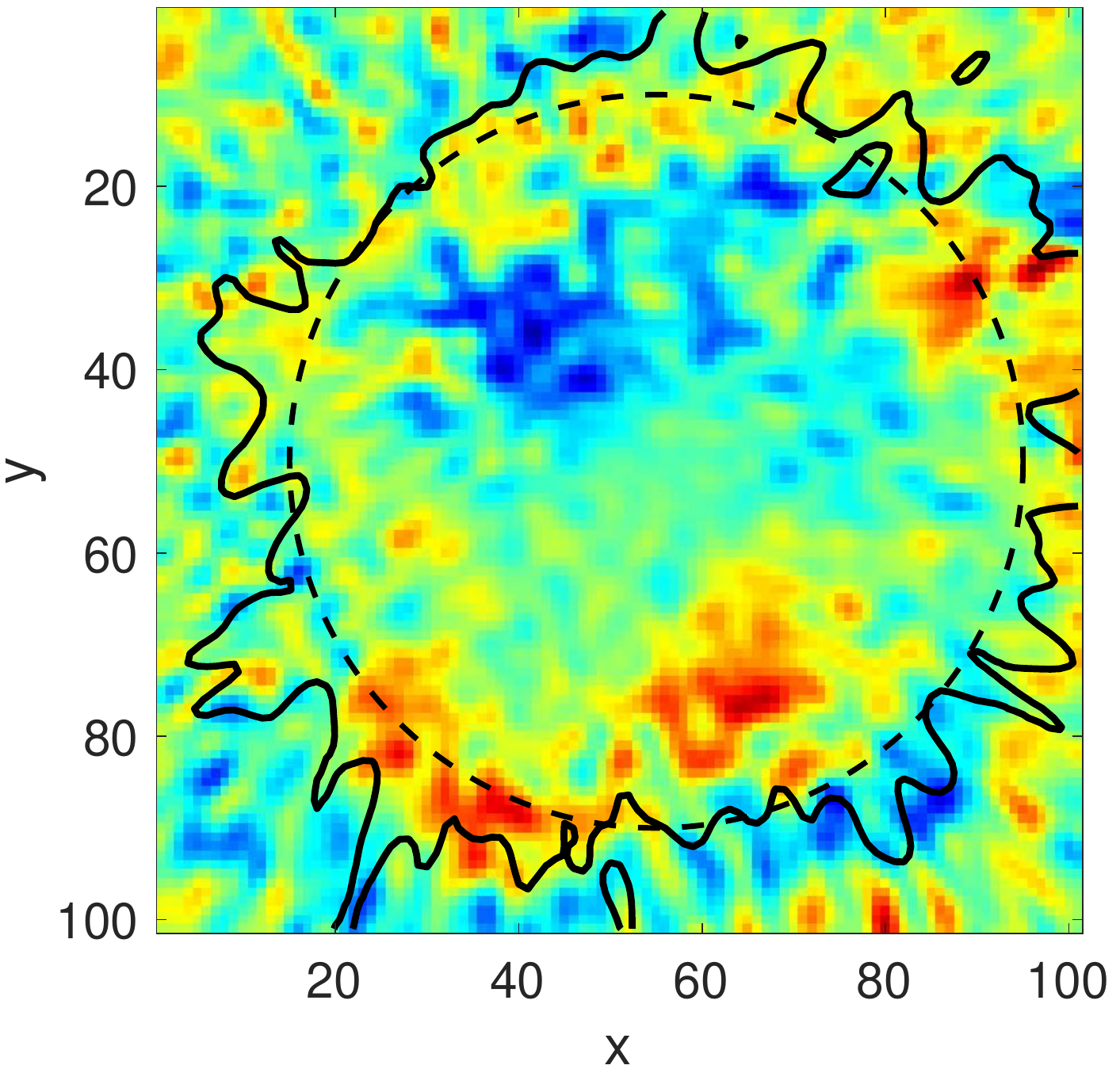}&
    
     \includegraphics[trim=1mm 0mm 0mm 0mm, clip, width=0.31\textwidth]{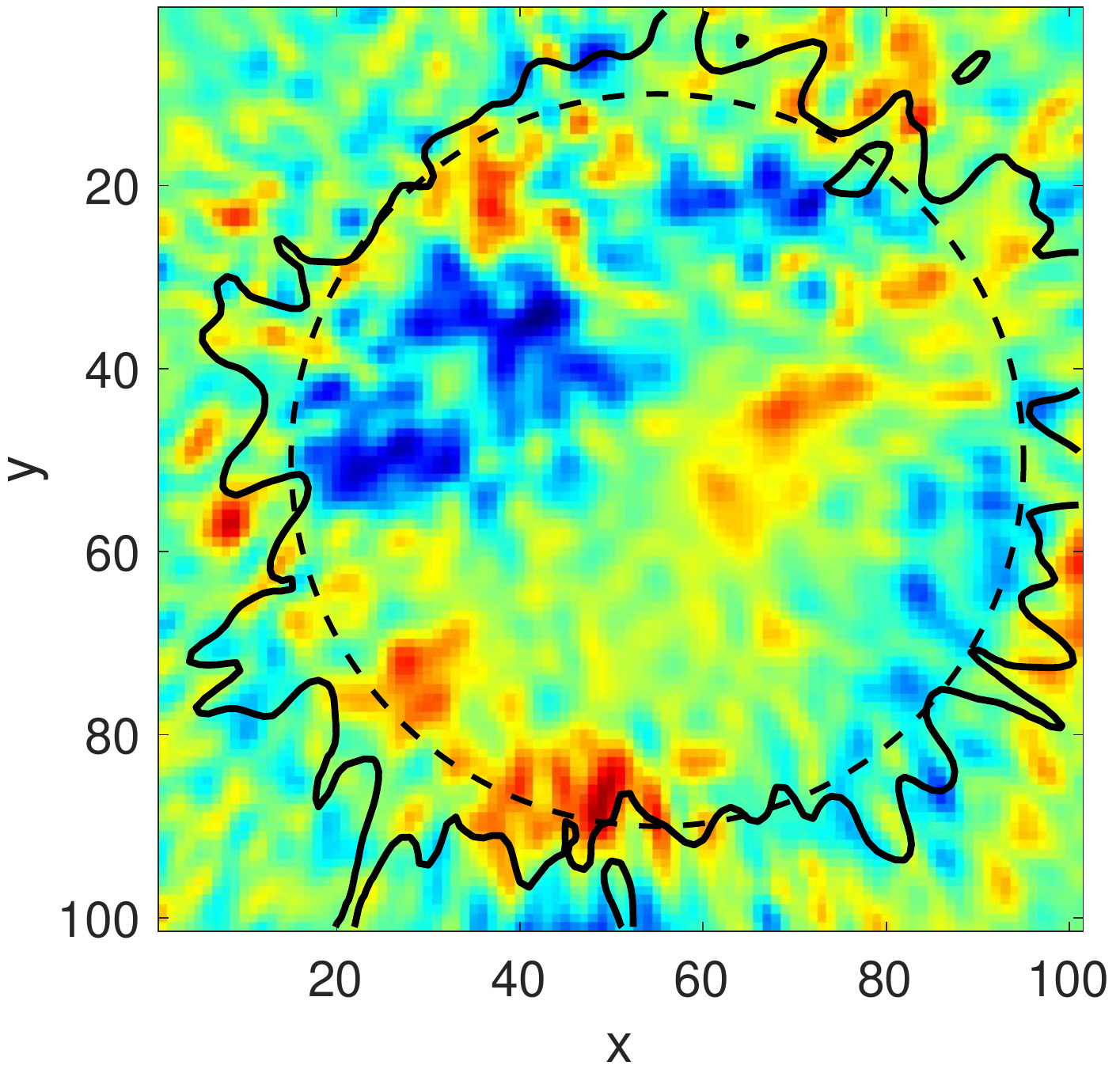}&

     \includegraphics[trim=1mm 0mm 0mm 0mm, clip, width=0.35\textwidth]{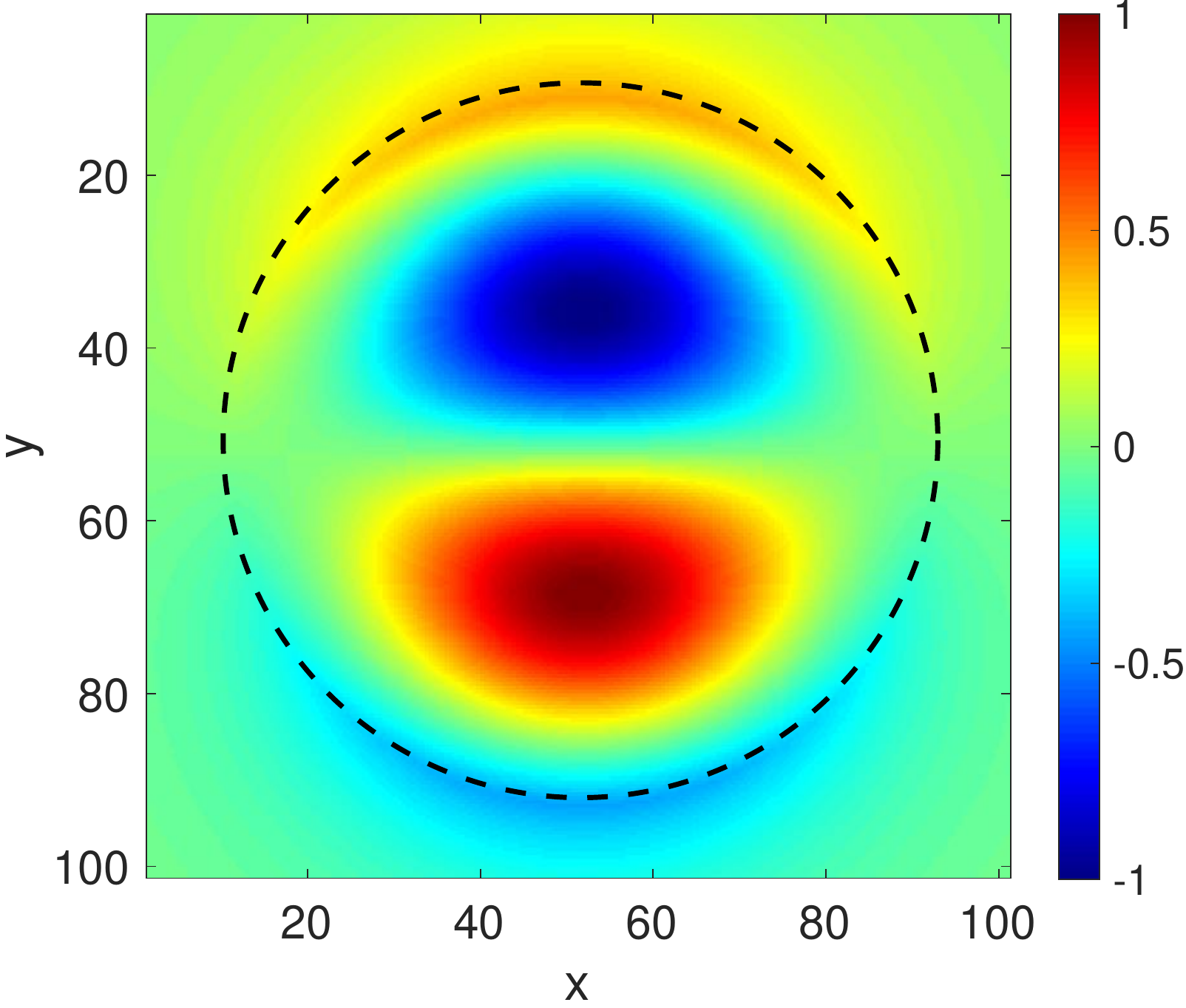}\\
\end{tabular}
\caption{The top and bottom rows show snapshots of the slow body sausage and kink modes, respectively. From left to right, the columns show the POD and DMD modes from HARDcam H$\alpha$ intensity observations of a sunspot \citep{2017ApJ...842...59J}, then the corresponding magnetic cylinder model modes. As shown in the color bar, the intensity oscillations are normalized between $-1$ and $1$, hence the blue and red regions are in anti-phase. The methods of POD and DMD provide a most promising approach to decompose MHD wave modes in pores and sunspots, even if their cross-sectional shapes are much more irregular than this example. Image adapted from \citet{2021RSPTA.37900181A}.}
\label{fig:POD_DMD}
\end{figure}

Both the POD and DMD produce 2D eigenfunctions as shown in the left and middle columns of  Figure \ref{fig:POD_DMD}, however, they achieve this using different approaches. Essentially, DMD identifies the spatial modes which best fit a constant sinusoidal behavior in time, as with a Fourier transform. POD ranks the spatial modes in order of contribution to the total variance, which DMD cannot do.

Since POD can produce as many modes as there are time snapshots, the challenge is to identify which modes are physical and which are not. Similarly, not all DMD modes may be physical. For practical purposes a physical model, such as the magnetic cylinder model (see Section~\ref{sec:mag cylinder} for discussion of MHD wave modes of a magnetic cylinder) can be used to select POD and DMD modes which most closely correspond to predicted MHD wave modes. For the approximately circular sunspot shown in Figure~\ref{fig:POD_DMD}, the predicted MHD cylinder modes which are in the strongest agreement with the selected POD and DMD modes are shown in the right column. These are the fundamental  slow body sausage (top row) and kink modes (bottom row).

In the case of the magnetic cylinder model, assuming a static background, the eigenmodes, e.g., kink, sausage and fluting, are orthogonal to each in other in space by definition. Furthermore, each mode can have a broadband signal in $\omega$ and $k$ as shown for a real sunspot in the right panel of Figure~\ref{fig:HARDcam_POD}.  Hence, POD can identify such modes in pores and sunspots, providing there is no significant background flow that will break the condition of orthogonality.  Furthermore, if a mode has a dominant power frequency, this can be identified with DMD as well as POD. Indeed, this was done by \citet{2021RSPTA.37900181A} for the 8 POD modes shown in the PSD plot in the right panel of Figure \ref{fig:POD_DMD} which have distinct power peaks between 4.3 mHz and 6.5 mHz. In such cases a combined POD/DMD approach is a most promising avenue for identifying physical modes. However, it must be highlighted, as initially introduced in Section~{\ref{sec:1Dfourieranalysis}}, that the characterization of waves using POD and DMD techniques must be treated with the same caution as traditional FFT approaches. For example, it is essential that the relative amplitudes of each eigenfunction are compared to noise and/or background sources to establish its true significance.

As will be shown in Section~\ref{sec:globaleigenmodes}, POD and DMD methods are especially useful for decomposing MHD wave modes in pores and sunspots of more irregular cross-sectional shapes than the example shown in Figure \ref{fig:POD_DMD}. This is because POD and DMD do not have the limitation of having their eigenfunctions pre-defined as they are with Fourier decomposition, where the basis functions are simply fixed as sinusoids. Even in the standard cylinder model, the eigenfunctions in the radial direction are Bessel functions not sinusoids. Hence, when it comes to identifying the spatial structure of individual MHD wave modes in pores and sunspots, the methods of POD and DMD are more suited to the job than Fourier decomposition. However, Fourier transforming the time coefficient of a POD mode is still necessary to calculate its PSD as shown in the right panel of Figure~{\ref{fig:HARDcam_POD}}.

\subsection{$B$-$\omega$ Diagrams}
\label{sec:Bomega}
Imaging spectropolarimetry offers the additional possibility to study the variations in the wave power spectrum as a function of magnetic flux. To this aim, \citet{Stangalini2021_bomega} have proposed a new visualization technique, called a $B$-$\omega$ diagram (see Figure~\ref{fig:BOmega}), which combines the power spectrum of a particular quantity (e.g., Doppler velocities) with its corresponding magnetic information. In this diagram, each column represents the average power spectrum of pixels within a particular magnetic field strength interval as inferred from polarimetry \citep[e.g., via spectropolarimetric inversions or center-of-gravity methods;][]{rees_line_1979}. The $B$-$\omega$ diagram therefore has the capability to help visualize changes in the oscillatory field as one transitions from quiet Sun pixels outside the magnetic tube to the inner (more concentrated) magnetic region. In Figure~\ref{fig:BOmega} we show an example of $B$-$\omega$ diagram taken from \citet{Stangalini2021_bomega}, which reveals unique wave information for a magnetic pore observed by IBIS in the photospheric Fe~{\sc{i}} 6173~{\AA} spectral line. Here, we clearly see that the the amplitude of five-minute ($\approx$3~mHz) oscillations in the quiet Sun is progressively reduced as one approaches the boundary of the magnetic pore (increasing $B$ values). On the other hand, immediately inside the boundary of the pore (highlighted using a dashed vertical line), a set of spectral features is observed in both Doppler velocity and CP (circular polarization) oscillations (i.e., magnetic field oscillations), which are interpreted as specific eigenmodes of observed magnetic cylinder.

\begin{figure*}[]
\begin{center}
\includegraphics[trim = 1cm 0cm 0cm 0cm, clip, width=\textwidth]{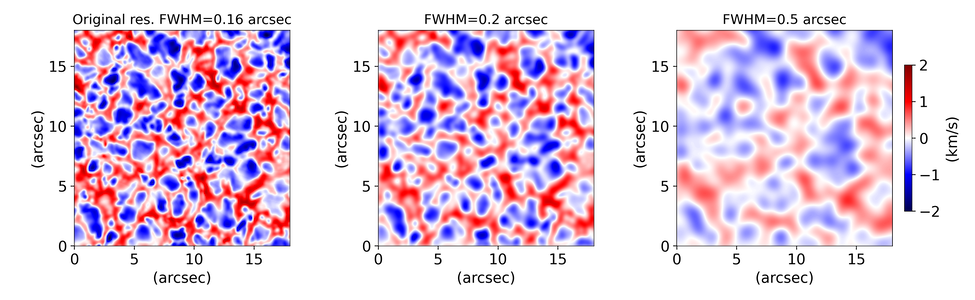}
\end{center}
\caption{Estimated effects of the spatial resolution (i.e., different FWHMs of the instrumental PSF; see Equation~{\ref{eqn:Rayleigh}}) on the observed Doppler velocity field. The original Doppler velocity field observed by the CRISP instrument at the SST in the Fe~{\sc{i}} 6301.5~{\AA} photospheric spectral line (left panel) is convolved with a Gaussian PSF with larger and larger FWHMs to mimic the effects of a lower spatial resolution (middle and right panels). The sign convention employed shows downflows (positive velocities) as red colors and upflows (negative velocities) as blue colors. It can be seen in the middle (${\mathrm{FWHM}}=0.2''$) and right (${\mathrm{FWHM}}=0.5''$) panels that progressively worsening seeing conditions results in lost velocity signals from primarily small-scale features (e.g., intergranular lanes).}
\label{fig:image_degradation} 
\end{figure*}

\subsection{Effects of Spatial Resolution}
\label{sec:spatial_resolution}
The solar atmosphere is highly structured, presenting features across a wide range of spatial scales down to the resolution limit of current instrumentation. Oscillations can be localized at particular spatial scales/features (see, e.g., the discussion in Section~{\ref{sec:globaleigenmodes}}). This means that, for instance, the Doppler velocity, or indeed any other diagnostic, is the average within the resolution angle of the observations. For this reason, the signal itself and its inherent temporal oscillations associated with features below (or close to) the resolution limit can be underestimated \citep{2021RSPTA.37900171M}. 

To illustrate this effect, we consider a case study based on CRISP observations acquired at the SST of a quiet Sun region, which were previously deconvolved using the MOMFBD code \citep{2005SoPh..228..191V} to reduce the effects of residual image aberrations. Here, for the seek of simplicity we consider the starting data as ``perfect data'' for the only purpose of illustrating the effects of spatial resolution of the final power spectra of the oscillations. In the left panel of Figure~\ref{fig:image_degradation} we show the original instantaneous Doppler velocity field obtained from the Fe~{\sc{i}} 6301.5~{\AA} photospheric spectral line. In order to mimic the effect of a lower spatial resolution, we convolve this data using a point spread function (PSF), assumed here to be Gaussian, with a larger full-width at half-maximum (FWHM). In order to simplify the process, we do not consider the effects of residual seeing aberrations present in the original and convolved images. Therefore, our PSF model only considers the effect of the instrumental PSF, which can be represented by the Houston diffraction limited criterion \citep{1927PhRv...29..478H}, 
\begin{equation}
\label{eqn:Rayleigh}
    {\mathrm{FWHM}} = \frac{1.03\lambda}{D} \ ,
\end{equation}
where $\lambda$ is the observed wavelength and $D$ is the diameter of the telescope. Local seeing effects in ground-based observations can further reduce the effective resolution, in addition to the seeing conditions themselves varying significantly throughout the observations, thus providing further (time varying) degradation to the data.
In the left panel of Figure~{\ref{fig:image_degradation}}, the photospheric velocity field is the result of two components: downflows in the intergranular lanes (red colors) and upflows in the granules (blue colors). Since the integranular lanes are much smaller and narrower with respect to the granules, the velocity signals associated with the integranular regions become more affected (i.e., reduced) by the lower spatial resolution induced by worsening seeing conditions. This effect is apparent in the middle and right panels of Figure~{\ref{fig:image_degradation}}, where the progressively worsening seeing conditions (${\mathrm{FWHM}}=0.2''$ middle panel; ${\mathrm{FWHM}}=0.5''$ right panel) result in lost fine-scale velocity information. 

\begin{figure*}[]
\begin{center}
\includegraphics[trim = 0cm 0cm 0cm 0cm, clip, width=8cm]{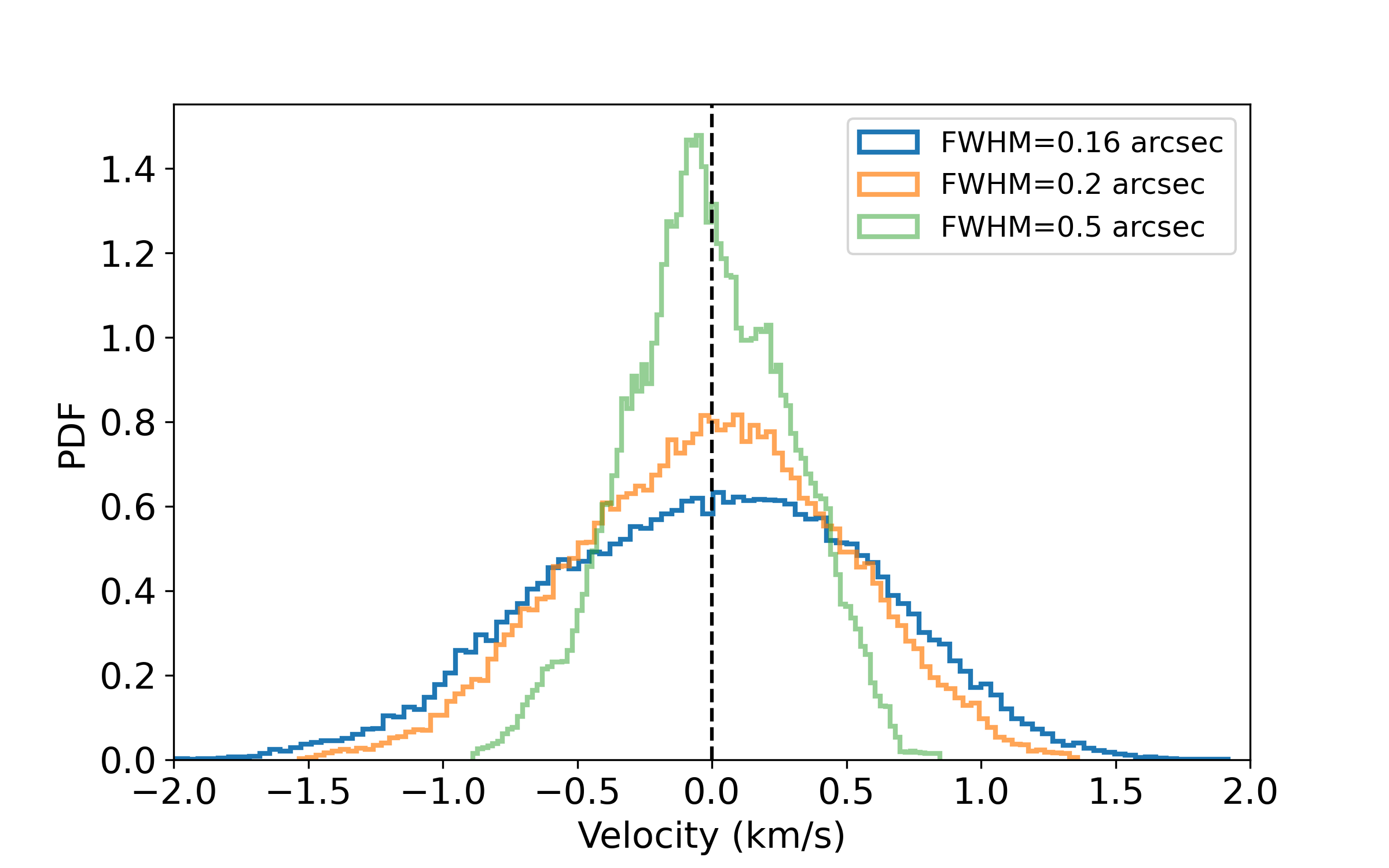}
\end{center}
\caption{Probability density functions (PDFs) of the instantaneous velocity fields shown in Figure~{\ref{fig:image_degradation}} as a function of spatial resolution. Here, the blue, orange, and green lines represent the PDFs for three different seeing conditions represented by a ${\mathrm{FWHM}}=0.16''$, ${\mathrm{FWHM}}=0.2''$, and ${\mathrm{FWHM}}=0.5''$, respectively. It can be seen that worse seeing conditions (e.g., the green line) produce more symmetric distributions and smaller velocity amplitudes due to the spatial averaging of the measured signals.}
\label{fig:image_degradation_PDFs} 
\end{figure*}

If the resolution angle is smaller than the angular size of the feature being studied, then the measured signal will approach the true value. This is due to the `filling factor' being equal to `1', whereby the feature of interest occupies the entirety of the resolution element on the detector. On the contrary, if the resolution element is larger than a particular spatial feature, then the signal measured will be a combination of both the feature of interest and plasma in its immediate vicinity. Here, the filling factor of the desired structure is $<1$, resulting in a blended signal that produces the measured parameters. In the specific case of integranular lanes (see, e.g., Figure~{\ref{fig:image_degradation}}), this means that if the resolution element is larger than their characteristic width, signal from the neighboring granules will be collected too. This effect is shown in Figure~{\ref{fig:image_degradation_PDFs}}, where the probability density functions (PDFs) of the instantaneous velocities for different spatial resolutions is shown. By lowering the spatial resolution, the original skewed distribution of the velocity, which is a consequence of the different spatial scales associated with the upflows (blueshifts) and downflows (redshifts), is transitioned into a more symmetric distribution that is characterized by smaller velocity amplitudes. 

\begin{figure*}[]
\begin{center}
\includegraphics[trim = 0cm 0cm 0cm 0cm, clip, width=7cm]{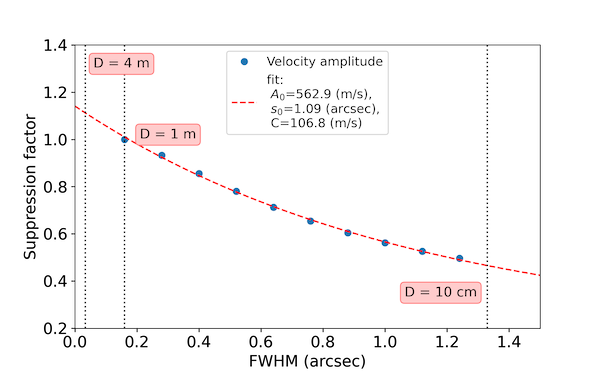}
\includegraphics[trim = 0cm 0cm 0cm 0cm, clip, width=7cm]{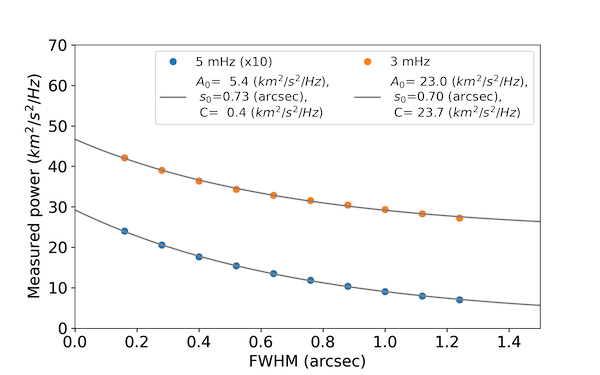}
\end{center}
\caption{Wave amplitude suppression factor (upper panel) and the resulting power spectral densities (lower panel) for observations acquired with different spatial resolutions. In the upper panel, the wave amplitude suppression factors (blue dots) are computed with respect to the velocity information displayed in Figure~{\ref{fig:image_degradation}}, with the vertical dotted lines highlighting telescope aperture sizes of 4m (DKIST), 1m (SST), and 0.1m. The dashed red line displays an exponential fit (using Equation~{\ref{eqn:suppressionfactor}}), with the fit parameters shown in the figure legend. The lower panel displays the resulting power spectral densities, as a function of spatial resolution, for two key frequencies commonly found in observations of the solar atmosphere, notably $2.5-3.5$~mHz (orange dots) and $4.5-5.5$~mHz (blue dots). Again, the power spectral densities are fitted using Equation~{\ref{eqn:suppressionfactor}}, with the corresponding fit parameters shown in the figure legend. These panels document the importance of spatial resolution when attempting to measure weak oscillatory processes, since poor spatial resolution (either through small telescope aperture sizes or poor local seeing conditions) may result in complete suppression of the observable signal. }
\label{fig:suppression_factor}
\end{figure*}

These effects, in turn, also translate into a reduction of the measured amplitudes of any oscillations present in the data. This effect can be seen in Figure~{\ref{fig:suppression_factor}}, where the suppression factor of the Doppler velocity amplitudes (upper panel) and the resulting power spectral densities in two distinct frequency bands, namely 3~mHz and 5~mHz (1~mHz bandwidth; lower panel), are shown as a function of the spatial resolution. The suppression factor gives an idea of the underestimation of the amplitudes of the embedded oscillations, and in the top panel of Figure~{\ref{fig:suppression_factor}} it is normalized to the value associated with the original SST/CRISP data used here (i.e., ${\mathrm{FWHM}}=0.16''$ provides a suppression factor equal to 1.0). From the upper panel of Figure~{\ref{fig:suppression_factor}} we can also predict the amplitudes of the velocity oscillations captured in forthcoming observations from the new 4m DKIST facility, which could be as large as $1.3 - 1.4$ times that of the velocity amplitudes measured with a 1m class telescope at the same wavelength (under similar local seeing conditions).

Both the suppression factor and the resulting power reduction, as a function of spatial resolution, are well modeled by an exponential decay of the form,
\begin{equation}
\label{eqn:suppressionfactor}
    A = A_{0} e^{-\frac{{\mathrm{FWHM}}}{{s_{0}}}} + C \ ,
\end{equation}
where $A_{0}$ is either the amplitude of the velocity signals or the wave power, $s_{0}$ is a characteristic spatial length, and $C$ is a constant. Equation~{\ref{eqn:suppressionfactor}} characterizes very nicely the impact spatial resolution has on the visible wave characteristics, whereby when the resolution element is larger than the characteristic physical scale of the observed process in the solar atmosphere (i.e., ${\mathrm{FWHM}}>s_{0}$), then the oscillatory signal is strongly suppressed. This may result in weak oscillatory amplitudes being lost from the final data products, a process that was recently discussed by \citet{2021NatAs...5....5J} in the context of sunspot oscillations.

Such amplitude suppression effects imply that when estimating the energy flux of waves, one needs to consider the specific spatial resolution achieved and correct the resulting estimates by a factor depending on the FWHM of the instrumental PSF and the local seeing effects. We note that this effect strongly depends on the characteristic spatial length of the processes observed in the solar atmosphere. In order to illustrate the problem we have made use of photospheric observations (i.e., Figures~{\ref{fig:image_degradation}} -- {\ref{fig:suppression_factor}}). However, due to the presence of narrow filamentary structures observed in the chromosphere, the power of the oscillations can be even more underestimated at those atmospheric heights.

\subsection{Identification of MHD wave modes}
\label{sec:Identification_of_MHD_wave_modes}

In this Section we will not review MHD wave theory in any great detail since this has been covered previously in many books and reviews \citep[see e.g.,][]{2004psci.book.....A, 2005LRSP....2....3N, 2014masu.book.....P, 2015SSRv..190..103J, roberts2019mhd}.  Instead, we would like to highlight the particular challenges of identifying MHD wave modes from observational data given what is known from MHD wave theory. 

\subsubsection{Homogeneous and unbounded plasma}
In most textbooks, for simplicity MHD waves are rightly introduced by assuming a homogeneous unbounded plasma with a straight and constant magnetic field. This highly idealized plasma configuration only permits propagating Alfv\'en, slow, and fast magnetoacoustic wave modes. In stark contrast, the Sun's atmosphere is actually very inhomogeneous and the newest high resolution instrumentation reveal the solar plasma to be ever more finely structured. But let us assume the wavelengths are large enough so that these MHD wave modes do not ``feel'' the effect of any plasma fine structure, hence allowing us to apply the unbounded homogeneous plasma model, as a zeroth order approximation, to observational data. How can we actually identify the Alfv\'en, slow, and fast magnetoacoustic MHD wave modes?  As we shall discuss, in practical terms, even in this simplest of plasma configurations, each MHD wave mode would actually be non-trivial to identify without ambiguity, even from excellent quality spectropolarimetric data. 

First, let us consider the Alfv\'en wave \citep{1942Natur.150..405A}. The only restoring force of this wave is magnetic tension, but since this wave is incompressible the magnetic field lines remain equidistant from each other as they are oscillating. Hence, although the direction of the magnetic field vectors will change with time as the field lines oscillate the magnitude of the vectors will remain constant. Therefore, this wave will not reveal itself through variations in the magnetic field strength using the Zeeman or Hanle effects. Also, due to its incompressibility the Alfv\'en wave would not reveal itself in intensity oscillations since the density is not perturbed. This only leaves the velocity perturbations associated with this wave, which could in principle be detected in Doppler measurements. However, to truly identify an Alfv\'en wave it would have to be established that the velocity perturbations were perpendicular to the magnetic field lines and that the wave vector was not perpendicular to the direction of the magnetic field. To add even more difficulty to the challenge of identifying an Alfv\'en wave, it is only approximately anisotropic, in the sense that the fastest propagation is along the direction of the magnetic field and only completely perpendicular propagation is forbidden, i.e., the more perpendicular the wave vector becomes relative to the magnetic field the slower the propagation will be. 

What about identifying the slow and fast magnetoacoustic modes? The allowed directions for the slow magnetoacoustic wave vector are very similar to that of the Alfv\'en wave, meaning that it is only approximately anisotropic and propagation perpendicular to the magnetic field direction is forbidden. However, unlike the Alfv\'en wave, the slow magnetoacoustic wave is compressible and should reveal itself in intensity oscillations if the amplitude of the perturbations are large enough relative to the background. However, to establish even more convincing evidence, a slow magnetoacoustic wave requires validation that the plasma and magnetic pressure perturbations are in anti-phase. Of course, this is not an easy task in observational data and would require both a fortuitous line-of-sight and an excellent signal-to-noise ratio to determine perturbations in both intensity and Zeeman/Hanle effect measurements. In contrast to the Alfv\'en and slow magnetoacoustic waves, the fast magnetoacoustic wave is more isotropic in nature since it can also propagate perpendicular to the magnetic field. A further key difference to the slow magnetoacoustic wave is that the plasma and magnetic pressure perturbations associated with a fast magnetoacoustic wave are in phase. To show this from observational data would provide compelling evidence that a fast magnetoacoustic wave mode has indeed been identified, but, as with showing the anti-phase behavior between plasma and magnetic pressures for a slow magnetoacoustic wave, this is not a trivial task, even with excellent quality spectropolarimetric data.

\begin{figure*}[!t]
\begin{center}
\includegraphics[width=0.8\textwidth]{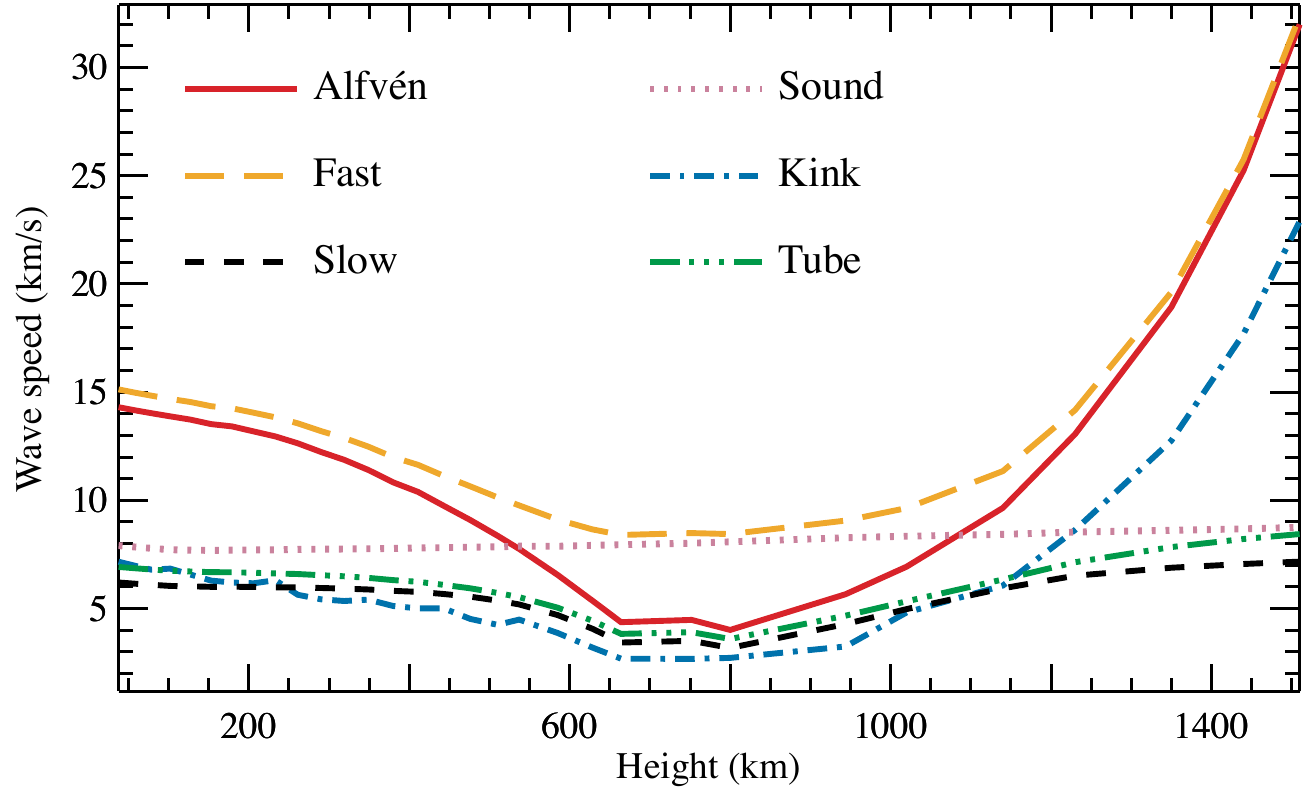}
\end{center}
\caption{Various wave speeds in a flux tube in the lower solar atmosphere, from the hot `NC5' flux tube model put forward by \citet{1993A&A...273..293B}, in combination with the surrounding cool VAL-A atmosphere \citep{1981ApJS...45..635V}.}
\label{fig:Bruls_model} 
\end{figure*}

There are also more subtle points in distinguishing between the Alfv\'en, slow, and fast magnetoacoustic wave modes depending on the value of plasma-$\beta$, which itself is difficult to determine from observational data. Importantly, for MHD wave modes the value of plasma-$\beta$ also indicates the relative values of the sound and Alfv\'en speeds. Especially problematic is the case when the sound speed is close to the Alfv\'en speed, since here the propagation speeds of the Alfv\'en, slow and fast magnetoacoustic waves along the direction of the magnetic field are practically indistinguishable. This effect is clearly demonstrated in Figure~{\ref{fig:Bruls_model}}, which is based on the `NC5' flux tube model presented by \citet{1993A&A...273..293B}, and clearly shows how the localized velocities associated with different wave modes can become difficult to disentangle in the lower solar atmosphere, hence providing some ambiguity when attempting to diagnose the true wave mode from the propagation velocity alone. But remember, the nuanced discussion we have had here on wave mode identification assumed that the solar plasma was both homogeneous and unbounded. In practical terms, it is more likely that the analysis of waves in the lower solar atmosphere will be directly related to their excitation in, and propagation through, large scale magnetic structures such as sunspot and pores (see Section~\ref{sec:largescalestructures}) or smaller scale structures such as spicules and fibrils (see Section~\ref{sec:smallscalestructures}).  In such cases the most applied model is that of the magnetic cylinder \citep[e.g.,][]{Wentzel1979, 1979A&A....71....9W, 1980A&A....87..121W, 1982SoPh...75....3S, 1983SoPh...88..179E}, which we shall discuss next.

\subsubsection{Magnetic cylinder model}
\label{sec:mag cylinder}
\begin{figure*}[!t]
\begin{center}
\includegraphics[trim=0mm 0mm 0mm 0mm, clip, width=0.42\textwidth]{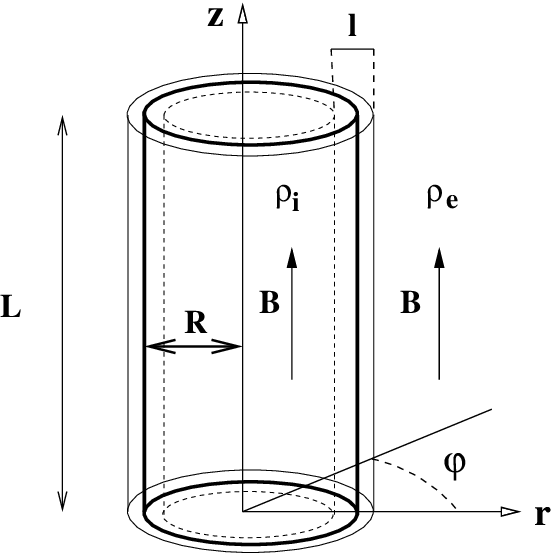}
\hfill \includegraphics[trim=0mm 0mm 0mm 0mm, clip, width=0.55\textwidth]{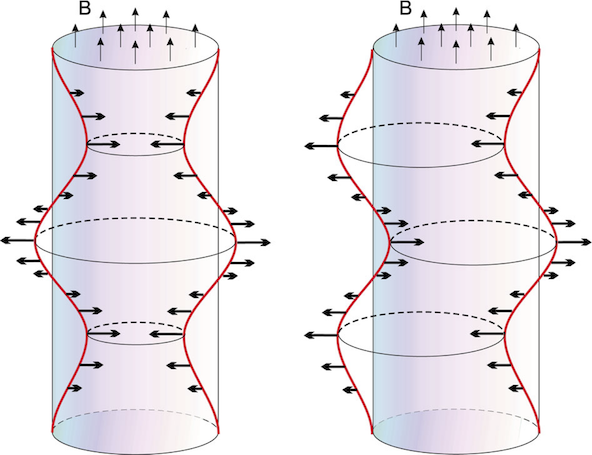}
\end{center}
\caption{A typical cylindrical flux tube model (left panel) represented by a straightened magnetic tube of length, $L$, and radius, $R$. The magnetic field, $B$, is uniform and parallel to the $z$-axis and the whole configuration is invariant in the azimuthal direction, $\theta$ (labeled as $\varphi$ in the diagram). In the schematic, the density varies in a non-uniform transitional layer of width, $l$, from a constant internal value, $\rho_{i}$, to a constant external value in the local plasma environment, $\rho_{e}$. The middle and right panels show the effects of $m=0$ (sausage) and $m=1$ (kink) wave perturbations, respectively, to the equilibrium flux tube. The sausage wave (middle) is characterized by an axi-symmetric contraction and expansion of the tube's cross-section. This produces a periodic compression/rarefaction of both the plasma and magnetic field. The kink wave (right) causes a transverse displacement of the flux tube. In contrast to the sausage wave, the kink wave displacement/velocity field is not axi-symmetric about the flux tube axis. The red lines show the perturbed flux tube boundary and thick arrows show the corresponding displacement vectors. The thin arrows labelled $B$ show the direction of the background magnetic field. Images reproduced from \citet[][left panel]{2005A&A...441..361A} and \citet[][middle \& right panels]{2012NatCo...3.1315M}.}
\label{fig:cylindermodel} 
\end{figure*}

The advantage of the magnetic cylinder model is that it allows for the key plasma parameters, e.g., magnetic field strength and plasma density, to differ inside and outside of the flux tube, allowing us to introduce inhomogeneity in the direction perpendicular to the cylinder axis. In this model, relative to the cylindrical coordinates $(r, \theta, z)$, where $r$, $\theta$, and $z$ are the radial, azimuthal, and axial directions, respectively, waves can either be standing or propagating in all three orthogonal directions (see the left panel of Figure~{\ref{fig:cylindermodel}}). If the wave is propagating in the radial direction this is a so-called ``leaky'' wave, which is not trapped by the cylindrical waveguide and damps due to MHD radiation. The so-called ``trapped'' modes are standing in the radial direction with the greatest wave energy density in the internal region of the cylinder. Outside of the cylinder the trapped mode is evanescent and decays with increasing distance from the tube. 

Beyond the basic descriptions of whether the mode is ``leaky'' or ``trapped'', the azimuthal integer wave number, $m$, defines whether the waves are the so-called ``sausage'', ``kink'', or ``fluting'' modes. The sausage mode has $m=0$ and is azimuthally symmetric, the kink mode has $m=1$ and is azimuthally asymmetric (see the middle and right panels of Figure~{\ref{fig:cylindermodel}}). The fluting modes are higher order in the azimuthal direction with $m \ge 2$. A further classification of wave types in a magnetic cylinder is ``body'' or ``surface'' modes. A body wave is oscillatory in the radial direction inside the tube and evanescently decaying outside. Because the body wave is oscillatory inside the tube, it has a fundamental mode in the radial direction and also higher radial harmonics. In contrast, a surface wave is evanescent inside and outside of the tube with its maximum amplitude at the boundary between the internal and external plasma. Since it is strictly evanescent inside the tube, the surface mode cannot generate higher radial harmonics.

\begin{figure*}[!t]
\begin{center}
\includegraphics[trim=2mm 20mm 5mm 0mm, clip, width=0.35\textwidth]{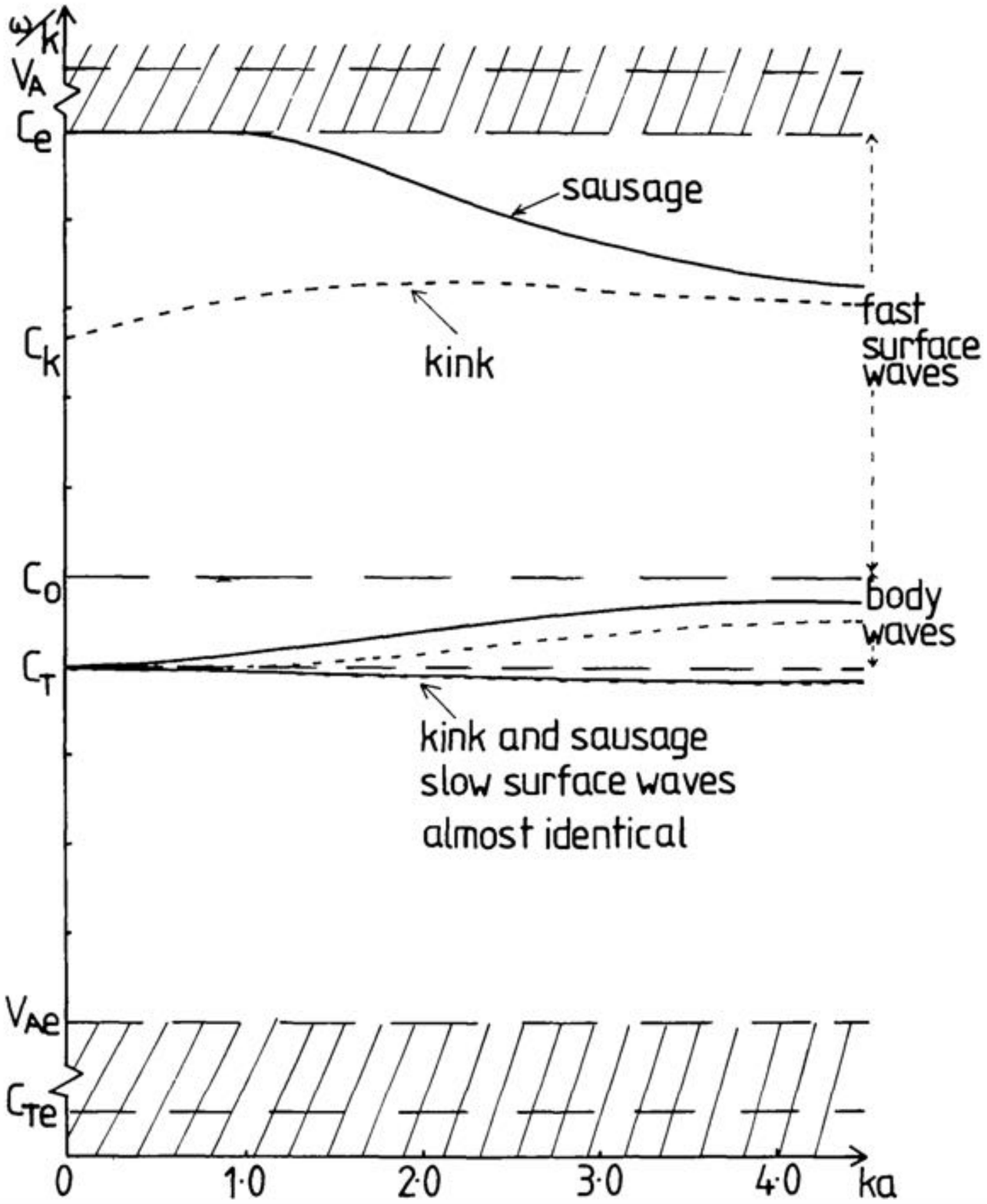}
\hfill\includegraphics[trim=0mm 0mm 0mm 0mm, clip, width=0.63\textwidth]{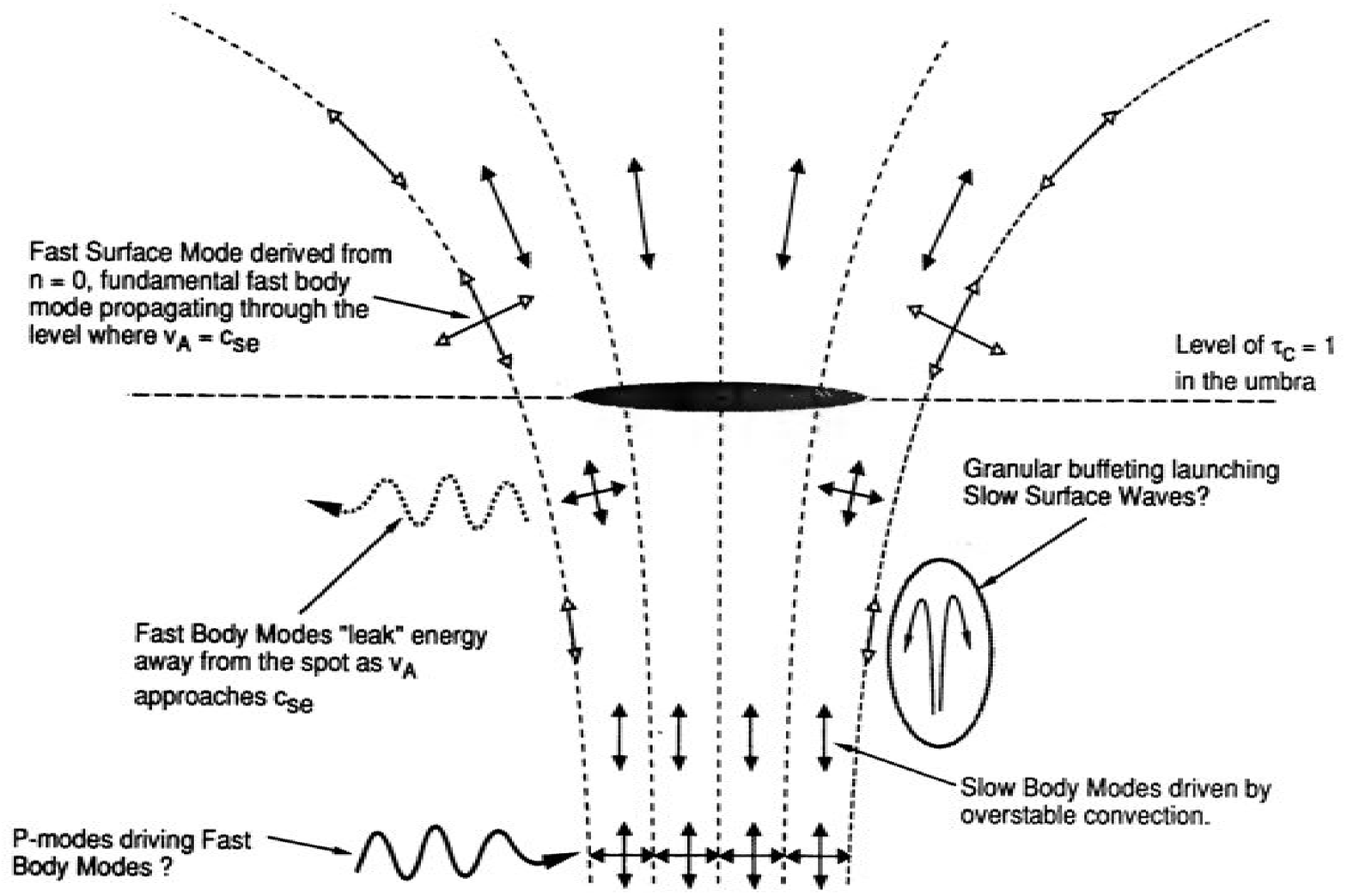}
\end{center}
\caption{Left panel: A dispersion diagram is shown for a representative photospheric magnetic cylinder. It can be seen that there are two distinct horizontal bands with slower and faster phase speeds. The fast band is bounded between $[c_0,c_e]$ and the slow band between $[c_T,c_0]$. The adjectives ``slow'' and ``fast'' here have a quite distinct meaning from the terms slow and fast when referring to the magnetoacoustic wave modes of a homogeneous and unbounded plasma. Right panel: A cartoon of theoretically predicted MHD wave modes in a sunspot, and their possible sources, based on the magnetic cylinder model of \citet{1983SoPh...88..179E}. Images adapted from \citet[][left panel]{1983SoPh...88..179E} and \citet[][right panel]{1990ApJ...348..346E}.}
\label{fig:Edwin_dispersion} 
\end{figure*}

At this point it will be worth explaining why confusion has arisen over the years since the seminal publication by \citet{1983SoPh...88..179E}, who also introduced the terms ``fast'' and ``slow'' to classify the propagation speeds of MHD wave modes along the axis of the magnetic cylinder. In the dispersion diagrams of a magnetic cylinder, distinct bands appear for a particular wave mode where the axial phase speed is bounded by characteristic background speeds. As an example, we can model a photospheric waveguide as being less dense than the surrounding plasma and having a stronger magnetic field internally than externally. This would be a reasonable basic model for, e.g., a pore or sunspot, where the internal density depletion is a result of the increased magnetic pressure \citep{1986ApJ...306..284M, 1992ApJ...399..300L, 2017ApJ...837L..11C, 2021RSPTA.37900172G, 2021A&A...648A..77R}. In this case, we can form the inequality of the characteristic background speeds as $v_A >c_e >c_0>v_{Ae}$, where $v_A$ is the internal Alfv\'en speed, $c_e$ is the external sound speed, $c_0$ is the internal sound speed, and $v_{Ae}$ is the external Alfv\'en speed. This results in a slower band with phase speeds between $[c_T, c_0]$, where the internal tube speed, $c_T$, is defined as, 
\begin{equation}
\label{eqn:tubespeed}
    c_T = \frac{c_0 v_A}{\sqrt{c_0^2+v_A^2}} \ .
\end{equation}
In addition, a faster band also exists with phase speeds between $[c_0, c_e]$. Wave modes with phase speeds below the ``slow'' band and above the ``fast'' band are not trapped modes (having real $\omega$ and $k_z$ values).  The ``slow'' and ``fast'' bands for these chosen photospheric conditions are shown in the dispersion diagram in the left panel of Figure~\ref{fig:Edwin_dispersion}.

Although \citet{1983SoPh...88..179E} used the perfectly apt adjectives, ``slow'' and ``fast'', to describe the phase speed bounds of these distinct bands of trapped MHD wave modes, they have quite a different physical meaning to the terms of the slow magnetoacoustic and fast magnetoacoustic waves from the homogeneous and unbounded plasma model. This is most clearly illustrated when comparing the same label ``fast'' in both scenarios. For a cylindrical waveguide any trapped fast MHD mode is strictly anisotropic since the propagating wave vector is restricted to being absolutely parallel to magnetic field direction, which is also aligned with the cylinder axis. However, a fast magnetoacoustic wave in a homogeneous plasma can propagate with any angle relative to the magnetic field orientation. 

There is a special class of incompressible Alfv\'en modes that can exist in a magnetic cylinder with any azimuthal wave number, $m$, the so-called torsional Alfv\'en waves \citep[see e.g.,][]{1982SoPh...75....3S}. Like the Alfv\'en wave in a homogeneous plasma, the only restoring force is magnetic tension. However, torsional Alfv\'en waves are strictly anisotropic since they can only propagate along the direction of the tube axis, whereas their counterpart in a homogeneous plasma can propagate at any angle (with the exception of perpendicular) relative to the magnetic field. The torsional Alfv\'en wave can only be excited if the driver itself is incompressible, meaning that the tube boundary is not perturbed at all in the radial direction. However, in reality it likely that the boundary of solar magnetic flux tubes are perturbed to some degree in the radial direction.  If the boundary is only slightly perturbed in the radial direction, and the dominant perturbations are in the axial direction, then this will excite a slow mode. If the radial perturbation dominates over the axial perturbation, resulting in a greater perturbation of the boundary, then this will excite a fast mode. The greater radial perturbation for a fast mode means that magnetic tension plays a larger role in the restoring force than for a slow mode, where the longitudinal compressive forces of plasma and magnetic pressure dominate. 

Understanding the phase relations between the restoring forces for MHD wave modes in a magnetic cylinder is not as straightforward as it is for the three possible MHD modes in a homogeneous plasma. This is because the phase relations between plasma pressure, magnetic pressure, and magnetic tension restoring forces depend on whether the wave is propagating or standing in each of the three orthogonal directions, i.e., radial $(r)$, azimuthal $(\theta)$, and axial $(z)$. Also, the radial spatial structuring of the plasma in a magnetic cylinder means that perturbed MHD variables, such as the magnetic field $(B_r,B_{\theta},B_z)$ and velocity $(v_r,v_{\theta},v_z)$ components, are related, not only by time derivatives, but spatial derivatives dependent on the variation of the background plasma properties. 

A simplified thin tube or ``wave on a string'' approximation was made by \citet{2009ApJ...702.1443F} to derive the phase relations between $v_r$ and $B_r$ for a kink mode, and $v_z$ and $B_z$ for a sausage mode. This was done for both propagating and standing waves in the axial direction, but caution should be taken in applying these results to structures of finite width. A more detailed investigation into the phase relations of these MHD variables was done for the sausage mode by \citet{2013A&A...551A.137M}, utilizing a magnetic cylinder of finite width under photospheric conditions. Like \citet{2009ApJ...702.1443F}, this model predicted the phase relations for both standing and propagating waves in the axial direction. A note of caution should be introduced here to state that both the models of \citet{2009ApJ...702.1443F} and \citet{2013A&A...551A.137M} assume the kink and sausage modes are ``free'' oscillations of the structure and are not being driven. To correctly derive the phase relations between the MHD wave variables in a driven system demands that system is solved as an initial value problem. However, currently the exact spatial and temporal structures of the underlying drivers of the waves observed in pores and sunspots are not universally understood. 

Although the phase relations between the perturbed variables for any MHD wave mode may be not simple to predict theoretically, at the least the spatial structure of these variables (independent of time), providing the cross-section of the wave guide is resolved (e.g., particularly in the case of larger magnetic structures such as pores and sunspots), should correlate in straightforward way. First, let us consider a fixed axial position, $z$, which for a vertical tube would correspond to a fixed height in the solar atmosphere. If the magnetic cylinder is oscillating with an eigenmode, then the variables related to compressible axial motion, i.e., $v_z$, $B_z$ and plasma pressure (also related to perturbations in temperature and plasma density), should have the same spatial structure in the radial $(r)$ and azimuthal $(\theta)$ directions. Likewise, the spatial structure of variables related to radial perturbations of the magnetic field, i.e., $v_r$ and $B_r$, should be consistent. The same is also true for the variables that relate to the torsional motions of the magnetic field, i.e., $v_{\theta}$ and $B_{\theta}$. Again, all these theoretical predictions assume free oscillations of the entire magnetic structure, e.g., a pore or sunspot. If the oscillations are being driven, then this is a more complicated and computationally expensive modeling problem to solve. Also, the spatial scale of the driver relative to the size of the magnetic structure is crucial. To excite the global eigenmodes of magnetic structures the driver has to be at least as large as the structure itself. If the driver is much smaller than the magnetic structure, it will still excite localized MHD waves, but these will not be global eigenmodes of the entire magnetic structure. This too requires a different modeling approach, see e.g., \citet{2006ApJ...653..739K}, who modeled $p$-mode propagation and refraction through sunspots.

\begin{figure*}[!t]
\begin{center}
\includegraphics[width=\textwidth]{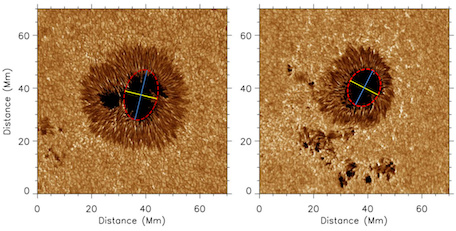}
\end{center}
\caption{Two active regions, NOAA AR12565 (left) and NOAA AR12149 (right), captured in the G-band by ROSA at the Dunn Solar Telescope. To show the departure from circular cross-sectional shape, ellipses are fitted to the sunspot umbrae. The eccentricity of the left umbra is $\epsilon=0.76$, while the right umbra is $\epsilon=0.58$. Image adapted from \citet{Aldhafeeri2021}.}
\label{fig:ellipse_spots}
\end{figure*}

\begingroup
\setlength{\tabcolsep}{0.25pt} 
\renewcommand{\arraystretch}{0.25}  
     \begin{figure}
     \centering
     \begin{tabular}{|c|}
                \hline
                \\
        \multicolumn{1}{|c|}{Fast body kink modes, odd $m=1$} \\
        \\
      \hline
       \includegraphics[width=.75\textwidth]{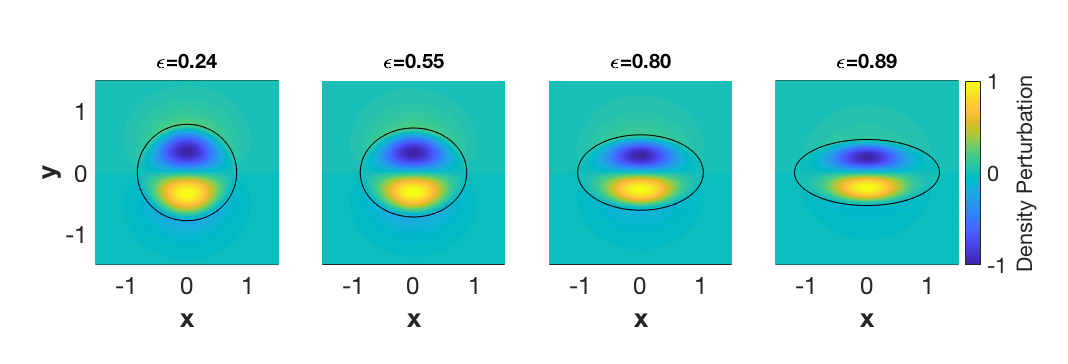}\\
      \hline\\
        \multicolumn{1}{|c|}{Fast body fluting modes, odd $m=2$} \\\\
      \hline
    \includegraphics[width=0.75\textwidth]{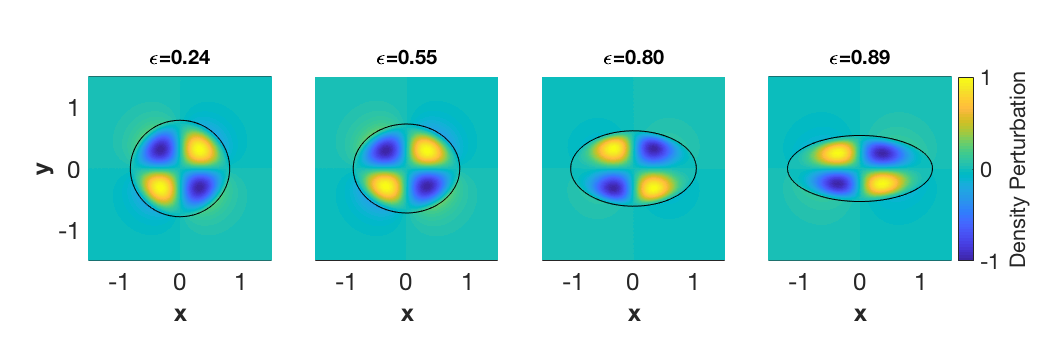}\\
           \hline\\
        \multicolumn{1}{|c|}{Fast body fluting modes, odd $m=3$} \\\\
      \hline
    \includegraphics[width=0.75\textwidth]{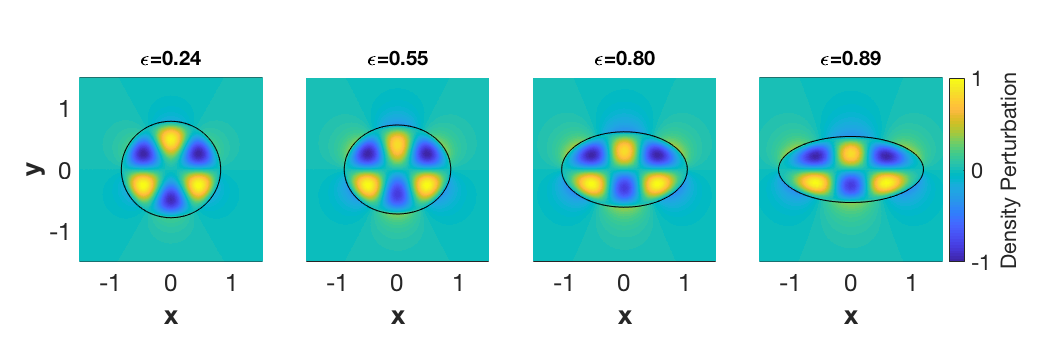}\\
           \hline   
      \end{tabular}
      \caption{The normalized density perturbations of fast body wave modes under representative coronal conditions for the different values of eccentricity $\epsilon$. Note that the eigenfunctions for slow body wave modes under photospheric conditions would have a very similar appearance. From top to bottom, the $m=1$ (kink) and $m=2,3$ (fluting) modes are shown which have an odd phase structure with respect to the major axis of the ellipse. Image adapted from \citet{Aldhafeeri2021}.}
            \label{fig:odd_modes}
      \end{figure}

\begingroup
\setlength{\tabcolsep}{0.25pt} 
\renewcommand{\arraystretch}{0.25}  
     \begin{figure}
     \centering
     \begin{tabular}{|c|}
                \hline
                \\
        \multicolumn{1}{|c|}{Fast body kink modes, even $m=1$} \\
        \\
      \hline
       \includegraphics[width=0.75\textwidth]{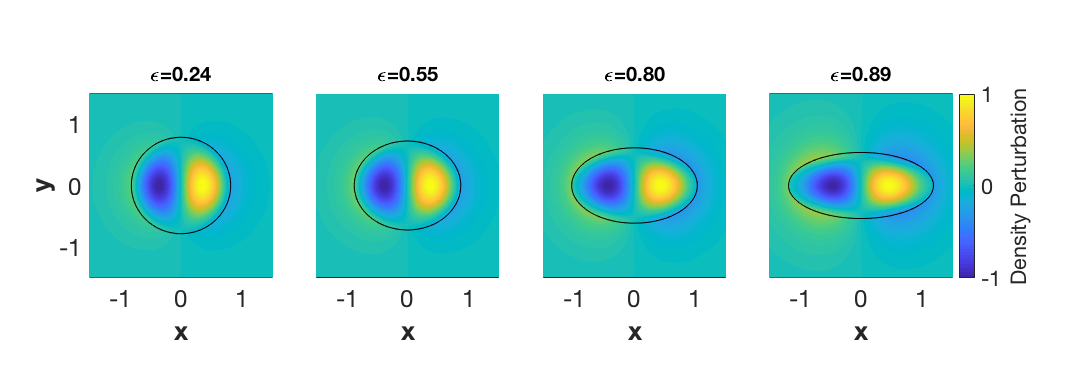}\\
      \hline\\
        \multicolumn{1}{|c|}{Fast body fluting modes, even $m=2$} \\\\
      \hline
    \includegraphics[width=0.75\textwidth]{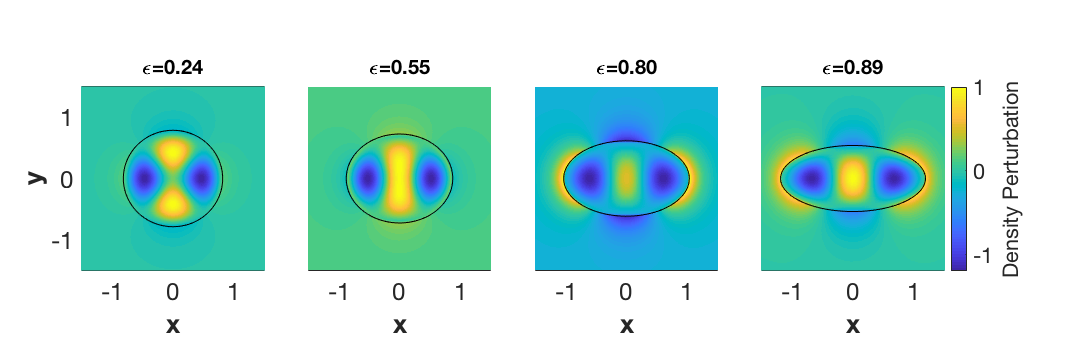}\\
           \hline\\
        \multicolumn{1}{|c|}{Fast body fluting modes, even $m=3$} \\\\
      \hline
    \includegraphics[width=0.75\textwidth]{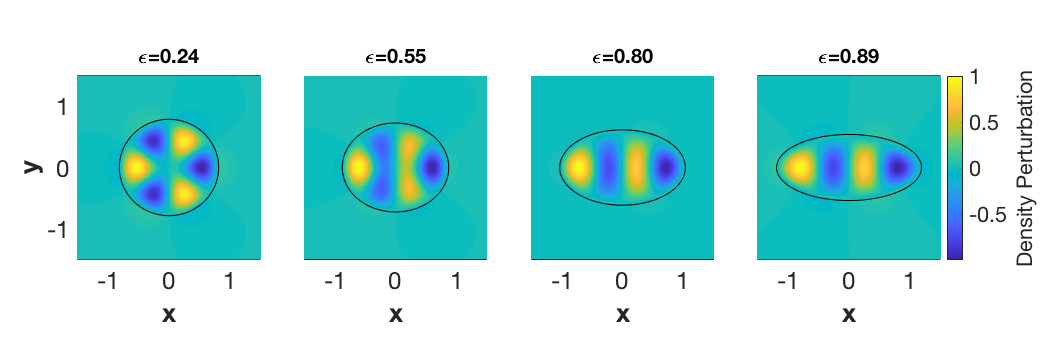}\\
           \hline
      \end{tabular}
      \caption{The same wave modes are shown as in Figure~\ref{fig:odd_modes} but their phase structure is even with respect to the major axis of the ellipse. Image adapted from \citet{Aldhafeeri2021}.}  
      \label{fig:even_modes}
      \end{figure}
\endgroup

High resolution images of sunspots, pores, magnetic bright points, and fibrillar structures are continually telling us that modeling these features using cylindrical flux tube geometries, while more mathematically simplistic, is far from realistic.  Even from basic membrane models, in which separation of variables is possible, the cross-sectional shape has a fundamental effect on the structure of the eigenfunctions. For elliptical magnetic flux tubes, \citet{Aldhafeeri2021} investigated the effect of eccentricity, $\epsilon=\sqrt{1-b^2/a^2}$, where $a$ and $b$ are the semi-major and semi-minor axes, respectively, on the spatial structure of eigenfunctions. See, for example Figure~\ref{fig:ellipse_spots}, which shows two sunspot umbrae fitted with ellipses with eccentricities $\epsilon=0.58$ and $\epsilon=0.76$. These are not negligible values since a circle has $\epsilon=0$. Figure~\ref{fig:odd_modes} shows $m=1$ (kink) and $m=2,3$ (fluting) fast body modes where the phase is odd with respect to the major axis as eccentricity increases, while Figure~\ref{fig:even_modes} shows the same modes where the phase is even with respect to the major axis. Although all MHD wave modes in flux tubes of elliptical cross-section have their spatial structure distorted when compared to their equivalent versions in flux tubes of circular cross-section, it can be seen that the fluting modes that have even phase with respect to the major axis (shown in Figure~\ref{fig:even_modes}) become notably different in character as eccentricity increases, since previously distinct regions of phase or anti-phase end up coalescing. This advancement from the cylindrical flux tube model demonstrates that more sophisticated modeling of magnetic flux tubes with more realistic, and hence more irregular, cross-sectional shapes is required to more accurately interpret what type of wave modes are present in pores and sunspots. Recently this was done by \citet{2022ApJ...927..201A} and \citet{2022NatCo..13..479S} to identify MHD wave modes in sunspot umbrae and this will be discussed in Section~\ref{sec:largescalestructures}.

In Section~\ref{sec:spatial_resolution} the crucial issue of spatial resolution was discussed. In smaller scale magnetic structures, such as off-limb spicules or on disc fibrils, it is not possible to observe the true cross-section of the wave guide (as is possible for larger on-disc features such as pores and sunspots) in order to identify eigenmodes. However, fast sausage and kink modes can still be identified in these smaller structures if the amplitude of the radial motion (i.e., transverse to the magnetic field direction) is large enough. The kink mode is the only cylinder mode which causes a transverse displacement of the axis. For smaller magnetic structures, such as fibrils, the kink mode will appear as a ``swaying'' motion. If the radial motion of the fast sausage mode is large enough, then this causes periodic changes in the width of the structure, which can be resolved. Wave mode identification in smaller magnetic structures is addressed in detail in Section~\ref{sec:smallscalestructures}. As for larger scale magnetic waveguides, where the cross-section can be resolved fully, such as in pores or sunspots, the right panel of Figure~\ref{fig:Edwin_dispersion} shows the wide variety of theoretically predicted MHD wave modes, including slow/fast and body/surface, that can exist in such structures based on the magnetic cylinder model of \citet{1983SoPh...88..179E}. Recent progress in the identification of such wave modes from observations is discussed in Section~{\ref{sec:globaleigenmodes}}.

Across Section~{\ref{sec:waveanalysistools}}, we have discussed the fundamental theoretical considerations of waves manifesting in the solar atmosphere (Section~{\ref{sec:Identification_of_MHD_wave_modes}}), we have provided an overview of the techniques used to characterize them (Sections~{\ref{sec:1Dfourieranalysis}} -- {\ref{sec:Bomega}}), and summarized the challenges faced in light of variable spatial resolution (Section~{\ref{sec:spatial_resolution}}). Regardless of these challenges, over the last number of decades the solar community has overcame many obstacles, which has allowed for the successful acquisition, extraction, and identification of many different types of wave modes across a wide variety of solar features. In the following section, we will overview recent discoveries in the field of waves in the lower solar atmosphere, as well as comment on the difficulties still facing the global community in the years ahead.

\section{Recent Studies of Waves}
\label{sec:recentstudiesofwaves}
In the past, review articles have traditionally segregated wave activity in the solar atmosphere into a number of sub-topics based on the specific wave types and structures demonstrating the observed behavior. For example, \citet{2015SSRv..190..103J} divided up the review content on a feature-by-feature basis, including sections related to compressible and incompressible waveforms, which were subsequently further sub-divided into quiet Sun, magnetic network, and active region locations. However, as modern observations and modeling approaches continue to produce data sequences with ever improving spatial resolutions, placing the physical boundary between two locations becomes even more challenging. Indeed, emerging (and temporally evolving) magnetic fields often blur the boundaries between magnetic network elements, pores, proto-sunspots, and fully developed active regions. Hence, it is clear that solar complexity continues to increase with each improvement in spatial resolution made. As a result, dividing the content between previously well-defined structures becomes inappropriate, which is even more apparent now that mixed MHD waves \citep[e.g., compressible {\it{and}} incompressible modes;][]{2012NatCo...3.1315M} are being identified in a broad spectrum of magnetic features. 

Hence, for this topical review we employ just three (deliberately imprecise) sub-section headings, notably related to `global wave modes', as well as `large-scale' and `small-scale' structures. This is to avoid repetition and confusion, and to allow the overlap between many of the observables in the Sun's atmosphere to be discussed in a more transparent manner. Importantly, while discussing the recent developments surrounding wave activity in the lower solar atmosphere, we will attempt to pinpoint open questions that naturally arise from the cited work. We must stress that closing one research door more often than not opens two (or more) further avenues of investigation. Therefore, discussion of the challenges posed is not to discredit the cited work, but instead highlight the difficult research stepping stones facing the solar physics community over the years and decades to come.  

\subsection{Global Wave Modes}
\label{sec:globalmodes}
The field of helioseismology has employed long-duration data sequences \citep[some spanning multiple continuous solar cycles;][]{2018A&A...619A..99L} to uncover the internal structure and dynamics of the Sun through its global oscillation properties. Pioneering observations by \citet{1968ZA.....68..345F} suggested, for the first time, the presence of dual oscillating modes in the solar atmosphere, something that contradicted previous interpretations where the observed oscillations were simply considered to be an atmospheric response to granular impacts. It was subsequently shown that a variety of global wavenumbers could be seen in the photospheric velocity field of the C~{\sc{i}}~$538$~nm line \citep{1975A&A....44..371D}. Importantly, the pioneering work of \citet{1975A&A....44..371D} revealed clear ridges in photospheric $k$-$\omega$ power spectra, which helped to highlight, for the first time, that the ubiquitous 5-minute $p$-mode oscillations are in-fact resonant eigenmodes of the Sun. Novel observations acquired during austral summer at the South Pole discovered global 5-minute global oscillations at a wide range of horizontal wavelengths, revealing the true extent of oscillation modes associated with global solar resonances \citep{1980Natur.288..541G, 1983Natur.302...24D}. Traditionally, in the field of helioseismolgy, the Sun is considered as an approximate spherically symmetric body of self-gravitating fluid that is suspended in hydrostatic equilibrium \citep{Christensen-Dalsgaard2000}. This results in the modes of solar oscillations being interpreted as resonant vibrations, which can be represented as the product of a function of radius and a spherical harmonic, $Y_{l}^{m}(\theta, \phi)$. Here, $l$ relates to the horizontal scale on each spherical shell (commonly referred to as the `angular degree'), while $m$ determines the number of nodes in solar longitude (commonly referred to as the `azimuthal order'). 

The specific modes of oscillation can be divided up into three main categories: (1) Pressure modes ($p$-modes), which are essentially acoustic waves where the dominant restoring force is pressure, providing frequencies in the range of $\sim1-5$~mHz and angular degrees spanning $0 \le l \le 10^{3}$ \citep{1997SoPh..175..287R, 2011LNP...832....3K, 2013ApJ...772...87K}, (2) Internal gravity modes ($g$-modes), where the restoring force is predominantly buoyancy (hence linked to the magnitude of local gravitational forces), which typically manifest in convectively stable regions, such as the radiative interior and/or the solar atmosphere itself \citep{2014ApJ...784...88S}, and (3) Surface gravity modes ($f$-modes), which have high angular degrees and are analogous to surface waves in deep water since they obey a dispersion relation that is independent of the stratification in the solar atmosphere \citep{2007nsc..conf...13M}. In the limit that the wavelength is much smaller than the solar radius these wave are highly incompressible. The main restoring force for $f$-modes is gravity, which acts to resist wrinkling of the Sun's surface.

The intricacies of helioseismology become even more complex once isolated magnetic features, such as sunspots, develop within the solar atmosphere, which impact the velocities and travel times of the embedded global wave modes \citep{2000SoPh..192..285B, 2001ApJ...563..410R, 2004SoPh..220..381R, 2019ApJ...871..155R, 2012SoPh..279..323K, 2013A&A...558A.130S, 2016A&A...595A.107S}. A complete overview of the progress in helioseismology is beyond the scope of the present review. Instead, we refer the reader to the vast assortment of review articles that focus on the widespread development of helioseismology over the last few decades \citep[e.g.,][]{1983SoPh...82..103D, 1983SoPh...82..487B, RevModPhys.74.1073, 2005LRSP....2....6G, 2010ARA&A..48..289G, 2017A&A...600A..35G, 2013SoPh..287....9G, 2016LRSP...13....2B,  10.3389/fspas.2019.00042}

Importantly, the magnetic field in the solar photosphere is inhomogeneous and found in discrete concentrations across all spatial scales \citep{1987ARA&A..25...83Z}. Outside of the magnetic concentrations, where plasma pressure and gravity are the dominant restoring forces, longitudinal acoustic waves (i.e., $p$-modes) are generated at the top of the convection zone from the turbulent motions constituting the convective motion  \citep{1967SoPh....2..385S, 1990ApJ...363..694G, 1993ApJ...407..316B}. The $p$-modes can propagate upwards and contribute to heating of the higher layers if their frequency is larger than the acoustic cut-off frequency \citep{1971A&A....14..275U, 1995ApJ...444..879W}. Thus, the acoustic waves can dissipate their energy in the solar chromosphere by forming shocks (as a result of gas-density decreases with height), which are manifested in intensity images as, e.g., intense brightenings \citep{1991SoPh..134...15R,1997ApJ...481..500C,2008A&A...479..213B, 2020A&A...644A.152E, 2021RSPTA.37900185E}, or drive, e.g., Type~{\sc i} spicules, in the so-called `magnetic portals' \citep{2006ApJ...648L.151J}. Moreover, \citet{2016ApJ...817..124S} showed that these shocks are associated with dynamic fibrils in an active region they exploited from observations with SST and the Interface Region Imaging Spectrograph (IRIS; \citealt{2014SoPh..289.2733D}) space telescope.

Properties of propagating acoustic/magnetoacoustic waves through the lower solar atmosphere have also been reported in a number of recent studies from both ground-based \citep[e.g.,][]{2016ApJ...826...49S,2020ApJ...890...22A,2020A&A...642A..52A} and space-borne observations \citep[e.g.,][]{2015ApJ...803...44M,2016ApJ...830L..17Z, 2021A&A...648A..28A}. From IRIS  observations of a quiet-Sun area in Mn~{\sc i}~2801.25{\,}{\AA}, Mg~{\sc ii}~k~2796.35{\,}{\AA}, and C~{\sc ii}~1334.53{\,}{\AA} spectral lines (sampling the photosphere, chromosphere, and transition region, respectively), \citet{2018MNRAS.479.5512K} found upwardly propagating $p$-modes with periods on the order of $1.6-4.0$~min, and downward propagation in the higher period regime (i.e., periods larger than $\approx$4.5~min). Furthermore, \citet{2020A&A...634A..63K} identified the propagation of slow magnetoacoustic waves (with 2-9~min periodicities), within a plage region, from the high-photosphere/low-chromosphere to the transition region, using SDO/AIA and IRIS observations. 

In addition, $g$-modes (i.e., internal gravity waves) can be produced within turbulent convective flows \citep{1981ApJ...249..349M,1982ApJ...263..386M} and propagate through the lower solar atmosphere, with frequencies shorter than $\approx2$~mHz \citep{2010MNRAS.402..386N,2011A&A...532A.111K,2017ApJ...835..148V,2019ApJ...884L...8J,2020A&A...633A.140V}. Their identifications in the solar atmosphere have, however, been in a challenging task since they become evanescent in the convection zone and their amplitudes at the surface are exceedingly small \citep{2018SoPh..293...95S, 2021RSPTA.37900178C}. The internal gravity waves can potentially carry a large amount of energy flux (of $\approx5$~kW/m$^2$; \citealt{2008ApJ...681L.125S}) to the chromosphere, thus contributing to its radiative losses \citep{2021RSPTA.37900177V}.

Fortunately, unlike $g$-modes, $f$-modes (i.e., surface gravity modes) have been detected in abundance and have provided valuable diagnostic information about flows and magnetic field in the near surface region \citep{1995ApJ...451..851G, 1995MNRAS.276.1003R, 1996Sci...272.1286C}. Furthermore, $f$-modes have been exploited to quantify the Sun's effective seismic (or acoustic) radius \citep{ 1997ApJ...489L.197S, 2000SoPh..192..459A,2001ApJ...553..897D,  2005ApJ...625..548D}, a relatively new concept driven by results from helioseismology, as opposed to the measuring the Sun's physical (or true) radius. Such studies have shown that $f$-mode frequencies, as well as being sensitive to the seismic radius, are also modified by changes in the magnetic field during the solar cycle.

\subsubsection{Global $p$-modes in the lower solar atmosphere}
Acoustic waves (i.e., $p$-modes) can propagate both outside and inside magnetic concentrations. Through a number of studies prior to the turn of the century, properties of their `global' oscillations (i.e., properties averaged over a relatively large field of view) became a ``basic fact'', describing the characteristic periodicity of $p$-modes as 5~minutes in the solar photosphere  \citep{1962ApJ...135..474L,1970ApJ...162..993U,1997ApJ...488..462R,2009ASPC..416...49S}, and 3~minutes in the chromosphere \citep{Evans1963, Orrall1966, 1978A&A....70..345C,1991A&A...250..235F,1991SoPh..134...15R,1992ApJ...397L..59C}. Standing acoustic waves have also been reported from multi-line observations in the solar chromosphere \citep{1994chdy.conf..103F}, though wave patterns (and power spectra) were found to be somewhat different in He~{\sc i}~1080~{\AA} observations \citep{1994IAUS..154...65F,1995itsa.conf..437F}, compared to those in other chromospheric diagnostics (e.g., Ca~{\sc ii}~H~\&~K, Ca~{\sc ii}~8542~{\AA}, and H$\alpha$). While the global $p$-modes are more purely acoustic in nature in the photosphere, they are more likely to manifest as magnetoacoustic waves in the upper atmosphere, where the magnetic forces dominate \citep{2013JPhCS.440a2048K}.

Many of the observations demonstrating the characteristic periodicities of the global $p$-modes have been based on wide-band imaging at low-spatial resolutions, with very large fields of view. A recent study by \citet{2021RSPTA.37900170F} highlighted the presence of ubiquitous 3-minute characteristic periodicities by exploring several advanced state-of-the-art numerical models. Even so, considerable differences between the various simulations were also reported, including the height dependence of wave power, in particular for high-frequency waves, varying by up to two orders of magnitude between the models \citep{2021RSPTA.37900170F}. Thus, although the numerical simulations provide us with important information regarding the physical processes embedded within observational data, they should be interpreted with caution since the numerical domains are too small to resolve the true physics driving large-scale global eigenmodes.

Development of modern instruments in recent years, resulting in relatively narrow-band (often spectrally-resolved) observations at high resolution, have further explored the highly dynamic nature of the lower solar atmosphere. These novel observations reveal that the physical properties and structure of the lower solar atmosphere may significantly vary over different solar regions (with different levels of magnetic flux and/or topology), as well as through different atmospheric layers. Therefore, chromospheric wide-band filtergrams, that often integrate over a significant portion of strong chromospheric lines (hence, sampling across a large range of heights), may result in mixing (or averaging) of observable information (e.g., the oscillatory power), which can largely vary within a short distance in the lower solar atmosphere. A large variation in the height of formation can also cause a strong temporal modulation that may consequently destroy the oscillatory signal. Furthermore, the effect of spatial resolution can be crucial, as information may be lost in lower resolution observations due to, e.g., smearing (see Section~{\ref{sec:spatial_resolution}} for more discussion related to resolution effects). Moreover, an average power spectrum over a very large field of view can predominantly be dominated by characteristics of quiet-Sun regions (which cover the majority of the solar surface at any given time).

An example of the influence of spatial resolution is the larger (total) energy flux of acoustic waves (larger by a factor of $\approx2$) found in a quiet-Sun region by the 1m {\sc Sunrise} telescope \citep{2010ApJ...723L.134B}, compared to that from the 0.7m VTT telescope \citep{2009A&A...508..941B}. However, the effect of seeing-free observations with {\sc Sunrise} could also play a role in that difference, highlighting again the importance of spatial resolution (as discussed in Section~{\ref{sec:spatial_resolution}}). Such variations in atmospheric seeing (that directly affect the spatial resolution achievable) can influence the measured periodicities, in particular the global $p$-modes that are ubiquitously visible across the photosphere and chromosphere.

\begin{figure*}[!t]
\begin{center}
\includegraphics[trim=0mm 0mm 0mm 0mm, clip, width=\textwidth]{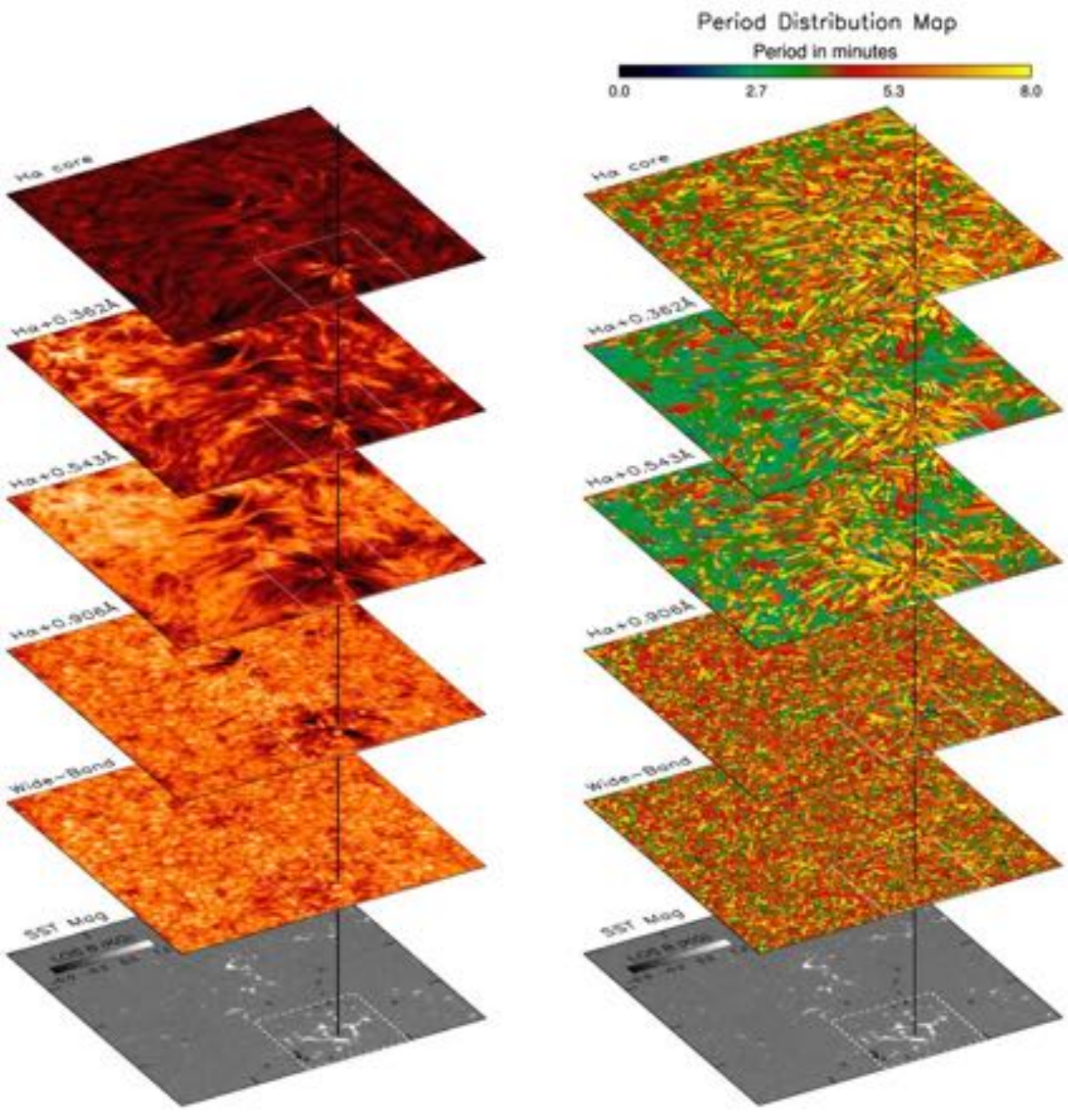}
\end{center}
\caption{Multi-layer observations of a quiet-Sun region from the low photosphere to the high chromosphere (left) whose dominant oscillatory periods are shown on the right. The green, red, and yellow colors in the dominant-period maps roughly represent periods around 3, 5, and 7 minutes, respectively. The bottom panels illustrate the corresponding line-of-sight magnetogram, from Stokes inversions of Fe~{\sc i}~630.2~nm spectral line. Images reproduced from \citet{2016ApJ...828...23S}.}
\label{fig:power_changes} 
\end{figure*}

In the presence of strong magnetic fields (e.g., in network or plage regions, where a group of concentrated small-scale magnetic features reside), the global acoustic power is enhanced at photospheric and low chromospheric heights \citep[known as a `power halo';][]{1992ApJ...394L..65B,2010A&A...510A..41K,2013SoPh..287..107R}, while it is suppressed in the high chromosphere \citep[so-called `magnetic shadows';][]{1962ApJ...135..474L,1992ApJ...393..782T, 2001ApJ...561..420M}. While the exact mechanisms behind such power variation are not yet fully understood, a number of suggestions have been provided in recent years, from both observations and simulations. 
In particular, models have shown that the power enhancement at lower heights can be due to the reflection of fast waves at the magnetic canopy, as a result of a large Alfv{\'e}n speed gradient \citep{2012ApJ...746...68K, 2016ApJ...817...45R}. From observations, both magnetic-field strength and inclination have been found to play an important role, with greater power in the stronger and more horizontal fields \citep{2011SoPh..268..349S,2019ApJ...871..155R}.
The power suppression of the acoustic waves in the high chromosphere has been suggested to be due to the mode conversion at the plasma-$\beta\approx1$ level \citep[i.e., as a result of interactions between $p$-mode oscillations and the embedded magnetic fields;][]{2007A&A...471..961M,2012A&A...542L..30N}, less efficient wave propagation under the canopy, or the wave-energy dissipation before it reaches the canopy \citep{1971A&A....12..297U, 2005ApJ...631L.155U, 2017ApJ...846...92S, 2020ApJ...889...95M, 2021JGRA..12629097S}. The power suppression and its spatial scale has found to be directly correlated with the magnetic-field strength and/or geometric height \citep{2012ApJ...744...98C,2014ApJ...796...72J,2016ApJ...823...45K}. From MHD simulations with the Bifrost code \citep{2011A&A...531A.154G}, \citet{2011ApJ...743..142H} showed that field inclination plays an important role in propagation of long-period waves (longer than 3 minutes) in the solar chromosphere. As such, they primarily found 3-minute periodicities in regions with weak or vertical magnetic fields (including the center of strong flux tubes), whereas 5-minute dominant waves in strong or inclined magnetic fields (such as the edges of flux tubes).

Power suppression of 3~minute oscillations in the upper solar chromosphere has been reported by \citet{2016ApJ...828...23S}, where almost no oscillatory power at this period was observed in time series of H$\alpha$ line-core intensity images from SST/CRISP. The authors, however, found a slightly larger number of pixels demonstrating 3~minute oscillations in Doppler velocity signatures of the same spectral line. In addition, they found power halos at lower atmospheric heights. In this study, the presence of ubiquitous chromospheric transient events (i.e., short-lived fibrillar structures) was speculated to be responsible for the power enhancements at lower heights. In addition, it was speculated that mode conversion was causing the magnetic shadows found around the 3~minute periodicity in the upper chromosphere. Figure~\ref{fig:power_changes} illustrates the multi-height observations studied by \citet{2016ApJ...828...23S} (on the left) along with their corresponding distribution of dominant periods of the oscillations (on the right), representing periods corresponding to the maximum power at each pixel. The lack of 3-minute oscillations (i.e., the green color) on the top layer is evident.
Using high-resolution H$\alpha$ line-core observations with SST, \citet{2007ApJ...655..624D} had found longer periods in regions where the field is supposedly more inclined. From spatial distribution of dominant periods (from a wavelet analysis) they showed that while sunspots and plage regions were dominated by 3-minute global $p$-modes, 5-minute and longer periodicities were found in adjacent to the dense plage regions and in more inclined-field areas, respectively. We should, however, note that such dominant-period maps demonstrated by \citet{2016ApJ...828...23S} and \citet{2007ApJ...655..624D} should be interpreted with great caution, since multiple peaks with comparable (or even equal) power may co-exist in a power spectrum. As such, the period associated to one absolute maximum of the power may not solely be representative of the oscillations in that pixel. In addition, we should note that the global wavelet spectrum is often considered a biased estimation of the true Fourier spectrum, with variable frequency resolution through the entire spectrum, that can also depend on the choice of wavelet function (see Section~{\ref{waveletanalyses}} for more details).

A recent investigation of such global oscillations (in brightness temperature) from millimeter observations with ALMA also revealed the lack of 3-minute oscillations in the solar chromosphere in datasets with relatively large amounts of magnetic flux \citep{2021RSPTA.37900174J}. Conversely, the same study showed the presence of dominating 3-minute oscillations in the most magnetically quiescent datasets employed. However, due to the uncertain nature of those millimeter observations, particularly, their exact heights of formation, further investigations are required. Furthermore, \citet{Norton2021} reported on global oscillations in the photosphere, from SDO/HMI data, in various regions, namely, the quiet-Sun, plage, umbra and the polarity inversion line of an active region. While the 5-minute periodicity, with a considerably large power,  was found in all four areas in Doppler velocity perturbations, much smaller power enhancements could be observed in intensity and line-width observations of the quiet and plage regions. 

Of particular importance is also the effect of magnetic topology in the chromosphere, with the multi-layer magnetic canopy whose strength and thickness depends on, e.g., the magnetic flux involved in their generation \citep{2017ApJS..229...11J}.
By exploring the formation and properties of various chromospheric diagnostics, \citet{2017A&A...598A..89R} showed that the dense canopies of long opaque fibrils in the upper chromosphere, seen in H$\alpha$ line-core intensity images, could act as an `umbrella', obscuring the dynamics underneath. Thus, this could perhaps explain the lack of 3-minute oscillations in the high chromosphere (in addition to the magnetic shadows effect). In the case of ALMA observations,  \citet{2017A&A...598A..89R} speculated that the same phenomena could also occur, though at those wavelengths the dense fibrillar structures might not be visible due to their reduced lateral contrast (i.e., an insensitivity to Doppler shifts; ALMA observes continuum emission, and as such cannot be used to derive Doppler velocities) We note that similarities between ALMA observations (at 3~mm) and H$\alpha$ line-width images have been shown by \citet{2019ApJ...881...99M}.

All in all, it is important to investigate the variation of the global $p$-modes with height, throughout the lower solar atmosphere, in greater details. This can hopefully clarify whether the characteristic periodicity reported in previous studies is constant through the photosphere and the chromosphere, or whether they vary with height and/or in various solar regions.

\subsection{Large-scale Magnetic Structures}
\label{sec:largescalestructures}
Large magnetic structures, in the form of sunspots and solar pores, are considered ideal laboratories for the study of the excitation and propagation of MHD waves. Modern high resolution observations have revealed an extremely complex physical scenario in which different wave modes simultaneously co-exist in the same magnetic structure, hampering an unambiguous identification of individual wave modes. This is even more the case for highly structured magnetic fields, where the wave propagation reflects the geometrical complexity of the field lines acting as waveguides. However, in recent years our understanding of MHD waves in large magnetic structures, and their corresponding role in the heating of the solar atmosphere, has dramatically changed thanks to the opportunity offered by high-resolution fast-cadence tomographic imaging and to the new spectropolarimetric diagnostic capabilities, which have progressively extended up to chromospheric heights thanks to the technological advances of modern instrumentation. In particular, the inference of the plasma and magnetic field parameters obtained by spectropolarimetric inversion techniques have enabled the investigation of the effects of the magnetic field geometry on the wave propagation itself. In addition, new spectropolarimetric diagnostics have started to provide additional information about the magnetic field fluctuations, which are expected from several MHD wave modes.

The first oscillatory phenomena in the umbra of sunspots were observed by \citet{1969SoPh....7..351B} and \citet{1972SoPh...27...61B}, where the spatially localized brightenings (so-called `umbral flashes') were immediately associated with locally excited magnetoacoustic waves propagating upwards along the field lines. Today, after more than 50~years from the first discovery of these oscillations, our view of wave excitation and propagation of MHD waves in large magnetic structures has changed dramatically. From the observational point of view, in addition to the localized wave phenomena and disturbances in sunspots and pores (e.g., umbral flashes), the aforementioned instrumental advances have also allowed the identification of global eigenmodes of the magnetic structure (e.g., sausage modes, kink modes). These are generally mixed with other local disturbances, requiring specific filtering techniques (e.g., $k$-$\omega$ filtering; see Section~{\ref{sec:3Dfourieranalysis}}) for their identification. Although most of the literature on the subject mainly reflects this apparent dichotomy with the two classes of waves (i.e., global and local oscillations) addressed independently, most recent observations have started suggesting a superposition of locally excited magnetoacoustic waves resulting, for example, from $p$-mode absorption or residual local magneto-convection \citep{2015ApJ...812L..15K}, and global resonances of the magnetic structure. These two components can coexist, both of which contribute to the physical complexity of the observed velocity patterns; an aspect that was highlighted by \citet{roberts2019mhd}. In the following we wish to strike a balance between local disturbances and global resonances, and we will summarize the results from the most relevant studies in recent literature.

\begin{figure*}[!t]
\begin{center}
\includegraphics[trim=0mm 0mm 0mm 0mm, clip, width=0.49\textwidth]{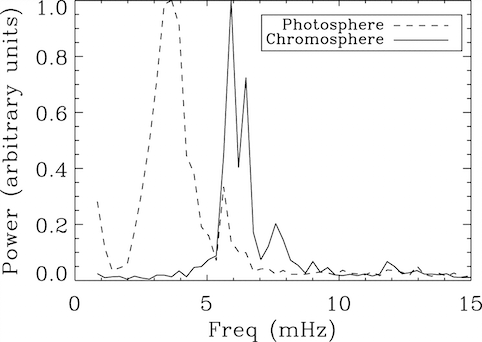}
\includegraphics[trim=0mm 0mm 0mm 0mm, clip, width=0.49\textwidth]{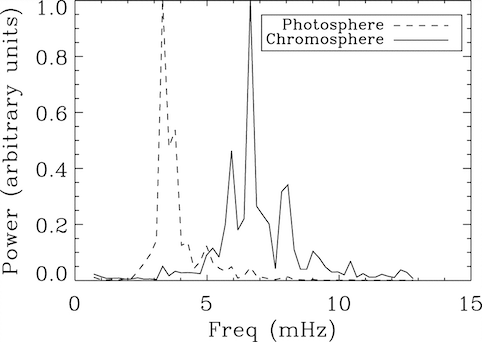}
\end{center}
\caption{Average umbral power spectra for two different sunspots. Solid lines indicate the power spectra of the chromospheric velocity oscillations averaged over each entire umbra, demonstrating a peak around 6~mHz. Dashed lines reveal the photospheric velocity power spectra averaged over each entire umbra, with a peak around 3.3~mHz and secondary peaks around 6~mHz. Image reproduced from \citet{2006ApJ...640.1153C}.}
\label{fig:Centeno_spectra} 
\end{figure*}

\subsubsection{Magnetoacoustic waves in large-scale magnetic structures}
\label{subsec:magnetoacoustic_waves}
Sunspots and other large magnetic structures, such as solar pores, typically display intensity and Doppler velocity power spectra that are dominated by 5-minute ($\sim$3~mHz) oscillations in the photosphere, and 3-minute ($\sim$5~mHz) oscillations in the chromosphere \citep[see for instance,][and references therein]{2006ApJ...640.1153C, 2009ApJ...692.1211C, 2010ApJ...722..131F, 2020NatAs.tmp..148F, 2020A&A...640A...4F}. Of course, it must be noted that the frequencies/periodicities found at photospheric and chromospheric heights are not universal values at precisely 3~mHz and 5~mHz, respectively. Indeed, windows of power are normally referred to when discussing the corresponding Fourier spectra \citep{2006ApJ...640.1153C, 2011ApJ...743..142H, 2013SoPh..282...67G, 2015LRSP...12....6K}, for example, $5\pm0.5$~minutes ($3.0-3.7$~mHz) and $3\pm0.5$~minutes ($4.8-6.7$~mHz) for the photosphere and chromosphere, respectively. These spectral features are depicted in Figure~\ref{fig:Centeno_spectra}, which clearly shows the frequency transition of peak power between the photospheric and chromospheric layers of two sunspots. While some authors have interpreted this to be the combined action of an acoustic cut-off \citep[$\omega_{c} \sim 5.3$~mHz, allowing the upward propagation of magnetoacoustic waves with $\omega > \omega_{c}$;][]{1984ARA&A..22..593D, 1991ApJ...373..308D, 1992A&A...266..532F, 1998MNRAS.298..464V} and the atmospheric density stratification resulting in the subsequent amplification of the wave amplitudes with height, others have explained the spectral features as the result of the presence of an acoustic resonator \citep{2021NatAs...5....5J, 2020NatAs...4..220J, 2020ApJ...900L..29F}. Power spectra similar to those shown in Figure~\ref{fig:Centeno_spectra} were also obtained by \citet{2016ApJ...831...24K} from observations of a sunspot with Hinode/SP (in Fe~{\sc i}~6301.5/6302.5{\,}{\AA}) and IRIS (in Si~{\sc iv}~1403{\,}{\AA}), corresponding to photospheric and transition-region heights, respectively. By comparing energy fluxes at the two atmospheric regions, \citet{2016ApJ...831...24K} speculated the three orders of magnitude energy decrease with height could suggest wave dissipation in the chromosphere. 

\begin{figure*}[!t]
\begin{center}
\includegraphics[width=\textwidth]{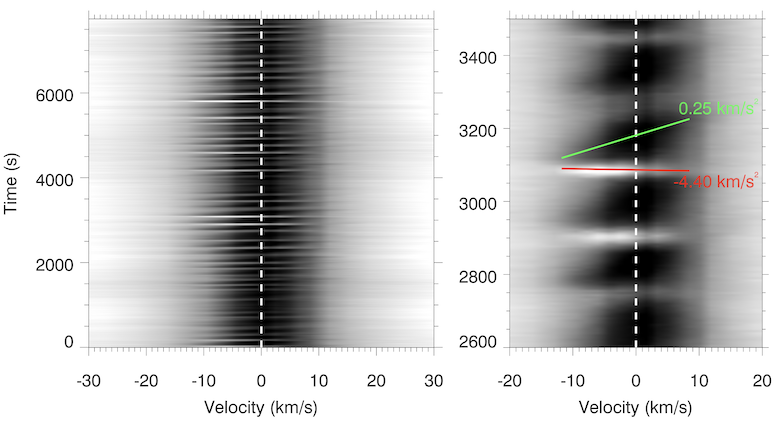}
\end{center}
\caption{A velocity--time graph extracted from IBIS observations of the sunspot umbral core on 24th August 2014 by \citet{2018NatPh..14..480G}. The horizontal axis represents Doppler line-of-sight velocity shifts from the rest wavelength, with the brightnesses displayed correlating to the Ca II 8542\AA~spectral profile of a single umbral pixel over the full time series. The left panel displays velocities of up to 30~km~s$^{-1}$ (or 0.85\AA), while the right panel zooms in to a smaller sub-set for a closer examination of the associated signatures. The red and green lines in the right panel denote the accelerations associated with the rising (blue-shifted) and falling (red-shifted) plasma, respectively. Image reproduced from \citet{2018NatPh..14..480G}.}
\label{fig:Grant_2018_shocks} 
\end{figure*}

When considering the umbrae of sunspots, the aforementioned umbral flashes dominated early observations after their initial detection. Umbral flashes (UFs) initially manifested as intensity brightenings in the core of the Ca~{\sc{ii}}~K spectral line, subtending multiple arcseconds across an umbra \citep{1969SoPh....7..351B}. Consistent with the dominant chromospheric frequencies mentioned above, UFs exhibited a 3-minute periodicity, however observed brightness increases of up to 150\% \citep{2000SoPh..192..373B} and line-of-sight velocity excursions of 10~km{\,}s$^{-1}$ \citep{1972SoPh...27...61B, 1975SoPh...41...71P} implied that these were not the signatures of linear MHD magneto-acoustic oscillations. Subsequent modelling efforts \citep[see the seminal works of][]{1997ApJ...481..500C, 2010ApJ...722..888B} established that UFs were the signature of shocks formed from the steepening of slow magneto-acoustic waves as they propagate through the large negative density gradients of the low chromosphere. When these non-linear shock fronts are formed, the intensity brightenings correspond to the dissipation of wave energy into plasma heat. At this stage, the plasma is no longer frozen into the magnetic field, and can propagate isotropically, however as the shocked plasma radiatively cools, gravitational effects will cause this overdense plasma to infall. The observational signatures of this morphology can be seen in Figure~\ref{fig:Grant_2018_shocks}, with periodic 3-minute intensity brightenings seen in concert with large velocity excursions consistent with the steepening of slow modes. The right panel of Figure~\ref{fig:Grant_2018_shocks} details the development of the shocked plasma, with the impulsive shock formation process, characterized by a notable blue-shifted velocity, leading to a more gradual red-shifted signature due to the infall of the plasma as it cools. This spectral morophology, known as a {\em{`saw-tooth'}} is distinctive in comparison to the sinusoidal behavior of linear MHD waves.

The nature of shock development in the solar atmosphere is deserving of its own dedicated review, as the three characteristic MHD wave speeds lead to a plethora of potential shock configurations \citep{2011JPlPh..77..207D}. Wave activity is also not the only driver, with magnetic reconnection capable of generating a range of shocks \citep{1964NASSP..50..425P, 2010RvMP...82..603Y}. The physical processes involved in shock dynamics entails that current modeling work still strives to replicate their behavior in realistic conditions \citep[e.g.,][]{2019A&A...626A..46S, 2021arXiv211102242S} and there have only been initial detections of other modes in the magnetic solar atmosphere \citep{2018NatPh..14..480G, 2020ApJ...892...49H}. In the context of UFs, it is more instructive to consider them as a dissipative process of waves, as opposed to wave behavior synonymous with the focus of this review. It is, however, useful to outline recent studies that characterize the effect of UFs on umbral plasma and their effectiveness as wave dissipators. It is also valuable to discuss the effect UFs have on observables, and their influence on attempts to extract linear MHD modes from sunspot umbrae.

\begin{figure*}[!t]
\begin{center}
\includegraphics[width=\textwidth]{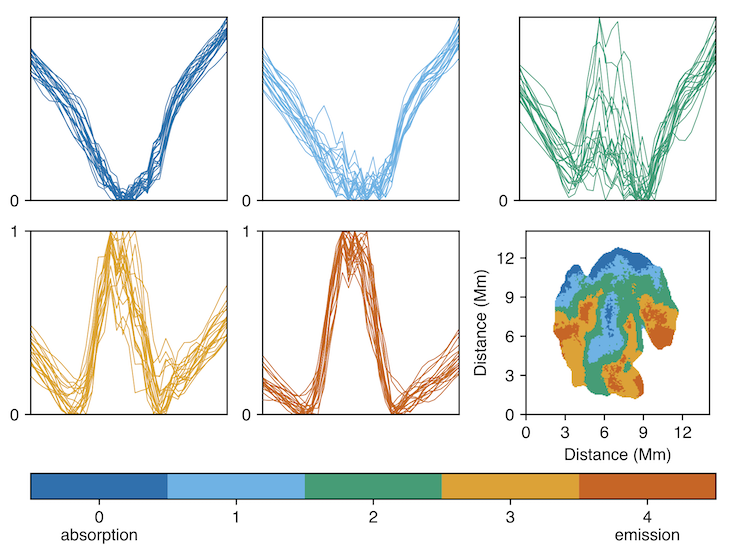}
\end{center}
\caption{Plots of stacked Ca~{\sc{ii}}~8542{\,}{\AA} umbral line spectra grouped by the neural network classification of \citet{2021RSPTA.37900171M}, where the intensity scale for each spectrum is normalized between `0' and `1' to aid visualization. A two-dimensional map (lower right) reveals the prominent neural network classifications for the Ca~{\sc{ii}}~8542{\,}{\AA} spectra present spatially across the umbra for a single IBIS spectral imaging scan. Image reproduced from \citet{2021RSPTA.37900171M}.}
\label{fig:Macbride_2021_class} 
\end{figure*}

At the turn of the 21$^{\mathrm{st}}$ century, as instrumental capabilities took a leap forward, initial studies into UF atmospheres still could not resolve any influence on the umbral magnetic field from shock fronts \citep{2003A&A...403..277R}. Instead, the temperature enhancements of UFs were characterized by \citet{2013A&A...556A.115D} by applying NICOLE inversion techniques on Ca~{\sc{ii}}~8542{\,}{\AA} data, with temperature excursions of up to 1000~K inferred. However, magnetic field perturbations were still unresolvable, likely due to the coarse spatial sampling of the data and small sample size of profiles inverted. The modification of the umbral magnetic field due to UFs was finally detected by \citet{2018ApJ...860...28H} using polarimetric He~{\sc{i}}~10830{\,}{\AA} observations. By sampling this high-chromospheric spectral line with the high spectral resolution of the FIRS instrument, HAZEL inversions \citep{2008ApJ...683..542A} revealed $\sim$200~G fluctuations in the vector magnetic field, and incremental changes in the inclination and azimuth of the field, approximately 8~degrees, implying that the magnetic field enhancement is predominantly along the direction of wave and shock propagation.  
\citet{2018ApJ...860...28H} also corroborated the temperature enhancements of shocks, though with smaller average values of $\sim$500~K, consistent with the higher atmospheric height sampled by He~{\sc{i}}~10830{\,}{\AA}, leading to observation of the shocked plasma as it enters into its cooling stage. Subsequently, magnetic field changes in Ca~{\sc{ii}}~8542{\,}{\AA} were reported \citep{2018A&A...619A..63J}, and the subsequent derivation of semi-empirical UF atmospheric models performed by \citet{2019A&A...627A..46B}. 

These results further reinforce the scenario where UFs perturb the magnetic field geometry of the umbra. However, the perturbation is always predominantly along the magnetic field vector, as the shock propagates, and the field returns to its unperturbed state once the shock has propagated through. Thus, shocks are not seen as candidates to further incline umbral fields to permit longer period waves to pass, or to greatly impact on adjacent waves as they propagate. Indeed, the perceived propagation of UF shock fronts horizontally across the umbra towards the penumbral boundary was instead interpreted as successive UFs developing along more inclined fields \citep{2015ApJ...800..129M}, further implying that the shocked plasma does not play a notable role in umbral morphology. Despite this, UFs have proved valuable in uncovering fine-scale umbral features and waves. \citet{2017ApJ...845..102H} utilized the brightenings of UFs to reveal small-scale horizontal magnetic fields across the umbra, revealing a complex `corrugated' structure to the field geometry in the chromosphere. UFs have been shown to generate a number of plasma flows with wave implications, notably \citet{2020A&A...642A.215H} detected downflows, upflows, and counter-flows before, after, and during the UFs, respectively. Recently, downflowing UFs have been found to be a signature of standing oscillations above sunspot umbrae \citep{2021A&A...645L..12F}. 

When considering the processes necessary to balance the chromospheric energy budget, shocks provide a macroscopic method for converting wave energy directly into local plasma heating. As discussed earlier, the intensity excursions of UFs are confirmed as signatures of heating, with between 500 - 1000K temperature increases observed as a result. When UFs are judged in terms of sole heaters of the chromosphere, \citet{2019ApJ...882..161A} derived a UF shock heating energy per unit mass of plasma that was insufficient to balance radiative losses. It is unsurprising that UFs alone are incapable of heating the chromosphere, particularly as they only occur in localized umbrae. However, the identification that slow-mode shocks impart heating energy is notable, given that it is proposed that such wave-driven shocks occur on a variety of scales across the solar atmosphere \citep{2021arXiv211102242S}, they present a viable method to potentially contribute to heating. In addition, as has been discussed, shock formation is not limited to slow-mode waves. \citet{2018NatPh..14..480G} observed shocks at the umbra-penumbra boundary of a sunspot that are inconsistent with the scenario of UF formation, as the inclined magnetic fields in this region does not produce a large density gradient to steepen magneto-acoustic waves. Instead, it was proposed that {\Alfven} waves were coupling to, and resonantly amplifying, magneto-acoustic waves in at the penumbral boundary to allow for shock formation, which was verified through the transverse velocity signatures in the shocks. These {\Alfven}-induced shocks produced local temperature enhancements of 5\%, less than simultaneous UFs, but providing dissipation of {\Alfven} waves in the chromosphere. This further highlights that the range of possible shock configurations are capable of dissipating a wide assortment of waves, including the elusive incompressible {\Alfven} mode, inferring that there is an tapestry of shock heating across the atmosphere that is yet to be fully uncovered.

When seeking to observe waves in sunspot umbrae, UFs must always be taken into account. The intensity excursions associated with UFs have noticable effects on the spectral profiles sensitive to the density and temperature perturbations of UFs, such as Ca~{\sc{ii}}~H/K, Ca~{\sc{ii}}~8542{\,}{\AA}, He~{\sc{i}}~10830{\,}{\AA}, and upper chromospheric/transition region channels from IRIS, such as C~{\sc ii}~1335.71{\,}{\AA}, Mg~{\sc ii}~2796.35{\,}{\AA}, and Si~{\sc iv}~1393.76{\,}{\AA} \citep{2014ApJ...786..137T,2021ApJ...906..121K}. The line-core emission from these brightenings causes non-trivial profile shapes to develop, and introduces opacity effects that inhibits velocity inference through profile fitting (as seen by classes 3 \& 4 of Figure~\ref{fig:Macbride_2021_class}). A method of recovering velocity information from these profiles is to use inversion methods such as NICOLE and HAZEL, however, these are computationally intensive, time consuming, and do not account for multi-component atmospheres. The seminal work of \citet{2000Sci...288.1396S} showed that in the observation of a column of shocked plasma, there is always an intermixing of active and quiescent atmospheres below the resolution limit of the observations. Thus, an observed spectral profile will in-fact be the result of a two-component atmosphere, where the filling factor is unknown. It is therefore possible to extract the linear oscillations embedded in a bright pixel if the atmospheres can be separated. This was investigated by \citet{2021RSPTA.37900171M} and \citet{2021JOSS....6.3265M}, who employed machine learning techniques to develop a neural network capable of classifying Ca~{\sc{ii}}~8542{\,}{\AA} as a function of line-core emission (see Figure~\ref{fig:Macbride_2021_class}). The authors were then able to separate the two individual atmospheres through model fitting to provide values for both the shocked plasma velocity and the associated quiescent component. For any observer looking into waves in the umbra of a sunspot, care must be taken to account for UF signatures, either through exclusion of these signatures, two-component fitting, or the use of spectral lines that are less sensitive to temperature changes, such as H$\alpha$ \citep{2008A&A...480..515C, 2009A&A...503..577C}.

The propagation of waves from the umbrae outward (i.e., along penumbral filaments) are known as running (penumbral) waves \citep[RPWs;][]{1972SoPh...27...71G, 1972ApJ...178L..85Z, 1997ApJ...478..814B, 2000A&A...354..305C,  2001A&A...375..617C, 2000A&A...363..306G, 2004A&A...424..671K, 2006RSPTA.364..313B, 2007ApJ...671.1005B, 2009A&A...505..791S, 2013ApJ...779..168J, 2015ApJ...800..129M, 2015A&A...580A..53L, 2016AN....337.1040L, 2018ApJ...869..110S}, which have also been attributed to magnetoacoustic wave modes \citep{1997ApJ...478..814B, 2004A&A...424..671K}. RPW phenomena are mostly prominent in the mid-to-upper chromosphere, though they have also been observed at photospheric heights \citep{2015A&A...580A..53L, 2015ApJ...809L..15Z}.

The origin of RPWs has long been debated as either a chromospheric phenomenon visible as a result of trans-sunspot wave interactions \citep[e.g.,][]{1992SoPh..138...93A, 1996SoPh..167...79T, 2000A&A...355..375T, 2006A&A...456..689T, 2006RSPTA.364..313B, 2017ApJ...850..206S, 2017Ap&SS.362...46Z, 2018ApJ...852...15P}, or as the chromospheric signature of upwardly propagating and magnetically guided $p$-mode waves from the sub-photospheric layers \citep[e.g.,][]{2000A&A...354..305C, 2001A&A...375..617C, 2000A&A...363..306G, 2003A&A...403..277R, 2007ApJ...671.1005B, 2012ApJ...756...35R, 2012ApJ...746..119R, 2013ApJ...779..168J, 2014A&A...561A..19Y, 2015ApJ...800..129M}. When viewed as a function of radial distance from the umbral center, RPW signatures manifest with large apparent phase speeds ($\sim40$~km/s) and relatively high frequencies ($\sim5$~mHz) at the umbra/penumbra boundary, decreasing to lower apparent phase speeds ($\sim10$~km/s) and reduced frequencies ($\sim1$~mHz) towards the outer penumbral edge \citep{2006SoPh..238..231K}. This effect can be also seen in Figures~\ref{fig:Jess_2013_power_vs_geometry1} \& \ref{fig:Jess_2013_power_vs_geometry2}, which are reproduced from \citet{2013ApJ...779..168J}. Here, both the amplitude and frequency of the captured wave modes is found to depend strongly on the magnetic field geometry at chromospheric heights. The dominant frequency of the waves progressively extends towards lower values (longer periods) as one moves from the center of the umbra and into the surrounding regions with more heavily inclined magnetic fields (see the discussions below involving the ramp effect).

\citet{2015ApJ...800..129M} examined datasets from both the SDO and IRIS spacecrafts and concluded that the apparent trans-sunspot motion associated with RPWs is not a real effect, but instead is a result of the waves (originating from the photospheric $p$-modes) traveling along magnetic field lines of increasing inclination angle away from the umbral core. On the other hand, \citet{2018ApJ...852...15P} examined high resolution observations from the Goode Solar Telescope \citep[GST;][]{2010AN....331..636C}. The authors found that oscillatory events in the sunspot umbra appeared to initiate from earlier occurring RPWs, which, in turn, caused the development of new RPW events. This was proposed to be evidence that many of the RPW signatures that are seen at high spatial resolutions may be entirely chromospheric in origin. However, the authors also suggest that complex, twisted magnetic field geometry can create a scenario where wave emergence seems to contradict \citet{2015ApJ...800..129M}. As a result, with next generation instrumentation and facilities imminent, close attention will need to be paid to multi-wavelength (i.e., multi-height) observations in order to compare the small- and large-scale characteristics of RPWs, which will help to unequivocally determine the underlying physics that underpins their visible signatures.

\begin{figure*}[!t]
\begin{center}
\includegraphics[width=\textwidth]{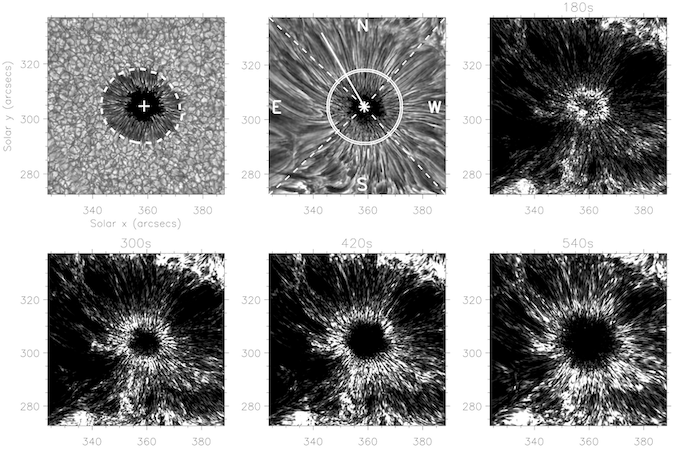}
\end{center}
\caption{Simultaneous images of the blue continuum (photosphere; upper left) and H$\alpha$ core (chromosphere; upper middle) acquired by the DST using the ROSA imaging instrument. A white cross marks the center of the sunspot umbra, while a white dashed line in the continuum image displays the extent of the photospheric plasma-$\beta = 1$ isocontour. The white concentric circles overlaid on the chromospheric image depict a sample annulus used to extract wave characteristics as a function of distance from the center of the umbra, while the solid white line extending into the north quadrant reveals the slice position used for the time–distance analysis displayed in Figure~{\ref{fig:Jess_2013_power_vs_geometry2}}. The dashed white lines isolate the active region into four distinct regions, corresponding to the North (N), South (S), East (E), and West (W) quadrants. The scale is in heliocentric coordinates where $1'' \approx 725$~km. The remaining panels display a series of chromospheric power maps extracted through Fourier analysis of the H$\alpha$ time series, indicating the locations of high oscillatory power (white) with periodicities equal to $180$, $300$, $420$, and $540$~s. As the period of the wave becomes longer, it is clear that the location of peak power expands radially away from the center of the umbra. This effect is synonymous with the presence of running penumbral waves (RPWs), which were first identified in solar images by \citet{1972SoPh...27...71G} and \citet{1972ApJ...178L..85Z}. Image reproduced from \citet{2013ApJ...779..168J}.}
\label{fig:Jess_2013_power_vs_geometry1} 
\end{figure*}

\begin{figure*}[!t]
\begin{center}
\includegraphics[width=8cm]{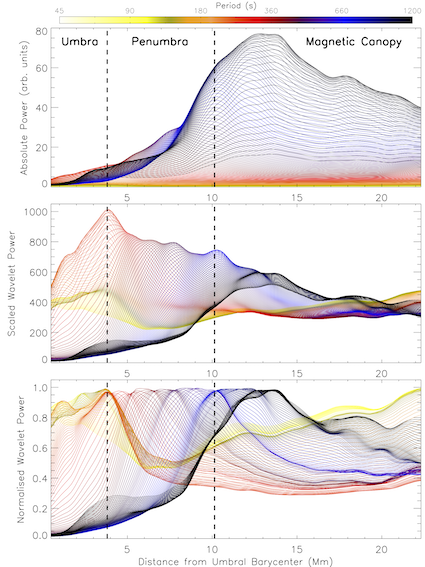}
\end{center}
\caption{{\it{Top:}} azimuthally averaged absolute Fourier power displayed as a function of radial distance from the center of the umbra. {\it{Middle:}} power spectra from the top panel normalized by the average power for that periodicity within the entire field of view. Thus, the vertical axis represents a factor of how much each period displays power above its spatially and temporally averaged background. {\it{Bottom:}} power spectra normalized to their own respective maxima. The vertical dashed lines represent the radial extent of the umbral and penumbral boundaries, while the graduated color spectrum, displayed in the color bar at the top, assigns display colors to a series of increasing periodicities between $45$ and $1200$~s. Image reproduced from \citet{2013ApJ...779..168J}.}
\label{fig:Jess_2013_power_vs_geometry2} 
\end{figure*}

\begin{figure*}[!t]
\begin{center}
\includegraphics[width=8cm]{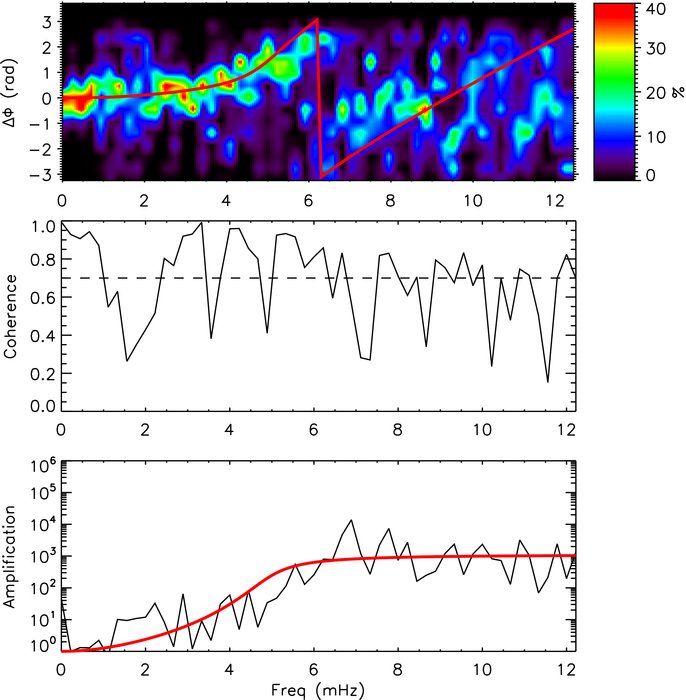}
\end{center}
\caption{Phase spectrum (upper), coherence (middle; see Section~{\ref{sec:calculatingconfidencelevels}}), and amplification (lower) of Doppler velocity oscillations observed in the photospheric (Si~{\sc{i}}) and chromospheric (He~{\sc{i}}) spectral lines in a sunspot atmosphere. The red line in the upper panel represents the best fit from a theoretical model. The horizontal dashed black line at a coherence value of 0.7 in the middle panel highlights the lower confidence threshold. The red line in the lower panel represents the best fit from a theoretical model of acoustic waves propagating in an isothermal and stratified atmosphere (see text for more details). Image reproduced from \citet{2010ApJ...722..131F}.}
\label{fig:Felipe_propagation} 
\end{figure*}

\begin{figure*}[!t]
\begin{center}
\includegraphics[width=6cm]{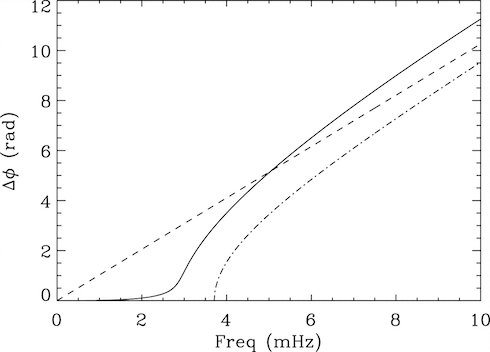}
\end{center}
\caption{Phase difference spectrum for acoustic oscillations sampled at two geometric heights. The dashed and dot-dashed lines represent the expected phase as a function of frequency for non-stratified and vertically stratified isothermal atmospheres, respectively. The solid line indicates the phase relationship in a vertically stratified atmosphere including radiative losses from Newton's cooling law. For each of the three cases depicted, the same plasma parameters are utilized ($T = 9000$~K, $\Delta{z} = 1600$~km, and $g = 274$~m{\,}s$^{-2}$). Image reproduced from \citet{2006ApJ...640.1153C}.}
\label{fig:Centeno_cutoff} 
\end{figure*}

In Figure~\ref{fig:Felipe_propagation} we show a typical phase diagram from \citet{2010ApJ...722..131F} that was obtained by simultaneously measuring the Doppler velocity at both photospheric and chromospheric heights in a sunspot umbra. Here, we see the clear effect of an embedded acoustic cut-off, with frequencies above $\sim$5.3~mHz having a positive phase lag, highlighting the upward propagation of these waveforms.  Consequently, these waves experience a rapid density drop as they propagate into the chromosphere, thus resulting in strong amplification of their amplitudes (see the lower panel of Figure~{\ref{fig:Felipe_propagation}}), which eventually results into shock formation. Following the methodology put forward by \citet{1958ApJ...127..459F} and \citet{2006ApJ...640.1153C}, the amplitude, $A$, of a monochromatic wave with frequency, $\omega$, in a plane-parallel isothermal atmosphere permeated by a uniform vertical magnetic field comes from the solution to the equation,
\begin{equation}
\label{eqn:amplitude_amplification}
    c_{s}^{2} \frac{d^{2}A(z)}{dz^{2}} - \gamma g \frac{dA(z)}{dz} + \omega^{2}A(z)=0 \ ,
\end{equation}
where $z$ is the vertical coordinate, $g$ is the acceleration due to gravity, ${c_{s}=\gamma g H_{0}}$ is the speed of sound, $H_{0}$ is the pressure scale height, and $\gamma$ is the ratio of specific heats, which equals $5/3$ for a monoatomic gas demonstrating adiabatic propagation. A solution to Equation~{\ref{eqn:amplitude_amplification}} is given by the trial function,
\begin{equation}
    A(z)=e^{i k_{z} z} \ ,
\end{equation}
where $k_{z}$ represents the vertical wavenumber. Solving for the vertical wavenumber, $k_{z}$, provides a dispersion relation of the form,
\begin{equation}
    k_{z}=\frac{1}{c_{s}} \left( -i \omega_{c} \pm \sqrt{\omega^{2} -\omega^{2}_{c}} \right) \ ,
\end{equation}
where $\omega_{c}=\gamma g/2c_{s}$ is the cut-off frequency. For $\omega<\omega_{c}$, $k_{z}$ takes imaginary values and the wave is evanescent. In the opposing regime (i.e., $\omega>\omega_{c}$), waves are able to propagate. This is illustrated in Figure~\ref{fig:Centeno_cutoff}, where the phase angle is displayed as a function of frequency for waves measured at two independent geometric heights for both non-stratified and stratified atmospheric models. The cut-off frequency appears as a natural consequence of the stratification itself. However, as also shown in Figure~{\ref{fig:Centeno_cutoff}}, a distinct cut-off frequency only exists in the limit of negligible radiative losses. In more realistic models, which include aspects of radiative cooling, a sharp separation between the propagating and evanescent regimes does not exist (solid line in Figure~{\ref{fig:Centeno_cutoff}}), with the resulting phase diagram displaying a smooth transition at around $3$~mHz. Of course, the above equations that represent the cut-off frequency are only valid in the limit of an isothermal atmosphere.

The strong vertical stratification of the atmospheric parameters in sunspots result in orders-of-magnitude changes to the propagation speeds of the embedded magnetoacoustic waves; namely the {\Alfven} speed, $v_{A}$, and the sound speed, $c_{s}$. This, together with the vertical and horizontal gradients of the background magnetic field, alongside other variations in local plasma parameters, constitutes important ingredients in the propagation characteristics of magnetoacoustic waves in these magnetic structures \citep{2022ApJ...938..154M}. In Figure~\ref{fig:sunspot_model} we show the variation of the sound and {\Alfven} speeds in a typical small sunspot model. To relate these variations to real observations we must employ approximations, which leads to slightly different but significant changes \citep{2020A&A...640A...4F}. The cut-off frequency has been found to significantly change as a function of atmospheric height, generating important implications for both the heating of the upper layers of the Sun's atmosphere and the wave propagation itself \citep{Wi_niewska_2016, 2018A&A...617A..39F}.

Further to changes with atmospheric height, it has been shown that the cut-off frequency depends on the magnetic field inclination, with more inclined fields allowing the upward propagation of frequencies below the $\sim$5.3~mHz threshold \citep{2007ApJ...671.1005B, 2013ApJ...779..168J} -- the so-called {\it{ramp effect}}. These waves are generally interpreted as longitudinal slow magnetoacoustic waves, with the general consensus on their origin being photospheric, through the absorption of externally driven $p$-modes \citep[see for instance,][]{1992ApJ...391L.109S, 2003SoPh..214..201C, 2012ApJ...757..160J, 2015ApJ...812L..15K}, with trans-sunspot oscillations at chromospheric heights potentially influencing the behaviour of these wave trains  \citep[e.g.,][]{2017ApJ...836...18C, 2020ApJ...888...84S}. Although these observational results are in agreement with the theoretical scenario of propagating slow magnetoacoustic modes, which in the low plasma-$\beta$ regime correspond to acoustic-like waves propagating along field lines with the magnetic pressure as the dominant restoring force, other magnetoacoustic modes exist in spatially uniform plasmas: namely an incompressible wave with magnetic tension as the sole restoring force ({\Alfven} wave), and an intermediate wave mode that can be thought of as a generalization of an acoustic wave with contributions from magnetic pressure (fast wave in the presence of a low plasma-$\beta$). In other words, in low plasma-$\beta$ environments the fast mode is an acoustic wave modified by the magnetic tension, capable of propagating isotropically with respect to the magnetic field. 

Interestingly, at the equipartition layer where the sound and {\Alfven} speeds are nearly equal, a fraction of the energy, $C$, can be either channeled from a fast mode in the high plasma-$\beta$ regime (which is mainly an acoustic-like wave) to a fast magnetoacoustic mode in the low plasma-$\beta$ regime, or converted into a slow mode, thus preserving its acoustic nature. If we consider the sound speed, $c_{s}=\sqrt{\gamma P_{0}/\rho_{0}}$, and the {\Alfven} speed, $v_{A}=B/(4 \pi \rho_{0})$, where $\gamma$ is the adiabatic index, $P_{0}$ is the gas pressure, $\rho_{0}$ is the density, and $B$ is the magnitude of the magnetic field strength, then the ratio between the two speeds squared can be given by,
\begin{equation}
\label{eqn:plasma_beta}
    \frac{c_{s}^2}{v_{A}^2}=\gamma \frac{4 \pi P_{0}}{B^{2}} \ .
\end{equation}
Substituting the magnetic pressure, $P_{B}=B^{2}/(8 \pi)$, into Equation~{\ref{eqn:plasma_beta}} we obtain, 
\begin{equation}
  {\frac{c_{s}^2}{v_{A}^2} = \frac{\gamma}{2} \beta \ . }
   \label{eqn:equipartition}
\end{equation}
This means that the equipartition layer is in practice close to the plasma-$\beta=1$ surface, and although they are conceptually different, they are often difficult to segregate from one another in observational data sequences \citep[e.g., see the discussion points raised by][]{2018NatPh..14..480G}.

\begin{figure*}[!t]
\begin{center}
\includegraphics[trim = 0cm 2cm 0cm 0cm, clip, width=\textwidth]{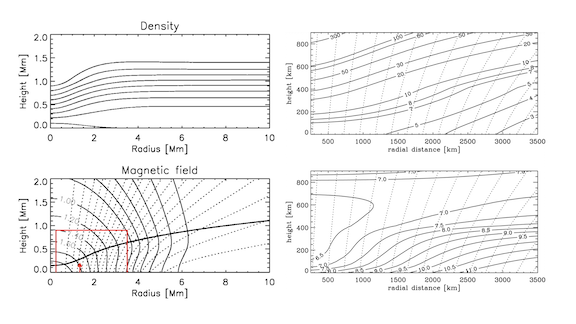}
\end{center}
\caption{Contours of constant density (upper left) and constant magnetic field strength (lower left) for a typical small sunspot. In the lower-left panel the labels indicate the magnetic field strength in units of kG, while the thick black line denotes the isosurface where $v_{A} = c_{s}$. The dotted lines indicate the geometries of the embedded magnetic field lines. The red box corresponds to the domain size displayed in the right hand panels, where contours of constant {\Alfven} speed ($v_{A}$) and constant sound speed ($c_{s}$), in units of km/s, are depicted in the upper-right and lower-right panels, respectively. Note the strong horizontal gradients of both $v_{A}$ and $c_{s}$ due to the Wilson depression. The dotted lines, as per the lower-left panel, indicate the geometries of the magnetic field lines. Image reproduced from \citet{2006ApJ...653..739K}.}
\label{fig:sunspot_model} 
\end{figure*}

The wave translation process from one form to another is generally referred to as one of two processes: {\it{mode conversion}} or {\it{mode transmission}} \citep{2005SoPh..227....1C, 2005ApJ...632L..49S, 2015ApJ...814..106C, 2017A&A...601A.107P}. Here, `mode conversion' refers to a wave that retains its original character (i.e., fast-to-fast or slow-to-slow), yet {\it{converts}} its general nature in the form of acoustic-to-magnetic or magnetic-to-acoustic. Contrarily, `mode transmission' refers to a wave that preserves its general nature (i.e., remains a `magnetic' or `acoustic' mode), yet changes character from fast-to-slow or vice versa. The fraction of energy that can be converted from fast to slow modes depends on the attack angle of the wave with respect to the magnetic field lines. The precise transmission coefficient, $T$, is defined as the proportion of incident wave energy flux transmitted from fast to slow acoustic waves \citep{2001ApJ...548..473C, Cally2007}, which is governed by, 
\begin{equation}
    T = e^{-\pi k h_{s} sin^{2}(\alpha)} \ , 
\end{equation}
where $k$ is the wavenumber, $h_{s}$ the thickness of the conversion layer, and $\alpha$ the attack angle itself. The fast-to-fast conversion coefficient, $C$, can then be obtained by invoking energy conservation: $T + |C|=1$, where $C$ is a complex energy fraction to take into account possible phase changes during the process of mode conversion \citep{2009SoPh..255..193H}. We note that the conversion coefficient, $C$, is larger when the frequency of the incident waves is higher and the attack angle is larger \citep{2014A&A...567A..62K}.

\begin{figure*}[!t]
\begin{center}
\includegraphics[width=\textwidth]{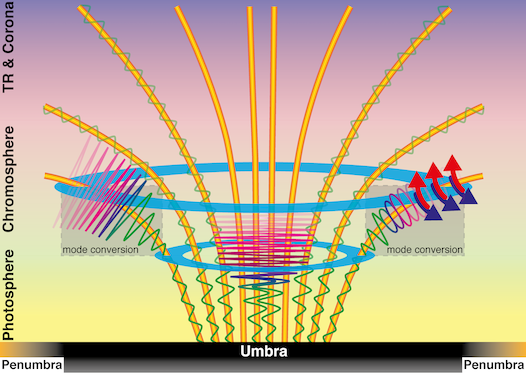}
\end{center}
\caption{A cartoon representation of a sunspot umbral atmosphere demonstrating a variety of shock phenomena. A side-on perspective of a typical sunspot atmosphere, showing magnetic field lines (orange cylinders) anchored into the photospheric umbra (bottom of image) and expanding laterally as a function of atmospheric height, into the upper atmospheric regions of the transition region (TR) and corona. The light blue annuli highlight the lower and upper extents of the mode-conversion region for the atmospheric heights of interest. The mode-conversion region on the left-hand side shows a schematic of non-linear {\Alfven} waves resonantly amplifying magnetoacoustic waves, increasing the shock formation efficiency in this location. The mode-conversion region on the right-hand side demonstrates the coupling of upwardly propagating magnetoacoustic oscillations (the sinusoidal motions) into {\Alfven} waves (the elliptical structures), which subsequently develop tangential blue- and red-shifted plasma during the creation of {\Alfven} shocks. The central portion represents the traditional creation of umbral flashes that result from the steepening of magnetoacoustic waves as they traverse multiple density scale heights in the lower solar atmosphere. The image is not to scale and is reproduced from \citet{2018NatPh..14..480G}.} 
\label{fig:Grant2018figure03} 
\end{figure*}

\citet{2006MNRAS.372..551S} have shown how the combination of mode conversion alongside the ramp effect can result in an acoustic flux which is strongly dependent on the magnetic field geometry. This was also confirmed by \citet{2011A&A...534A..65S}, who found a strong dependence of the wave flux between the photosphere and chromosphere on the magnetic field geometry inferred from spectropolarimetric inversions.

\begin{figure*}[!t]
\begin{center}
\includegraphics[width=\textwidth]{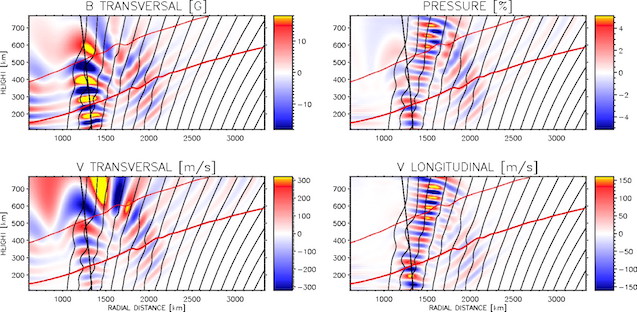} \\
\vspace{3mm} \includegraphics[width=\textwidth]{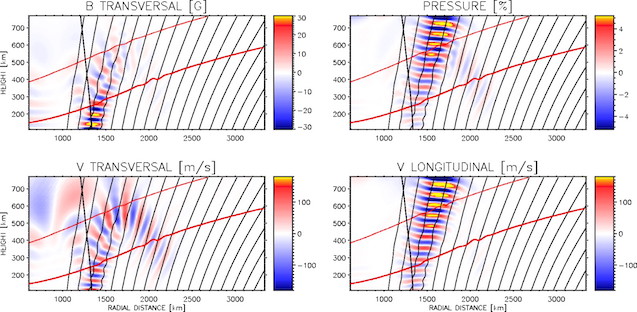}
\end{center}
\caption{Variations of the magnetic field (upper-left), pressure (upper-right), and velocity (transverse, lower-left; longitudinal, lower-right) at an elapsed time, ${t = 100}$~s, after the beginning of the simulations for a vertical longitudinal driver with a 10~s (100~mHz) periodicity. The lower 4 panels are identical to the upper 4 panels, only now show the variations for a horizontal transverse driver with a 10~s (100~mHz) periodicity. In each panel, the horizontal axis represents the radial distance from the center of the sunspot, while the inclined black lines highlight the magnetic field orientation. The two red lines in each panel indicate contours of constant $c_{s}/v_{A}$, with the thicker red line corresponding to $v_{A} = c_{s}$, and the thinner red line to $c_{s}/v_{A} = 0.1$. The thick black lines inclined towards the left (with increasing atmospheric height) of each panel indicate the direction of $\nabla v_{A}$, starting at the location of the pulse. Here, the $\nabla v_{A}$ line represents the boundary that separates waves refracting to the right from those refracting to the left, which is perpendicular to the contours of constant $v_{A}$ at every geometric height. Image reproduced from \citet{2006ApJ...653..739K}.}
\label{fig:sunspot_model_longitudinal_pulse} 
\end{figure*}

\citet{2018NatPh..14..480G} have also shown, by exploiting unique high-resolution observations and magnetic field extrapolations, combined with thermal inversions and MHD wave theory, that magnetoacoustic waves can couple with {\Alfven} waves at the equipartition layer in a sunspot, resulting in {\Alfven}-driven shocks that can efficiently contribute to the overall energy budget of the chromosphere (see Figure~\ref{fig:Grant2018figure03}).

Mode conversion and propagation of magnetoacoustic wave modes in sunspot atmospheres has also been investigated through numerical two-dimensional simulations \citep[e.g.,][]{2008ApJ...689.1379K} incorporating realistic sunspot atmospheres (see Figure~\ref{fig:sunspot_model}). In particular, the atmospheric response to both longitudinal and transverse pulses (with respect to the magnetic field lines) has been investigated. Figure~{\ref{fig:sunspot_model_longitudinal_pulse}} reveals the velocity, magnetic field, and pressure fluctuations for a wave with input frequency above the cut-off value (10~s periodicity or 100~mHz) following 100~s of simulation run time for both longitudinal and transverse pulses, respectively. It is clear from Figure~{\ref{fig:sunspot_model_longitudinal_pulse}} that the specific input pulse type results in different mixtures of transverse and longitudinal wave modes, which undergo mode conversion at the {\Alfven}/acoustic equipartition layer, and may also be reflected or refracted by the vertical and horizontal gradients. 


From theoretical studies, it was suggested that the enhanced 3-minute wave power observed at chromospheric heights in sunspot umbrae may come from the presence of an acoustic resonance cavity, which is established by the temperature gradients at both the photospheric and transition region boundaries \citep[see for instance,][]{1979SoPh...62..227H, 2011ApJ...728...84B, 2015A&A...580A.107S, 2019A&A...627A.169F}. Recently, \citet{2020NatAs...4..220J} exploited multi-height high spatial and temporal resolution observations, spectropolarimetric inversions, and numerical modeling to provide an observational confirmation of this physical mechanism. The authors examined the Fourier power spectra originating within a sunspot umbra and compared this to high-precision simulations encompassing a variety of different atmospheric stratifications. It was found that once steep temperature gradients were introduced into the simulation, the resulting cavity produced resonant amplification of the 3-minute oscillations (see Figure~{\ref{fig:Jess_IBIS_stack}}). Following on from the study by \citet{2020NatAs...4..220J}, \citet{2020ApJ...900L..29F} independently confirmed the presence of an acoustic resonator for another sunspot structure, and highlighted the potential importance of such findings for future helioseismic investigations of the solar atmosphere.

\begin{figure*}[!t]
\begin{center}
\includegraphics[width=0.27\textwidth]{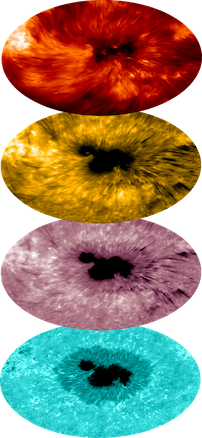}
\includegraphics[width=0.64\textwidth]{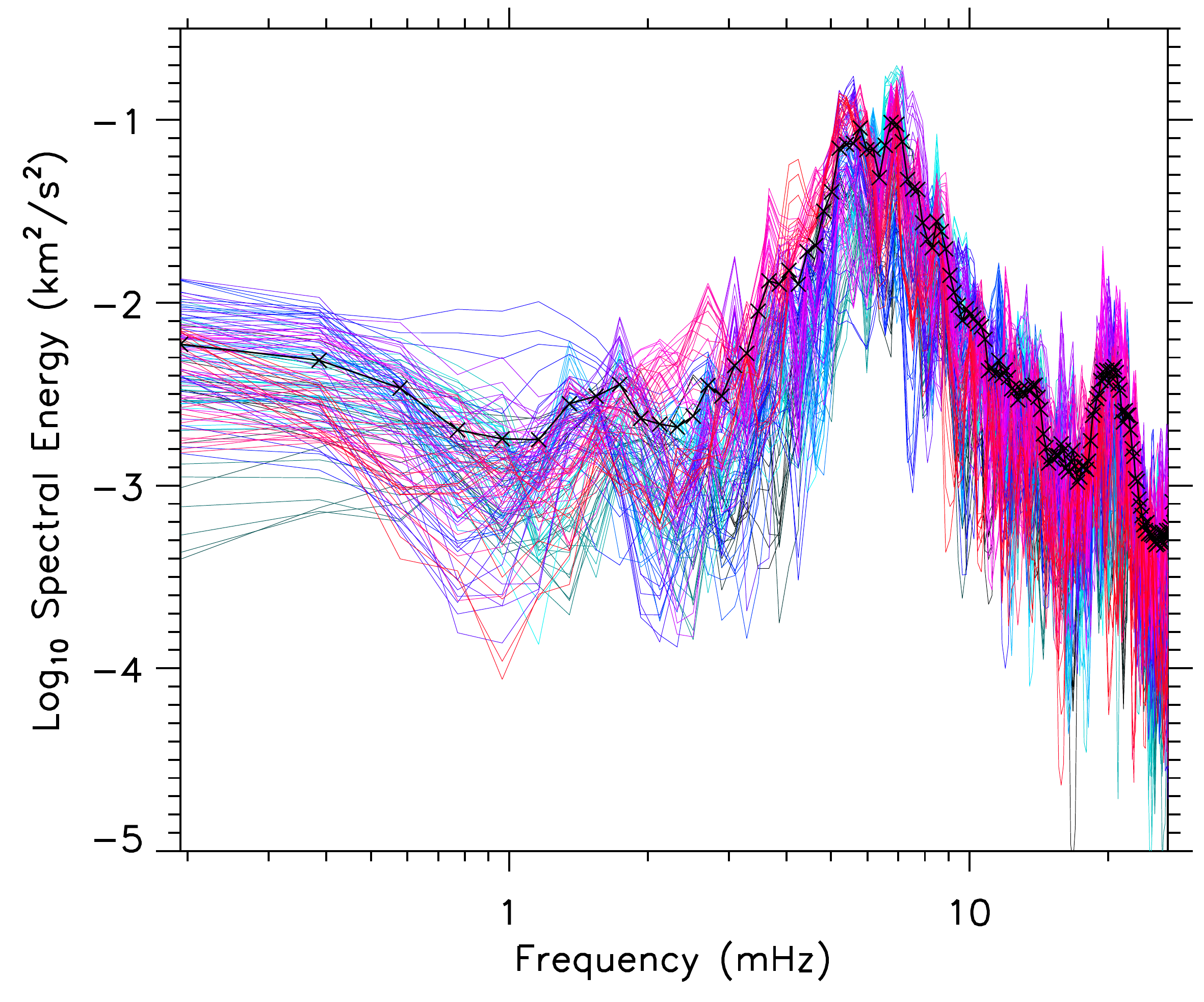}
\end{center}
\caption{A vertical stack of narrowband images taken across the Ca~{\sc{ii}}~8542{\,}{\AA} spectral line by the IBIS Fabry-P{\'{e}}rot instrument at the DST, revealing the photospheric (blue) to chromospheric (red) stratification of the sunspot atmosphere (left panel). This sunspot was found to display characteristics consistent with the presence of a resonance cavity, which caused the manipulation of spectral energies across the frequency domain (right panel), in particular providing a resonant enhancement at $\approx$20~mHz. Image adapted from \citet{2020NatAs...4..220J}.}
\label{fig:Jess_IBIS_stack} 
\end{figure*}

\subsection{Eigenmodes of Large-scale Magnetic Structures}
\label{sec:globaleigenmodes}
MHD theory applied to cylindrical magnetic flux tubes predicts a series of wave modes \citep{1983SoPh...88..179E}, which represent the intrinsic overall response of the magnetic structure to the external forcing. These wave modes manifest over all scales of magnetic flux tube, however, due to their irregular cross sectional shapes and corrugated boundaries with surrounding plasma, sunspots and pores may significantly alter the characteristics of their resonant modes and their azimuthal properties \citep{2021RSPTA.37900181A}. Although as in any natural system one may also expect resonant modes in large scale magnetic structures, such as sunspots and pores, they are still poorly investigated in these structures. One reason is likely coupled to the fact that the amplitudes of the oscillations belonging to these modes are inherently weaker than that of the intensity and velocity excursions associated with local magnetoacoustic fluctuations such as umbral flashes, requiring particular filtering techniques for their identification.

\begin{figure*}[!t]
\begin{center}
\includegraphics[width=\textwidth]{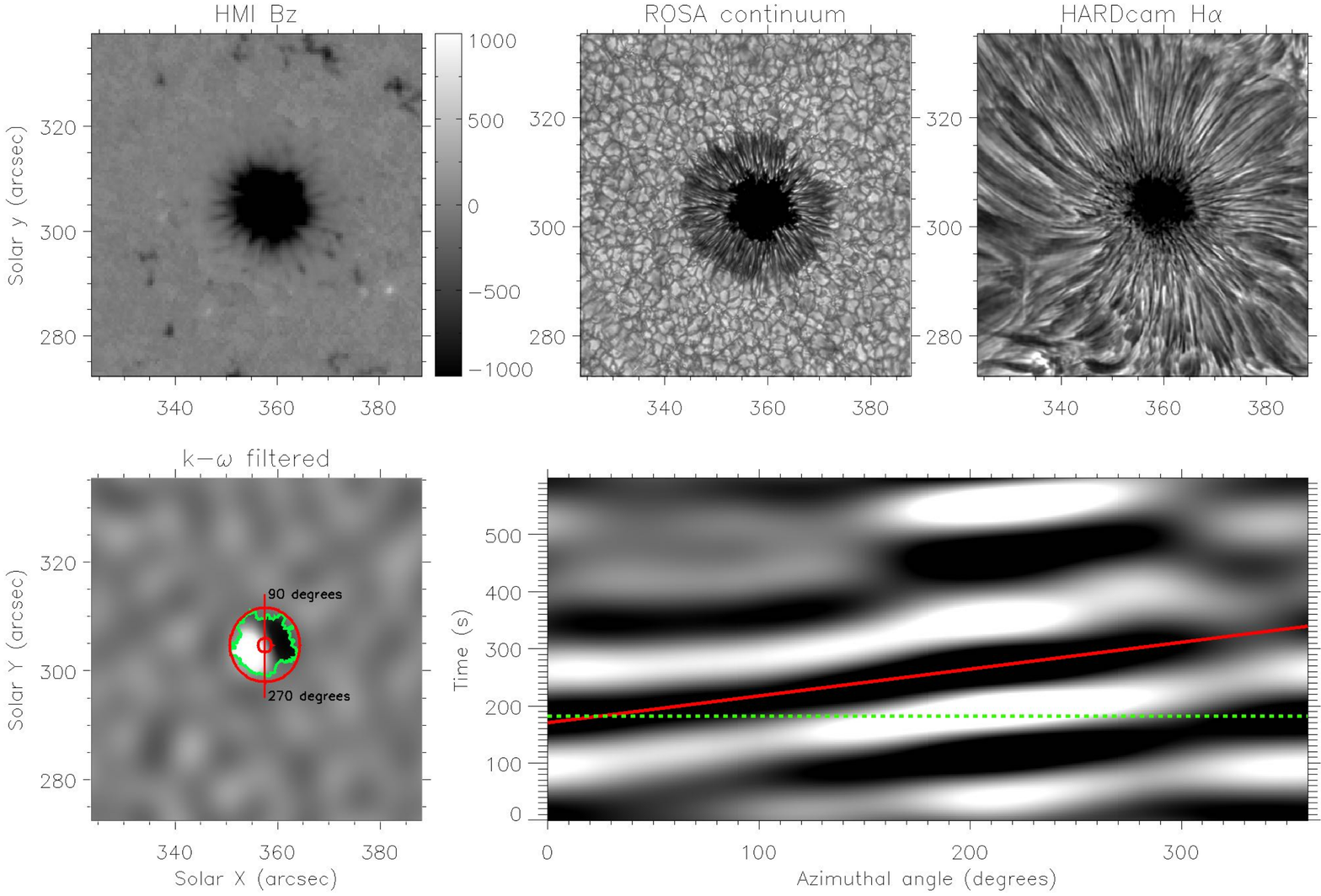}
\end{center}
\caption{Sample images of a near circularly symmetric sunspot, including the vertical component of the magnetic field ($B_{z}$; upper left), the photospheric continuum (upper middle), and the chromospheric H$\alpha$ line core (upper right). The color bar corresponding to the strength of the magnetic field is saturated at $\pm$1000~G for visual clarity. The lower left panel displays a snapshot of H$\alpha$ intensities following $k$-$\omega$ filtering (i.e., in both temporal and spatial domains; see Section~{\ref{sec:threedimensionalFourierfiltering}}). The solid green contour outlines the time-averaged umbra/penumbra boundary, while the red annulus depicts the extent of the region used for examining azimuthal wave motion within the umbra, where the center of the annulus is placed at the center of the umbra. The lower right panel is a time-azimuth diagram following the polar transformation of the signals contained within the red annulus in the lower left panel, which allows the circular nature of the wave rotation to be investigated in a similar way to traditional time-distance diagrams. The horizontal dashed green line highlights the azimuthal intensity signal corresponding to the filtered image shown in the lower left panel, while the solid red line represents the fitted angular frequency (i.e., degrees per second) of the rotating wave amplitudes. Image reproduced from \citet{2017ApJ...842...59J}.}
\label{fig:Jess_2017_fig01} 
\end{figure*}

\begin{figure*}[!t]
\begin{center}
\includegraphics[trim=0mm 3mm 0mm 0mm, clip, width=0.43\textwidth]{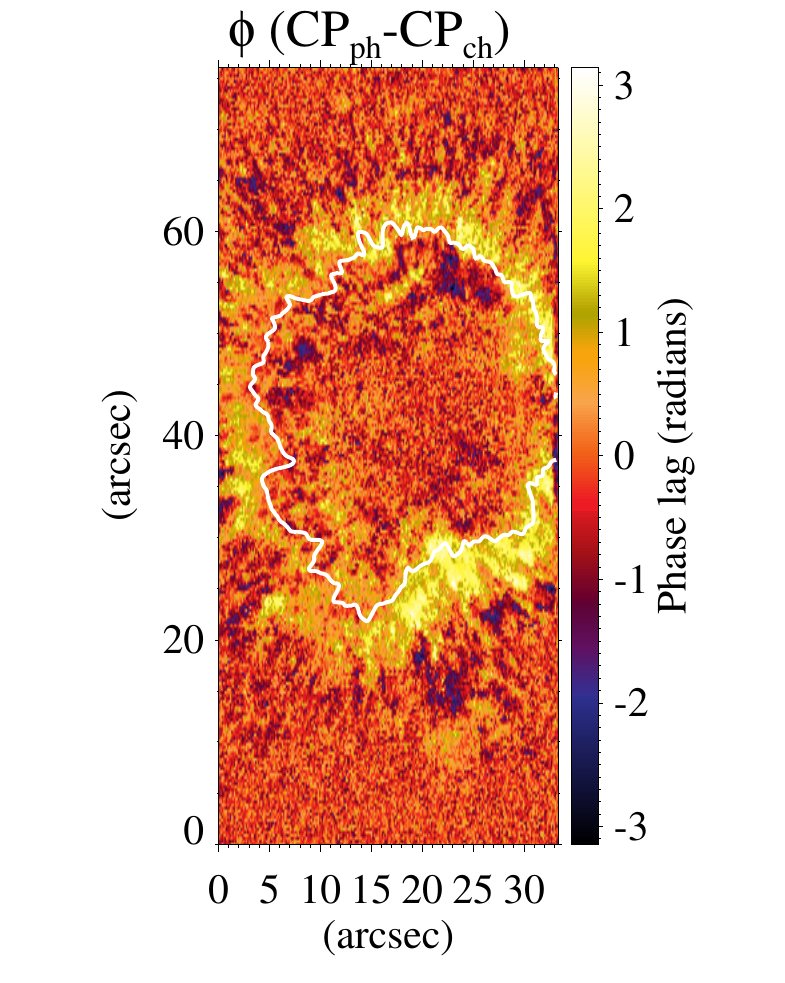}
\includegraphics[trim=0mm 4mm 0mm 0mm, clip, width=0.55\textwidth]{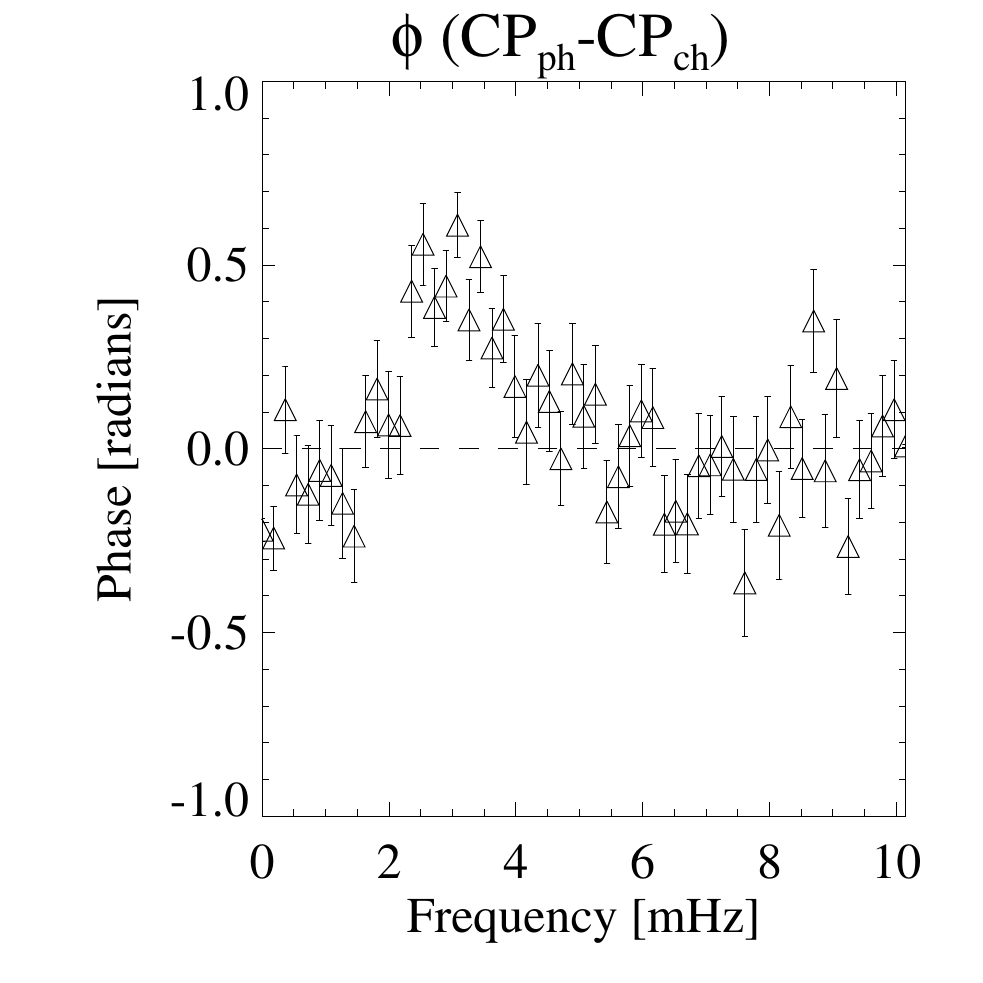}
\end{center}
\caption{{\em{Left:}} Phase lag map of the amplitude of circular polarization (CP) fluctuations at 3~mHz (across a bandwidth of $0.7$~mHz) between the photosphere and chromosphere. The solid white contour depicts the umbra/penumbra boundary. The positive phase lags towards the edges of the sunspot umbra reveal the presence of upwardly propagating magnetic waves. {\em{Right:}} Phase lag diagram obtained in umbra-penumbra boundary region showing the presence of a distinct positive peak (downward propagation) in the spectrum. Plots reproduced from \citet{2018ApJ...869..110S}.}
\label{fig:magnetic_waves_sunspotIBIS} 
\end{figure*}

In this regard, a first attempt to disentangle the signal associated with resonant modes from the rest of the spatially incoherent oscillations was made by \citet{2017ApJ...842...59J}, who applied three-dimensional Fourier (i.e., $k$-$\omega$; see Section~{\ref{sec:threedimensionalFourierfiltering}}) filtering to study the intensity oscillations of the sunspot (shown in Figure~{\ref{fig:Jess_2017_fig01}}) at chromospheric heights. The $k$-$\omega$ diagram of the intensity oscillations displays several horizontal ridges, which are evident in the right panel of Figure~{\ref{fig:komega_azimuthal_average}}. These spectral features are associated to the spatial scale of the entire sunspot (i.e., at wavelengths corresponding to the size of the overall sunspot umbra), highlighting the presence of a spatially coherent oscillations that encompass the entire sunspot umbra itself. It is worth stressing here that, in order for these modes to be readily identified, one requires sufficient spectral resolution in both the temporal and spatial frequency domains. As discussed in Section~{\ref{sec:commonmisconceptionsinvolvingfourierspace}}, to maximize the temporal frequency resolution requires the acquisition of long duration data sequences. In an analogous manner, large field-of-view sizes are required to provide sufficient frequency resolution in the spatial domain (i.e., a small wavenumber resolution, $\Delta{k}$). 

After filtering at the wavelengths and frequencies corresponding to the horizontal spectral ridges shown in the $k$-$\omega$ diagram in the right panel of Figure~{\ref{fig:komega_azimuthal_average}}, \citet{2017ApJ...842...59J} were able to detect a coherent rotational wave within the sunspot umbra. The observed rotational motion is visible in the lower panels of Figure~{\ref{fig:Jess_2017_fig01}}, where the lower-left panel shows the reconstructed intensities following $k$-$\omega$ filtering within the solid black box depicted in the right panel of Figure~{\ref{fig:komega_azimuthal_average}}. The lower-right panel of Figure~{\ref{fig:Jess_2017_fig01}} reveals the time-azimuth map following the polar transformation of the circular sunspot wave patterns, where the straight diagonal trends highlight coherent bulk rotations of the MHD wave phenomena. Thanks to numerical MHD modeling, it was possible to interpret these results as the first detection of an $m = 1$ slow magnetoacoustic mode in the chromospheric umbra of a sunspot. Building upon the early work of \citet{1970ApJ...162..993U} and \citet{1975A&A....44..371D}, which has subsequently been improved through the application of more sensitive instrumentation and modern techniques \citep[e.g.,][to name but a few]{1997SoPh..170...43K, 1997SoPh..175..287R, 1999ApJ...515..832H, RevModPhys.74.1073, 2004ApJ...608..562H, 2006ApJ...638..576G}, significant Fourier power at $\sim5$~mHz (i.e., consistent with the generalized $p$-mode spectrum) has demonstrated coherency down to wavenumbers on the order of $k \sim 0.05$~arcsec$^{-1}$, corresponding to spatial wavelengths of approximately $140''$ ($\sim100{\,}000$~km), with radio observations from missions such as GOLF \& BiSON observing global wavenumbers \citep[e.g.,][]{2004ApJ...604..969J,2007ApJ...659.1749C}. This may suggest that the elevated Fourier power bands shown in the right panel of Figure~{\ref{fig:komega_azimuthal_average}} may be linked to large-scale, sub-surface drivers. While the HARDcam H$\alpha$ observations are chromospheric in nature, being formed at a geometric height of $\sim1500$~km \citep{1981ApJS...45..635V, 2012ApJ...749..136L}, the highly magnetic composition of sunspot structures may enable direct and efficient coupling with the solar layers below \citep[see, e.g., the recent review by][]{2016GMS...216..489C}.

\begin{figure*}[!t]
\begin{center}
\includegraphics[width=\textwidth]{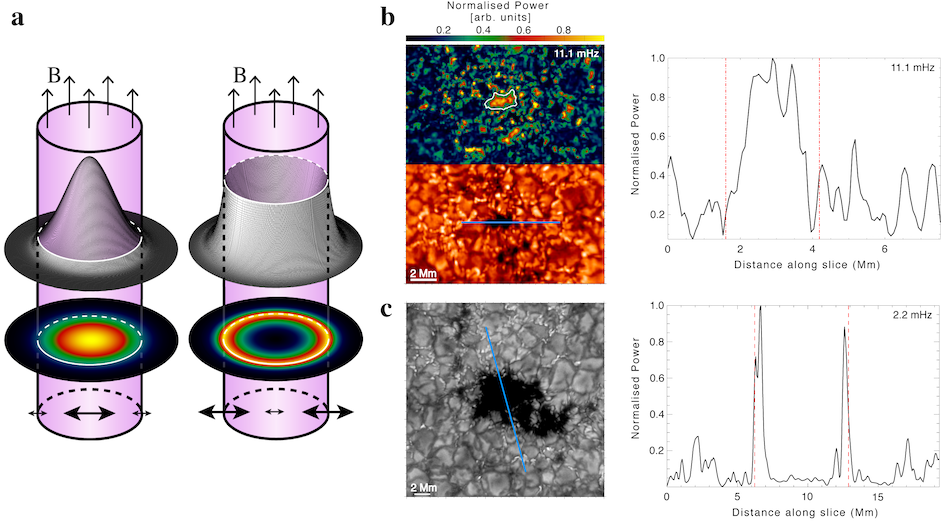}
\end{center}
\caption{Panel a shows a schematic of the spatial structure of the pressure perturbation for the body mode ({\em{left}}) and the surface mode ({\em{right}}) in both two and three dimensions. Arrows at the bottom of each schematic show the sausage oscillation in the flux tube. Panel b shows an example of a body mode in an elliptical pore. The lower left image shows a G-band intensity image of the pore, while the image above shows the two-dimensional power plot for the pore filtered at a central frequency of 11.1~mHz. The white contour here indicates the location of the pore. The blue cross-cut on the intensity image indicates the region taken for the one-dimensional power plot across the pore shown on the right. Panel c shows a similar example for a surface mode. Again, the blue cross-cut on the intensity image shows the location for the corresponding one-dimensional power plot shown to the right in this panel. In this instance, the power plot is produced for the pore filtered at a central frequency of 2.2~mHz, consistent with the timescale of granular evolution. Red dashed lines in the one-dimensional power plots in both panels b and c indicate the pore boundary in both instances. The body mode is characterized by a central peak decaying to the pore boundary, while the surface mode is characterized by peaks in power at the pore boundary decaying to zero in the center of the pore. Image adapted from \citet{2018ApJ...857...28K}.}
\label{fig:Keys_2018_surface_body} 
\end{figure*}

In addition, sausage modes \citep[see][]{2008IAUS..247..351D, 2011ApJ...729L..18M, 2014A&A...563A..12D, 2015ApJ...806..132G, 2016ApJ...817...44F, 2022ApJ...938..143G} have also been identified in magnetic pores through variations in their cross-sectional area of the magnetic structures and associated out-of-phase intensity oscillations. This highlights the interesting possibility to simultaneously exploit several diagnostics for the identification of global resonances in magnetic structures. Besides velocity and intensity perturbations, MHD waves are also expected to be characterized by magnetic perturbations. However, their identification has been long debated, as opacity effects can also mimic magnetic perturbations \citep[for further details, see][and the references therein]{2015LRSP...12....6K}. However, the recent advances in multi-height spectropolarimetric imaging observations have enabled the use of phase lag analyses between different layers of the solar atmosphere, which can be used to robustly identify real magnetic oscillations and disentangle them from spurious effects (e.g., changes in opacity). By doing this, \citet{2018ApJ...869..110S} were able to identify propagating magnetic fluctuations at the umbra-penumbra boundaries of a large sunspot observed by IBIS, which constitutes a spatially coherent oscillation that was interpreted as the signature of a surface mode of the sunspot flux tube. By studying the phase relationship between circular polarization (CP) and intensity signals, it was also argued that the oscillations were not consistent with opacity effects (see Figure~{\ref{fig:magnetic_waves_sunspotIBIS}}).

\begin{figure}[t!]
 \centering
\includegraphics[width=8cm]{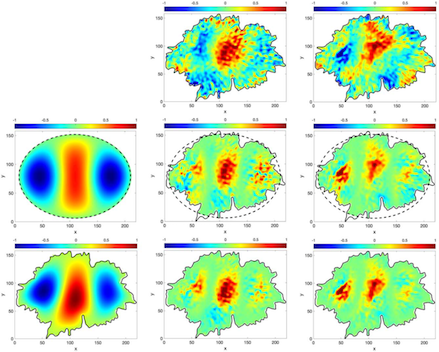}
\caption{The first row displays the spatial structure of the modes that were detected from the observational HARDcam data in an approximately elliptical sunspot by \citet{2022ApJ...927..201A}: the first POD mode (middle) and the DMD mode that
corresponds to the frequency of 5.6 mHz. In the first column, the theoretical spatial structure of the fundamental slow body
fluting  mode ($m=2$), even with respect to the major axis, in the elliptical magnetic flux tube model (middle) and the corresponding model using the exact umbral shape (bottom). The
rest of the panels show the Pearson correlation between the theoretical and POD/DMD modes. The positive/negative numbers in the color bars denote regions of phase/anti-phase. The dashed line show the boundary of the theoretical elliptical tube, and the solid black line shows the actual umbra/penumbra boundary. Image reproduced from \citet{2022ApJ...927..201A}.}
\label{fig:ellipse_flute}
 \end{figure}

Studying sausage modes in numerous pore data sets obtained with ROSA, \citet{2018ApJ...857...28K} were able to identify signatures of surface and body waves associated with the oscillating pores. Surface and body modes can be identified by the spatial distribution of the amplitude across the flux tube \citep{1983SoPh...84...99R, 2010SoPh..263...63E}. For surface modes, all perturbations have a maximum at the tube boundary and will be zero at the center of the tube. For body modes, the kinetic gas pressure, $v_z$ and $B_z$ will have maximum amplitude at the center of the tube, decreasing to the tube boundary. In the case of body modes, higher harmonics in the radial direction may result in nodes between the axis of symmetry and the boundary of the flux tube. For both surface and body modes in a homogeneous ambient plasma, the external wave power should decrease exponentially as a function of distance from the flux tube boundary. Figure~\ref{fig:Keys_2018_surface_body} shows a schematic diagram (adapted from \citealt{2018ApJ...857...28K}) of the expected spatial distribution of power for both body (left image in panel a) and surface (right image of panel a) waves in a cylindrical flux tube. To detect these signatures, the authors identified oscillations in the pore data sets by looking for oscillations in area and intensity in the pores, which would indicate the presence of a sausage mode. Periodicities were found to range from $90-700$~s, with the most common periods in area and intensity occurring at $\sim 300 \pm 45$~s. To determine if the wave was a surface or body mode, the spatial distribution of the power was then analyzed. This was performed by employing Gaussian filtering of the data for the dominant oscillatory frequencies within the data. This was limited to sections of the time series where significant power was found from the wavelet and EMD analysis of the area and intensity signals for the pores.

\begin{figure}[t!]
 \centering
\includegraphics[width=8cm]{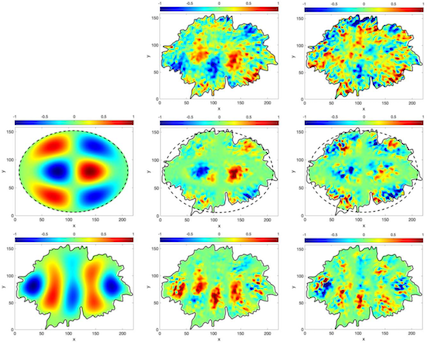}
\caption{This is the equivalent display for the same approximately elliptical sunspot as shown in Figure~\ref{fig:ellipse_flute} for the first overtone of the slow body kink mode which is odd with respect to the major axis. Here the DMD mode frequency is 5.3 mHz. Image reproduced from \citet{2022ApJ...927..201A}.}
\label{fig:ellipse_kink_overtone}
\end{figure}

By looking at the spatial distribution of the power, \citet{2018ApJ...857...28K} were able to identify surface and body modes associated with their data sets. The right panels of Figure~\ref{fig:Keys_2018_surface_body} shows examples of the power distribution observed by \citet{2018ApJ...857...28K} for both the body (upper plot) and surface (lower plot) waves. The surface mode was determined to be the most frequently occurring of the two. The authors suggest that this could be due to either the size or the magnetic field strength of the pore, as smaller weaker pores were more likely to display signatures of body modes. The authors suggest that it is possible that with a stronger field strength, there is a larger magnetic field gradient between the pore and the ambient plasma, possibly resulting in surface modes when the field strength is larger. The authors are clear to note though that the number of samples they have is small and, so, this relation is something that needs to be analyzed further. Estimates were also made for the energies associated with the observed surface and body modes using the framework of \citet{2015A&A...578A..60M}. Surface modes had an energy flux estimate of $22\pm 10$~kW\,m$^{-2}$, while the body modes had an observed energy flux of $11\pm5$~kW\,m$^{-2}$. 

Recently, \citet{2022NatCo..13..479S, 2022ApJ...927..201A} have found multiple slow body modes in sunspot umbrae, from low to high order in both the photosphere and chromosphere (see Fig. \ref{fig:ellipse_flute}, \ref{fig:ellipse_kink_overtone}, \ref{fig:stangalini_2022}). These works have shown that higher order modes (which have smaller spatial regions of phase and anti-phase) are particularly sensitive to the cross-sectional shape of the umbra. It can be seen for the $m=2$ slow body mode shown in Figure~\ref{fig:ellipse_flute} detected in an approximately elliptical sunspot by \citet{2022ApJ...927..201A} that the Pearson correlation between the POD/DMD modes and the the modes predicted by the elliptical and exact shape models are very similar (the regions of red show a strong in phase correlation). However, the first kink overtone, which has smaller regions of phase and anti-phase compared with the $m=2$ fluting mode, shown in Figure~\ref{fig:ellipse_kink_overtone}, demonstrates that model with the exact umbral shape has a much stronger correlation to the observed POD/DMD modes. This shows that higher order modes ``feel'' the irregularities in umbral cross-sectional shapes more than the lower order modes \citep[see][for more examples]{2022ApJ...927..201A}.

\begin{figure}[t!]
 \centering
\includegraphics[width=11.5cm]{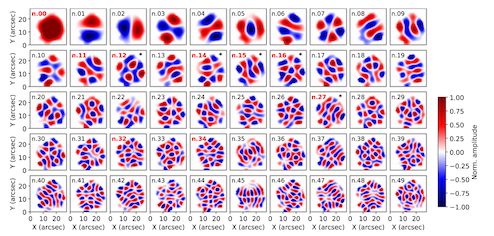}
\caption{The first fifty slow body mode eigenfunctions calculated for the vertical component of velocity using the observed sunspot umbra shape by \citet{2022NatCo..13..479S}. The dominant nine mode numbers are highlighted in red and these were used to reconstruct the observed Doppler velocity signal to high accuracy as shown in the top panel of Figure~\ref{fig:stangalini_2022}. Plot reproduced from \citet{2022NatCo..13..479S}.}
\label{fig:model_basis}
\end{figure}

Furthermore, if the observed sunspot umbra is far from circular or elliptical such simple models cannot be applied and the actual cross-sectional shape must be used, even to get an accurate representation of low order modes. Such an example from \citet{2022NatCo..13..479S} is shown in Figure~\ref{fig:model_basis} where first fifty eigenfunctions are modelled for the vertical velocity component of slow body modes in a large sunspot umbra (40 Mm across). The nine most dominant modes are highlighted in Figure ~\ref{fig:model_basis} and were used to reconstruct the observed Doppler signal to high accuracy as shown in the top panel of Figure~\ref{fig:stangalini_2022}. This would not have been possible by simply approximating the observed irregular umbra cross-sectional shape with circular or even elliptical cylinder MHD wave models.

From theory, each individual MHD wave mode can be broadband in both $\omega$ and $k$, as is shown for the magnetic cylinder model dispersion diagram in the left panel of Figure~\ref{fig:Edwin_dispersion}, i.e., each MHD mode on the dispersion diagram forms a continuous line, meaning that for every $\omega$ there is a unique $k$ that will satisfy the phase speed relation for a particular mode.  This broadband behaviour has now actually been observed, as can be seen in the bottom two panels of Figure~\ref{fig:stangalini_2022} taken from \citet{2022NatCo..13..479S}, where the B-$\omega$ diagram of a sunspot in the photosphere shows that, within the umbra, there are multiple strong frequency peaks, other than the usual dominant p-mode frequency, excited at the same time. It must be emphasised that the actual fine structure of the frequency power spectrum inside the umbra was not predicted at all from any model and therefore potentially opens up a whole new field of lower solar atmospheric MHD waves research.

The limitation of studies by \citet{2018ApJ...857...28K}, \citet{2022ApJ...927..201A} and \citet{2022NatCo..13..479S} is that they only looked at either the photospheric or chromospheric signatures of MHD waves in pores and sunspots. Multi-height studies of MHD waves in pores and sunspots are needed to determine the variation of properties and energy flux with height to give a clearer understanding of their contributions and relative importance to heating.\\
It also must be noted that the identification of global eigenmodes in large-scale magnetic wave guides is synonymous with the identification of coherent wave motions in a particular flux tube, be it a pore or sunspot umbra. Nevertheless, the limited number of works in this direction \citep{2015RAA....15.1449Y, 2017ApJ...842...59J, 2018ApJ...857...28K, 2021RSPTA.37900181A, 2022ApJ...927..201A, 2022NatCo..13..479S} testifies to the intrinsic difficulty of this task. As mentioned above, a potential reason for that could be due to the intrinsically small amplitudes of the associated oscillations, compared to the other omnipresent spatially incoherent fluctuations. However, filtering techniques represent a viable solution, but these require very high spatial and temporal resolution data with enough temporal and spatial coverage to reach the necessary frequency and wavelength resolutions in $k$-$\omega$ space.

\begin{figure*}[!t]
\begin{center}
\includegraphics[trim=0mm 20mm 0mm 160mm, clip, width=\textwidth]{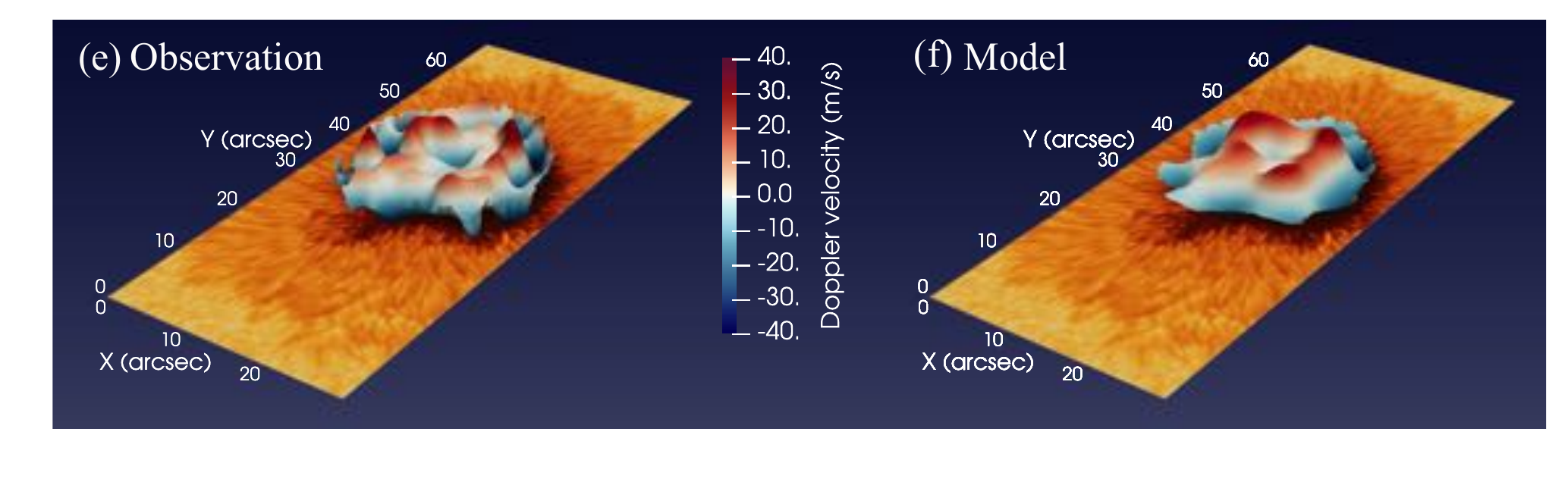}
\end{center}
\caption{Top: Comparison between the spatially coherent wave pattern observed in a sunspot after properly filtering the data (left), and the one expected from numerical modeling (right, see Figure~\ref{fig:model_basis} for the eigenfunctions that were used). The filtered wave pattern displays a high order oscillation in the umbra that agrees very well with numerical predictions of eigenmodes. Bottom: B-$\omega$ diagram of the same sunspot showing a rich variety of frequencies excited within the umbra and consistent with resonant modes of the magnetic structure.  Plots reproduced from \citet{2022NatCo..13..479S}.}
\label{fig:stangalini_2022} 
\end{figure*}

\subsection{Small-scale Magnetic Structures}
\label{sec:smallscalestructures}
Concentrations of intense magnetic fields at small scales are mostly found in intergranular regions (in the photosphere), where strong downflows occur \citep{1973SoPh...32...41S,2010ApJ...719L.134J,2014A&A...568A..13R,2017SSRv..210..275B}. In such small magnetic elements, in addition to gravity and pressure, the magnetic field also acts as a restoring force \citep[e.g.,][]{2010ASSP...19..166S}. Therefore, in addition to the longitudinal (compressible) acoustic and gravity waves, other MHD wave modes may also propagate along the flux tubes \citep{1978SoPh...56....5R,1983SoPh...88..179E,1987MNRAS.228..427H,1993ASPC...46..522F,1998ApJ...495..468S,2003ApJ...599..626B,2003A&A...400.1057M,2008SoPh..251..589K,2011ApJ...727...17F}. Such waves can be generated as a result of, e.g., (1) buffeting of the flux tubes (i.e., bundles of magnetic-field lines; \citealt{1976SoPh...50..269S,1981NASSP.450..385S,1996A&A...310L..33S}) by the surrounding granules (and intergranular turbulence), i.e., kink modes \citep{2003A&A...406..725M}, (2) compression and contraction of the flux tubes by convective forces from opposite directions (sausage modes), or (3) twisting the flux tubes by rotating flows around the tubes (torsional Alfv{\'e}n waves; \citealt{1982SoPh...75....3S,1993SSRv...63....1S}). It is, however, more complex in real situations where wave modes of various eigenmodes may co-exist in the same magnetic elements. Excitation of various wave modes in a magnetic cylinder has been reviewed in detail by \citet{2005LRSP....2....3N}.
Such waves may occur in either propagating or standing states \citep{2002ApJ...564..508R,2014A&A...563A..12D}. The propagating magnetoacoustic (or MHD) waves are channeled from the base of the photosphere to the chromosphere and beyond along the magnetic field lines (where the magnetic field acts as a guide; \citealt{2008ApJ...676L..85K}) whose strength and inclination plays a role in their leakage to the upper solar atmosphere \citep{1973SoPh...30...47M,2006ApJ...648L.151J,2006MNRAS.372..551S,2011A&A...534A..65S}. In addition, the magnetoacoustic waves may propagate as `fast' or `slow', traveling at speeds faster or slower compared to the ratio of the Alfv{\'e}n and sound speeds. The various wave modes may be converted from one form to another (i.e., mode conversion) and/or only switch the fast and slow labels (i.e. mode transmission) at the plasma-$\beta\approx1$ level, where the Alfv{\'e}n and sound speeds nearly coincide (\citealt{2019ApJ...883..179K}; see Sections \ref{sec:Identification_of_MHD_wave_modes} and \ref{subsec:magnetoacoustic_waves} for greater details).

As a (thin) flux tube extends into the solar atmosphere, it expands with height (while the gas density and pressure decrease). In addition, many of such flux tubes bend over at lower atmospheric heights (compared to larger and stronger magnetic-field structures, thus producing the multi-height magnetic canopy; \citealt{1982SoPh...79..267G,1991A&A...250..220S,2002ApJ...564..508R,2017ApJS..229...11J}). The (inclined) flux tubes are thought to be observed throughout the solar chromosphere as (dark or bright) thread-like structures in intensity images \citep{2007PASJ...59S.655D,2009A&A...502..647P,2009ApJ...705..272R,2012ApJ...759...18P,2017ApJS..229....6G,2017ApJS..229...11J,2020A&A...637A...1K}. Only recently, has it been possible to identify the MHD wave modes in small-scale structures through the entire lower solar atmosphere, thanks to high-resolution observations provided by modern facilities. Such small structures are often very dynamic and short lived (with observational timescales on the order of a few seconds to a few minutes), therefore, not only are high-spatial resolution observations required to resolve them, but any study of their rapid evolution also needs a high temporal resolution. Furthermore, multi-line (i.e., multi height) observations are also essential in order to trace waves as they propagate through the atmosphere. In this regards, narrow-band observations in spectral lines (with a relatively high spectral resolution) would reduce mixing information from different atmospheric heights, but ambiguities may still exist.

In the following, we summarize the most recent advances on detection and analysis of various MHD wave modes in small-scale magnetic elements and fibrillar structures in the solar photosphere and the chromosphere.

\subsubsection{Excitation, propagation, and dissipation of MHD waves in small-scale magnetic structures}
\label{sec:waves_at_small_scales}
Due to different (distinct) kinematics of various solar photospheric regions (with different levels of magnetic flux; \citealt{2011ApJ...743..133A,2014A&A...561L...6S,2014A&A...566A..99K,2017ApJS..229....8J}), characteristics of waves and oscillations in small-scale magnetic elements may depend on the environment in which they are embedded. Small magnetic elements are able to laterally move within a supergranular body (i.e., the internetwork). In concert with surrounding granules interacting, due to expansions and explosions, and effects of intergranular turbulence, a variety of MHD wave modes can be generated across a range of frequencies. Of particular interest, from the detection point of view, is also a lower number density of these elements in the internetwork, hence, they are more isolated compared to those found in network/plage regions. In the latter, while such interactions also occur between the plasma and magnetic elements, they are often found in groups of concentrated field structures trapped in sinks (stagnation points) where inflows from surrounding supergranules prevent them from moving in a preferred direction, but rather they experience random walks within a relatively small area \citep{1998ApJ...509..435V, 2003ApJ...587..458N, 2010A&A...511A..39U, 2011A&A...531L...9M, 2012ApJ...752...48C, 2014A&A...563A.101J, 2014ApJ...788..137G}. 

Oscillations (of different properties) in the low photosphere, in both network and internetwork magnetic elements as well as those in plage areas (in the vicinity of large magnetic structures), have been identified over the past decades \citep[e.g.,][to name but a few]{1963ApJ...138..631N, 1964ApJ...140.1120S, 1967SoPh....2....3H, 1970ApJ...162..993U, 1972SoPh...27...71G, 1973SoPh...33...33C, 1974ARA&A..12..407S, 1977ApJ...211..934G, 1982MNRAS.198..141C, 1988ApJ...333..427N, 1992A&A...257..287T, 1993ApJ...405..787F, 1999A&A...341..617Z, 2003Natur.421...43G, 2006A&A...446..669V}. Thanks to simultaneous multi-height (multi spectral-line) observations of the entire solar photosphere and chromosphere (although at different resolution/band width), propagation of the various types of magnetoacoustic waves have also been investigated. However, it is worth noting that the highest chromospheric layer to which such waves have been traced can depend on the properties of the lines in which the observations were made and/or the magnetic topology of the observed area. Thus, the relatively wide-band observations of the chromospheric lines include a wide range of chromospheric heights. In contrast, narrow-band observations have revealed more filamentary structures of the chromosphere, while their number density and thickness tend to increase with height through the entire solar chromosphere. Therefore, formation heights at which waves are identified/traced should be interpreted with caution. In addition, the presence or density number of fibrillar structures at chromospheric heights may depend on the level of magnetic flux within (and/or in the immediate vicinity of) the observed photospheric field of view. Using the Ca~{\sc ii}~H filter (with a width of 0.1~nm) onboard the {\sc Sunrise} balloon-borne solar observatory, \citet{2017ApJS..229...11J} illustrated a field of view of an active region filled with slender fibrils where almost no magnetic bright points could be observed. On the contrary, they presented a quiet-Sun region (taken with the same filter) where no fibrillar structures appeared.

Various kinds of MHD modes have been identified at small scales in high-resolution observations. Often they are identified in intensity images of features such as magnetic bright points (MBPs) or in fibrillar structures at chromospheric heights, utilizing various spectral lines, thus, sampling different atmospheric layers. As a result, their propagation from the solar photosphere to the chromosphere, or in a particular region within these layers, has been characterized in a number of studies \citep[e.g.,][to name a few]{1979ApJ...231..570L, 1997ApJ...486L.145K, 2002ESASP.505..305M, 2005ApJ...624L..61D, 2011ApJ...736L..24O, 2012ApJ...746..183J, 2012ApJ...750...51K, 2017ApJS..229...10J, 2022ApJ...930..129B}. Such waves include transverse oscillations (often interpreted as MHD kink or Alfv{\'e}nic waves), oscillations in intensity and/or Doppler velocity (characterizing longitudinal magnetoacoustic waves), twist perturbations (describing torsional Alfv{\'e}n waves), as well as fluctuations in the size/width of small magnetic structures, as a signature of (compressible) MHD sausage modes, particularly, when they are in anti-phase relationships with intensity oscillations \citep{1983SoPh...88..179E,2006ASPC..358..465C,2009A&A...494..295E,2013SSRv..175....1M,2016GMS...216..449J}. Kink modes and Alfv{\'e}n waves are incompressible and their dissipation in the solar atmosphere requires a large gradient in the Alfv{\'e}n speed \citep{2009A&A...503..213G}. Alfv{\'e}n waves also result in Doppler velocity perturbations, while the longitudinal magnetoacoustic waves result in fluctuations in both intensity and Doppler velocity. Therefore, spectral observations with sufficiently high wavelength resolution for fine doppler studies should be interpreted by taking the nature of various wave modes, as well as mode-coupling/mixing, into consideration.


It is thought that rapid (greater than $\approx2$~km/s) pulse-like kicks to small magnetic elements, as a result of, e.g., granular explosions, can excite transverse kink waves along the flux tubes \citep{1981A&A....98..155S, 1993SoPh..143...49C, 1993ApJ...413..811C, 1998ApJ...495..468S, 1999ApJ...519..899H, 2003A&A...406..725M, 2006SoPh..237...13M}. Such impulsively excited waves, as a result of rapid continuous jostling of the flux tube by granules \citep{2000ApJ...535L..67H} can upwardly propagate into the upper solar chromosphere, with their amplitudes increasing exponentially. The kink waves may become nonlinear in the upper chromosphere where their propagating speeds are comparable to tube speeds \citep{1997ApJ...486L.145K}. Such waves may, however, couple to the longitudinal magnetoacoustic waves in the low-to-mid chromosphere and dissipate by forming shocks \citep{1991A&A...241..625U,1995A&A...300..302Z}. \citet{1994A&A...283..232M} examined the pulse-like excitation mechanism and estimated an energy flux of 2000~W/m$^2$ that could be carried by kink waves in network MBPs. A smaller energy flux on the order of 440~W/m$^2$ was later reported by \citet{1998A&A...335..323W}, based on horizontal motion of chromospheric Ca~{\sc ii}~K bright points in network areas. However, the magnetic nature of such small-scale brightenings were not determined.

Thanks to the high spatial resolution provided by {\sc Sunrise}, \citet{2013A&A...549A.116J} were able to identify such jerky rapid (sometimes supersonic) pulse-like motions in small internetwork MBPs observed in the upper photosphere/low chromosphere. Such waves were found to be energetic enough (with a net energy flux of $\approx300$~W/m$^2$) to potentially heat the outer solar atmosphere. The somewhat large difference between the energy fluxes found by different authors could be due to, e.g., different geometric heights, network versus internetwork (the latter hosts considerably smaller number of MBPs, that can move around more freely, compared to the former), spatial resolution (influencing the number of detected elements, their sizes, and horizontal-motion measurements), and their nature (magnetic versus non-magnetic features).

Incompressible horizontal convective motions, on the solar surface, are generally thought to be the prime excitation mechanism of the transverse MHD waves in magnetic flux tubes, by being either perpendicular or tangential to the surface of magnetic elements (resulting in kink or torsional Alfv{\'e}n modes, respectively; e.g., \citealt{1996SSRv...75..453N, 2007Sci...318.1572E}). In addition, turbulent convective downflows have been suggested to generate transverse displacements inside the magnetic concentrations \citep{2011ApJ...736....3V}, occurring on smaller length and time scales, compared to those from the granular motions.

\citet{2014ApJ...784...29M} exploited observations from multiple instruments (i.e., from the Swedish Solar Telescope \citep[SST;][]{2003SPIE.4853..341S}, Hinode/SOT \citep{2007SoPh..243....3K, 2008SoPh..249..167T}, DST/ROSA, and Coronal Multi-channel Polarimeter, CoMP) to study the generation and transport of energy by kink waves in small scale structures through the entire quiescent solar atmosphere. They found similar power spectra for transverse oscillations of photospheric MBPs (and granular flows) and the chromospheric H$\alpha$ fibrils, suggesting the granular motions have excited the kink waves identified in the small structures. In addition, \citet{2014ApJ...784...29M} found that the higher-frequency wave energy was significantly diminished in the corona's power spectra, thought to be a signature of energy dissipation at those frequencies. However, the authors give no consideration to the attenuation effects of the contribution function and transmission of spectral lines in the solar atmosphere at high frequencies. \citet{1976A&A....51..189D} found notable reductions in wave power at evenly spaced frequencies above 10~mHz, and proposed that the spacing equated to wavelengths that were integers of the length of the atmospheric region contributing to the spectral line. Subsequently, it was conclusively shown that the extent of the atmospheric column that contributes to a spectral line directly impacts the transmission of high frequency waves in the atmosphere \citep{1979A&A....76..208D, 1979ApJ...234..768C, 1980A&A....84...96M}. As a result, the transmission of frequencies is dependent on both the wavelength, and the spectral window that observations are integrated over \citep{2005ApJ...625..556F}. In relation to energy transport and dissipation, differential transmission has been shown to impact the potential of acoustic waves in particular to propagate energy in higher frequency modes beyond the photosphere \citep[e.g.,][]{2009A&A...508..941B,2010ApJ...723L.134B}. In the case of \citet{2014ApJ...784...29M}, a variety of different spectral lines are studied, including $H\alpha$ imaging with a spectral window of 0.25~\AA~and sampling a range of formation heights. Without consideration of the relative transmission functions of the multi-wavelength observations, it cannot be verified whether the damping observed in \citet{2014ApJ...784...29M} is dissipative, or due to transmission effects. This must also be taken into account whenever damping is observed across different spectral line observations, either in mitigation, or through transmission function analysis.

Nonlinear propagation of transverse waves to the solar chromosphere at small scales, previously predicted by theoretical models \citep{1986A&A...166..291M,1991A&A...241..625U}, was identified by \citet{2015A&A...577A..17S}, where the authors exploited transverse perturbations in several MBPs simultaneously recorded in both the photosphere and the low chromosphere at high resolution with SST. They found the identified kink waves to nonlinearly propagate upward above a cut-off frequency of $\approx2.6$~mHz. The nonlinearity was concluded due to remarkable differences between the photospheric and chromospheric power spectra (i.e., considerably different patterns of peaks in the power spectra).

Using a relatively long time series (of $\approx4$~hours) from Hinode/NFI \citep{2008SoPh..249..167T}, \citet{2013A&A...559A..88S} provided the full power spectra of transverse (kink) oscillations in small photospheric magnetic elements (limited to the spatial resolution of the 0.5~m telescope). They found a wide range of frequencies of $1-12$~mHz, of which, the lower frequencies would only be reliably identifiable by exploiting such a long image sequence, which is rare for ground-based observations. However, on the higher frequency end, the measurements were limited to the 30~s cadence of the observations, hence, a Nyquist frequency of 16.7~mHz. Therefore, detection of higher frequencies would only be possible with a higher temporal resolution. In addition, it is worth noting that spatial resolution is also an important factor for detecting power at high frequencies \citep{2007ASPC..368...93W}. Therefore, the higher spatial and temporal resolutions are, the higher frequency oscillations can be detected (if they exist).

\begin{figure*}[!t]
\begin{center}
\includegraphics[trim=0mm 0mm 0mm 0mm, clip, width=10cm]{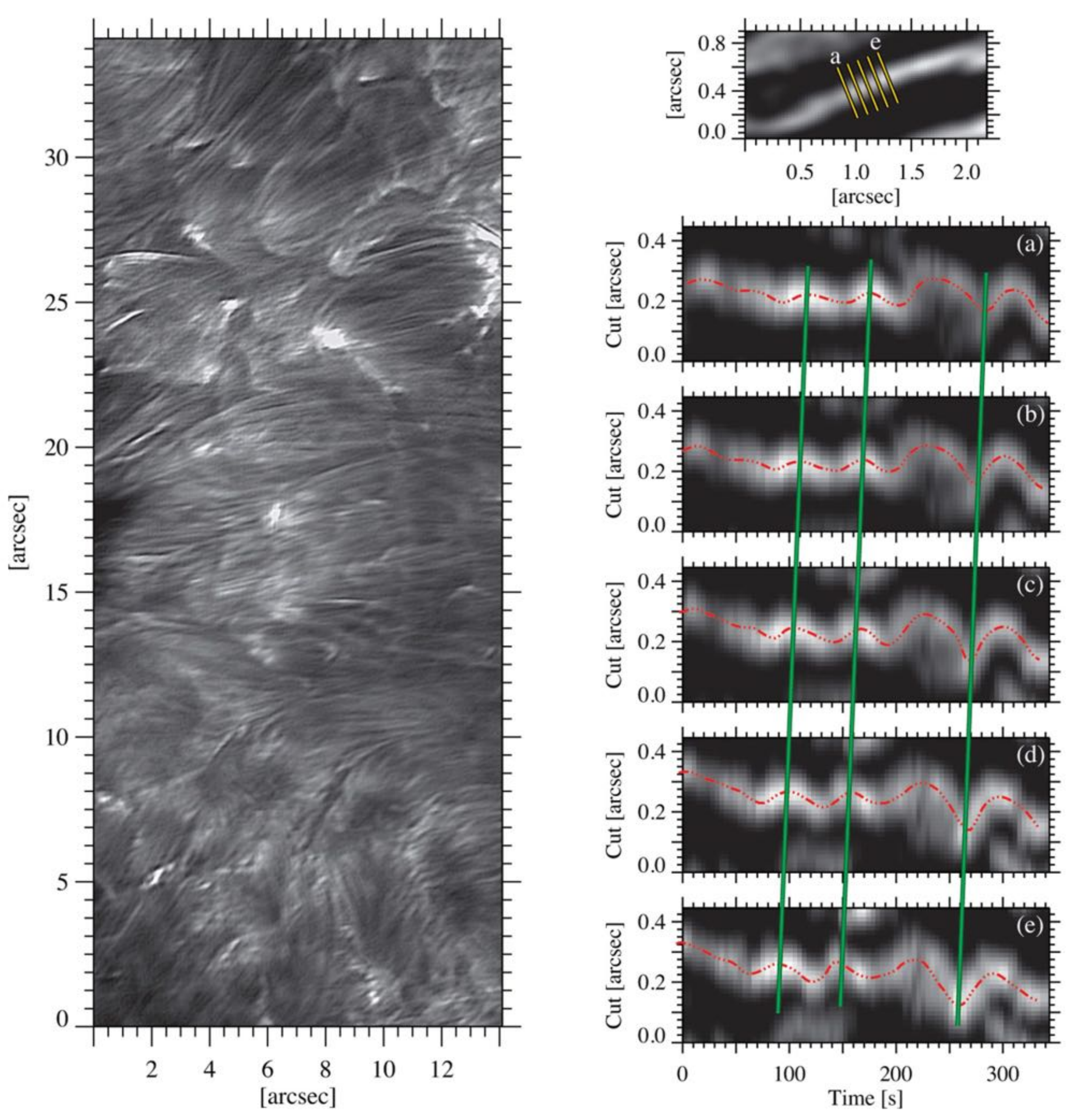}
\end{center}
\caption{Small-scale (slender) Ca~{\sc ii}~H fibrils in the low chromosphere (left). Transverse oscillations in one fibril are illustrated on the right panel in space-time plots, at multiple cuts in different locations along the fibril (shown on the top). The solid (green) lines connect the extrema of the fluctuations, indicating wave propagation from right to left along the fibril. Images reproduced from \citet{2017ApJS..229...11J} and \citet{2017ApJS..229....9J}.}
\label{fig:SCFs_kink}
\end{figure*}

Following earlier works by \citet{2003ApJ...585.1138H} and \cite{2005ApJ...631.1270H}, \citet{2008ApJ...680.1542H} employed numerical simulations to propose that the excess brightness in the network Ca~{\sc ii}~H \& K MBPs (in the solar chromosphere) could be due to (1) high-frequency (higher than 10~mHz) transverse oscillations at the base of the magnetic flux tubes, and (2) absorption of acoustic waves from the surrounding medium. These result in temperature perturbations (of up to 900~K) due to shock dissipation at chromospheric heights, following the upwardly propagating (slow) magnetoacoustic waves along the flux tubes. Such propagating high-frequency transverse waves (up to 23~mHz) as well as (longitudinal) intensity oscillation (up to 30~mHz) in small MBPs were detected by \citet{2017ApJS..229...10J} from high temporal and spatial resolution observations with {\sc Sunrise}. These authors studied the MBPs at two atmospheric heights, corresponding to the low photosphere and the low chromosphere, with an approximate height difference of 450~km on average (estimated using two independent approaches). Together with phase differences between the intensity oscillations at the two atmospheric layers, a wide range of propagating velocities were determined. Of which, phase speeds larger than 30~km/s could not satisfy expected propagating speeds (from theoretical models) at these heights. Uncertainties in $I-I$ phase analysis can be introduced through radiative damping, particularly in atmospheric regions where the radiative relaxation time is equivalent to, or less than, the wave period \citep[e.g.,][]{1972A&A....17..458S,1978SoPh...57..245S, 1990A&A...236..509D}. Estimates predict that $3-5$~minute oscillations are on the order of the relaxation time in the low chromosphere \citep{1982ApJ...263..386M, 2013SoPh..284..297S}, thus the unexpectedly large phase speeds will be influenced by non-adiabatic atmospheric evolution. In addition, refraction of the propagating path of these waves may influence phase speed estimations. Using numerical simulations, \citet{2012A&A...538A..79N} studied such possible modifications of wave propagation (and wave travel time) in small magnetic features in the solar network atmosphere. They found that the travel time (and hence the propagating speed) is strongly influenced by mode conversion, sometimes at multiple plasma-$\beta$=1 levels which are placed on top of each other as a result of the highly dynamic atmosphere. In addition, they found that the measured wave travel-time could significantly be reduced as a result of the fast waves being refracted above the magnetic canopy due to the large gradient of the Alfv{\'e}n speed. Thus, the two mechanisms, i.e., the fast waves due to (multiple) mode conversion inside the magnetic canopy and the refraction of the propagation path above the canopy may lead to observations of wave travel time that is too short (i.e., propagating speeds which are too large) between two atmospheric layers. 
Short time delays of 4~s and 29~s were also reported by \citet{2013MNRAS.428.3220K} between horizontal velocity variations within 500~km from simulated data and within heights sampled by the G-band and Ca~{\sc ii}~K MBPs from DST/ROSA, respectively. The authors interpreted such short time intervals as the results of oblique granular shock waves in the simulations, and of a semi-rigid flux tube in the observations.

Incompressible kink and compressible sausage modes at high frequencies (of $\approx12$ and 29~mHz, respectively) were also detected in slender Ca~{\sc ii}~H fibrils (located in the low-to-mid chromosphere) from high-resolution observations with {\sc Sunrise} \citep{2017ApJS..229....9J,2017ApJS..229....7G}. Figure~\ref{fig:SCFs_kink} shows such slender fibrillar structures, filling the entire field of view (left), with an example of the detected transverse oscillations at 5 locations along one fibril (right). The fibril and locations of the artificial slits (marked with a-e) are illustrated on the top of the right panel. The transverse oscillations are identified in space-time plots, where time variations of the location of fibrils has been inspected at each `cut' perpendicular to the fibril's axis. Slope of the lines (green), connecting the same peaks/troughs of the oscillations at different locations, indicate the propagation of the transverse (kink) waves from right to left in the top panel. To quantify the propagating speeds (and periods), \citet{2017ApJS..229....9J} employed a wavelet analysis to compute phase differences between oscillations at different locations (whose distances are known). The energy flux transported by the kink waves along these slender fibrils was found to be $\approx15$~kW/m$^2$, on average.

\begin{figure*}[!t]
\begin{center}
\includegraphics[trim=0mm 0mm 0mm 0mm, clip, width=\textwidth]{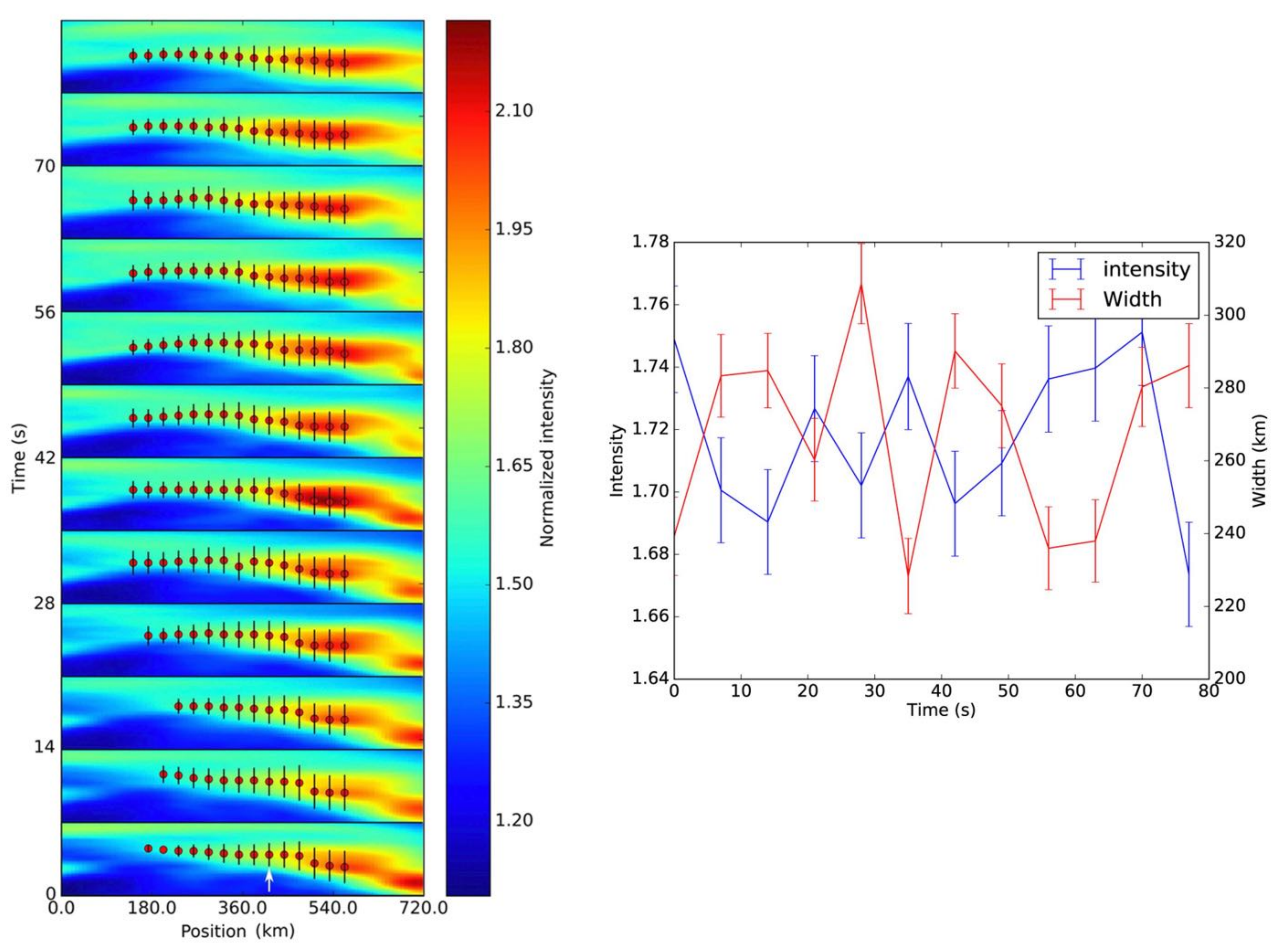}
\end{center}
\caption{Measurements of intensity and width in different locations along one slender Ca~{\sc ii}~H fibril is illustrated on the stack of the fibril at different times (left). Oscillations of the two quantity at one location (marked with the arrow) is illustrated on the right panel. Images reproduced from \citet{2017ApJS..229....7G}.}
\label{fig:SCFs_sausage}
\end{figure*}

Furthermore, \citet{2017ApJS..229....7G} identified sausage modes (with periods on the order of 32-35~s) by measuring intensity and size at various locations (cuts) along these small-scale fibrils. The left panel in Figure~\ref{fig:SCFs_sausage} illustrates one example where the measurements of both intensity and width (by fitting a Gaussian function perpendicular to the fibril's axis) at multiple locations are shown on the stack of one fibril (on top of each other) at different times. The vertical lines, depicted at the different spatial locations, indicate the width of the fibril at those locations (marked with a small circle). The fluctuations of intensity and width at one location, marked with the arrow on the left panel, is presented on the right, where a clear anti-correlation between the two oscillations is evident. The authors also measured the wave properties by means of wavelet analysis, resulting in a propagating speed of $11-15$~km/s, interpreted as fast sausage modes. These phase speeds are considerably smaller than those found in chromospheric H$\alpha$ fibrils, propagating with $\approx67$~km/s, on average, for the fast sausage modes \citep{2012NatCo...3.1315M}, also with longer periods on the order of 197~s. \citet{2012NatCo...3.1315M} also detected transverse oscillation in the same H$\alpha$ fibrils with periods and propagating speeds of 232~s and $\approx80$~km/s, respectively. The increase in propagating speed with height is expected from theoretical models, due to the height stratification of the physical parameters or, possibly, these waves are close to the cut-off frequency and are close to being evanescent resulting in the observed high speeds.

Generation and propagation of both kink and sausage modes in chromospheric fibrillar structures (i.e., on-disc Type~{\sc i} spicules) were studied in detail by \citet{2012ApJ...744L...5J} from both observations (from DST/ROSA) and MHD simulations, where the mode conversion at the lower solar atmosphere was found to be the main driver of the MHD waves. \citet{2012ApJ...744L...5J} showed that the longitudinal waves in the photospheric MBPs (with periods on the order of $130-440$~s) could be converted to kink modes at higher frequencies (higher by a factor of $\approx2$), which were concluded to be the result of a 90 degrees phase difference encompassing opposite sides of the photospheric driver. Indeed, they found these waves are energetic enough (with an energy flux of $\approx300$~kW/m$^2$) to heat the outer solar atmosphere (or accelerate the solar wind). 
\citet{2014A&A...569A.102S} provided observational evidence for excitation of kink modes in small photospheric magnetic elements as a result of granular buffeting. Their relatively long time-series of images (of about 4 hours) from Hinode/NFI, and the use of EMD technique, allowed  \citet{2014A&A...569A.102S} to reveal hints about the mechanisms of excitation of low frequency kink oscillations in small-scale magnetic tubes through their sub-harmonic response. Indeed, the probability density function of the periodicities of horizontal oscillations in a large sample of magnetic tubes, revealed several peaks in the statistical distribution corresponding to sub-harmonic oscillations with periods multiple of a fundamental one of $\approx7.6$~min, which is comparable with evolution time of granular cells. Furthermore, the application of EMD approach on horizontal-velocity fluctuations of small (low) chromospheric MBPs, seen in SST Ca~{\sc ii}~H images, led \citet{2017ApJ...840...19S} to find an elliptic polarization of the velocity vector associated to the low-frequency (smaller than $5-6$~mHz) kink oscillations. The Ca~{\sc ii}~H MBPs were showed to more freely move around, in a helical motion (while fluctuating transversely) compared to their photospheric counterparts (bounded to the granular flows). The left panel of Figure~\ref{fig:polarized_kink} schematically illustrates such a polarized kink wave where the superposition of both helical motion and transverse kink waves co-exist in the same flux tube. The power spectra of the $x$ and $y$ components of the horizontal velocity of a Ca~{\sc ii}~H MBP, as well as coherence spectrum between the two components, are plotted on the top-right panel (with solid red, dashed blue, and open circles, respectively), indicating the presence of higher frequencies, in addition to the larger peaks in the lower end. The helical motion, which was characterized from a phase relationship between the two components of the horizontal velocity, can also be visualized through the plot of the vector velocity on the bottom-right panel of Figure~\ref{fig:polarized_kink}, where a rotation in the displacement direction of the MBP is observed.

\begin{figure*}[!t]
\begin{center}
\includegraphics[trim=0mm 0mm 0mm 0mm, clip, width=\textwidth]{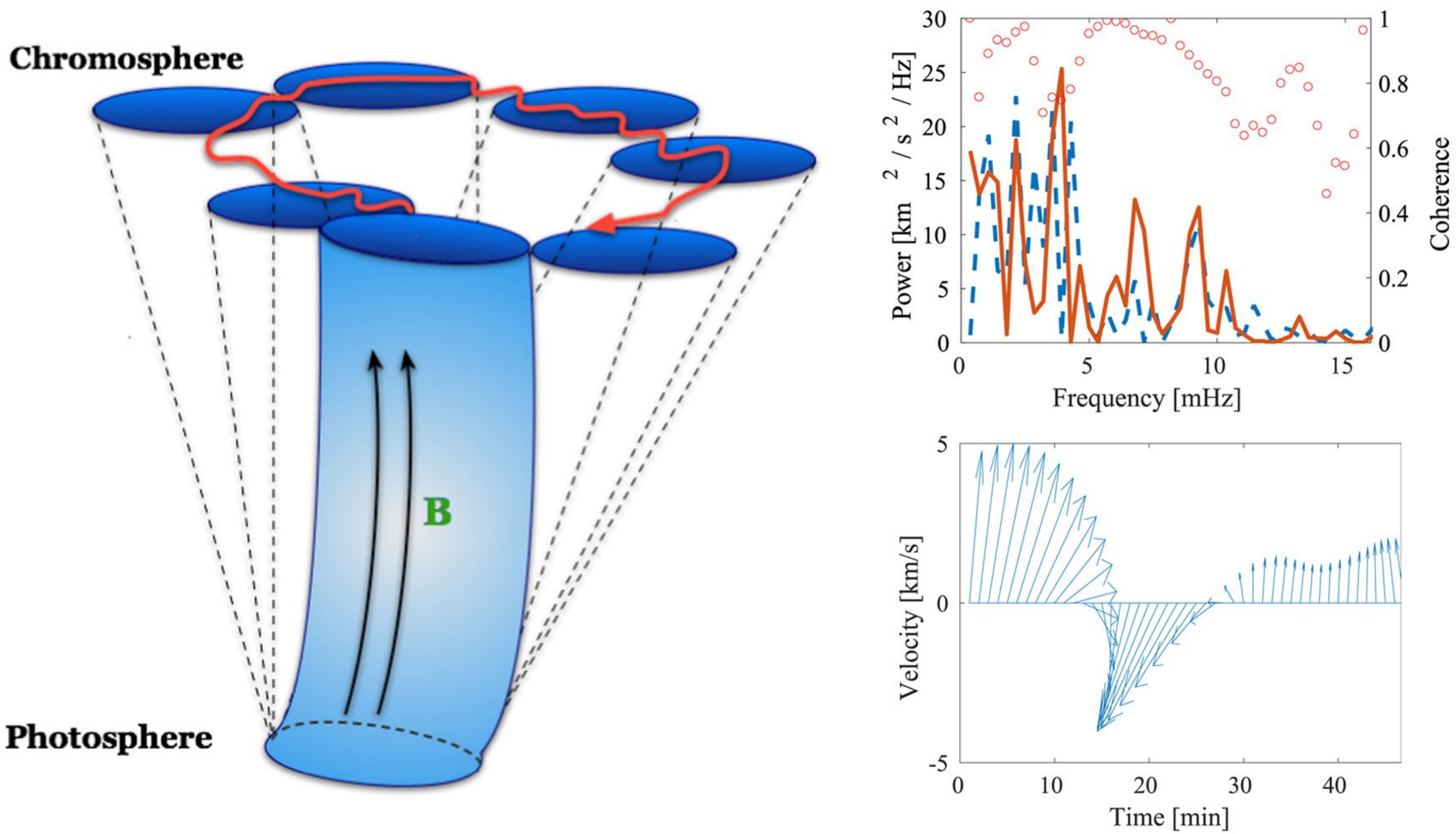}
\end{center}
\caption{Left: A cartoon illustrating a low-frequency helical displacement superimposed on a high-frequency kink wave in the solar chromosphere. Top right: Power spectra of the $x$ and $y$ components of the horizontal velocity of a Ca~{\sc ii}~H MBP (solid red and dashed blue lines, respectively). The coherence spectrum of the two components are also plotted with the red open circles. Bottom right: Vector horizontal-velocity (both direction and magnitude) as a function of time. The helical motion (as the rotation of the velocity direction) is evident. Images reproduced from \citet{2017ApJ...840...19S}.}
\label{fig:polarized_kink}
\end{figure*}

Transverse kink waves have also been studied in small-scale structures in sunspots. \citet{2011ApJ...739...92P} identified kink waves in dynamic fibrils (in the immediate vicinity of a sunspot; from Ca~{\sc ii}~8542~{\AA} observations with SST) with periods of $\approx135$~s. More recently, \citet{2021RSPTA.37900183M} used high-resolution observations in Ca~{\sc ii}~8542~{\AA} spectral line from SST/CRISP to demonstrate that transverse waves also pervaded the sunspot super-penumbral fibrils (in the solar chromosphere). They interpreted the oscillations as MHD kink modes with periods and propagation speeds on the order of 754~s and 25~km/s, on average, respectively. The velocity amplitudes (with an average of $0.76\pm0.47$~km/s) were found to increase with distance from the umbral center by about 80\%, as illustrated in Figure~\ref{fig:sunspot_fibrils}. \citet{2021RSPTA.37900183M} speculated this variation as, possibly, a result of a density decrease along the fibrils as the super-penumbra is extending to higher atmospheric heights while moving away from the umbra until reaching its highest magnetic-canopy point and returning to the surface. Thus, considering the field topology (in the chromosphere) is an important key when interpreting the observations, particularly, in intensity images in which the projection effects cannot be directly realized.
\citet{2021RSPTA.37900183M} also discussed a number of possible excitation mechanisms (for the transverse oscillations), namely, convection driven, reconnection, and mode conversion, of which, they found the latter to be more convincing. Oscillations in these sunspot's small-scale structures may be different when compared to other chromospheric features due to a number of reasons, e.g., the very strong magnetic fields of the sunspots.

As has been discussed at length, observed frequencies in the lower solar atmosphere center around a range between $2-10$~mHz. Observations with the Vacuum Tower Telescope (VTT) led \citet{1995AA...304L...1V} to find relatively high-frequency oscillations in horizontal motions (with frequencies of $\approx10$~mHz), consistent with kink modes, and in Doppler velocity (with frequencies that peaked at $\approx8$~mHz) in small-scale structures in a plage region in the photosphere. The Doppler velocities were computed from Stokes-$V$ profiles of the Fe~{\sc i}~630.15~nm spectral line. High-resolution observations from SST led \citet{2007SoPh..246...65L} to identify the signature of propagating kink oscillations (in both intensity and Doppler velocity) along  numerous thin, thread-like structures in a H{$\alpha$} filament. The $3-9$~min perturbations found to travel along the small-scale structures with an average phase speed of $12$~km/s. Using high-spatial resolution with DST/ROSA, and in agreement with numerical simulations, \citet{2012ApJ...746..183J} reported upwardly propagating longitudinal magnetoacoustic waves in photospheric MBPs with periods in the range $100-600$~s. They also found standing waves at shorter periods in about 27\% of their MBPs. By employing time-series of slit-jaw images from IRIS (in 279.6~nm, 133~nm, and 140~nm channels), \citet{2020JApA...41...18Z} found $2-5.5$~min intensity oscillations in small MBPs, propagating from the chromosphere to transition region with phase speeds ranging from $30-200$~km/s.

\begin{figure*}[!t]
\begin{center}
\includegraphics[trim=0mm 0mm 0mm 0mm, clip, width=\textwidth]{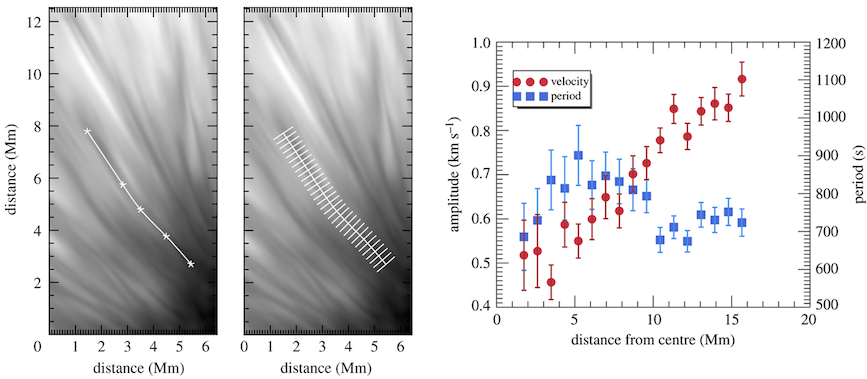}
\end{center}
\caption{Left: a sample sunspot super-penumbral fibril, along which transverse kink waves have been identified. Right: average velocity amplitudes and periods of the transverse oscillations as a function of distance to the umbral center. The error bars indicate the standard deviation of distributions of the parameters. Images reproduced from \citet{2021RSPTA.37900183M}.}
\label{fig:sunspot_fibrils}
\end{figure*}

More recently, \citet{2021RSPTA.37900184G} identified both kink and sausage modes in small, likely magnetic, bright points from pioneering observations with ALMA (at 3~mm), with periods on the order of 60~s, on average, for the transverse oscillations, and periodicities of about 90~s for the brightness temperature and size fluctuations. Although the exact heights of formation of these observations are still unclear, there have been indications to suggest that the ALMA Band~3 observations represents a wide range of heights, mostly from the mid-to-high chromosphere, but there may also be contributions from the lower chromosphere, and possibly the upper atmosphere \citep{2020A&A...635A..71W,2021RSPTA.37900174J}. Thus, it is difficult to conclude at this moment where these small structures reside, although it is highly likely to be in the chromosphere. The high-frequency oscillation reported by  \citet{2021RSPTA.37900184G} are comparable to those previously found in the low-to-mid chromosphere.

The high frequencies observed in the photosphere and the lower/middle chromosphere ($>$10~mHz) have been reported less frequently in the upper chromospheric fibrillar structures \citep{2007ApJ...655..624D,2012ApJ...750...51K,2012A&A...543A...6M,2013ApJ...768...17M,2014ApJ...784...29M}, which however, were observed at different resolutions (and with different properties) compared to those seen at lower heights. These could be speculated as the result of wave energy dissipation (associated to those high frequencies) through the chromosphere. However, no clear observational evidence for such energy release has been found to date. We note that frequencies higher than 10~mHz were also observed by \citet{2013ApJ...768...17M} and \citet{2014ApJ...784...29M} in fibrillar structures, however their mean values lie in lower frequencies. One exception is the high-frequency (on the order of 22~mHz) transverse oscillations that \citet{2011ApJ...736L..24O} found in type~{\sc ii} spicules, however, this was postulated by the authors to be a result of the method they employed in their study. Direct evidence of kink wave damping in the solar chromosphere (i.e., in a Ca~{\sc ii}~H spicule from Hinode/SOT) was provided by \citet{2014A&A...566A..90M} when an initial rapid increase in the oscillation's amplitude with height was followed by an amplitude decrease in the upper chromosphere. The conclusion of wave damping was reached by combining the amplitude variations with changes in width (of the spicule) and phase speed with height, while also comparing to theoretical models.

Oscillations in the chromospheric thread-like structures, including the off-limb Type~{\sc{i}} and Type~{\sc{ii}} spicules, and the on-disk counterparts of the latter, so-called rapid blue-/red-shifted events (RBEs/RREs; \citealt{2009ApJ...705..272R}), have also been reported in a number of studies from both ground-based and space-born observing facilities. By exploiting joint observations of the lower solar atmosphere with SST and IRIS \citep{2020A&A...641A.146R} and the help of MHD simulations, \citet{2017Sci...356.1269M} described the generation of spicules as a result of magnetic tension and ion-neutral interactions. These authors found that impulsive release of the magnetic tension to excite Alfv{\'e}n waves in these small-scale thread-like structures. \citet{2013ApJ...769...44S} identified longitudinal, transversal, and torsional oscillations in numerous RBEs/RREs from H$\alpha$ and Ca~{\sc ii}~8542~{\AA} observations with SST. The three types of oscillations were found to propagate with velocity amplitudes on the order of $50-100$~km/s, $15-20$~km/s, and $25-30$~km/s, respectively. Later, \citet{2015ApJ...799L...3R} speculated that bright features around their extended network regions observed in IRIS 1330{\,}{\AA} and 1400{\,}{\AA} slit-jaw images could be heating signatures associated to (waves in) H$\alpha$ RBEs and/or RREs from their coordinated observations with SST.

Another important interaction between small-scale magnetic concentrations and the convective motion is in the form of vortices at the solar surface \citep{2012ASPC..456....3S}. The presence and and properties of vortex motions in the solar photosphere and in the chromosphere have been studied from both observations and numerical simulations \citep[e.g.,][]{2008ApJ...687L.131B,2009A&A...507L...9W,2010ApJ...723L.180S,2011A&A...526A...5S,2016A&A...586A..25P,2019ApJ...881...83S,2020ApJ...894L..17Y,2020ApJ...898..137S, 2021RSPTA.37900176K}. Of particular interest is that the vortex flows can excite a variety of MHD wave modes, including torsional (Alfv{\'e}n) waves, at small scales, as the magnetic field lines are frozen in the plasma in the lower photosphere \citep{2011AnGeo..29.1029F,2020A&A...643A.166T}.

\begin{figure*}[!t]
\begin{center}
\includegraphics[trim=0mm 0mm 0mm 0mm, clip, width=\textwidth]{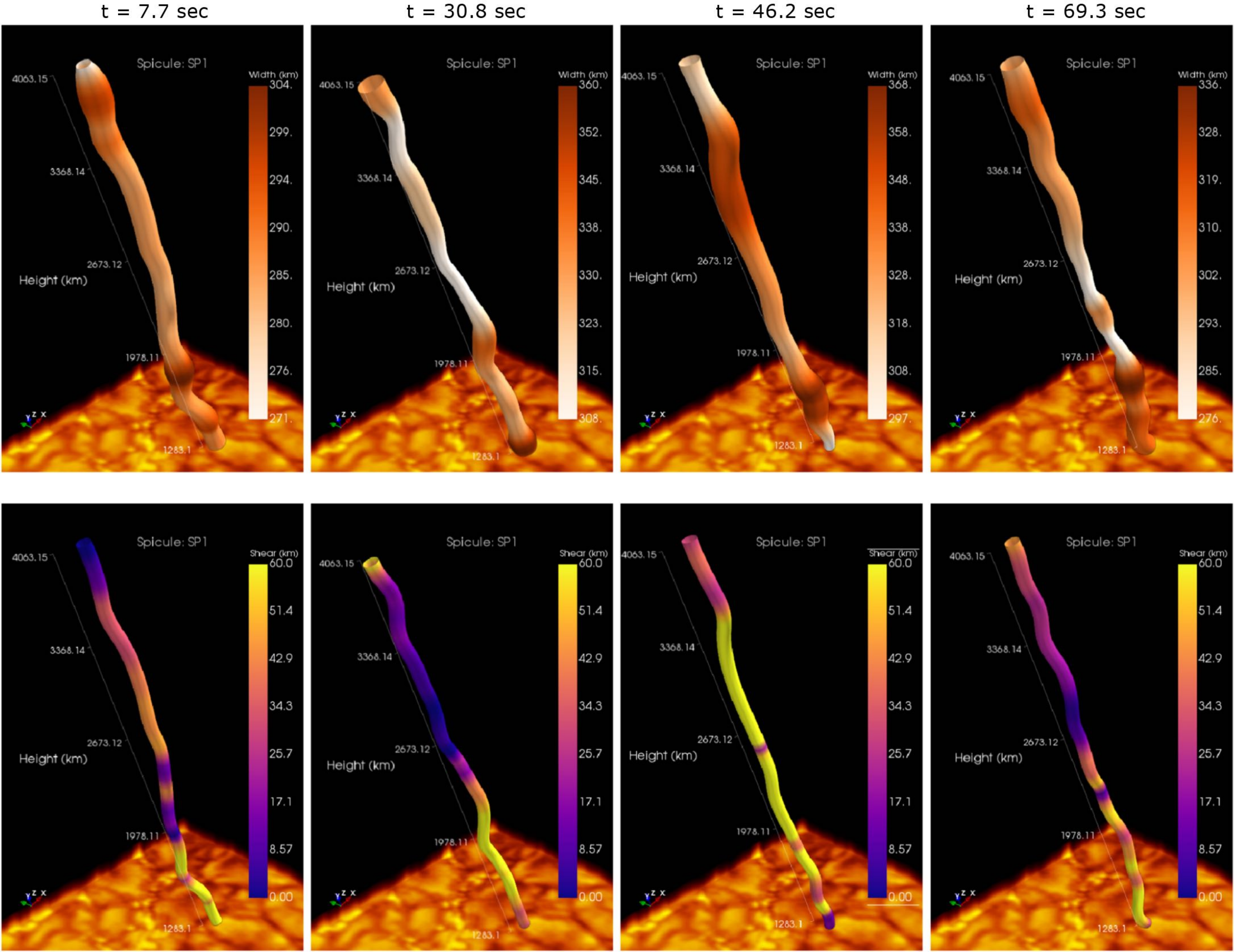}
\end{center}
\caption{Visualization of the coupled MHD wave modes in a spicule, constructed from high-resolution observations with SST/CRISP. The four columns illustrate the 3D structure at different time steps indicated on the top. Top row: coupled transverse and width with intensity. Bottom row: transverse and azimuthal shear components. Image reproduced from \citet{2018ApJ...853...61S}.
An \href{https://WaLSA.team/animations/apjaaa07ff5_video.mp4}{animation} of this figure is also available.}
\label{fig:spicule_3D}
\end{figure*}

\citet{2009Sci...323.1582J} provided the first observational evidence of the torsional (Alfv{\'e}n) waves, detected as full-width half-maximum oscillations in a small MBP through the lower solar atmosphere (with periods on the order of $126-700$~s). They estimated an energy flux of $\approx15000$~W/m$^2$ carried by these waves. Later, \citet{2013ApJ...768...17M} demonstrated the excitation of incompressible kink modes by vortex motions of strong photospheric magnetic concentrations whose chromospheric counterparts showed quasi-periodic torsional motions. In addition, they identified transverse waves in the chromospheric fibrillar structures, connected to the magnetic concentrations, to be driven by the torsional motion. Using MURaM radiation-MHD simulations \citep{2005A&A...429..335V}, \citet{2021A&A...645A...3Y} discussed the formation mechanism of the small-scale vortices and showed how they can heat the solar chromosphere, though propagation of torsional (Alfv{\'e}n) waves. Particularly, they showed that small-scales vortices are produced as a result of cascading in the relatively larger scales (residing in the interganular lanes in the photosphere) due to the turbulent nature of the plasma. That is, the twisted flux tubes create turbulence in the chromosphere, where the magnetic-field pressure dominates that of gas, by co-rotating the surrounding plasma. It is worth noting that some of the small features they found in their simulations (with diameters of $50-100$~km in the photosphere; $100-200$~km in the chromosphere) cannot yet be resolved in observations from currently available instruments. 

According to MHD wave theory, an infinite number of wave modes may co-exist in the same magnetic structure, where phase mixing and mode coupling may also occur \citep{2016GMS...216..431V}. However, many fundamental wave modes (specifically the higher order modes) and their coupling/interaction, particularly at small scales, have been difficult to identify in observations.
\citet{2013A&A...554A.115S} reported interaction between transverse and longitudinal waves in small magnetic elements from both observations (with {\sc Sunrise}/IMaX) and MHD simulations (with MURaM). They particularly found a 90 degree phase difference between transverse oscillations (with frequencies larger than 10~mHz) and longitudinal (velocity) perturbations characterized by frequencies smaller than $7-8$~mHz. The interaction between the two type of MHD waves were, however, found to take place with a high confidence level at periods shorter than 200~s.

High spatial and temporal resolution H$\alpha$ observations from SST/CRISP provided \citet{2017ApJ...840...96S} three-dimensional velocity vectors to identify MHD kink modes in spicules from two approximately perpendicular angles. Furthermore, \citet{2018ApJ...853...61S} used the same dataset to inspect coupling between various MHD wave modes in the off-limb thread-like structures. In this regard, they also explored time variations of longitudinal, cross-sectional width, photometric, and azimuthal shear/torsion parameters at selected spicules, that were concluded to be coupled over the period scale, supported by mutual phase relationships. In particular, they found that the nonlinear kink waves (identified as the displacement of the spicule's axis in both the plane-of-sky and Doppler directions) were coupled with the longitudinal (field-aligned) flows. These led \citet{2018ApJ...853...61S} to explain the coupling of the independent wave modes in the spicules as a result of a single pulse-like driver following a twist. Figure~\ref{fig:spicule_3D} visualizes a 3D structure of a spicule studied by \citet{2018ApJ...853...61S}, where the coupled transverse and width oscillations (top row) as well as transverse and azimuthal shear components (bottom row) are shown in four time steps.

While ubiquitous rapid (supersonic) and high-frequency intensity fluctuations in the chromosphere and transition region, observed with the rocket-borne Chromospheric Lyman-Alpha Spectropolarimeter (CLASP; \citealt{2012SPIE.8443E..4FK}) instrument, were explained as MHD fast-mode waves \citep{2016ApJ...832..141K}, they were later attributed to both waves and jets (i.e., small-scale transient features) from joint observations with CLASP and IRIS \citep{2020ApJ...889..112S}. In the latter study, the authors found non-linear wave propagation in the core of plages and linear propagation of fluctuations as a result of non-recurrent jet-like features. Moreover, using an unprecedented high temporal cadence of 0.3~s with CLASP in hydrogen~Ly$\alpha$~1216{\,}{\AA}~line, \citet{2019ApJ...887....2Y} found high-frequency oscillations (of the Doppler velocity) in the early phase of a spicule evolution, with period and propagating speed on the order of 30~s and 470~km/s, respectively.

Other excitation mechanisms have been proposed for the observation of magnetoacoustic waves at small scales. Of particular interest is small magnetic reconnection which have been thought to be the driver of kink modes \citep{2009ApJ...705L.217H,2014Ap&SS.353...31E}.
Furthermore, through numerical studies, \citet{2011ApJ...730L..24K} proposed a new mechanism called ``magnetic pumping'' to excite upwardly propagating slow modes in magnetic flux concentrations. They showed that the convective downdrafts around a flux tube can eventually result in pumping downflows inside the tube, hence, creating magnetoacoustic oscillations

Table~\ref{table:summary} summarizes the average properties of some of the various MHD waves in small-scale magnetic structures, reviewed in this section. As evidenced, the mean periods and phase speeds measured in different studies have wide ranges on the order $34-754$~s and $13-270$~km{\,}s$^{-1}$, respectively. We note that comparison of these wave characteristics, from different studies, should be performed with great caution. These may include results obtained for the same wave types/modes and/or in similar structures (e.g., in magnetic bright points or fibrillar structures). Such values may not always be one-to-one comparable due to various reasons, such as structures with different spatial/temporal scales residing in different geometric heights and/or different solar environments, as well as some measurement effects. Estimating accurate formation heights is a challenging task, even when the observations are made at similar wavelengths. For instance, the spectral resolution or the width of filters employed, and/or the level of magnetic flux contained can play important roles in observations of different geometric heights, not only on average, but also across the field of view or along the magnetic structures. The choice of analysis approaches is another important factor in the reported (often) mean values (due to, e.g., some selection effects). Furthermore, the spatial and temporal resolutions of the observations as well as the length of the time series can limit, e.g., the identified structures (which are found in a variety of spatial and temporal scales, though with similar names), and the range of detectable frequencies. Hence, the wave characteristics reported in the literature (using different observations) may not necessarily represent waves in the same structures, same geometric heights, and/or same solar regions.

\begin{table*}[!htp]
\centering
\begin{threeparttable}
\small
\caption{Average (or median) period ($T$), phase speed ($v_{\mathrm{ph}}$), and energy flux ($F_E$) of MHD waves in small-scale magnetic structures observed in the lower solar atmosphere. If no mean value is available, the range is indicated instead.}
\label{table:summary}
\setlength{\tabcolsep}{.29em}
\renewcommand{\arraystretch}{1.3}
\begin{tabular}{llllllllll}
\hline\hline

{Wave} & {Event\tnote{\emph{a}}} & {Reg.\tnote{\emph{b}}} & {Diagnostic} & {$\Delta\lambda$\tnote{\emph{c}}} & {Telescope} & {$T$} & {$v_{\mathrm{ph}}$} & $F_E$ & {Ref.\tnote{\emph{d}}} \\
{mode} &  &  &  & {(\AA)} &  & {(s)} & {(km/s)} & (kW/m$^2$) &  \\

\hline

Kink / & MBP & AR & Fe~{\sc i}~6302~\AA &  & VTT & 100 &  & 16--23 & 1 \\ 

Alfv{\'e}nic & Sp.II & QS & Ca~{\sc ii}~H & 3.0 & Hinode & 150--350 &  & 4--7 & 2 \\ 

 & Filament & QS & H$\alpha$ &  & SST & 180--540 & 12 &  & 3 \\

 & Sp.II & QS & Ca~{\sc ii}~H & 3.0 & Hinode & 45 & 270 & 0.25 & 4 \\ 
  
 & Fibril & AR & Ca~{\sc ii}~8542~\AA & 0.11 & SST & 135 & 190 &  & 5 \\
  
 & SP.I & QS & H$\alpha$ & 0.25 & DST & 56--220 &  & 300 &  6 \\
  
 & Mottle & QS & H$\alpha$ & 0.25 & DST & 70--280 & 40--110 &  & 7 \\

 & Fibril & QS & H$\alpha$ & 0.25 & DST & 232 & 40--130 & 4.3 & 8 \\

 & RBE & QS & Ca~{\sc ii}~8542~\AA & 0.11 & SST & 54 &  &  & 9 \\
 
 & Fibril & AR & H$\alpha$ & 0.25 & DST & 130 &  &  & 10 \\

 & Fibril & AR & Ca~{\sc ii}~H & 1.1 & {\sc Sunrise} & 89 & 15 & 15 & 11 \\

 & MBP & AR & 3~mm &  & ALMA & 60 &  &  & 12 \\

 & Fibril & AR & Ca~{\sc ii}~8542~\AA & 0.11 & SST & 754 & 25 & 0.08--1.2 & 13 \\

 & Fibril & QS & H$\alpha$ & 0.06 & SST & 120 & 446 &  & 22 \\ 
 
 & MBP & QS & Ca~{\sc ii}~H & 0.10 & SST & 67--333 & 6 &  & 21 \\
 
 & MBP & AR & 3~mm &  & ALMA & 66 & 96 & 3.8 & 19 \\
 
 & Spicules & AR & H$\alpha$ &  & DST & 54 & 128--147\tnote{\emph{e}} & 40--80\tnote{\emph{e}} & 14 \\

\hline 

Sausage  & Fibril & QS & H$\alpha$ & 0.25 & DST & 197 & 67 & 11.7 & 8 \\

 & Fibril & AR & Ca~{\sc ii}~H & 1.1 & {\sc Sunrise} & 34 & 13 &  & 15 \\

 & MBP & AR & 3~mm &  & ALMA & 90 &  &  & 16 \\
 
\hline

Torsional & MBP & QS & H$\alpha$ & 0.25 & DST & 126--700 & 22 & 15 & 17 \\

(Alfv{\'e}n) & Fibril & QS & H$\alpha$ & 0.25 & DST & 120--180 &  &  & 18 \\ 

 & Spicules & QS & H$\alpha$ & 0.06 & SST & 24--83 &  & 100 & 20 \\ 

\hline\hline
\vspace{-1mm}
\end{tabular}
\hspace{1mm}
\vspace{-0.6cm}
\begin{tablenotes}
      \footnotesize
      \item[(\emph{a})]{Name of the observed structure according to the authors (MBP: Magnetic Bright Point; RBE: Rapid Blueshifted Events; Sp.I: Type~{\sc i}~Spicules; Sp.II: Type~{\sc ii}~Spicules).}
      \item[(\emph{b})]{Regions: AR: Active Region; CH: Coronal Hole; QS: Quiet Sun.}
      \item[(\emph{c})]{Spectral resolution, or Full Width at Half Maximum (FWHM), of transmission profile of the passband.}
      \item[(\emph{d})]{References:
      1: \citet{1995AA...304L...1V}, 
      2: \citet{2007Sci...318.1574D}, 
      3: \citet{2007SoPh..246...65L}, 
      4: \citet{2011ApJ...736L..24O}, 
      5: \citet{2011ApJ...739...92P}, 
      6: \citet{2012ApJ...744L...5J}, 
      7: \citet{2012ApJ...750...51K}, 
      8: \citet{2012NatCo...3.1315M}, 
      9: \citet{2013ApJ...764..164S}, 
      10: \citet{2014ApJ...784...29M}, 
      11: \citet{2017ApJS..229....9J}, 
      12: \citet{2021RSPTA.37900184G}, 
      13: \citet{2021RSPTA.37900183M}, 
      14: \citet{2022ApJ...930..129B}. 
      15: \citet{2017ApJS..229....7G}, 
      16: \citet{2021RSPTA.37900184G}, 
      17: \citet{2009Sci...323.1582J}, 
      18: \citet{2013ApJ...768...17M}, 
      19: \citet{2022A&A...665L...2G},
      20: \citet{2017NatSR...743147S}, 
      21: \citet{2015A&A...577A..17S},
      22: \citet{2017A&A...607A..46M}.
      }
      \item[(\emph{e})]{Mean values corresponding to various geometric heights between 4890--7500~km (off the limb) for upward wave propagation. While the phase speed increases with height, the energy flux decreases. The values for downward propagation are 75--128~km{\,}s$^{-1}$ and $\approx$40~kW{\,}m$^{-2}$.}
\end{tablenotes}
\end{threeparttable}
\end{table*}

\subsubsection{Magnetic-field Perturbations in Small-scale Magnetic Structures}
\label{sec:magnetic_field_perturbations}
Measurement of the magnetic fields at small scales has been a challenge due to various reasons, particularly, due to the fact that most of them are spatially unresolved. Furthermore, the photospheric linear polarization signals are often weak (i.e., on the order of the photon noise level) in the quiet Sun where the small-scale magnetic structures reside. Therefore, computations of the full vector magnetic field at these structures are rare \citep{2019LRSP...16....1B} and may lead to incorrect field parameters, such as field inclination angles, when, e.g., traditional Stokes inversions, are employed \citep{2011A&A...527A..29B,2012A&A...547A..89B,2014A&A...569A.105J}. Such measurements at chromospheric heights are even more challenging, due to the smaller magnetic flux \citep{2017SSRv..210...37L}. 

As a result, it is often difficult, if not impossible, to accurately constrain the vector magnetic fields of small-scale magnetic elements as they weave their way from the base of the photosphere through to the chromosphere and beyond. A novel way of uncovering the magnetic field information associated with such small-scale features is to examine them off-limb, where the background polarimetric signals contaminate the Stokes profiles to a lesser extent. Hence, viewing structures, like spicules, against the black background of space is a compelling way of measuring their small-amplitude magnetic signals \citep[e.g.,][]{2005A&A...436..325L, 2005ApJ...619L.195S, 2005ApJ...619L.191T, 2010ApJ...708.1579C}. In particular, Utilizing high-resolution Ca~{\sc{ii}}~8542~{\AA} limb observations acquired with the SST, \citet{2020A&A...642A..61K} employed the weak field approximation to radiative transfer equations to infer the line-of-sight component of the magnetic field in a multitude of different spicule types. Magnetic field strengths on the order of 100~G were found ubiquitously along the spicule structures, with little difference found between spicules embedded close to active regions and those associated with the quiet Sun \citep{2020A&A...642A..61K}.

Moving on from the weak field approximation, \citet{2021ApJ...908..168K} harnessed a new version of the non-LTE NICOLE inversion code to invert a prominent off-limb spicule captured in the Ca~{\sc{ii}}~8542~{\AA} spectral line by the SST. By considering true geometry effects through the inclusion of vertical stratification, \citet{2021ApJ...908..168K} were able to provide a semi-empirical model for the specific spicule structure examined, which consisted of a uniform temperature of 9560~K, coupled with an exponential density decrease as a function of atmospheric height, providing a density scale height on the order of $1000-2000$~km. These results are consistent with those previously deduced by \citet{1968SoPh....3..367B}, \citet{1973SoPh...32..345A}, and \citet{1976SoPh...46...93K}, albeit with more modern non-LTE considerations invoked. However, the spicule studied by \citet{2021ApJ...908..168K} is an interesting structure with atypical characteristics. Specifically, the spicule demonstrated a clear inverted Y-shaped base, consistent with anemone jets driven by magnetic reconnection in the lower solar atmosphere \citep[e.g.,][]{1995Natur.375...42Y, 2007Sci...318.1591S, 2009ApJ...705L.217H}. In addition, the spicule structure reached atmospheric heights of $\sim$10~Mm above the surface and was clearly visible in the far blue-wing of the Ca~{\sc{ii}}~8542~{\AA} spectral line, suggesting the feature may be similar to dynamic type~{\sc{ii}} spicule events \citep{2018ApJ...860..116M}. However, the lifetime of the structure examined by \citet{2021ApJ...908..168K} was $>$20~minutes, which is not consistent with the shorter duration lifetimes ($\sim50-150$~s) of traditional type~{\sc{ii}} spicules \citep{2012ApJ...759...18P, 2016ApJ...824...65P, 2012ApJ...752..108S, 2013ApJ...764..164S}. As such, the spicule feature examined by \citet{2021ApJ...908..168K} may be more closely related to the macrospicules initially observed as cool plasma in the He~{\sc{ii}}~304~{\AA} spectral line by Skylab \citep{1975ApJ...197L.133B}, and later as vibrant O~{\sc{v}} emission by the Solar Ultraviolet Measurements of Emitted Radiation \citep[SUMER;][]{1995SoPh..162..189W, 2000A&A...360..351W} spectrograph onboard the Solar and Heliospheric Obervatory \citep[SOHO;][]{1995SoPh..162....1D} spacecraft.

Small-scale magnetism can also be studied in the upper-chromospheric He~{\sc{i}}~10830~{\AA} spectral line. Here, it is possible to make use of the Hanle effect to deduce magnetic field information since it is more sensitive (compared to the traditionally employed Zeeman effect) to the weaker magnetic fields present in small magnetic elements \citep{1994ASSL..189.....S, 1998ApJ...493..978L}. Similar to the work of \citet{2020A&A...642A..61K} and \citet{2021ApJ...908..168K}, many studies have attempted spectropolarimetric inversions of off-limb spicular features to uncover the magnetic field variations away from the solar disc \citep[e.g.,][]{2005ApJ...619L.191T, 2010ApJ...708.1579C, 2015ApJ...803L..18O}. Utilizing He~{\sc{i}}~10830~{\AA} diagnostics, \citet{2005ApJ...619L.191T} uncovered spicule magnetic fields as low as 10~G. This highlights the difficulties when attempting to uncover magnetic field perturbations arising from propagating wave phenomena. Even with a relatively large 10\% amplitude variation, this only equates to a $\pm$1~G fluctuation in the associated magnetic field strength of the spicule. As a result, it becomes statistically challenging to reliably quantify such minuscule variations in the magnetic field strength, especially with these plasma parameters inferred from a multitude of spectral lines, often with weak Stokes~$Q/U$ components. Hence, measurements of oscillations in individual components of the polarization signals has predominantly been limited to the (often dominant) circular polarization (Stokes~$V$) component.

Only recently, small-scale kilogauss magnetic elements could spatially be fully resolved \citep{2010ApJ...723L.164L}, not only because the high-spatial resolution provided by the 1-m {\sc Sunrise} balloon-borne solar telescope, but also due to the seeing-free observations \citep[and the high precision of the IMaX spectropolairemeter;][]{2011SoPh..268...57M} which in turn resulted in higher polarization signal-to-noise compared to those normally achieved with similar ground-based instruments currently available. Thus, fluctuations of polarization signals could also be detected in small magnetic elements observed by {\sc Sunrise} \citep{2013A&A...549A.116J}.
With a relatively lower spatial resolution from Hinode, but also from seeing-free data, \citet{2013A&A...554A..65U} were also able to measure the kilogauss field strength at small magnetic elements. However, it should be noted that the definition of `small' features may vary from one study to another, with various spatial sizes reported on.

\citet{2011ApJ...730L..37M} presented magnetic-field oscillations (from {\sc Sunrise}/IMaX observations) in a very quiet photospheric area whose field strength did not exceed 500~G. They concluded that the oscillations were not associated to oscillatory modes of magnetic concentrations, but rather, to buffeting of the magnetic-field lines by granular flows. It is worth noting that they found two different prominent period ranges, one corresponding to magnetic flux density patches of $10^{16}-10^{17}$~Mx (with a period range of $4-11$~min) and one associated with more intense patches of 10$^{18}$~Mx (with periods on the order of $3-5$~min). Thus, the small magnetic patches studied by \citet{2011ApJ...730L..37M} were representative of relatively weak magnetic flux concentrations that preferentially emerge within granules \citep{2007A&A...469L..39M,2007ApJ...666L.137C,2008A&A...481L..33O,2009ApJ...700.1391M,2010ApJ...723L.149D,2018SoPh..293..123K}. On the other hand, \citet{2011ApJ...730L..37M} found clear evidence of $p$-modes oscillations (with a 5~min periodicity) in the Doppler velocity, as well as both continuum and line-core intensities (of the Fe~{\sc i}~525.02~nm spectral line). Interestingly, they found a 180 degree phase difference between fluctuations in the continuum and line-core intensities, as well as an anti-phase between Doppler velocity and continuum intensity perturbations. Such anti-correlations between Doppler velocity and intensity in small-scale magnetic structures were also found by \citet{2021A&A...647A.182C} in high-resolution observations with GREGOR \citep{2012AN....333..796S}.

Recently, \citet{Norton2021} examined the identification of magnetic-field perturbations from SDO/HMI observations, in various solar regions. They found that except in the umbra, almost no oscillatory power could be detected in the field-strength oscillations in, e.g., the quiet-Sun and plage regions. No oscillatory signature could also be identified in the field inclination and azimuth.

Using high spatial and temporal resolution observations with SST/CRISP, \citet{2020A&A...633A..60K} detected rapid (within $33-99$~s) magnetic-field amplification (by a factor of $\approx2$ on average) in numerous MBPs whose field strengths followed a bimodal distribution \citep{2019MNRAS.488L..53K}. With the help of numerical simulations, \citet{2020A&A...633A..60K} found that the field amplification could possibly be explained as a result of convective collapse \citep{1979SoPh...61..363S}, as the most frequent process, thus the decrease in size of the MBPs is accompanied by amplification of the field strength. This process is similar to that responsible for excitation of sausage modes. Furthermore, the authors find evidence, albeit less frequently, for granular compression of the inter-granular lanes leading to magnetic field amplification in the MBPs. Although the ,mechanisms leading to field amplification are somewhat similar to convective collapse with both displaying a decrease in size, the authors pick out differences that imply a distinct process occurring in this case for amplification due to granular expansion into the lanes. Like the case with convective collapse, the authors note that this process is similar to that responsible for excitation of sausage modes.

In addition to the level of polarization signals (compared to the noise level) which is necessary for a reliable detection, inferring physical parameters from Stokes inversions seems to be challenging when waves pass through the solar atmosphere. \citet{2021RSPTA.37900182K} did an experiment with a synthesized dataset for the Fe~{\sc i}~6301~{\AA} and Fe~{\sc i}~6302~{\AA} line pair and found that, e.g., the Doppler velocities, could not be returned accurately by the inversion codes, after synthesis with NICOLE \citep{2015A&A...577A...7S} and inversion with SIR \citep{1992ApJ...398..375R} (compared to their initial values in the numerical simulations), at the presence of an upwardly propagating wave in a thin flux tube. This was explained as waves perturb the atmosphere over a smaller height range compared to that sampled by the spectral lines. Thus, development of inversion codes in this regard is crucial, otherwise, the prevalence of waves and oscillations in the solar atmosphere may largely influence the inferred parameters from inversions. It is worth nothing that some recent advancements in development of powerful (multi-line) non-LTE inversion codes \citep[e.g.,][]{2017ApJS..229...16R, 2018A&A...617A..24M, 2019A&A...623A..74D, RuizCobo2021_DeSIRe}, and other similar approaches \citep{2018ApJ...866...89C,2019A&A...626A.102A}, will significantly improve these measurements, particularly in the quiet regions through the entire lower solar atmosphere.

Although, the current pioneering instruments, with enhanced spatial resolutions and spectropolarimetric sensitivities over the past decade, have advanced our understanding around the magnetic-field oscillations at small scales (and in the quiet-Sun regions), a clear detection of such perturbations remains difficult to date. It is foreseen that the next generation spectropolarimetric instruments, including those on DKIST and {\sc Sunrise}~III, together with new generations of Stokes inversion codes, will revolutionize our vantage points of magnetic-field oscillations in the lower solar atmosphere.

\section{Future Directions and Progress}
\label{sec:unansweredquestionsarising}
Section~{\ref{sec:recentstudiesofwaves}} highlights key advancements and accomplishments made by the global solar physics community over the last number of years. Importantly, these achievements also reveal areas where our collective understanding is currently lacking. Here, we will pinpoint specific topics that we believe are within scientific reach following the realization of next-generation observatories, modeling techniques and analysis tools.

It has been shown in a multitude of studies that wave signatures manifesting in the lower solar atmosphere may be the result of upward wave propagation from sub-photospheric layers \citep[e.g.,][to name but a few recent examples]{2015A&A...580A..53L, 2017ApJ...836...18C, 2019ApJ...883...72C, 2018AdSpR..61..720G, 2018MNRAS.479.5512K, 2020ApJ...896..150Y, 2020JApA...41...18Z}. However, while it seems that global eigenmodes may be responsible for many of the observed wave signatures, they are unable (by themselves) to account for all of the perceived signals found in lower atmospheric wave guides, which often require additional (often unknown) perturbations, pulses, and/or excitation sources to explain \citep[e.g.,][]{2016ApJ...830L..17Z, 2018ApJ...869..110S, 2021RSPTA.37900216S}. Therefore, what other mechanisms are responsible for the complex wave signatures captured in high-resolution photospheric and chromospheric data sequences? This is where Fourier filtering techniques (see, e.g., Section~{\ref{sec:threedimensionalFourierfiltering}}) can play an important role, since they have the ability to disentangle small-scale wave perturbations from macroscopic flows and dominant MHD modes. Indeed, wavelet filtering and pixelized wavelet techniques can also provide useful diagnostic potential, especially if transient and/or rapidly developing wave signatures are present \citep{antoine2002application, 2008SoPh..248..395S, 2021RSPTA.37900180S}. Therefore, tailored Fourier and/or wavelet filtering algorithms will be of paramount importance to extract the smallest amplitude fluctuations, particularly those associated with higher order MHD modes, from the dominant (with often orders-of-magnitude larger power) wave signals. 

The power of novel visualization tools, such as $B$-$\omega$ diagrams (see Section~{\ref{sec:Bomega}}), cannot be overstated. In Figure~{\ref{fig:BOmega}}, the suppression of $p$-modes towards the boundary of a magnetic structure is clearly evident, as is the growth of distinct eigenmodes once magnetic fields synonymous with the solar pore are isolated. Such techniques go well beyond traditional one-dimensional Fourier power spectra, and allow not only the wave behavior to be more closely scrutinized as a function of other local plasma properties, but also provide impressive visualizations that allow the work to be more readily disseminated. We expect such powerful techniques to be widely exploited by the solar physics community in years to come. In particular, $B$-$\omega$ diagrams may be harnessed in longer-duration synoptic studies to see whether driving mechanisms in magnetic features (sunspots, pores, etc.) vary over the course of a solar cycle. As such, long-duration data sequences will provide excellent frequency resolutions that may help isolate true eigenmodes trapped within the boundaries of such magnetic structures.

Observations have shown systematic differences in the chemical abundances found in the corona and in the photosphere. In particular, in closed-loop systems, low ($<10$~eV) first ionization potential (FIP) elements are occasionally more abundant in the corona than in the photosphere, by a factor ranging from $2-4$. In contrast, the plasma composition in open magnetic field regions remains practically unfractionated between the lower and upper solar atmosphere \citep[][to mention but a few examples]{1995ApJ...440..884S, 1996ApJ...469..423S, 2013ApJ...778...69B}. Interestingly, the FIP bias is also observed in the solar wind \citep{1999ApJ...521..859S, 2019ApJ...879..124L} and could, therefore, be used to infer the magnetic connectivity in the heliosphere, and possibly help to identify the sources of the solar wind itself \citep{2015NatCo...6.5947B}. This is of primary importance to investigate and identify the solar wind acceleration mechanisms, which is one of the fundamental questions that the Solar Orbiter mission is being specifically designed to address through a combination of both remote sensing and in-situ instruments. It was predicted from theoretical modeling that the FIP bias in the solar corona could be due to the presence of  magnetic perturbations which, through the ponderomotive force, are responsible for the plasma fractionation \citep{2015LRSP...12....2L}. This was recently confirmed thanks to a combination of ground-based and space-borne observations, which identified magnetic oscillations at chromospheric heights magnetically linked to the coronal locations where the FIP bias was observed \citep{2021ApJ...907...16B, 2021RSPTA.37900216S}.

Theoretical and numerical modeling work has continually speculated that the lower solar atmosphere should be replete with the entire assortment of MHD wave modes: slow and fast magnetoacoustic modes, plus {\Alfven} waves \citep[e.g.,][to name but a few examples]{1983SoPh...88...77C, 1994ApJ...437..505C, 2006ApJ...653..739K, 2008SoPh..251..251C, 2011A&A...527A.132T, 2012ApJ...751...31H, 2012ApJ...746...68K, 2013MNRAS.435.2589C, 2015A&A...578A..60M, 2016ApJ...817...94A, 2016GMS...216..489C, 2017MNRAS.466..413C, 2018MNRAS.480.2839L, 2019ApJ...885...58C, 2019ApJ...870...94G, 2019ApJ...881L..21P, 2019SoPh..294..147R}. Indeed, observations have shown evidence for ubiquitous slow mode waves \citep{2015ApJ...806..132G, 2016ApJ...831...24K, 2016SoPh..291.3349T, 2017ApJ...842...59J, 2018ApJ...868....5C, 2019ApJ...877L...9K}, and indeed even (more challenging to detect) {\Alfven} modes \citep{2009Sci...323.1582J, 2014Sci...346D.315D, 2017NatSR...743147S}. Unfortunately, the universal existence of fast magnetoacoustic modes in the lower solar atmosphere is less well documented. \citet{2012NatCo...3.1315M} revealed evidence for fast incompressible MHD wave modes in chromospheric fibrils and mottles, which may be driven by photospheric flows and vortices \citep{2013ApJ...768...17M, 2018MNRAS.474...77M, 2019NatCo..10.3504L, 2020A&A...639A..59M}. However, the unequivocal ubiquity of fast mode waves in the lower solar atmosphere has still not been verified. 

Spectropolarimetric inversion processes are powerful techniques that allow key plasma parameters (magnetic fields, densities, temperatures, velocities, etc.) to be established as a function of optical depth. Examples of such software have allowed crucial plasma conditions associated with solar structures exhibiting wave activity to be uncovered \citep[e.g.,][]{2013A&A...553A..73B, 2013ApJ...776...56R, 2018ApJ...860...28H, 2020ApJ...892...49H, 2018NatPh..14..480G, 2019A&A...621A..43F, 2020NatAs...4..220J}. Next-generation inversion routines, including the Stockholm inversion code \citep[STiC;][]{2019A&A...623A..74D}, the Spectropolarimetic NLTE Analytically Powered Inversion \citep[SNAPI;][]{2018A&A...617A..24M} code, and the Departure coefficients Stokes Inversion based on Response functions (DeSIRe; \citealt{RuizCobo2021_DeSIRe}) code offer powerful new approaches for accurately constraining the derived plasma profiles, including the simultaneous use of multiple spectral lines. This is especially important when lines may not strictly be formed under LTE conditions, for example, the Fe~{\sc{i}} photospheric absorption lines in the presence of UV overionization \citep{2020A&A...633A.157S}. Recently, \citet{2019A&A...622A..36R} showed that many-line inversions of photospheric spectropolarimetric data, particularly at short wavelengths (where the photon noise is considerably higher than that at longer wavelengths) can significantly improve the outputs compared to inversions of a few spectral lines only. In addition, \citet{2012A&A...548A...5V} and \citet{2015A&A...577A.140A} have recently been developing spatially-coupled inversion routines, whereby the authors found that inclusion of the point spread function of the telescope, alongside the degree of spatial correlation between neighboring pixels, respectively, helps to minimize associated errors of the inversion outputs. As such, utilizing numerous spectral lines with differing magnetic field sensitivities (i.e., different Land{\'{e}} $g$-factors) spanning the base of the photosphere through to the upper extremities of the chromosphere will be required to converge spectropolarimetric inversion outputs to trustworthy values, which will be of paramount importance when attempting to benchmark small-scale seismological fluctuations throughout the lower solar atmosphere. 

There are a myriad of complexities involved in inversion studies involving oscillatory phenomena. One such issue is that most inversion codes return the plasma parameters as a function of optical depth, which is distinctly different from the geometric atmospheric height. It is a challenging endeavor to convert optical depth to atmospheric height, since this conversion hinges upon the reliability of the input model atmosphere. Indeed, \citet{2018SoPh..293...74I} recently showed the influence of different input atmospheric models on the inverted diagnostic outputs. This highlights the importance of ensuring the stationary (background) model is representative of the structure being investigated. Indeed, if wave-based perturbations are superimposed on top of this stationary input model, then it becomes difficult to disentangle what is a true fluctuation from what is just a simple deviation from the simplified input model. This complexity is even more pronounced when systematic effects associated with spectral fitting routines contaminate the inversion outputs \citep{2016A&A...590A..87A}. The next logical step is to try and generate background model atmospheres that already contain stratified periodic fluctuations in, e.g., density, temperature, velocity, etc., which are synonymous with the wave features wishing to be evaluated. Of course, such {\it{a-priori}} knowledge of the dominant embedded wave modes may not always be available to the researcher before the inversions are performed. As such, making use of inversion processes based on machine learning and neural networks \citep[e.g.,][]{2019A&A...626A.102A, 2020A&A...644A.129M, 2021arXiv210111445S} can help expedite the inversion process, especially when dealing with large degrees of freedom. Ultimately, this will allow Stokes profiles that contain wave perturbations to be reliably inverted, hence minimizing uncertainties and providing the first step in converting optical depths into true geometric heights once compared to the accurate background model that is reflective of the observations acquired.

Another possible solution to this issue is the MHD-assisted Stokes Inversion \citep[MASI;][]{2017ApJS..229...16R} code, which is based on the spectral syntheses of state-of-the-art MHD simulations and is a first step to being able to directly output plasma parameters as a function of true geometric scales. This work builds upon the legacy codes provided and documented by \citet{1999A&A...345..618M}, \citet{2001A&A...366..686T}, \citet{2005A&A...430..679B}, \citet{2008A&A...481L..37C}, and \citet{2013A&A...549A..24B, 2015ApJ...798..100B}. As highlighted by \citet{2017ApJS..229...16R}, the MASI code provides a platform for inversions that give results consistent with the underlying MHD equations, which is likely to be of huge benefit to researchers working exclusively in wave perturbations in the lower solar atmosphere. 

Yet another solution for inferring the magnetic-field vector throughout the lower solar atmosphere as a function of geometric height is the non-force-free fields extrapolations introduced by \citet{2015ApJ...815...10W,2017ApJS..229...18W}. These extrapolations use high-spatial resolution photospheric field vector as the boundary condition for a magnetohydrostatic (MHS) model. This extrapolation code was specifically designed for accurate approximation of magnetic-field vector at heights below $\approx2000$~km (i.e., the solar photosphere and the chromosphere) where the non-vanishing Lorentz force exists. Thus, to account for such a mixed plasma-$\beta$ environment, the MHS model self-consistently considers the pressure gradients and gravity forces through these regions of the atmosphere. We note that such MHS extrapolations are different from the traditional force-free field extrapolations which are mostly accurate in the solar corona where the Lorentz force vanishes \citep{2006SoPh..233..215W,2008SoPh..247..249W,2012LRSP....9....5W}. By combining MHS constraints with Stokes inversions, \citet{2021A&A...647A.190B} developed a new inversion code which is capable of retrieving the physical parameters in the solar photosphere as a function of geometric height. The new code makes use of an MHS solver and the Firtez-DZ code (a solver of the polarized radiative transfer equation in geometrical scale; \citealt{2019A&A...629A..24P,2019A&A...632A.111B}) which are based on three-dimensional MHD simulations of sunspots \citep{2012ApJ...750...62R}.

Recent work by \citet{2021RSPTA.37900182K} showed the difficulties in constraining atmospheric parameters in an oscillating waveguide using commonly used inversion techniques on a simulation with propagating MHD waves. The authors utilized simple two-dimensional MHD models with known driver and atmospheric parameters. From this, the authors synthesized Stokes profiles for the 6301~{\AA} and 6302~{\AA} line pair and then employed the Stokes Inversion based on Response functions \citep[SIR;][]{1992ApJ...398..375R} code to establish if the atmospheric parameters of the oscillation could be accurately returned. Results from a study by \citet{2011SoPh..273...15V} on the same dataset showed the necessity of high-spatial resolution in resolving asymmetries in the Stokes parameters, therefore, \citet{2021RSPTA.37900182K} degraded the spatial resolution to that typical of DKIST and the upcoming 4m European Solar Telescope \citep[EST;][]{2010AN....331..615C, 2013MmSAI..84..379C, 2022A&A...666A..21Q}. The results highlighted that inversion codes such as SIR, could return the atmospheric parameters fairly accurately within typical height of formation regions for the lines inverted. However, the authors note that the Doppler velocity is less accurately fit for perturbed spectra. This is an issue that could possibly be improved upon in the future. 

This is a potential difficulty for inversion algorithms that  utilize a nodal approach in minimizing the differences between the post radiative transfer profiles to the input models. Due to the fact that a spline interpolation is generally used to calculate the atmospheric parameters at grid points between nodes, there is a trade-off in selecting nodes (free parameters) in the inversion. Too few and it is unlikely that the perturbation will be accurately fit in the atmosphere. Too many nodal points could lead to potentially over-fitting the atmosphere when the spline interpolation is calculated \citep{2000ApJ...535..475B}. Therefore, there is a risk in inversion algorithms that employ the nodal approach that an over-fit atmosphere is misinterpreted as a perturbation in the atmosphere.  Furthermore, inversion algorithms generally assume the atmosphere is in hydrostatic equilibrium to make the computation manageable. Perturbations and oscillatory phenomena in the atmosphere will affect the validity of this assumption could affect the resultant output from the inversion. However, initial studies have shown that inversion methods can extract reliable wave behaviour. For instance, \citet{2021A&A...654A..50N} estimated the magnitude of the magnetic field of a pore with SIR inversions and the strong-field approximation method. Both methods detected magnetic perturbations of similar amplitude and frequency, consistent with sausage modes. Despite this, it is an uncertainty that deserves more comprehensive study, particularly when impulsive, non-linear wave modes are under consideration \citep[e.g.,][]{2020ApJ...892...49H}.

To an extent, these issues could be alleviated by multi-line studies using inversion codes such as STiC and DeSIRe to invert the atmosphere at multiple heights. Studies ascertaining the success of such approaches in the context of wave propagation need to be performed, however, to establish any potential issues with this approach. Another solution to this problem was suggested by \citet{2021RSPTA.37900182K}, whereby the authors suggest adapting an inversion code such as CAISAR (or MASI) to take spectra from MHD simulations archives with propagating wave phenomena with known wave properties to return the atmospheric parameters in observations with wave phenomena present. A similar idea has already been implemented with IRIS data by \citet{2019ApJ...875L..18S}. Here, the authors employed machine and deep learning techniques using a database of representative IRIS profiles of various solar features to ascertain the thermodynamic state in the upper photosphere and chromosphere. This method was found to be both accurate and significantly faster to process over more traditional techniques. A similar approach utilising MHD simulations as suggested by \citet{2021RSPTA.37900182K} is not without issues, as it is constrained by the accuracy and diversity of profiles generated by the MHD simulations. This is highlighted further by \citet{2021RSPTA.37900170F}, who find substantial differences in wave properties between the four commonly used MHD codes. Even considering these limitations, and keeping in mind all inversion algorithms will have limitations, it may be desirable to explore the possibilities of adapting existing tools,  to perhaps provide a somewhat robust method of inverting data with potential wave phenomena present. Regardless, the issue of accurately constraining atmospheric parameters from inversions in an oscillating atmosphere is still an open question in the field. It is highly likely that either a dedicated inversion code is needed or an existing code is adapted to invert datasets with suspected wave activity for a more accurate inversion of the observations. This is an even greater requirement for small-scale dynamic features, such as MBPs.

Further issues may arise due to the dataset being inverted as well. That is, instrumental limitations may make it harder to retrieve accurate inversion results in the presence of upwardly propagating waves. One such example is with the use of Fabry-P{\'{e}}rot interferometers. With these instruments there will be a time lag as the instrument scans across a given line, which can result in a scan time of at least around 10~s for an individual line after data reduction. Obviously this will have implications on wave studies. One study by \citet{2018A&A...614A..73F} looked at the effect of scanning time in umbral flashes appearing in simulations obtained with the MANCHA code. The authors synthesized (and inverted) the 8542~{\AA} line with the NLTE code NICOLE \citep{2015A&A...577A...7S}. The authors performed the inversions on the instantaneous synthesized Stokes parameters (with a single time step) as well as `synthetic scanned' Stokes parameters, where line scanning similar to instrument scans was introduced to the profiles to simulate actual observations. The authors report that, for profiles simulating a line scan, the inversions reasonably infer the atmosphere prior to and after the umbral flash is fully developed. However, during the early stage of the umbral flashes, issues appeared due to the short time for changes in the Stokes profiles in comparison to the temporal cadence of the scans. The Stokes-$V$ profiles in this case were similar to those for the instantaneous profiles, but with a remarkably different intensity profile associated to them. Such a difference can complicate the interpretation of the spectropolarimetric data, with the authors estimating that approximately 15\% of profiles in flashing regions are affected by this issue.

A subsequent study by \citet{2019A&A...632A..75F} looked at the effect of wavelength sampling on inversion results for wave studies. Using a similar approach to previous work, the authors synthesized the 8542~{\AA} line from a simulated umbral flash event, and degraded the spectral resolution (i.e., the wavelength sampling) before inverting the data with NICOLE and analyzing. The authors find that the vertical magnetic field inferred from the inversions is more accurate with finer wavelength sampling. However, finer sampling means an increase in scan time, which will miss sudden changes in the profiles given the dynamic nature of the umbral flash. The authors conclude that sampling positions should be selected with observing goal in mind: studies of the magnetic field should prioritize spectral resolution while studies of the temperature and Doppler velocity should prioritize faster scan times. 

In essence, the work of \citet{2019A&A...632A..75F} is intrinsically linked to the line scan time study \citep{2018A&A...614A..73F}, in that spectral resolution will affect the scan time for a line. However, the study highlights some key challenges facing the wave community in considering and interpreting the results from inversions. Key information can be lost and/or misinterpreted depending on the dataset and instrument employed. In terms of the limitations investigated in these studies linked to Fabry-P{\'{e}}rot interferometers, the next generation of instruments and facilities should alleviate the issues in obtaining reliable inversion results from the data. For example, instruments employing detectors with lower noise levels should attain faster scan times. Likewise, instruments such as integral field units can circumvent some of these issues as they render scanning superfluous. These solutions are not without their issues, however, as faster scan times from better detectors will likely lead to more complex, multi-line scans being used by observers (thus making the scan time issue emerge again). Also, integral field units have a limited field of view, which may impact certain wave studies.

It should be noted as well, that these studies focused on umbral flashes in the 8542~{\AA} line. Therefore, the focus is on a specific phenomena in the chromosphere. Further study is likely warranted for the effect of scan time and spectral resolution on inversion results for perturbations in the photosphere as well for other waveguides and oscillations. It is likely that similar issues will be reported, however, such studies would still be beneficial to determine optimal observing parameters for different features within the context of inversions. Realistically, a combination of instrument improvements and observing sequences with more modern detectors may be needed to fully understand the results from inversions in the context of wave studies in the lower solar atmosphere. Dedicated inversion algorithms for wave studies would likely complement such spectropolarimteric datasets and, therefore, give a deeper understanding and clearer understanding of the atmospheric parameters in the presence of propagating disturbances across a range of structures.

\begin{figure*}[!t]
\begin{center}
\includegraphics[width=\textwidth]{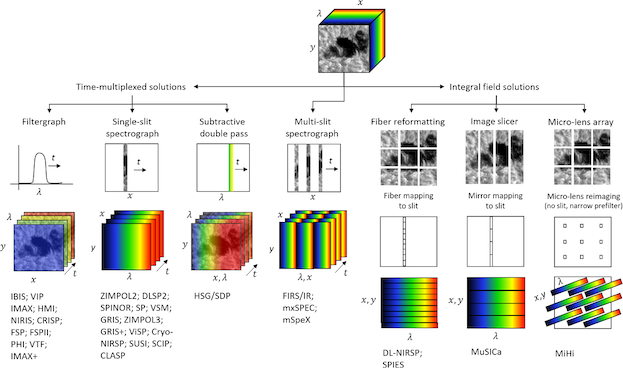}
\end{center}
\caption{Six different spectral mapping techniques used in solar spectropolarimetry. The input cube ($x,y,\lambda$)
at the top can be mapped on to the detector (bottom row) through a number of different mechanisms. Towards the left-hand side of the figure are more traditional methods, including scanning one-dimensional slit-based spectrographs and Fabry-P{\'{e}}rot interferometers, while the right-hand side of the panel depicts the three different types of IFU technology currently available. A multi-slit configuration (middle of the figure) can be considered an in-between solution. Image reproduced from \citet{2019OptEn..58h2417I}.}
\label{fig:IFUs} 
\end{figure*}

Of course, in order to obtain high-precision full Stokes spectropolarimetric signals that are suitable for the study of wave activity, instrumentation is required to capture such spectra with high cadences and low noise levels. Traditional instrumentation, including slit-based spectropolarimeters (e.g., the Facility Infrared Spectropolarimeter [FIRS] at the DST, \citealt{2010MmSAI..81..763J};  the Visible Spectropolarimeter [ViSP] on DKIST, \citealt{2012SPIE.8446E..6XD}; and the TRI-Port Polarimetric Echelle-Littrow [TRIPPEL] spectrograph at the SST, \citealt{2011A&A...535A..14K}) and Fabry-P{\'{e}}rot spectral imagers (e.g., CRISP, IBIS, and the GREGOR Fabry-P{\'{e}}rot Interferometer [GFPI], \citealt{2012AN....333..880P, 2013OptEn..52h1606P}; and the Near Infrared Imaging
Spectropolarimeter [NIRIS] at Big Bear Solar Observatory's New Solar Telescope, \citealt{2012ASPC..463..291C}), have long been the main acquisition systems for such observations. However, these instruments often run into difficulties when attempting to achieve the `trifecta' of high acquisition cadences, spectral precision, and spatial resolutions, all at the same time. For example, it is difficult to maintain high cadences for slit-based spectrographs when scanning their one-dimensional slit across a large field-of-view. Similarly, the simultaneity of spectral profile shapes can be lost if the wavelength scanning time is not fast enough in the case of Fabry-P{\'{e}}rot interferometers \citep{2018A&A...614A..73F}.

The next generation of solar instrumentation is attempting to address the inherent weaknesses of previous observing systems. Many are turning their attention to Integral Field Units (IFUs), which can help provide simultaneous high-precision spectra across a two-dimensional field of view \citep{2019OptEn..58h2417I}. IFUs typically come under three distinct guises: (1) fiber-fed bundles, (2) image slicers, and (3) microlens arrays, which are summarized in Figure~{\ref{fig:IFUs}}. At present, one drawback of IFUs is the reduced field-of-view sizes, which is a natural consequence of requiring the imaging detector to record spectra across two dimensions (i.e., $x$ and $y$ domains) simultaneously. Multi-slit variations (e.g., operating the FIRS instrument at the DST in a multiple-slit configuration), which are also shown in Figure~{\ref{fig:IFUs}}, can be viewed as an in-between solution that offers better cadences that single slit spectrographs, but lacks the true simultaneity of IFUs. 

Much progress has been achieved in recent years with regards to IFU development. The fiber-fed Diffraction Limited Near Infrared Spectropolarimeter \citep[DL-NIRSP;][]{2014SPIE.9147E..07E} is currently in the commissioning stage at the DKIST, providing spatially resolved spectra of infra-red features such as Ca~{\sc ii}~8542{\,}{\AA} and He~{\sc i}~10830{\,}{\AA}. Prototypes of image slicers \citep[e.g., the Multi-Slit Image Slicer based on collimator-Camera, MuSICa;][]{2013JAI.....250009C}, and microlens-fed spectrographs \citep[MiHi;][]{2019AdSpR..63.1389J} have also been trialed with great success. In addition, a new hyper-spectropolarimetric imager is being developed for the Indian National Large Solar Telescope \citep[NLST;][]{2010AN....331..628H, 2014MNRAS.437.2092D}, which is due to be built close to the Chinese border in the Merak region of northern India. This prototype instrument, named the Fibre-Resolved opticAl and Near-infrared Czerny-Turner Imaging Spectropolarimeter ({\sc{francis}}), is specifically designed to probe the lower solar atmosphere within the wavelength range of $350-700$~nm. A plethora of photospheric and chromospheric lines are available, including the Ca~{\sc{i}} 423~nm and Ca~{\sc{ii}} 393~nm spectral lines, so is optimally designed to contribute to the future goals of lower atmospheric wave studies through high-cadence polarimetry across multiple ionization states to help with subsequent spectropolarimetric inversions. {\sc{francis}} utilizes 400 optical fibers in a $20\times20$ configuration that are each only 40~$\mu$m in diameter. By focusing on the blue portion of the electromagnetic spectrum, the cladding thickness can be significantly reduced over its infrared counterparts (e.g., DL-NIRSP), giving each fiber a total overall diameter of 55~$\mu$m (40~$\mu$m diameter core plus 7.5~$\mu$m cladding). When arranged in an offset pattern to maximize the filling factor of the fiber inlet aperture, this results in the $20\times20$ fiber array occupying a $1.10\times1.13$~mm$^{2}$ surface area on the entrance ferrule, which can be seen in the upper-left panel of Figure~{\ref{fig:NLST_fibres}}. Each of the 400 fibers are mapped on to a one-dimensional linear array that is approximately 22~mm in length (upper-right panel of Figure~{\ref{fig:NLST_fibres}}), before being passed into the spectrograph entrance slit. The lower panels of Figure~{\ref{fig:NLST_fibres}} show a sample active region captured in the blue continuum at 390~nm (left) and in the core of the Ca~{\sc{ii}}~K spectral line at 393~nm (right), with the yellow circles depicting an example sampling provided by the $20\times20$ input fiber array. Here, each fiber covers an approximate $2''$ diameter on the solar surface, providing a total field of view spanning $40\times40$~arcsec$^{2}$. In this configuration, approximately 35 fibers cover the sunspot umbra, 170 fibers cover the penumbra, with the remaining 195 fibers sampling the surrounding quiet Sun. By adjusting the size of the imaged beam incident on the fiber entrance ferrule, it becomes possible to easily adjust the spatial field-of-view achievable with this instrument. Furthermore, since the inlet fiber ferrule is polished, it becomes possible to utilize a slitjaw camera to accurately coalign the resulting hyper-spectropolarimetric images with other instruments in operation at the same time. Installation and commissioning of {\sc{francis}} prototype instrument took place at the DST during the summer of 2022, where it will remain as a common-user instrument until such times as the Indian NLST becomes close to operations.

\begin{figure*}[!t]
\begin{center}
\includegraphics[width=\textwidth]{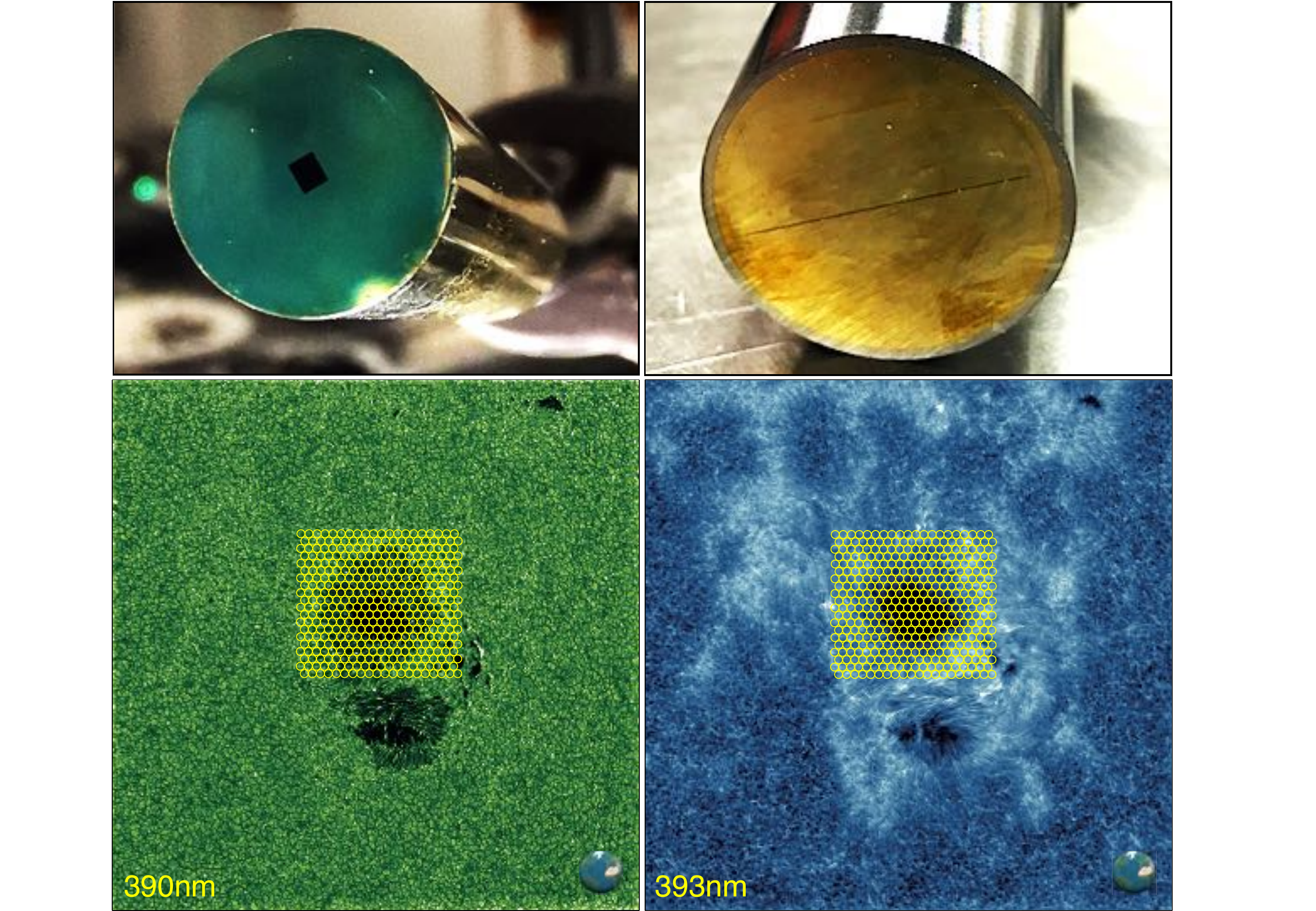}
\end{center}
\caption{The inlet fiber ferrule (upper left) of the hyper-spectropolarimetric imager being developed for the Indian NLST facility currently planned for construction. The dark rectangle ($1.10\times1.13$~mm$^{2}$) towards the center of the ferrule contains a $20\times20$ array of optical fibers, which are mapped into a linear configuration (upper right) allowing the light to be passed into the spectrograph housing. The $20\times20$ fiber grid is arranged in an alternating offset configuration to maximize the filling factor of the fibers in the inlet aperture, which can be visualized by the yellow circles in the lower two panels. The lower left and right panels show images of a solar active region captured in the blue continuum (390~nm) and core of the Ca~{\sc{ii}}~K (393~nm) spectral line, respectively. The yellow circles depict an example spacing (approximately $2''$ across each fiber), providing two-dimensional spectropolarimetric information at rapid cadences across a relatively large field of view. Here, approximately 35, 170, and 195 fibers sample the sunspot umbra, penumbra, and surrounding quiet Sun, respectively. An image of the Earth is provided in the lower-right corner of each solar image to provide a sense of scale. Images courtesy of the Astrophysics Research Centre at Queen's University Belfast.}
\label{fig:NLST_fibres} 
\end{figure*}

Of course, one cause for concern with any instrument attempting to acquire Stokes~$I/Q/U/V$ modulation states for the purposes of spectropolarimetric inversions is the introduction of time delays while the modulators re-orientate themselves in order to cycle between the various polarization states. Common hardware to perform this task include the use of liquid crystal variable retarders (LCVRs) and rotating waveplates, which can often operate at rates exceeding $10$~Hz to try and mitigate any timing delays \citep[e.g.,][]{1987ApOpt..26.3838L, 2010ApOpt..49.4278T, 2012SPIE.8446E..70H, 2017SPIE10563E..2ZA, 2018PASJ...70..102A, 2020A&A...642A..10A}. Even with the ability to progress rapidly through modulation states, the non-simultaneity of the resulting exposures means that the resulting Stokes~$I/Q/U/V$ spectra will not have identical noise values at each wavelength, making inversions challenging in the limit of low photon counts. One method to combat this involves the use of a charge-caching camera, which is able to utilize fast charge transfer in the detector's silicon chip to allow the use of a modulator running at kHz rates. By operating at (or above) the atmospheric turbulence timescales, such technology allows residual seeing effects to be largely eliminated. As a result, all Stokes modulation states can be accumulated at kHz rates before readout is performed, thus ensuring each Stokes modulation state has identical noise properties. An early development in this technique revolved around the Zurich Imaging Polarimeters \citep[ZIMPOL;][]{1995OptEn..34.1870P}, in which a CCD detector array was alternately divided into photosensitive rows and storage rows that were shielded from light by a mask. By shifting the accumulated photons into neighboring storage rows, the acquisition process could be repeated over many modulation cycles until sufficient charge had been accumulated, hence ensuring high signal-to-noise values. Building upon the earlier work of \citet{1995OptEn..34.1870P} and harnessing the more versatile and photosensitive back-illuminated CMOS technology, \citet{2004SPIE.5171..239K} discusses the Charge Caching CMOS Detector for Polarimetry (C$^{3}$Po) concept, whereby a typical multiplexed CMOS pixel would provide a well depth exceeding $6\times10^{6}$~electrons, providing approximately 500{\,}000 electrons per modulation state, helping to achieve polarimetric precision down to the $10^{-5}$ level \citep{1995OptEn..34.1870P, 2013A&ARv..21...66S}. 

By employing a kHz custom-made camera (using a frame-transfer, back-illuminated pn-type CCD sensor), \citet{2016A&A...590A..89I} introduced a novel Fast SpectroPolarimeter (FSP) based on a polarization modulator with ferroelectric liquid crystals. The FSP could result in high temporal- and spatial-resolution full-Stokes measurements with a high sensitivity in the visible wavelength range (i.e., $400-700$~nm), providing almost 100\% duty cycle. The $1024\times1024$ detector was shown to reach a high frame rate of 400~fps and low readout noise on the order of 5\,e$^-$~rms (i.e., considerably smaller than that of, e.g., CRISP and IBIS, by a factor of 4). This implies that the FSP can capture the full Stokes parameters in 10~ms, hence, resulting in a high signal-to-noise polarimetric measurements with a high-cadence of, e.g., 2~s, after image restoration. While the FPS, which can spatially resolve scattering polarization signals \citep{2018A&A...619A.179Z}, is an optimal instrument for studying high frequency waves through the solar photosphere and chromosphere, it has not yet been trialed with suitable hardware (e.g., a Fabry-P{\'{e}}rot) to allow sequential tuning and recording across multiple spectral lines. The latter is necessary for accurately constraining the plasma parameters through the multi-line inversions codes discussed above. 

Next-generation instruments onboard the next (third) flight of the {\sc Sunrise} balloon-borne solar observatory have been designed for such multi-line spectropolarimetric observations. In particular, two grating-based spectropolarimeters (with slit scanning and context imaging with slit-jaw cameras) are being employed \citep{2018cosp...42E.215B,2018cosp...42E3285S}. The {\sc Sunrise} UV Spectropolarimeter and Imager (SUSI) will sample the rich near-UV wavelength range of $300-430$~nm, which includes thousands of photospheric and over 150 chromospheric lines, poorly observable from the ground. Simultaneously, the {\sc Sunrise} Chromospheric Infrared spectroPolarimeter (SCIP) will explore two spectral windows in the near-infrared, within the $765-855$~nm wavelength range, which contains many magnetically sensitive lines sampling various heights in the lower solar atmosphere. The compromise for such many (full-Stokes)  lines measurements with the slit-scanning instruments is, however, a trade-off between size of the field of view and the cadence of observations, all of which are important for tracing wave dynamics in the 3D atmosphere.  Finally, of particular interest is the Polarimetric and Helioseismic Imager \citep[PHI;][]{2020A&A...642A..11S} instrument on Solar Orbiter, which examines the Zeeman and Doppler effects linked to the photospheric Fe~{\sc{i}} 617.3~nm spectral line. PHI employs two telescopes: the first is a full-disk view of the Sun designed to capture information across all phases of the orbit, while the second is a high-resolution telescope that will allow structures as small as 200~km ($\approx$$0{\,}.{\!\!}{''}27$) on the surface of the Sun to be examined at closest perihelion. This capability has the potential to provide long-duration, seeing-free polarimetric observations of the photosphere, which is particularly exciting for the examination of waves in developing magnetic features, such as network bright points,  pores, and sunspots.

Importantly, as can be seen from the text above, new types of instruments (e.g., IFUs) and acquisition techniques (e.g., charge caching cameras) are paving the way for higher precision spectropolarimetry with improved spatial resolutions and temporal cadences, i.e., converging towards the `trifecta' once thought impossible. Over the coming years, we expect the instrumentation described above to play a pivotal role in investigations of waves and oscillations in the lower solar atmosphere by providing the solar physics community with high precision data sequences necessary to take advantage of the cutting-edge inversion software concurrently being developed. What is clearly required at this stage is a `Level 2' data repository for observing sequences that have already been processed with a specific inversion scheme. This would be similar to what is commonly available from space-borne observatories, such as the Joint Science Operations Center\footnote{\href{http://jsoc.stanford.edu/}{http://jsoc.stanford.edu/}} (JSOC), which provides calibrated Level~1 observations, complete with Level~2 data products, such as disambiguated vector magnetograms from the HMI/SDO instrument \citep{2012SoPh..275...79M, 2016SoPh..291.1887C}. Providing a similar data repository scheme for current- and next-generation ground-based facilities would unquestionably boost the accessibility of these key scientific data products for the global solar physics community. 
We therefore look forward to additional proposals to increase the compute and storage capabilities of such a Level~2 data center, particularly if it can incorporate observations from other ground-based solar facilities, including the DST, SST, NST, GREGOR, ALMA, and once commissioned, the NLST and EST.

\section{Conclusions}
\label{sec:conclusions}
In this review, we have attempted to overview (see Section~\ref{sec:waveanalysistools}) the main wave analyses techniques currently harnessed by the global solar physics community. The reason for this is two-fold: (1) We would like to provide the next generation of researchers studying oscillatory phenomena a concise, yet helpful introduction to the techniques that underpin current research efforts, and (2) We wish to establish an analysis tool framework from which researchers can build upon. All too often analyses techniques are reinvented by researchers who do not have easy access to existing algorithms, which unfortunately delays time frames associated with the dissemination and publication of results. Hence, as part of this review, we wish to encourage researchers to visit the Waves in the Lower Solar Atmosphere\footnote{\href{www.WaLSA.team}{www.WaLSA.team}} (WaLSA) dedicated website repository, where a collection of the most readily used codes are available to download and employ. As a natural part of the feedback process, if researchers update and improve the existing wave analyses software, then their updated codes can be hosted on the WaLSA platform for others to avail of, with appropriate references provided to document the underlying improvements. 

Sections {\ref{sec:recentstudiesofwaves}} \& {\ref{sec:unansweredquestionsarising}} document a plethora of high-impact wave studies that have come to fruition over the last number of years. While these studies have unquestionably improved our understanding of the tenuous solar atmosphere, they naturally pose yet more unanswered questions. Thankfully, we are entering an era of discovery with the current and upcoming commissioning of high value (monetary {\it{and}} scientifically) ground-based and space-borne facilities such as DKIST, Solar Orbiter, and Solar-C. This places the community in an ideal position to explore new high-resolution observations with unrivaled accuracy. Many of the instruments associated with such observing facilities are well suited to wave studies, with multi-wavelength capabilities, rapid cadences, high polarimetric precisions, and unprecedented spatial resolutions planned from the very beginning. 

In addition to the imminent next-generation observational facilities we will have at our disposal, researchers will also be able to capitalize on continually updated high performance computing (HPC) infrastructures to model and simulate the observed wave signatures with unprecedented resolution. Globally, the Distributed Research utilising Advanced Computing\footnote{\href{https://dirac.ac.uk/}{https://dirac.ac.uk/}} (DiRAC), ARCHER\footnote{\href{http://www.archer.ac.uk/}{http://www.archer.ac.uk/}}, the Norwegian academic e-infrastructure\footnote{\href{https://www.sigma2.no/systems}{https://www.sigma2.no/systems}}, La Red Espa{\~{n}}ola de Supercomputaci{\'{o}}n\footnote{\href{https://www.res.es/en}{https://www.res.es/en}} (RES, Spanish Supercomputing Network), and the NASA Pleiades\footnote{\href{https://www.nas.nasa.gov/hecc/resources/pleiades.html}{https://www.nas.nasa.gov/hecc/resources/pleiades.html}} HPC supercomputing systems (to name but a few examples) provide tens of petaflops of compute capability. Such evolving HPC facilities are crucial for the accurate replication of physics, particularly down to the spatial and temporal scales imminently visible by the newest observing facilities. 

As a community, we therefore look forward in anticipation to the new era of understanding that will be brought to fruition by the newest researchers, observatories, and computing facilities we will have at our disposal over the decades to come.

\begin{acknowledgements}
D.B.J. and S.D.T.G. wish to thank Invest NI and Randox Laboratories Ltd for the award of a Research \& Development Grant (059RDEN-1), in addition to the UK Science and Technology Facilities Council (STFC) for the award of a Consolidated Grant (ST/T00021X/1). D.B.J. and S.D.T.G. also wish to thank the UK Space Agency for a National Space Technology Programme (NSTP) Technology for Space Science award (SSc~009).
S.J. acknowledges support from the European Research Council under the European Union's Horizon 2020 research and innovation programme (grant agreement no. 682462) and from the Research Council of Norway through its Centres of Excellence scheme (project no. 262622). 
M.S. is grateful for funding received from the European Research Council under the European Union's Horizon 2020 Framework Programme for Research and Innovation, grant agreements H2020 PRE-EST (no. 739500) and H2020 SOLARNET (no. 824135), in addition to support from INAF Istituto Nazionale di Astrofisica (PRIN-INAF-2014). 
We wish to acknowledge scientific discussions with the Waves in the Lower Solar Atmosphere (WaLSA; \href{https://www.WaLSA.team}{https://www.WaLSA.team}) team, which is supported by the Research Council of Norway (project no. 262622) and the Royal Society \citep[Hooke18b/SCTM;][]{2021RSPTA.37900169J}. Finally, all authors wish to thank the reviewers for offering suggestions that improved the quality and readability of the final review. 
\end{acknowledgements}

%
%

\bibliographystyle{spbasic}      


\end{document}